\documentclass[useAMS,usenatbib]{mn2e}
\usepackage{times}
\usepackage{graphics}
\usepackage{graphicx}
\usepackage{color}
\usepackage{verbatim}
\usepackage{amsmath}
\usepackage{mathtools}
\usepackage{subfigure}
\usepackage{float}
\usepackage{tabularx}
\usepackage{supertabular}
\usepackage{rotating}
\usepackage{longtable}
\usepackage{dpfloat}
\usepackage{booktabs}
\usepackage{color}
\usepackage{multicol}
\usepackage{wrapfig}
\usepackage{subfig}
\usepackage{soul}   
\definecolor{Mygrey}{gray}{0.75}

\title[Molecular gas in the arms and inter-arms of NGC~6946]{Molecular
  Gas Properties of the Giant Molecular Cloud Complexes in the Arms
  and Inter-arms of the Spiral Galaxy NGC~6946}
\author[S.\ Topal et al.]
{Sel\c{c}uk Topal$^1$\thanks{E-mail: selcuk.topal@astro.ox.ac.uk.},
  Estelle Bayet$^1$, Martin Bureau$^1$, Timothy A.\ Davis$^2$, Wilfred Walsh$^3$\\
$^1$Sub-department of Astrophysics, University of Oxford, Denys
Wilkinson Building, Keble Road, Oxford, OX1~3RH, U.K.\\
$^2$European Southern Observatory, Karl-Schwarzschild-Str.\ 2,
85748, Garching bei Muenchen, Germany\\
$^3$Harvard Smithsonian Center for Astrophysics, 60 Garden Street,
Cambridge, MA~02138, U.S.A.}
\begin{document}
\date{Accepted . Received ; in original form }
\pagerange{\pageref{firstpage}--\pageref{lastpage}} \pubyear{2012}
\maketitle
\label{firstpage}
%
% Abstract
%
\begin{abstract} 
  Combining observations of multiple CO lines with radiative transfer
  modeling is a very powerful tool to investigate the physical
  properties of the molecular gas in galaxies. Using new observations
  as well as literature data, we provide the most complete CO ladders ever
  generated for eight star-forming regions in the spiral arms and
  inter-arms of the spiral galaxy NGC~6946, with observations of the CO(1-0),
  CO(2-1), CO(3-2), CO(4-3), CO(6-5), $^{13}$CO(1-0) and
  $^{13}$CO(2-1) transitions. For each region, we use the large
  velocity gradient assumption to derive beam-averaged molecular gas
  physical properties, namely the gas kinetic temperature ($T_{\rm
    K}$), H$_2$ number volume density ($n$(H$_2$)) and CO number
  column density ($N$(CO)). Two complementary approaches are used to compare the
  observations with the model predictions: $\chi^2$ minimisation and
  likelihood. The physical conditions derived vary greatly from one
  region to the next: $T_{\rm K}=10$--$250$~K,
  $n$(H$_2$)$=10^{2.3}$--$10^{7.0}$~cm$^{-3}$ and
  $N$(CO)$=10^{15.0}$--$10^{19.3}$~cm$^{-2}$. The spectral line energy 
  distribution (SLED) of some of these extranuclear regions indicates a star-formation 
  activity that is more intense than that at the centre of our own Milky Way. The
  molecular gas in regions with a large SLED turnover transition
  ($J_{\rm max}>4$) is hot but tenuous with a high CO column density,
  while that in regions with a low SLED turnover transition ($J_{\rm
    max}\le4$) is cold but dense with a low CO column density. We
  finally discuss and find some correlations between the physical
  properties of the molecular gas in each region and the presence of
  young stellar population indicators (supernova remnants, H\,{\small
    II} regions, H\,{\small I} holes, etc). 
\end{abstract}
%
% Introduction
%
\section{Introduction}
\label{sec:intro}
The interstellar medium (ISM) of galaxies has many phases, each one
with its own temperature, density, chemical composition, etc. At high
densities, molecular hydrogen (H$_2$) dominates and provides the
reservoir for star formation (SF). Since its detection in the ISM by
e.g.\ \citet{w70}, \citet{s71} and \citet{p71}, the ground-state
rotational transition ($J$=1-0) of the CO molecule (second most
abundant in the ISM) has become the most widely used tracer of H$_2$,
with an excitation temperature of $\approx5$~K and a wavelength of
$2.6$~mm readily observable from the ground.

Stars form in the densest parts of giant molecular clouds
(GMCs), spanning tens of parsecs and weighting
$\sim10^4$--$10^6$~$M_{\sun}$ \citep{k11}. The exact conditions under
which star forms are however still poorly understood, and must
ultimately depend on the local physical conditions in the dense
gas. In this paper, we thus aim to constrain the physical conditions
in the star-forming regions of one particular galaxy, NGC~6946, in the
hope of futhering our understanding of SF processes.

\begin{figure*}
  \includegraphics[height=8.25cm,clip=]{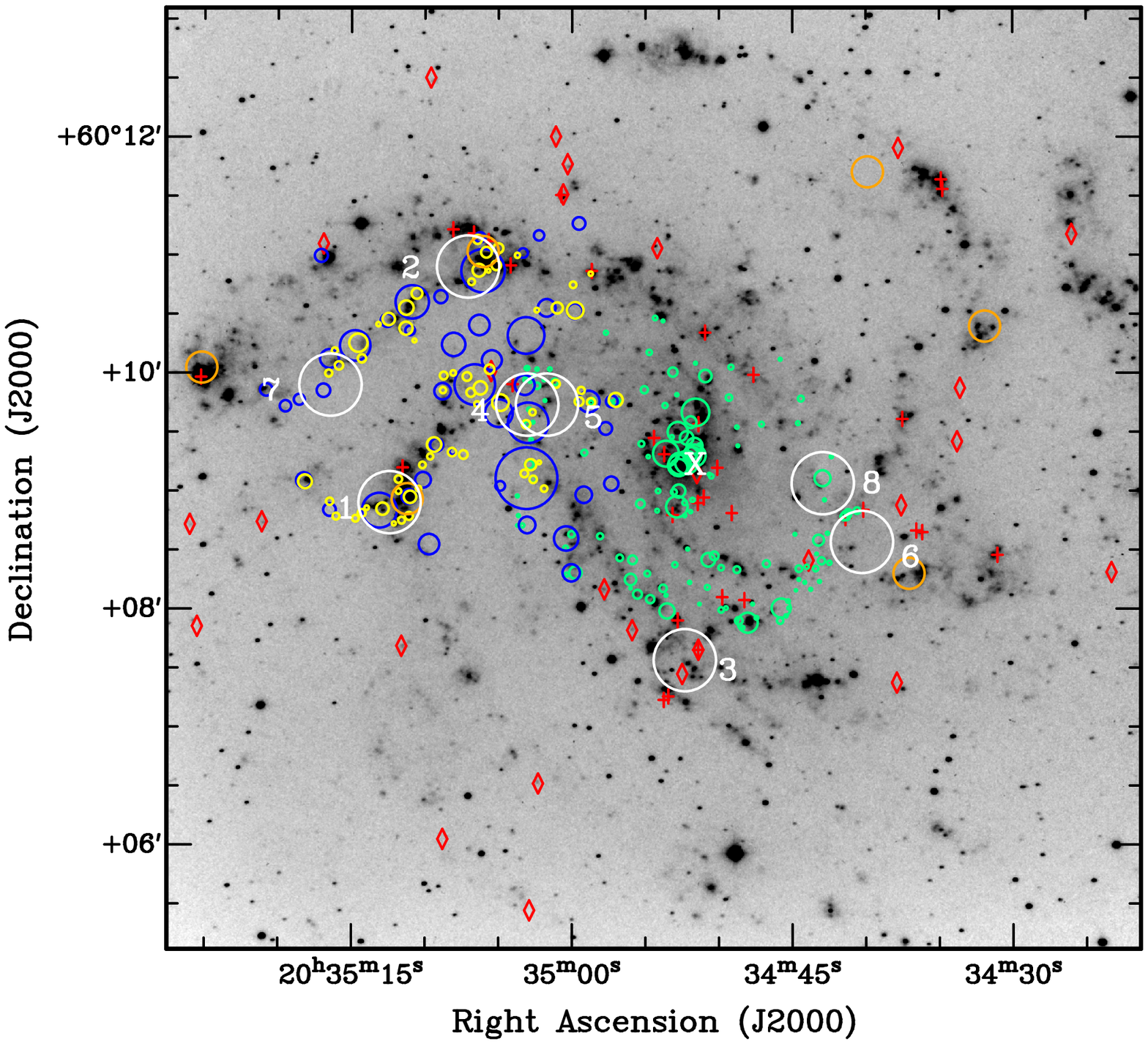}
  \includegraphics[height=8.25cm,clip=]{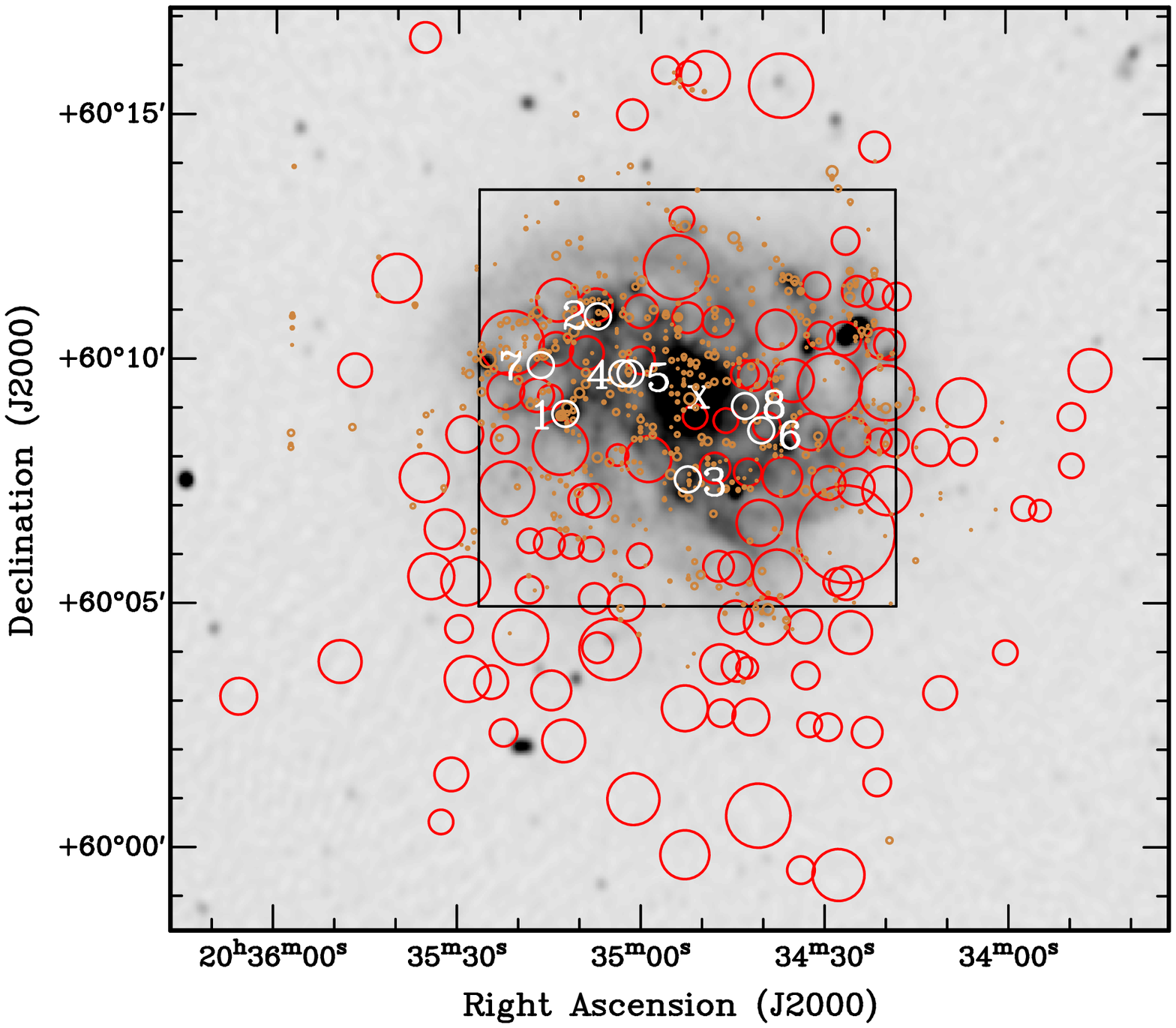}
  \caption{Regions studied and star formation-related features in
    NGC~6946. Left: H$\alpha$ image of NGC~6946 \citep{k04}. Overlaid
    green, yellow, blue and orange circles indicate respectively GMCs
    in the central $5$~kpc \citep{do12}, GMCs in the eastern part
    \citep{reb12}, CO(1-0) complexes in the eastern part
    \citep{reb12}, and giant H\,{\small II} complexes \citep{v77}. Red
    crosses and diamonds indicate supernova remnants identified in the
    radio \citep{ld01} and optical \citep{mf97}, respectively. Right:
    H\,{\small I} map of NGC~6946 \citep{br07}. Overlaid red circles
    and brown dots represent H\,{\small I} holes \citep{bo07} and
    H\,{\small II} regions \citep{bbm86}, respectively. White circles
    with numbers in both images represent the $8$ regions studied in
    this paper, with our final common beam size of $31\farcs5$, while
    the white ``X'' identifies the centre of the galaxy. The black
    square on the larger H\,{\small I} map shows our CO(2-1)
    pointings, covering a contiguous area of
    $510\arcsec\times510\arcsec$ centred on the galaxy. North is up
    and East to the left in all images.}
  \label{fig:regions}
\end{figure*}

NGC~6946 is an almost face-on, nearby ($5.5$~Mpc; \citealt{t88})
spiral galaxy whose basic properties are listed in
Table~\ref{tab:ngc6946}. It has been intensively studied at all
wavelengths, making it a particularly good target to study SF
processes using a multitude of tracers. Figure~\ref{fig:regions} shows
H$\alpha$ and H\,{\small I} images of NGC~6946, along with the
location of a number of features discussed below. In addition, the $8$
key regions on which our analysis will focus are identified there.

The star formation rate (SFR) of NGC~6946 is higher than that of the
Milky Way \citep{dgs84,sgc98,kktf05}. Extensive CO(1-0)
\citep{ty89,wbtwwd02}, CO(2-1) \citep{l09} and CO(3-2)
\citep{wbtwwd02} observations of NGC~6946 are available in the
literature, and we will present here new data comprising one of the
most extensive CO(2-1) maps. More limited mapping of the molecules
HCN, HCO and CO also exist in the central regions
\citep{bssls85,wcc88,bbg88,iki90,rw95,ndbw99,l08}, and the molecular
gas physical properties have been studied in the centre
\citep{wbtwwd02,bgpc06} and one position in the arms
\citep{wbtwwd02}. Here, we will extend this analysis to all $8$
regions defined in Figure~\ref{fig:regions}.
%
% Table: General properties of NGC6946
%
\begin{table}
  \begin{center}
    \caption{General properties of NGC~6946}
    \begin{tabular}{@{}lrc@{}}
      \hline
      Property & Value & Reference\\
      \hline
      Type & SABcd & 1\\
      RA (J2000)& $20^{\rm h}34^{\rm m}52.\!^{\rm s}3$ & 2\\
      Dec (J2000) & $60^\circ09\arcmin13\arcsec$ & 2\\
      Distance & $5.5$~Mpc & 3\\
      Major-axis diameter & $11\farcm5$ & 1\\
      Minor-axis diameter & $9\farcm8$ & 1\\
      Position angle & $243\degr$ & 4\\
      Inclination & $33\degr$ & 4\\
      $M_{{\rm H}_2}$ & $1.2\pm0.1\times10^{10}$~$M_\odot$ & 5\\
      $V_\odot$ & $40$~km~s$^{-1}$ & 6\\
      SFR & $3.12$~$M_\odot$~yr$^{-1}$ & 7\\
      \hline
    \end{tabular}
    \label{tab:ngc6946}
  \end{center}
  References: (1) NASA/IPAC Extragalactic Database (NED);
  (2) \citet{st06};
  (3) \citet{t88};
  (4) \citet{db08};
  (5) \citet{ty89};
  (6) \citet{ep08};
  (7) \citet{kktf05}. 
\end{table}

%
% Figure: NGC6946: SF-related features and definition of regions
%

As shown in Figure~\ref{fig:regions}, NGC~6946 has prominent
East/North-East spiral arms and more flocculent West/South-West arms
\citep{a66}, allowing us to probe possible differences in the gas
physical properties as a function of radius in two physically
different parts of the galaxy. Interestingly, NGC~6946 has hosted $9$
recorded supernovae in the last $100$ years ($3$ in the past $10$
years alone), the highest number in any galaxy \citep{s94,f08}. The
rate of supernova explosions at the centre of NGC~6946 is as high as
$6\times10^{-3}$~yr$^{-1}$ \citep{errl96}. Radio-bright ($35$;
\citealt{ld01}) and optically-selected ($27$; \citealt{mf97})
supernova remnants (SNRs) have also been detected. Some of the regions
probed in our study are associated with some of these SNRs (see
Fig.~\ref{fig:regions}). This SNR richness implies the presence of a
young stellar population, and it suggests an environment very rich in
cosmic rays (CRs) produced during supernova explosions. This in turn
allows us to probe for the first time in this source the possible
relationship between supernovae and molecular gas physical conditions.

A catalogue of $643$ H\,{\small II} regions was published by
\citet{bbm86}, who found that the H\,{\small II} regions on the
eastern side of NGC~6946 are $\approx30$~per cent larger than those on
western side. Many of these H\,{\small II} regions are associated with
regions studied here, including two giant H\,{\small II} complexes
(Fig.~\ref{fig:regions}). Ten star-forming regions were studied in
details by \citet{mhc10}, who detected $33$~GHz free-free emission and
showed that it is as useful a tool to derive SFRs as any other
standard diagnostic.

Over $100$ H\,{\small I} holes have also been found in NGC~6946, most
located in the inner regions where the gas density and SFR are high
\citep{bo08}. H\,{\small I} holes are thought to be created by
supernovae and stellar winds \citep{tik86,mk87,mm88,tb88}, and some of
the regions we study directly coincide with holes, while others are
located in their vicinity (see Fig.~\ref{fig:regions}).

As for the fuel for star formation, \citet{do12} found more than $100$
GMCs within the central $5$~kpc of NGC~6946, with sizes ranging from
$40$ to $200$~pc. Similarly, \citet{reb12} studied a
$6\times6$~kpc$^{\rm 2}$ area in the eastern part of NGC~6946 (where
regions~1, 2 and 7 of this work are located) and found $45$ CO
complexes and $64$~GMCs (see Fig.~\ref{fig:regions}).

As the physical conditions and thus the SF processes of external
galaxies may differ dramatically from those in the Milky Way, we
specifically aim here to probe these SF processes in the presence of
SNRs, stellar winds, H\,{\small I} holes, etc. All have the potential
to significantly perturb both the dynamics and the thermal balance of
GMCs. In this paper, we therefore study in details $8$
carefully-selected regions in the spiral arms and inter-arms of
NGC~6946 (see Fig.~\ref{fig:regions}). We probe the molecular gas
physical properties in these regions through the large velocity
gradient (LVG) method \citep{gk74,j75}, using the non-local
thermodynamic equilibrium (non-LTE) radiative transfer code RADEX
\citep{vbsjv07}, and exploiting the flux ratios of multiple CO
transitions. Our specific goals are first to probe the gas physical
conditions in various key regions of the galaxy, second to compare the
conditions in the various regions (galaxy centre, spiral arms and
inter-arms), and last to identify the main drivers of the gas physical
conditions.

This paper is set up as follows. Section~\ref{sec:data} describes our
own CO observations, the literature data we exploit, and their common
data reduction. Section~\ref{sec:models} presents the radiative
transfer code adopted to model the molecular gas properties, and the
two complementary methods used to identify the best models. In
Section~\ref{sec:results}, we present the outcome of our own
observations and the physical parameters of the best-fit models. A
discussion of the derived parameters in relation to the
characteristics and environment of each region is presented in
Section~\ref{sec:disc}. We conclude briefly in Section~\ref{sec:conc}.
%
% Observations
%
\section{OBSERVATIONS AND DATA REDUCTION}
\label{sec:data}
%
% New observations
%
\subsection{Observations}
\label{sec:data_new}
\begin{figure}
  \includegraphics[width=8.3cm,clip=]{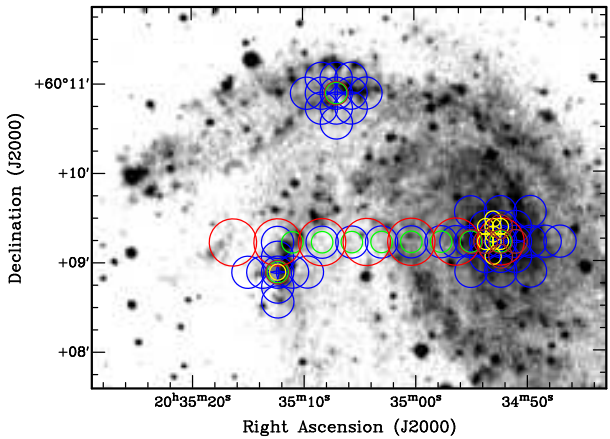}
  \caption{NGC~6946 optical image (DSS), overlaid with the grid of CSO
    pointings for the CO(3-2) (blue), CO(4-3) (brown), CO(6-5)
    (yellow) and C\,{\small I}\,($^3$P$_1$-$^3$P$_0$) (green)
    lines. The circles have a diameter equal to the beam size (FWHM)
    of each line. The CO(2-1) pointings cover a much larger contiguous
    area of $510\arcsec\times510\arcsec$ centred on the galaxy (see
    Fig.~\ref{fig:regions}), so only a few representative pointings
    are shown in red along the major-axis.}
  \label{fig:coverage}
\end{figure}

Caltech Submillimeter Observatory (CSO) 10.4-m telescope observations
of the CO(2-1), CO(3-2), CO(4-3), CO(6-5) and C\,{\small
  I}\,($^3$P$_1$-$^3$P$_0$) lines were performed in 2003-2005 in
atmospheric conditions varying from good to excellent
($\tau_{225\,{\rm GHz}}\la0.1$). The spatial coverage of the
observations in each line is shown in Figure~\ref{fig:coverage},
overlaid on an optical image (Digitized Sky Survey, DSS). The CO(2-1)
observations cover a $510\arcsec\times510\arcsec$ contiguous region
centred on the galaxy, with a full-width at half-maximum (FWHM) beam
size of $31\farcs5$. The CO(3-2) observations consist of clusters of
pointings around the galaxy centre, region~1 and region~2, and a
series of pointings along the major-axis toward the East (beam size
$21^{\prime\prime}$). The CO(4-3) observations consist of a cluster of
pointings around the galaxy centre, and a single pointing around
region~1 and region~2 (beam size $15\farcs7$). The same applies to
CO(6-5), although region~2 was not observed (beam size
$10\farcs5$). The C\,{\small I}\,($^3$P$_1$-$^3$P$_0$) observations
have only a few pointings along the major-axis toward the East (beam
size $14\farcs7$).
%
% Figure: Line coverage
%

The main observational parameters of the CSO observations are listed
in Table~\ref{tab:CSO}. All observations used the superconducting
tunnel junction receivers operated in double side-band mode. We
adopted a $3\arcmin$ chopping throw for the CO(2-1), CO(3-2), CO(4-3)
and C\,{\small I}\,($^3$P$_1$-$^3$P$_0$) lines, but a throw of only
$1\arcmin$ for the CO(6-5) transition as its emission is more
compact. The spectra were obtained with two acousto-optic
spectrometers with effective bandwidths of $1000$ and $500$~MHz and
spectral resolutions of $\approx1.5$ and $2$~MHz, respectively. There
was no sign of contamination by emission in the off beams in any
spectrum. The pointing was checked using planets (Jupiter, Mars and
Saturn) and evolved stars (IRC~10216 and R~Hya), yielding a pointing
accuracy of $\approx5\arcsec$. The overall flux calibration is good to
$\approx20$~per cent.
%
% Table: CSO observational parameters
%
\begin{table}
  \setlength{\tabcolsep}{5pt}
  \caption{CSO observational parameters.}
  \begin{tabular}{lcccc}
    \hline
    Line & Rest frequency & Beam size & Linear distance & $\eta_{\rm mb}^{\rm a}$\\
    & (GHz) & (arcsec) & (kpc) &\\
    \hline
    CO(2-1) & $230.538$ & $31.5$ & $0.84$ & $0.70$\\
    CO(3-2) & $345.795$ & $21.0$ & $0.56$ & $0.75$\\
    CO(4-3) & $461.040$ & $15.7$ & $0.42$ & $0.50$\\
    CO(6-5) & $691.473$ & $10.5$ & $0.28$ & $0.50$\\
    C\,{\small I} ($^3$P$_1$-$^3$P$_0$) & $492.160$ & $14.7$ & $0.39$ & $0.53$\\
    \hline 
  \end{tabular}
  \label{tab:CSO}
  $^{\rm a}$ Main beam efficiency (http://www.submm.caltech.edu/cso/).
\end{table}

\begin{figure}
  \includegraphics[height=7.0cm,clip=]{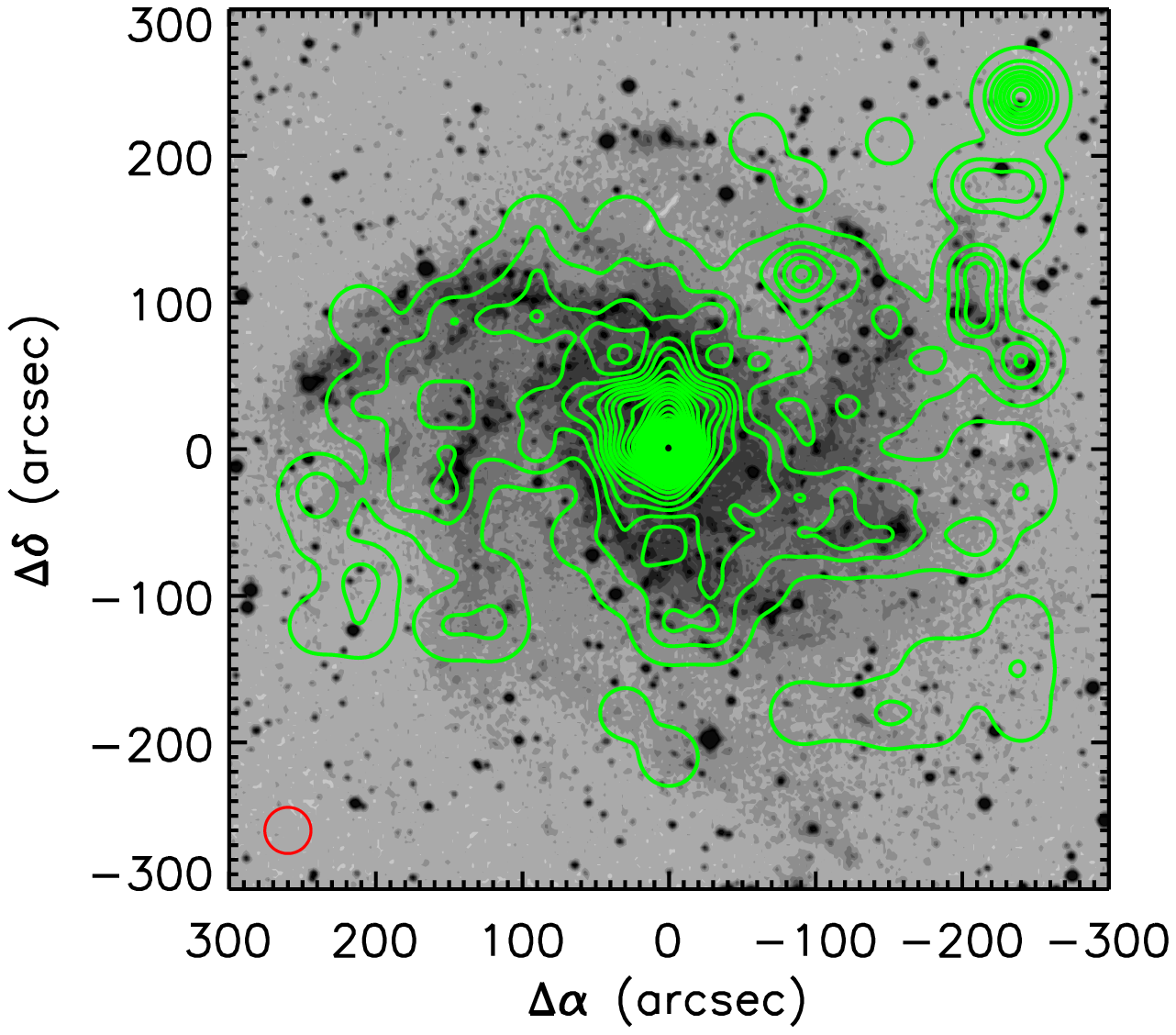}\\
  \includegraphics[height=7.0cm,clip=]{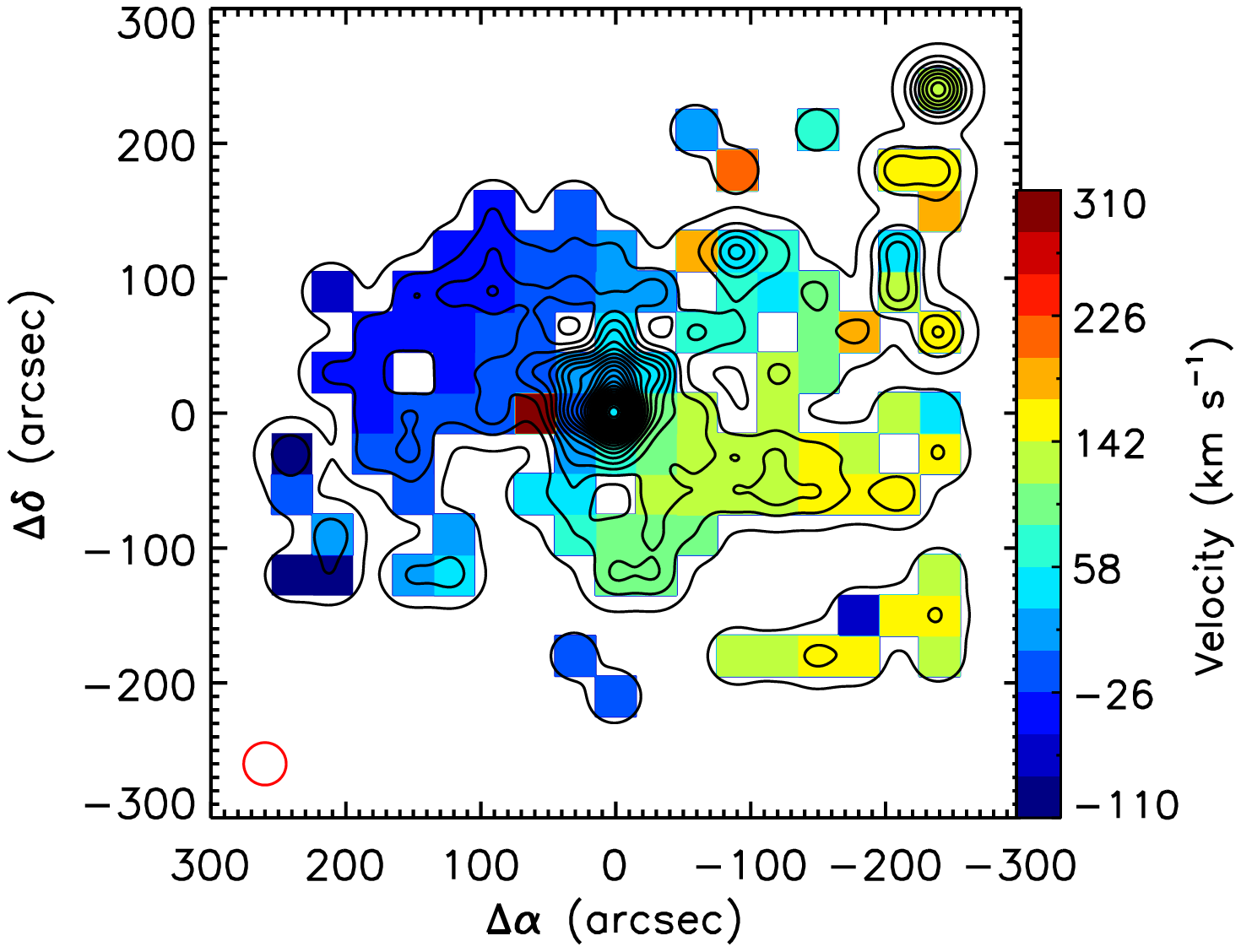}
  \caption{Top: NGC~6946 CO(2-1) integrated line intensity map (green
    contours), overlaid on an optical image (DSS). Bottom: Same
    CO(2-1) integrated line intensity map (black contours), overlaid
    on the CO(2-1) mean velocity field. Contours are from $1$ to
    $80$~per cent of the peak integrated intensity of
    $100$~K~km~s$^{-1}$, in steps of $5$~per cent. The CO(2-1) beam of
    $31\farcs5$ is shown in red in the bottom-left corner of each
    panel.}
  \label{fig:CO2-1}
\end{figure}

\begin{figure}
  \includegraphics[height=7.0cm,clip=]{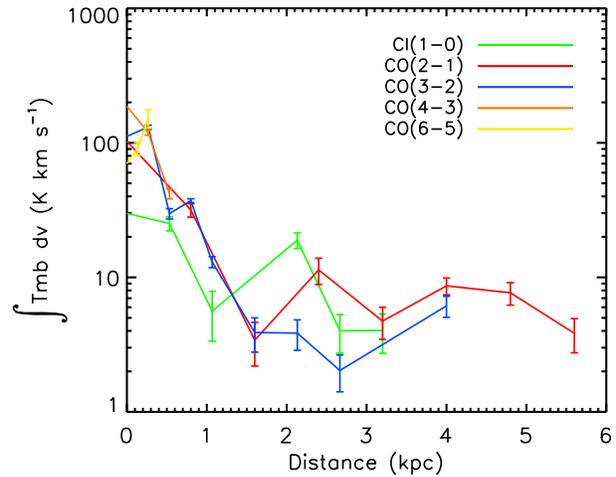}
  \caption{Radial distribution of the molecular gas in NGC~6946. The
    surface density profiles were obtained by taking a straight linear
    cut along the major-axis toward the East for each line. All
    surface density profiles decrease rapidly within $\approx1$~kpc
    and then remain nearly flat. Colours are as in
    Figure~\ref{fig:coverage}.}
  \label{fig:radial}
\end{figure}

We list our CSO detections (with a peak signal-to-noise ratio
$S/N\geq3$) in Table~\ref{tab:ourdata1} of Appendix~\ref{sec:cso} and
show the spectra in Figure~\ref{fig:spec12} of
Appendix~\ref{sec:spec}. The CO(2-1) integrated intensity and mean
velocity maps are presented in Figure~\ref{fig:CO2-1}, overlaid on an
optical image (DSS). These are some of the largest CO(2-1) maps of
NGC~6946 ever presented, with an average RMS noise level 
of T$_{\rm mb}$ = $45$~mK.
Finally, we also show in Figure~\ref{fig:radial} the radial distribution of the various
molecular lines observed. As we do not have full spatial coverage for
all lines (see above), standard azimuthal averages on ellipses are not
possible, and we simply took linear cuts along the major-axis toward
the East. We however also note a detection of CO(6-5) at a
distance of $\approx4$~kpc from the centre toward the East (region~1;
see Table~\ref{tab:intensities}).
%
% Figure: NGC6946 CO(2-1) observations
%

%
% Figure: Radial profiles
%

%
% Literature data
%
\subsection{Literature data}
\label{sec:data_lit}
In addition to our own CSO observations described above (CO(2-1),
CO(3-2), CO(4-3) and CO(6-5)), we use complementary data from
\citet{wbtwwd02}: CO(1-0) in region~1; CO(1-0), $^{13}$CO(1-0) and
$^{13}$CO(2-1) in region~2; CO(1-0), CO(3-2) and $^{13}$CO(2-1) in
regions~3, 4, 5, 6 and 8; and CO(1-0), CO(3-2) and CO(4-3) in
region~7. The CO(1-0) and $^{13}$CO(1-0) observations were obtained at
the Institut de Radioastronomie Millimetrique (IRAM) 30-m telescope
with beam sizes of $21\farcs8$ and $22\farcs8$, respectively. The
CO(3-2), CO(4-3) and $^{13}$CO(2-1) observations were obtained at the
Heinrich Hertz 10-m Telescope (HHT) with beam sizes of $21\farcs8$,
$16\farcs4$ and $34\farcs3$, respectively. We refer the reader to
\citet{wbtwwd02} for further information on these observations.
%
% Data reduction
%
\subsection{Data reduction}
\label{sec:data_reduction}
Both our own CSO data and the literature data from the IRAM 30-m
telescope and HHT were consistently (re-)reduced using the Continuum
and Line Analysis Single-Dish Software (CLASS) software package in the
Grenoble Image and Line Analysis System (GILDAS). A baseline fit
(polynomial of order $0$ or $1$, occasionally $2$) was removed from
each scan before averaging all integrations. Then, for each line in
each region, a Gaussian was fit to the integrated spectrum to derive
the velocity-integrated line intensity $S$ (in K~km~s$^{-1}$). The
fits are shown in Figure~\ref{fig:spec12} of Appendix~\ref{sec:spec},
overlaid on the spectra. The integrated line intensity was then
transformed from the antenna temperature scale ($T_{\rm A}^*$) to the
main beam brightness temperature scale ($T_{\rm mb}$) by dividing by
the main beam efficiency ($\eta_{\rm mb}$). The main beam efficiencies
for the IRAM 30-m telescope and HHT observations were taken from
\citet{wbtwwd02}. The velocity-integrated line intensities and velocity
widths (full width at half maximum, FWHM) are listed in Table~\ref{tab:ourdata1}. 

To estimate the beam-averaged total intensities per unit area $I$ and
fluxes $F$, we used the standard expressions below (see also
\citealt{b04}):
\begin{equation}
  \frac{I}{{\rm W}\,{\rm m}^{-2}\,{\rm
      sr}^{-1}}=1.02\times10^{-18}\,\bigg(\frac{\nu}{{\rm
        GHz}}\bigg)^3\,\bigg(\frac{S}{{\rm K}\,{\rm km}\,{\rm s}^{-1}}\bigg)\,, 
\end{equation}
\begin{equation}
  \frac{F}{{\rm W}\,{\rm m}^{-2}}=\bigg(\frac{I}{{\rm W}\,{\rm
        m}^{-2}\,{\rm sr}^{-1}}\bigg)\,\bigg(\frac{\Omega_{\rm B}}{\rm sr}\bigg)\,,
\end{equation}
\begin{equation}	
  \frac{\Omega_{\rm B}}{\rm sr}=1.133\,\bigg(\frac{\theta_{\rm B}^2}{{\rm arcsec}^2}\bigg)\,\bigg(\frac{1}{206265^2}\bigg)\,,
\end{equation}
\begin{equation}
  \frac{\theta_{\rm B}}{\rm rad}=1.22\,\,\frac{\lambda}{D}\,, 
\end{equation}
where $\nu$ and $\lambda$ are respectively the observed line frequency
and wavelength, $\Omega_{\rm B}$ is the solid angle sustained by the
beam, $\theta_{\rm B}$ is the beam size (FWHM) and $D$ is the diameter
of the telescope.

After calculating the integrated line intensity for each region and
line on which our analysis focuses, using both our and literature data, 
we applied beam dilution
corrections to all measured quantities to express them for a single,
common spatial resolution, chosen here to be the CO(2-1) beam size of
$31\farcs5$. Under the assumption that all the emission in each region
comes from the same cloud, we have corrected our values for beam
dilution assuming a fixed source size for each region. Similarly to
\citet{wbtwwd02} and \citet{bgpc06}, we used the highest spatial
resolution map available for the size measurements. As no molecular
gas map exists at sufficiently high resolution, we used instead the
$24$~$\micron$ image from the Spitzer Infrared Nearby Galaxies Survey
(SINGS; \citealt{k03}).
%
% Table: Beam dilution-corrected line fluxes
%
\begin{table*}
  \setlength{\tabcolsep}{4pt}
  \caption{Beam-corrected line quantities.}
  \begin{tabular}{cccrcrccc}
    \hline
    Region & Position & Radius$^{\rm a}$ & Line & $k$ & $\int T_{\rm mb}$ d$v^{\rm b}$ &
    Intensity & Flux & FWHM \\
    & ($\Delta\alpha(\arcsec)$, $\Delta\delta(\arcsec)$) & (kpc) & & &
    (K~km~s$^{-1}$) & (W~m$^{-2}$~sr$^{-1}$) & (W~m$^{-2}$) & (km~s$^{-1}$)\\
    \hline \\
    1 & $(150, -20)$ & $4.04$ & CO(1-0) & $0.51$ & $8.73\pm0.37$ &
    $1.36\,\pm\,0.06\times10^{-11}$ &
    $3.60\,\pm\,0.15\times10^{-19}$ & $24.9\,\pm\,\phantom{0}1.1$\\
      &              &        & CO(2-1) & $1.00$ & $9.16\pm1.40$ &
    $1.14\,\pm\,0.18\times10^{-10}$ &
    $3.02\,\pm\,0.46\times10^{-18}$ & $27.5\,\pm\,\phantom{0}4.6$\\
    &   	&	 & CO(3-2) & $0.48$ & $5.12\pm0.70$ & 
    $2.16\,\pm\,0.29\times10^{-10}$ & 
    $5.71\,\pm\,0.78\times10^{-18}$ & $26.8\,\pm\,\phantom{0}4.3$\\
    &   	&	 & CO(4-3) & $0.29$ & $3.90\pm0.55$ & 
    $3.89\,\pm\,0.55\times10^{-10}$ & 
    $1.03\,\pm\,0.14\times10^{-17}$ & $27.5\,\pm\,\phantom{0}5.1$\\
    &  		&	 & CO(6-5) & $0.16$ & $2.27\pm0.78$ & 
    $7.63\,\pm\,2.63\times10^{-10}$ & 
    $2.01\,\pm\,0.70\times10^{-17}$ & $20.2\,\pm\,\phantom{0}8.3$ \\
    \\
    2& $(110, 100)$ & $3.96$ & CO(1-0) & $0.52$ & $8.25\pm0.42$ & 
    $1.29\,\pm\,0.07\times10^{-11}$ & 
    $3.41\,\pm\,0.18\times10^{-19}$ & $28.7\,\pm\,\phantom{0}1.7$ \\
    &  		& 	& CO(2-1) & $1.00$ & $8.74\pm1.48$ & 
    $1.09\,\pm\,0.19\times10^{-10}$ & 
    $2.89\,\pm\,0.48\times10^{-18}$ & $44.6\,\pm\,\phantom{0}8.2$ \\
    &  		& 	& CO(3-2) & $0.49$ & $5.49\pm0.36$ & 
    $2.31\,\pm\,0.15\times10^{-10}$ & 
    $6.12\,\pm\,0.40\times10^{-18}$ & $34.1\,\pm\,\phantom{0}2.4$  \\
    &  		& 	& CO(4-3) & $0.31$ & $2.89\pm0.39$ & 
    $2.89\,\pm\,0.39\times10^{-10}$ & 
    $7.64\,\pm\,1.02\times10^{-18}$ & $27.1\,\pm\,\phantom{0}3.9$ \\
    &		&	& $^{13}$CO(1-0) & $0.56$ & $1.05\pm0.10$ & 
    $1.42\,\pm\,0.13\times10^{-12}$ & 
    $3.77\,\pm\,0.35\times10^{-20}$ & $29.3\,\pm\,\phantom{0}3.1$ \\
    &		&	& $^{13}$CO(2-1) & $0.20$ & $0.45\pm0.06$ & 
    $4.96\,\pm\,0.63\times10^{-12}$ & 
    $1.31\,\pm\,0.17\times10^{-19}$ & $28.1\,\pm\,\phantom{0}4.0$ \\
    \\
    3& $(0, -100)$ & $2.67$ & CO(1-0) & $0.63$ & $7.54\pm0.45$ & 
    $1.18\,\pm\,0.07\times10^{-11}$ & 
    $3.11\,\pm\,0.19\times10^{-19}$ & $26.2\,\pm\,\phantom{0}1.9$ \\
    &    	&	& CO(2-1) & $1.00$ & $7.80\pm1.78$ & 
    $9.74\,\pm\,2.22\times10^{-11}$ & 
    $2.57\,\pm\,0.59\times10^{-18}$ & $36.2\,\pm\,11.3$ \\
    &    	&	& CO(3-2) & $0.63$ & $6.76\pm1.12$ & 
    $2.85\,\pm\,0.47\times10^{-10}$ & 
    $7.53\,\pm\,1.24\times10^{-18}$ & $32.4\,\pm\,\phantom{0}6.9$ \\
    &    	&	& $^{13}$CO(2-1) & $1.13$ & $1.43\pm0.26$ & 
    $1.56\,\pm\,0.28\times10^{-11}$ & 
    $4.12\,\pm\,0.75\times10^{-19}$ & $30.6\,\pm\,\phantom{0}7.5$ \\
    \\
    4& $(80, 30)$ & $2.28$ & CO(1-0) & $0.58$ & $4.75\pm0.45$ & 
    $7.42\,\pm\,0.70\times10^{-12}$ & 
    $1.96\,\pm\,0.18\times10^{-19}$ & $15.9\,\pm\,\phantom{0}1.7$ \\
    &    	&	& CO(2-1) & $1.00$ & $11.14\pm1.76$ & 
    $1.39\,\pm\,0.22\times10^{-10}$ & 
    $3.68\,\pm\,0.58\times10^{-18}$ & $39.4\,\pm\,\phantom{0}8.3$ \\
    &    	&	& CO(3-2) & $0.58$ & $2.53\pm0.53$ & 
    $1.07\,\pm\,0.23\times10^{-10}$ & 
    $2.82\,\pm\,0.59\times10^{-18}$ & $13.2\,\pm\,\phantom{0}3.6$ \\
    &    	&	& $^{13}$CO(2-1) & $1.15$ & $1.61\pm0.39$ & 
    $1.75\,\pm\,0.43\times10^{-11}$ & 
    $4.63\,\pm\,1.12\times10^{-19}$ & $17.3\,\pm\,\phantom{0}5.2$ \\
    \\
    5& $(70, 30)$ & $2.03$ & CO(1-0) & $0.65$ & $5.73\pm0.34$ & 
    $8.95\,\pm\,0.53\times10^{-12}$ & 
    $2.36\,\pm\,0.14\times10^{-19}$ & $15.6\,\pm\,\phantom{0}1.1$ \\
    &   	& 	& CO(2-1) & $1.00$ & $13.07\pm1.98$ & 
    $1.63\,\pm\,0.25\times10^{-10}$ & 
    $4.32\,\pm\,0.65\times10^{-18}$ & $53.5\,\pm\,10.1$ \\
    &    	&	& CO(3-2) & $0.65$ & $3.90\pm0.49$ & 
    $1.65\,\pm\,0.21\times10^{-10}$ & 
    $4.35\,\pm\,0.55\times10^{-18}$ & $19.2\,\pm\,\phantom{0}3.0$ \\
    &    	&	& $^{13}$CO(2-1) & $1.12$ & $0.72\pm0.12$ & 
    $7.89\,\pm\,1.35\times10^{-12}$ & 
    $2.08\,\pm\,0.36\times10^{-19}$ & $20.9\,\pm\,\phantom{0}3.9$ \\
    \\
    6& $(-90, -40)$ & $2.63$ & CO(1-0) & $0.49$ & $4.77\pm0.43$ & 
    $7.46\,\pm\,0.67\times10^{-12}$ & 
    $1.97\,\pm\,0.18\times10^{-19}$ & $24.1\,\pm\,\phantom{0}2.5$ \\
    &    	&	& CO(2-1) & $1.00$ & $7.95\pm1.66$ & 
    $9.93\,\pm\,2.08\times10^{-11}$ & 
    $2.63\,\pm\,0.55\times10^{-18}$ & $27.6\,\pm\,\phantom{0}7.3$ \\
    &    	&	& CO(3-2) & $0.49$ & $3.21\pm0.51$ & 
    $1.35\,\pm\,0.21\times10^{-10}$ & 
    $3.58\,\pm\,0.56\times10^{-18}$ & $37.9\,\pm\,\phantom{0}5.6$ \\
    &    	&	& $^{13}$CO(2-1) & $1.18$ & $1.24\pm0.20$ & 
    $1.35\,\pm\,0.22\times10^{-11}$ & 
    $3.58\,\pm\,0.58\times10^{-19}$ & $25.4\,\pm\,\phantom{0}4.3$ \\
    \\
    7& $(180,40)$ & $4.92$ & CO(1-0) & $0.52$ & $3.67\pm0.43$ & 
    $5.73\,\pm\,0.66\times10^{-12}$ & 
    $1.51\,\pm\,0.17\times10^{-19}$ & $22.3\,\pm\,\phantom{0}2.9$ \\
    &    	&	& CO(2-1) & $1.00$ & $8.42\pm1.84$ & 
    $1.05\,\pm\,0.23\times10^{-10}$ & 
    $2.78\,\pm\,0.61\times10^{-18}$ & $34.9\,\pm\,\phantom{0}9.9$ \\
    &    	&	& CO(3-2) & $0.52$ & $2.36\pm0.61$ & 
    $9.95\,\pm\,2.59\times10^{-11}$ & 
    $2.63\,\pm\,0.68\times10^{-18}$ & $17.7\,\pm\,\phantom{0}4.3$ \\
    &   	&	& CO(4-3) & $0.33$ & $1.07\pm0.22$ & 
    $1.07\,\pm\,0.22\times10^{-10}$ & 
    $2.83\,\pm\,0.59\times10^{-18}$ & $16.7\,\pm\,\phantom{0}4.9$ \\
    \\
    8& $(-70,-10)$ & $1.89$ & CO(1-0) & $0.62$ & $7.32\pm0.44$ & 
    $1.14\,\pm\,0.07\times10^{-11}$ & 
    $3.02\,\pm\,0.18\times10^{-19}$ & $27.8\,\pm\,\phantom{0}1.9$ \\
    &   	&	& CO(2-1) & $1.00$ & $6.45\pm1.22$ & 
    $8.06\,\pm\,1.52\times10^{-11}$ & 
    $2.13\,\pm\,0.40\times10^{-18}$ & $22.6\,\pm\,\phantom{0}4.0$ \\
    &    	&	& CO(3-2) & $0.62$ & $3.87\pm0.58$ & 
    $1.63\,\pm\,0.25\times10^{-10}$ & 
    $4.31\,\pm\,0.65\times10^{-18}$ & $24.4\,\pm\,\phantom{0}4.5$ \\
    &   	&	& $^{13}$CO(2-1) & $1.07$ & $0.51\pm0.12$ & 
    $5.54\,\pm\,1.25\times10^{-12}$ & 
    $ 1.46\,\pm\,0.33\times10^{-19}$ & $18.7\,\pm\,\phantom{0}4.9$ \\ \\
    \hline   
  \end{tabular}
  \label{tab:intensities}
  \parbox[t]{0.92\textwidth}{$^{\rm a}$Radii are calculated with
    respect to the galaxy centre. $^{\rm b}$The CO(2-1) integrated intensities and the line widths
    were obtained by interpolating the two closest detections of the
    large-scale mapping.}
\end{table*}

Specifically, to calculate each source size, we first considered a
circular zone equal to the adopted beam (diameter $31\farcs5$) centred
on each region. We then calculated the total area with a flux above
$50$~per cent of the peak within this zone, and took the source size
as the diameter of a circle with that same area. We thus measured
source sizes of $7.8$, $9.3$, $22.1$, $15.6$, $5.9$, $19.9$, $9.7$ and
$19\farcs5$ for regions~1 to 8, respectively. Although the area 
with a flux above $50$~per cent of the peak is not contiguous in regions~3, 
4 and 8, the secondary peaks are much smaller in both area and flux and should
therefore have little impact on our modelling results. In any case, although this method of
measuring source sizes is only approximate, all the lines in a given
region are corrected to the same source size, so the method adopted
only has a small effect on the final line ratios used in the modeling
(see Section~\ref{sec:models}). Indeed, reducing the source sizes by
$50$~per cent (thus the source areas by a factor of $4$) or increasing
the source sizes by $50$~per cent (thus the source areas by a factor
of $\approx2$) only changes the line ratios by $\approx12$~per cent on
average.
 
A beam dilution correction factor $k$ was calculated for each line and
each region using the source sizes calculated and a method similar to that
described in \citet{wbtwwd02} and \citet{bgpc06}:
\begin{equation}
  k=(\theta_{\rm line}^{2}+\theta_{\rm source}^{2})\,/\,(\theta_{\rm
    common}^{2}+\theta_{\rm source}^{2})\,,
  \label{eq:beam_correction}
\end{equation}
were $\theta_{\rm line}$ is the beam size of the observations for the
line considered, $\theta_{\rm common}$ is the common beam size adopted
(here $31\farcs5$) and $\theta_{\rm source}$ is the source size of the
region considered. The correction factors obtained for all lines and
all regions are listed in Table~\ref{tab:intensities}.

Following this, the beam dilution-corrected integrated line
intensities (K~km~s$^{-1}$), beam-averaged total intensities
(W~m$^{-2}$~sr$^{-1}$) and total fluxes (W~m$^{-2}$) for the
8 regions studied were calculated
and are also listed in Table~\ref{tab:intensities}. These are the
values we will refer to and use in the rest of this paper, in
particular for the line ratio analysis discussed in
Section~\ref{sec:models}. The line ratios computed using these beam
dilution-corrected values are listed in Table~\ref{tab:line}.
%
% Table: Beam dilution-corrected line ratios
%	
\begin{table}
\caption{Beam-corrected line ratios.}
\centering
\begin{tabular}{ccc}
 \hline
Region & Line Ratio & Value  \\
 \hline \\
 1 & $^{12}$CO(2-1) / $^{12}$CO(1-0) & 1.05 $\pm$ 0.17 \\
   & $^{12}$CO(3-2) / $^{12}$CO(1-0) & 0.59 $\pm$ 0.08 \\
   & $^{12}$CO(4-3) / $^{12}$CO(1-0) & 0.45 $\pm$ 0.07 \\
   & $^{12}$CO(6-5) / $^{12}$CO(1-0) & 0.26 $\pm$ 0.09 \\ 
 \\
 2 & $^{12}$CO(2-1) / $^{12}$CO(1-0) & 1.06 $\pm$ 0.19 \\
    & $^{12}$CO(3-2) / $^{12}$CO(1-0) & 0.66 $\pm$ 0.06 \\
    & $^{12}$CO(4-3) / $^{12}$CO(1-0) & 0.35 $\pm$ 0.05 \\
    & $^{13}$CO(1-0) / $^{12}$CO(1-0) & 0.13 $\pm$ 0.01 \\ 
    & $^{13}$CO(2-1) / $^{12}$CO(2-1) & 0.05 $\pm$ 0.01 \\ 
 \\
 3 & $^{12}$CO(2-1) / $^{12}$CO(1-0) & 1.03 $\pm$ 0.24 \\		
    & $^{12}$CO(3-2) / $^{12}$CO(1-0) & 0.90 $\pm$ 0.16 \\
    & $^{13}$CO(2-1) / $^{12}$CO(2-1) & 0.18 $\pm$ 0.05 \\ 
 \\
 4 & $^{12}$CO(2-1) / $^{12}$CO(1-0) & 2.34 $\pm$ 0.43 \\
    & $^{12}$CO(3-2) / $^{12}$CO(1-0) & 0.53 $\pm$ 0.12 \\
    & $^{13}$CO(2-1) / $^{12}$CO(2-1) & 0.14 $\pm$ 0.04 \\ 
\\
 5 & $^{12}$CO(2-1) / $^{12}$CO(1-0) & 2.28 $\pm$ 0.37 \\
    & $^{12}$CO(3-2) / $^{12}$CO(1-0) & 0.68 $\pm$ 0.09 \\
    & $^{13}$CO(2-1) / $^{12}$CO(2-1) & 0.06 $\pm$ 0.01 \\ 
 \\
 6 & $^{12}$CO(2-1) / $^{12}$CO(1-0) & 1.67 $\pm$ 0.38 \\			
    & $^{12}$CO(3-2) / $^{12}$CO(1-0) & 0.67 $\pm$ 0.12 \\
    & $^{13}$CO(2-1) / $^{12}$CO(2-1) & 0.16 $\pm$ 0.04 \\ 
\\
 7 & $^{12}$CO(2-1) / $^{12}$CO(1-0) & 2.30 $\pm$ 0.57 \\
    & $^{12}$CO(3-2) / $^{12}$CO(1-0) & 0.64 $\pm$ 0.18 \\
    & $^{12}$CO(4-3) / $^{12}$CO(1-0) & 0.29 $\pm$ 0.07 \\ 
 \\
 8 & $^{12}$CO(2-1) / $^{12}$CO(1-0) & 0.88 $\pm$ 0.17 \\		
    & $^{12}$CO(3-2) / $^{12}$CO(1-0) & 0.53 $\pm$ 0.09 \\
    & $^{13}$CO(2-1) / $^{12}$CO(2-1) & 0.08 $\pm$ 0.02 \\ \\
   
\hline
\end{tabular}
\label{tab:line}
\end{table}		
%
% Modeling
%
\section{Modeling}
\label{sec:models}
%
% RADEX
%
\subsection{Radiative transfer code}
\label{sec:radex}
The radiative transfer equations need to be solved to probe the
physical conditions of the molecular gas. This is achieved here applying
RADEX \citep{vbsjv07}, a non-LTE radiative transfer code using the LVG
approximation \citep{gk74,j75} and yielding line intensities as a
function of a set of user-specified parameters: gas kinetic
temperature $T_{\rm K}$, molecular hydrogen number volume density
$n$(H$_2$) and CO number column density per unit line width
$N$(CO)/$\Delta v$. Here we take the FWHM of the Gaussian fit of each
line as the line width $\Delta v$ (see Table~\ref{tab:intensities}).

The LVG approximation assumes a one-dimensional (1D) isothermal ISM
with a velocity gradient sufficiently large to ensure that the source
function is locally defined. RADEX computes the first level
populations using statistical equilibrium in the optically thin limit
and considering a background radiation field (taken here to be a black
body with temperature $2.73$~K), and then calculates the optical
depths of the lines. RADEX further calculates both the
internally-generated radiation and the integrated intensity of the
lines as background-subtracted Rayleigh-Jeans equivalent radiation
temperatures. We assume here a uniform spherical geometry.

In our models, the range of parameters considered is: $T_{\rm K}=10$
to $250$~K in steps of $5$~K, $n$(H$_2$)$\,=10^2$ to
$10^{7}$~cm$^{-3}$ in setps of $10^{0.25}$~cm$^{-3}$ and
$N$(CO)$\,=10^{15}$ to $10^{21}$~cm$^{-2}$ in steps of
$10^{0.25}$~cm$^{-2}$. The collisional rate coefficients between the
molecules H$_2$, CO and $^{13}$CO have been measured by several
authors \citep[e.g.][]{gt76,sebmd85,fl85} and we use here the Leiden
Atomic and Molecular Database (LAMDA; \citealt{svvb05,yang10}) for the
most recent data. The ortho- to para-H$_2$ ratio is assumed to be
thermalised at the given kinetic temperature. The CO/$^{13}$CO
isotopic abundance ratio is taken to be $40$ \citep{hm93,ib01,bgpc06}
for the regions where $^{13}$CO observations are available.
%
% Best-fit models
%	 
\subsection{Best-fit models} 
\label{sec:bestfit}
We use a standard $\chi^2$ approach to identify the best-fit model in
each region. For each region and each set of model parameters ($T_{\rm
  K}$, $n$(H$_2$), $N$(CO)), the $\chi^2$ is defined as
\begin{equation}
  \chi^{2}\equiv\sum\limits_{i}\bigg(\frac{R_{i,{\rm
      mod}}-R_{i,{\rm obs}}}{\Delta R_{i,{\rm obs}}}\bigg)^2\,, 
  \label{eq:chi2}
\end{equation}
where $R_{\rm mod}$ is the modeled line ratio, $R_{\rm obs}$ is the
observed line ratio with uncertainty $\Delta R_{\rm obs}$ (see
Table~\ref{tab:line}), and the summation is over all independent line
ratios $i$ for that region (one fewer than the number of lines
available for that region). The best-fit model parameters for each
region are listed in Table~\ref{tab:results} and are taken as the set
of parameters yielding the smallest $\chi^2$ ($\chi_{\rm min}^2$).
%
% Table: Model results
%
\begin{table*}
 \setlength{\tabcolsep}{4pt}
 \caption{Model results for the two best-model identification methods.}
 \begin{tabular}{clllclll}
   \hline
   Region & Parameter & $\chi^{2}$ minimisation & Likelihood &
   Region & Parameter & $\chi^{2}$ minimisation & Likelihood\\
   \hline\\
   1 & $T_{\rm K}$ & $250^{\ast}$~K & $153^{+69}_{-138}$~K & 
   2 & $T_{\rm K}$ & $250^{\ast}$~K & $211^{+27}_{-42}$~K \\ 
     & $\log$($n$(H$_2$)) & $3.0$~cm$^{-3}$ & $3.0^{+2.4}_{-0.6}$~cm$^{-3}$ &
     & $\log$($n$(H$_2$)) & $2.3$~cm$^{-3}$ & $2.3^{+0.2}_{-0.2}$~cm$^{-3}$ \\
     & $\log$($N$(CO)) & $18.5$~cm$^{-2}$ & $19.2^{+0.9}_{-0.5}$~cm$^{-2}$ &
     & $\log$($N$(CO)) & $18.8$~cm$^{-2}$ & $18.8^{+0.2}_{-0.2}$~cm$^{-2}$ \\\\
     & $J_{\rm max}$ & $6$ & &
     & $J_{\rm max}$ & $5$ & \\
     & CO cooling & $3.30\times10^{-9}$~W~m$^{-2}$~sr$^{-1}$ & &
     & CO cooling & $1.76\times10^{-9}$~W~m$^{-2}$~sr$^{-1}$ &\\
     & $\tau_{\rm CO_{1-0}}$ & $0.3$ & &
     & $\tau_{\rm CO_{1-0}}$ & $4.7$ \\
     & $\tau_{\rm CO_{2-1}}$ & $4.3$ & &
     & $\tau_{\rm CO_{2-1}}$ & $13.7$ \\
     & $\tau_{\rm CO_{3-2}}$ & $8.2$ & &
     & $\tau_{\rm CO_{3-2}}$ & $22.9$ \\
     & $\tau_{\rm CO_{4-3}}$ & $10.3$& &
     & $\tau_{\rm CO_{4-3}}$ & $24.5$ \\
     & $\tau_{\rm CO_{6-5}}$ & $8.1$ & &
     & $\tau_{^{13}\rm CO_{1-0}}$ & $0.7$ \\
     & &&&
     & $\tau_{^{13}\rm CO_{2-1}}$ & $1.9$ \\
     &&\\\\
   3 & $T_{\rm K}$ & $175$~K & $139^{+76}_{-80}$~K&
   4 & $T_{\rm K}$ & $15$~K & $105^{+94}_{-71}$~K \\
    & $\log$($n$(H$_2$)) & $3.3$~cm$^{-3}$ & $2.9^{+0.9}_{-0.6}$~cm$^{-3}$ & 
    & $\log$($n$(H$_2$)) & $7.0^{\ast}$~cm$^{-3}$ & $4.4^{+1.8}_{-1.7}$~cm$^{-3}$ \\
    & $\log$($N$(CO)) & $19.3$~cm$^{-2}$ & $19.5^{+0.6}_{-0.6}$~cm$^{-2}$ & 
    & $\log$($N$(CO)) & $18.3$~cm$^{-2}$ & $19.6^{+0.7}_{-0.8}$~cm$^{-2}$ \\\\
    & $J_{\rm max}$ & $7$ & &
    & $J_{\rm max}$ & $2$ & \\
    & CO cooling & $6.77\times10^{-9}$~W~m$^{-2}$~sr$^{-1}$ & & 
    & CO cooling & $1.09\times10^{-9}$~W~m$^{-2}$~sr$^{-1}$ & \\
    & $\tau_{\rm CO_{1-0}}$ & $1.5$ & & 
    & $\tau_{\rm CO_{1-0}}$ & $8.5$ \\
    & $\tau_{\rm CO_{2-1}}$ & $6.8$ & & 
    & $\tau_{\rm CO_{2-1}}$ & $8.0$ \\
    & $\tau_{\rm CO_{3-2}}$ & $14.5$ & & 
    & $\tau_{\rm CO_{3-2}}$ & $22.1$ \\
    & $\tau_{^{13}\rm CO_{2-1}}$ & $1.1$ & & 
    & $\tau_{^{13}\rm CO_{2-1}}$ & $0.4$\\
    &&\\\\
   5 & $T_{\rm K}$ & $15$~K & $41^{+108}_{-28}$~K & 
   6 & $T_{\rm K}$ & $15$~K & $111^{+95}_{-77}$~K \\
    & $\log$($n$(H$_2$)) & $4.0$~cm$^{-3}$ & $3.7^{+1.9}_{-0.5}$~cm$^{-3}$ & 
    & $\log$($n$(H$_2$)) & $4.3$~cm$^{-3}$ & $3.5^{+1.8}_{-0.9}$~cm$^{-3}$ \\
    & $\log$($N$(CO)) & $17.8$~cm$^{-2}$ & $17.9^{+0.6}_{-0.4}$~cm$^{-2}$ & 
    & $\log$($N$(CO)) & $17.3$~cm$^{-2}$ & $18.6^{+1.1}_{-2.4}$~cm$^{-2}$ \\\\
    & $J_{\rm max}$ & $2$ & &
    & $J_{\rm max}$ & $3$ & \\
    & CO cooling & $5.24\times10^{-10}$~W~m$^{-2}$~sr$^{-1}$ & & 
    & CO cooling & $3.66\times10^{-10}$~W~m$^{-2}$~sr$^{-1}$ & \\
    & $\tau_{\rm CO_{1-0}}$ & $3.1$ & & 
    & $\tau_{\rm CO_{1-0}}$ & $0.6$ \\
    & $\tau_{\rm CO_{2-1}}$ & $2.4$ & & 
    & $\tau_{\rm CO_{2-1}}$ & $1.4$ \\
    & $\tau_{\rm CO_{3-2}}$ & $5.5$ & & 
    & $\tau_{\rm CO_{3-2}}$ & $0.9$ \\
    & $\tau_{^{13}\rm CO_{2-1}}$ & $0.2$ & & 
    & $\tau_{^{13}\rm CO_{2-1}}$ & $\ll~1$ \\
    &&\\\\	
   7 & $T_{\rm K}$ & $35$~K & $65^{+99}_{-43}$~K & 
   8 & $T_{\rm K}$ & $35$~K & $140^{+76}_{-81}$~K \\
    & $\log$($n$(H$_2$)) & $3.8$~cm$^{-3}$ & $4.1^{+1.8}_{-0.8}$~cm$^{-3}$ & 
    & $\log$($n$(H$_2$)) & $3.5$~cm$^{-3}$ & $2.9^{+0.4}_{-0.7}$~cm$^{-3}$ \\
    & $\log$($N$(CO)) & $15.0^{\ast}$~cm$^{-2}$ & $17.8^{+2.5}_{-2.06}$~cm$^{-2}$ & 
    & $\log$($N$(CO)) & $17.8$~cm$^{-2}$ & $18.2^{+0.7}_{-0.7}$~cm$^{-2}$ \\\\
    & $J_{\rm max}$ & $3$ & &
    & $J_{\rm max}$ & $4$ & \\
    & CO cooling & $3.82\times10^{-10}$~W~m$^{-2}$~sr$^{-1}$ & & 
    & CO cooling & $5.58\times10^{-10}$~W~m$^{-2}$~sr$^{-1}$ & \\
    & $\tau_{\rm CO_{1-0}}$ & $\ll~1$ & & 
    & $\tau_{\rm CO_{1-0}}$ & $0.6$ \\
    & $\tau_{\rm CO_{2-1}}$ & $\ll~1$ & & 
    & $\tau_{\rm CO_{2-1}}$ & $3.5$ \\
    & $\tau_{\rm CO_{3-2}}$ & $\ll~1$ & & 
    & $\tau_{\rm CO_{3-2}}$ & $4.2$ \\
    & $\tau_{\rm CO_{4-3}}$ & $\ll~1$ & & 
    & $\tau_{^{13}\rm CO_{2-1}}$ & $0.2$ \\\\		
   \hline\\
 \end{tabular}
 \parbox[t]{1.0\textwidth}{\textit{Notes:} $\chi^2$ minimisation
   results list the best-fit models, defined as having the smallest
   $\chi^2$. Likelihood results list the median value and
   $68$~per cent ($1\sigma$) confidence levels of the marginalised
   probability distribution function of each model
   parameter. A star ($^{\ast}$) indicates a value lying at the edge
   of the model grid. See Section~\ref{sec:models} for more details.}
 \label{tab:results}
\end{table*}

\begin{figure*}
  \includegraphics[width=5.5cm,clip=]{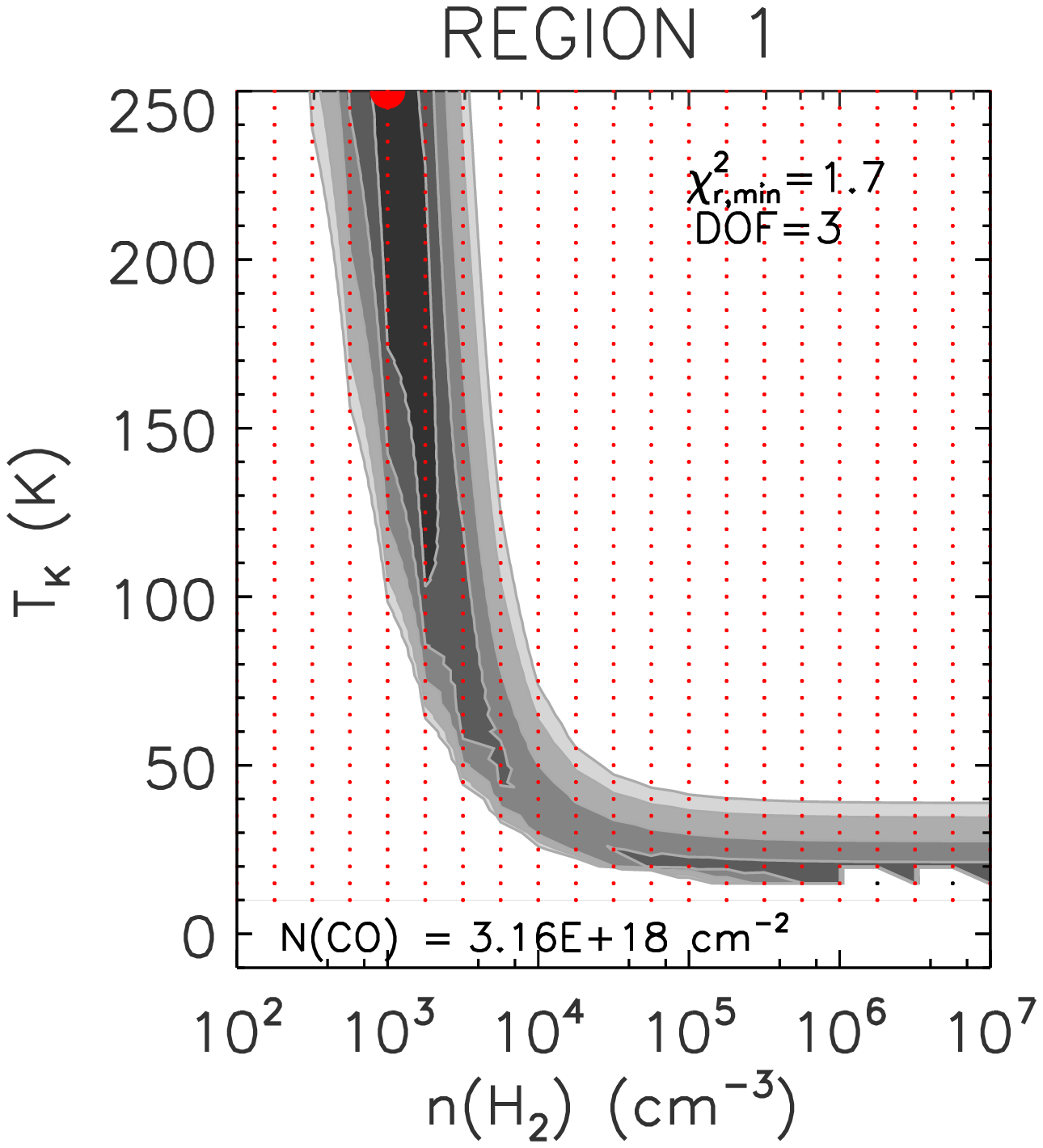}
  \includegraphics[width=5.5cm,clip=]{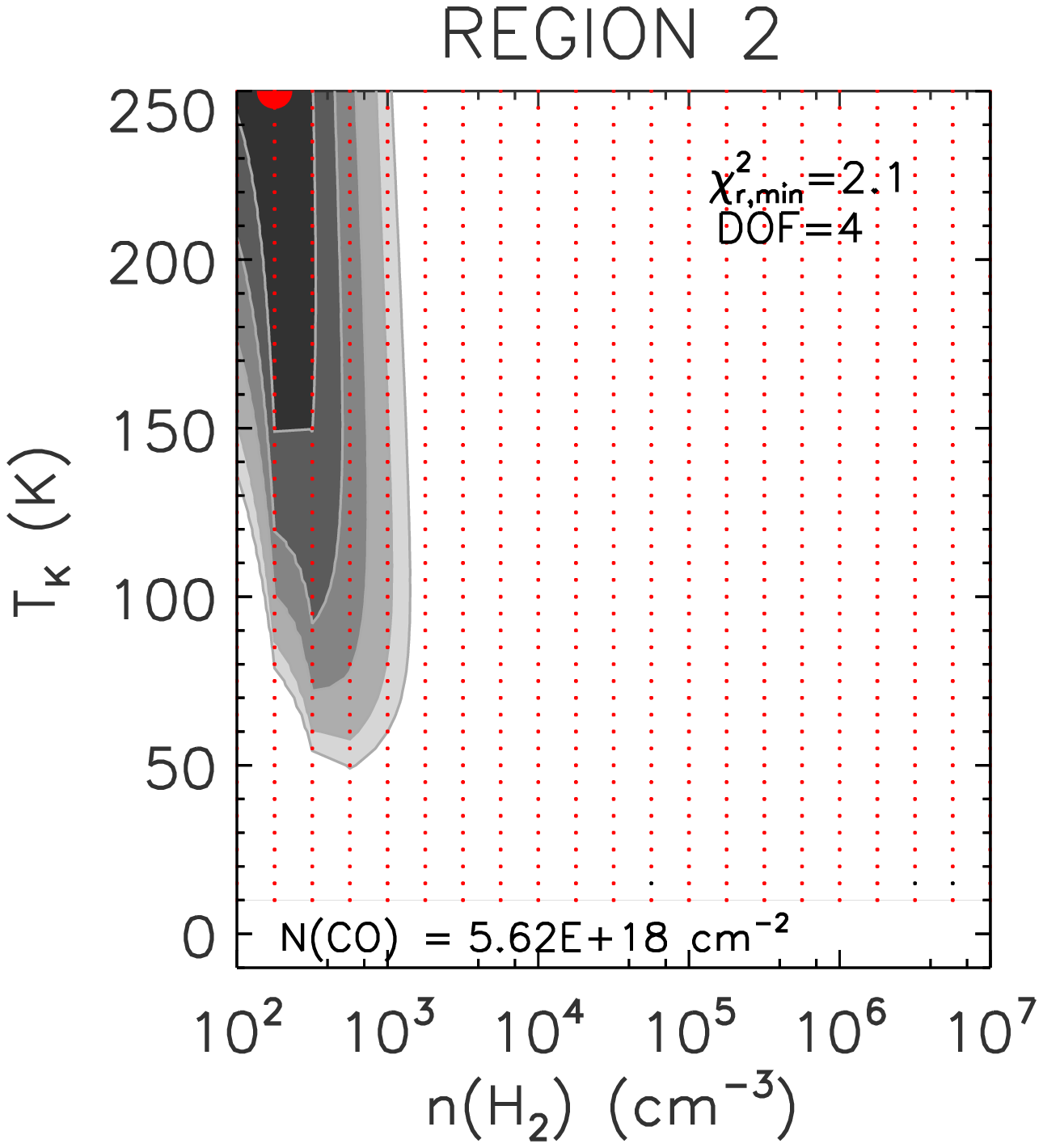}
  \includegraphics[width=5.5cm,clip=]{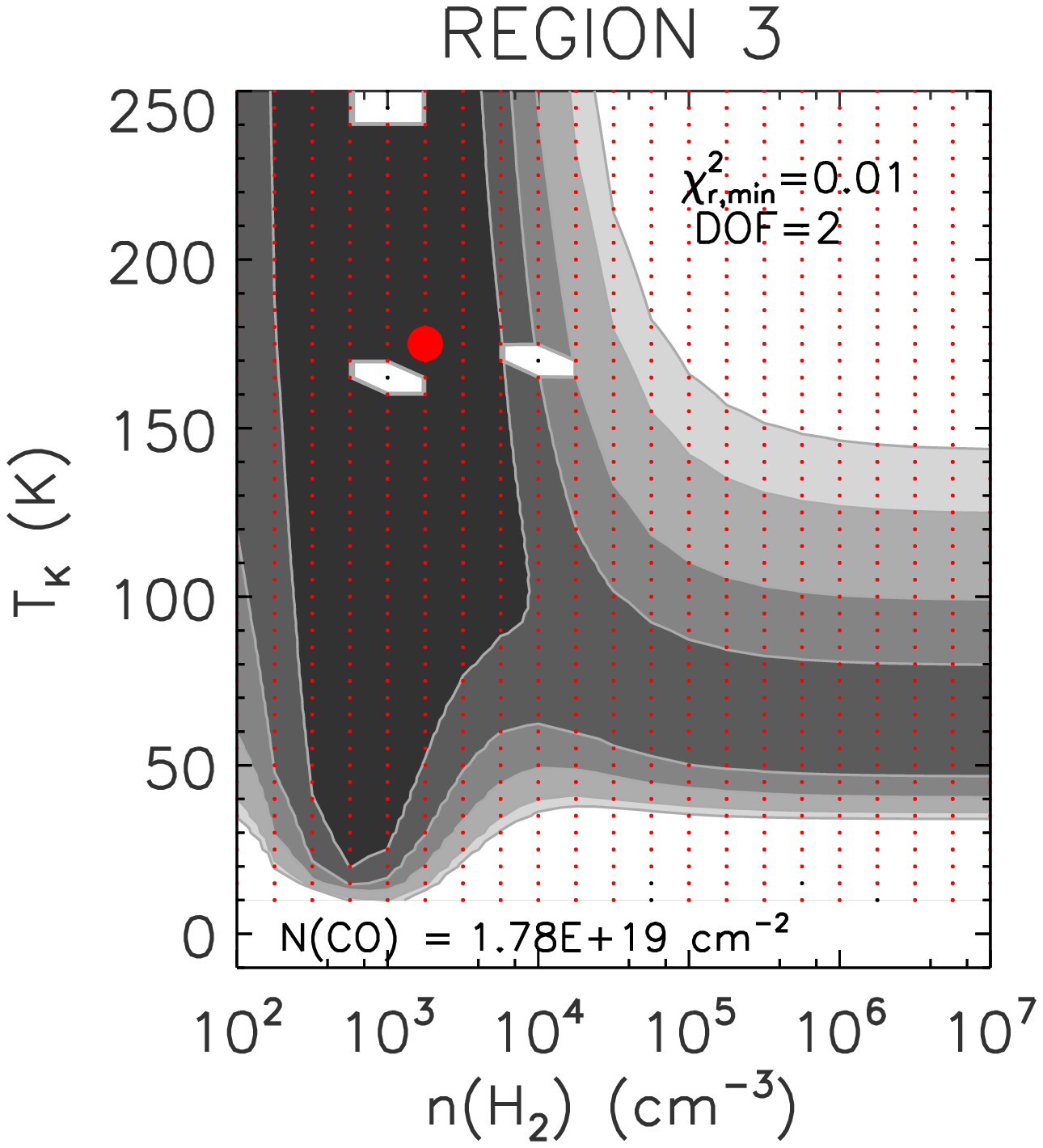}\\
  \includegraphics[width=5.5cm,clip=]{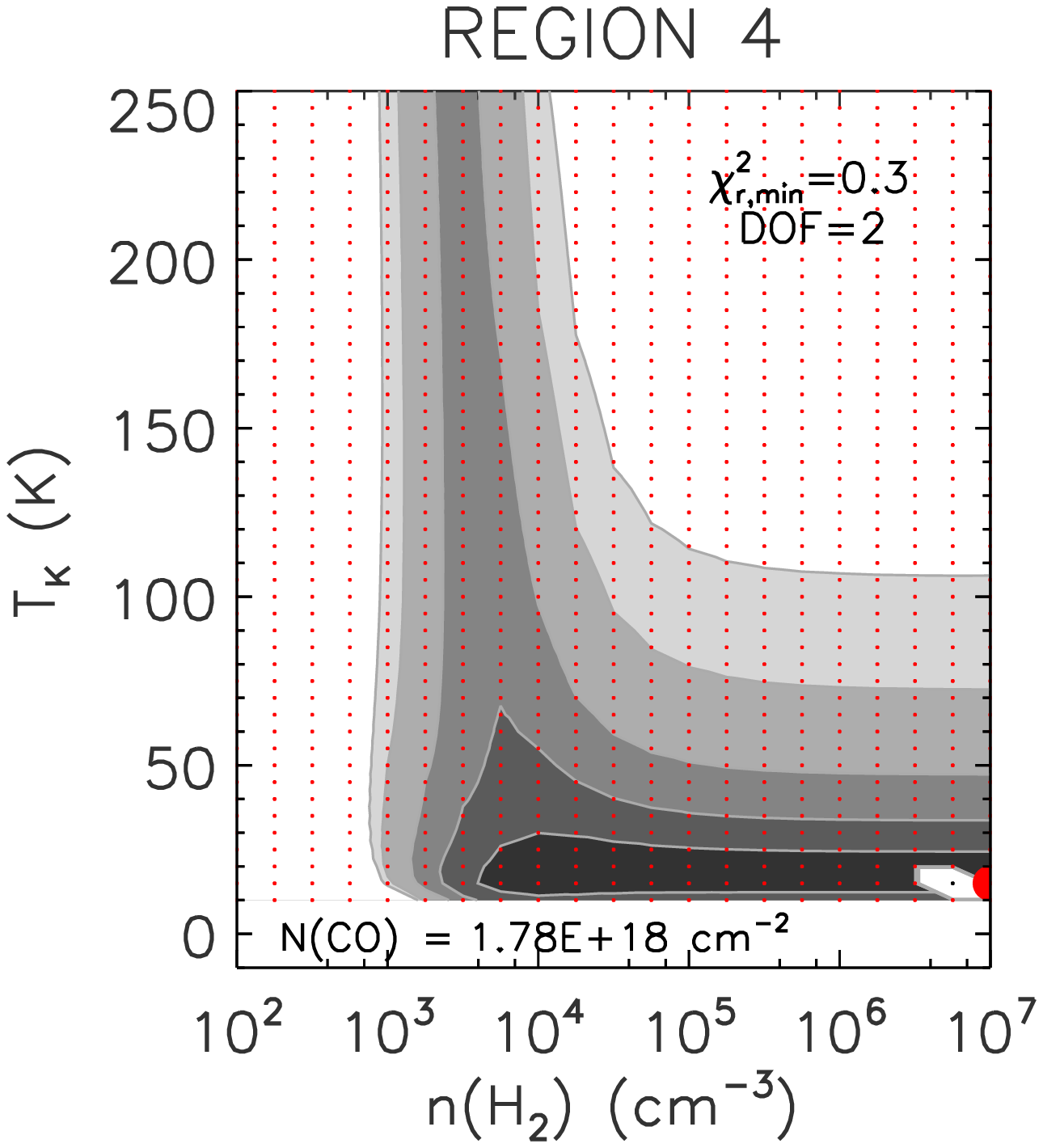}
  \includegraphics[width=5.5cm,clip=]{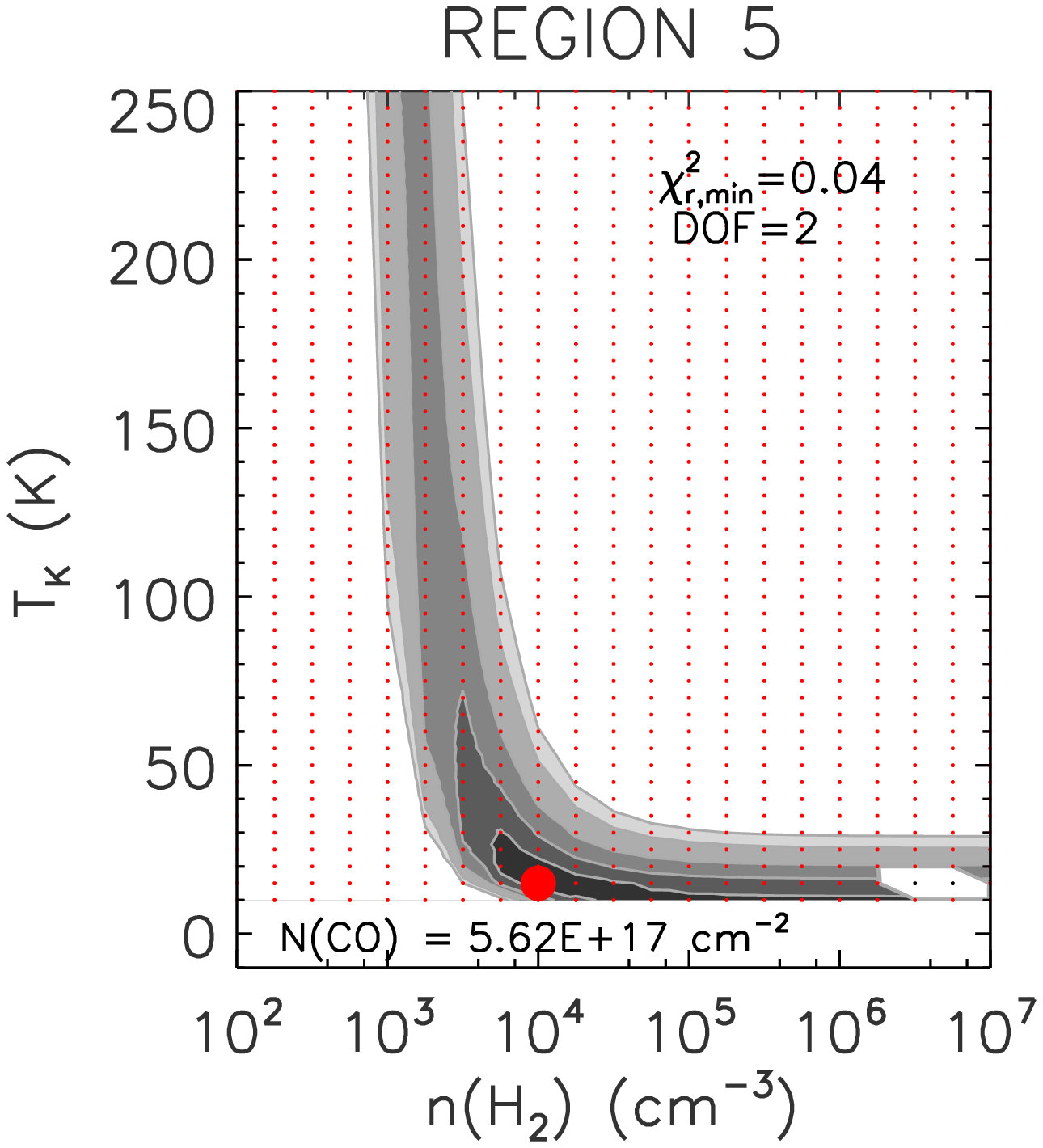}
  \includegraphics[width=5.5cm,clip=]{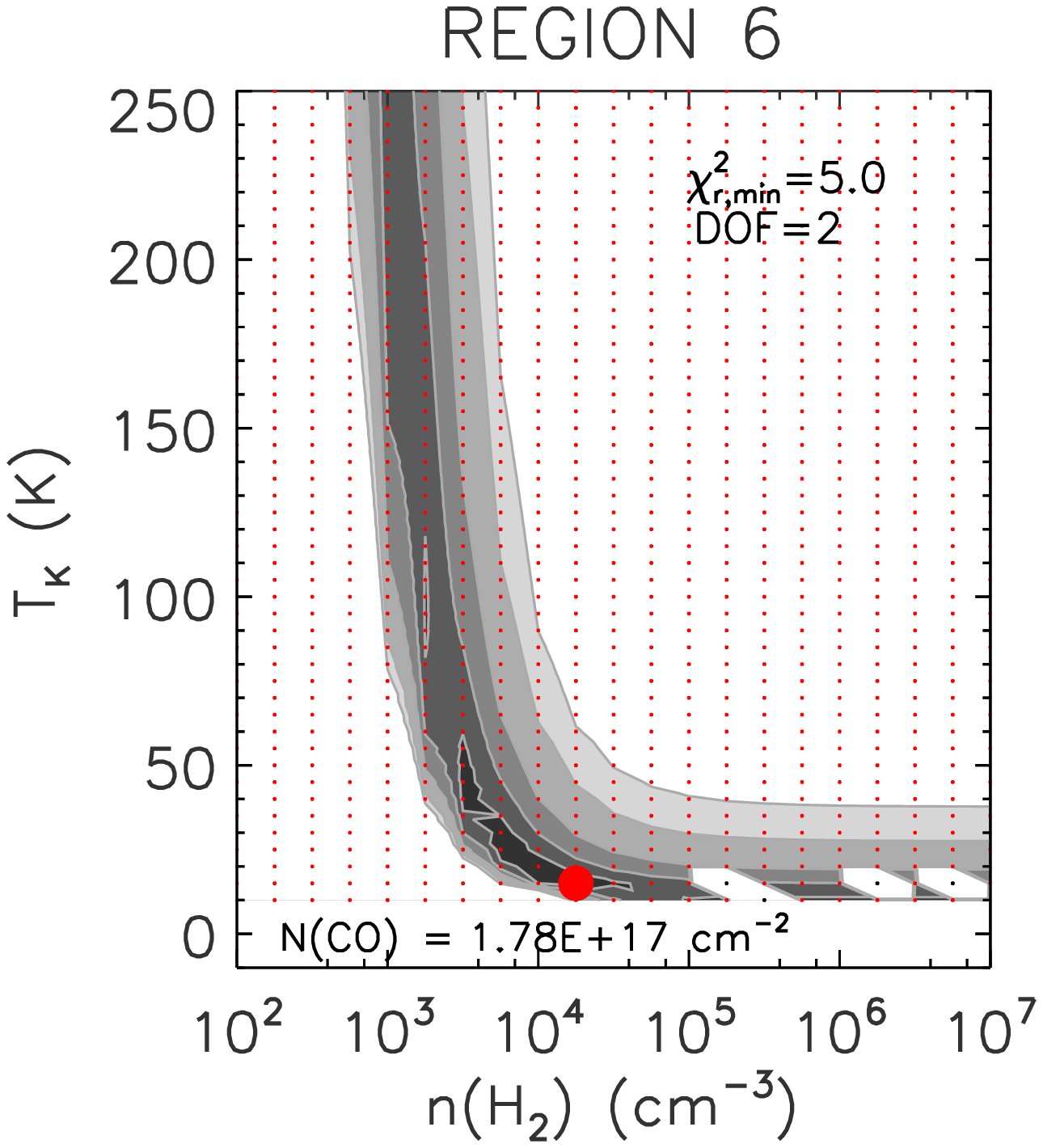}\\
  \includegraphics[width=5.5cm,clip=]{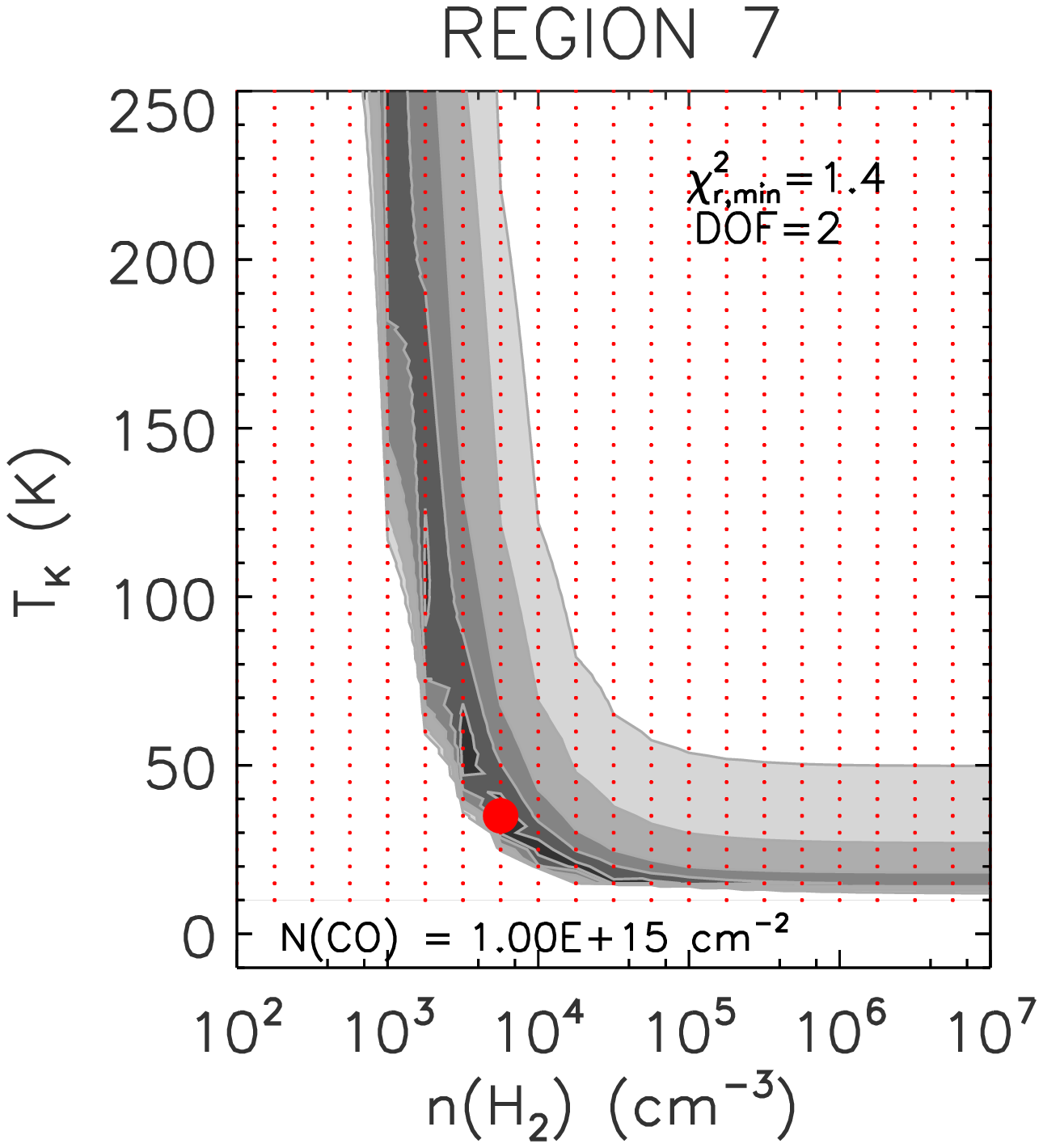}
  \includegraphics[width=5.5cm,clip=]{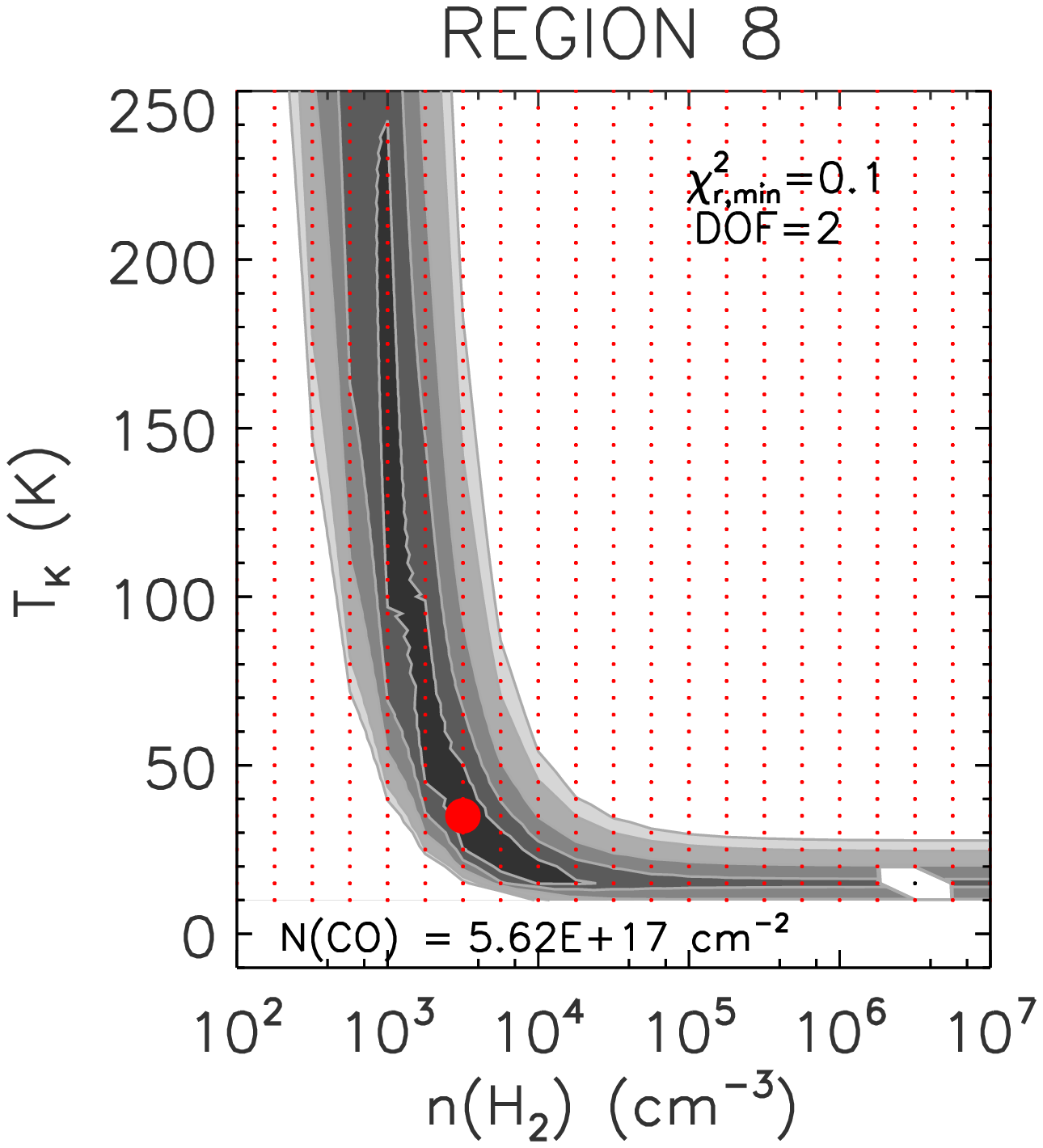}
  \caption{$\Delta\chi^2$ maps (contours and greyscales; see
    Section~\ref{sec:bestfit}). For each region, $\Delta\chi^2$ is
    shown as a function of $T_{\rm K}$ and $n$(H$_2$) for the best-fit
    $N$(CO) (indicated in the bottom-left corner of each panel). The
    model grid is indicated with red dots and the best-fit model with
    a red filled circle. Black dots represent bad models (e.g.\
    unacceptably low opacity; see \citealt{vbsjv07}). $\Delta\chi^2$
    contours indicate the $1$~$\sigma$ (darkest zone) to $5$~$\sigma$
    (lightest zone) confidence levels in steps of $1$~$\sigma$. For
    $3$ line ratios (all regions except regions~1 and 2, $2$ degrees
    of freedom), the levels are $2.3$, $6.2$, $11.8$, $19.3$ and
    $24.0$. For $4$ line ratios (region~1, $3$ degrees of freedom),
    the levels are $3.5$, $8.0$, $14.2$, $22.1$ and $28.0$. For $5$
    line ratios (region~2, $4$ degrees of freedom), the levels are
    $4.7$, $9.7$, $16.3$, $24.5$ and $32.0$. The reduced $\chi_{\rm
      min}^2$ ($\chi_{\rm r,min}^2$) values together with their
    corresponding degrees of freedom are also shown in each panel.}
  \label{fig:chi2}
\end{figure*}

We also plot in Figure~\ref{fig:chi2} the contours of
$\Delta\chi^2\equiv\chi^2-\chi_{\rm min}^2$ in ($T_{\rm K}$,
$n$(H$_2$), $N$(CO)) space, to illustrate the uncertainties of the
best-fit models and the usual degeneracies between the model
parameters. The only way to break those degeneracies is to add
observations of other lines coming from the same (or similar) gas
phase as the current CO lines, e.g.\ low-$J$ HCN lines (with similar
critical densities; see e.g.\ \citealt{j95}).
%
% Figure: chi2
%
\subsection{Likely models} 
\label{sec:likely}
Following \citet{b13}, we additionally applied the likelihood method
to our models, calculating for each region the probability
distribution function (PDF) of each model parameter. That is, for each possible
value of a model parameter, we calculated the sum of
the $e^{-\Delta\chi^{2} / 2}$ for all possible values of the other two
parameters. These PDFs are shown in Figure~\ref{fig:likelihood}
along with their peak (most likely) and median values within the model
grid, as well as the $68$~per cent ($1\sigma$) confidence levels
around the median (following \citealt{k07}). The latter values are
listed in Table~\ref{tab:results}.
%
% Figure: Likelihood
%
\begin{figure*}
  \includegraphics[width=5.5cm,clip=]{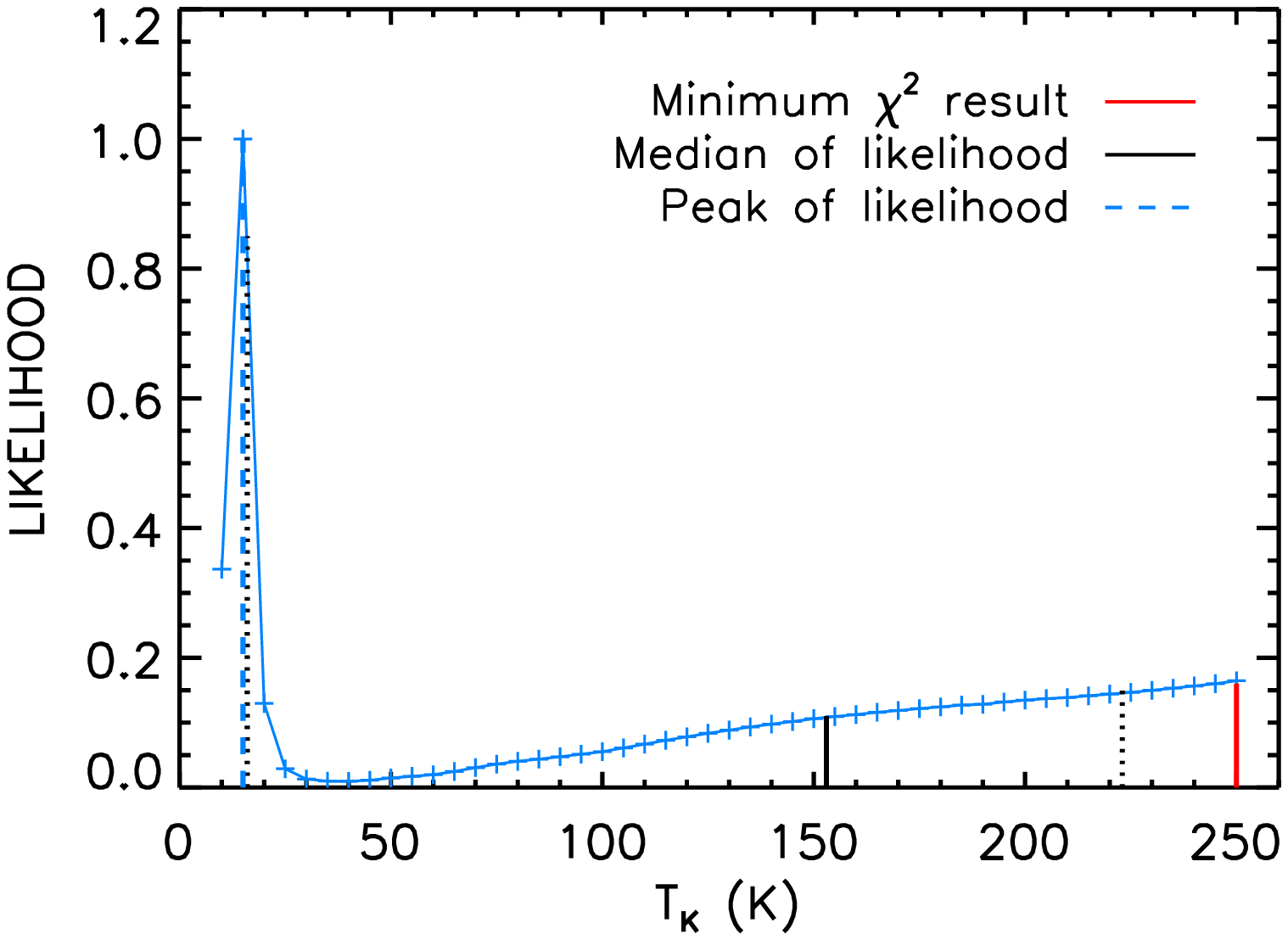}
  \includegraphics[width=5.5cm,clip=]{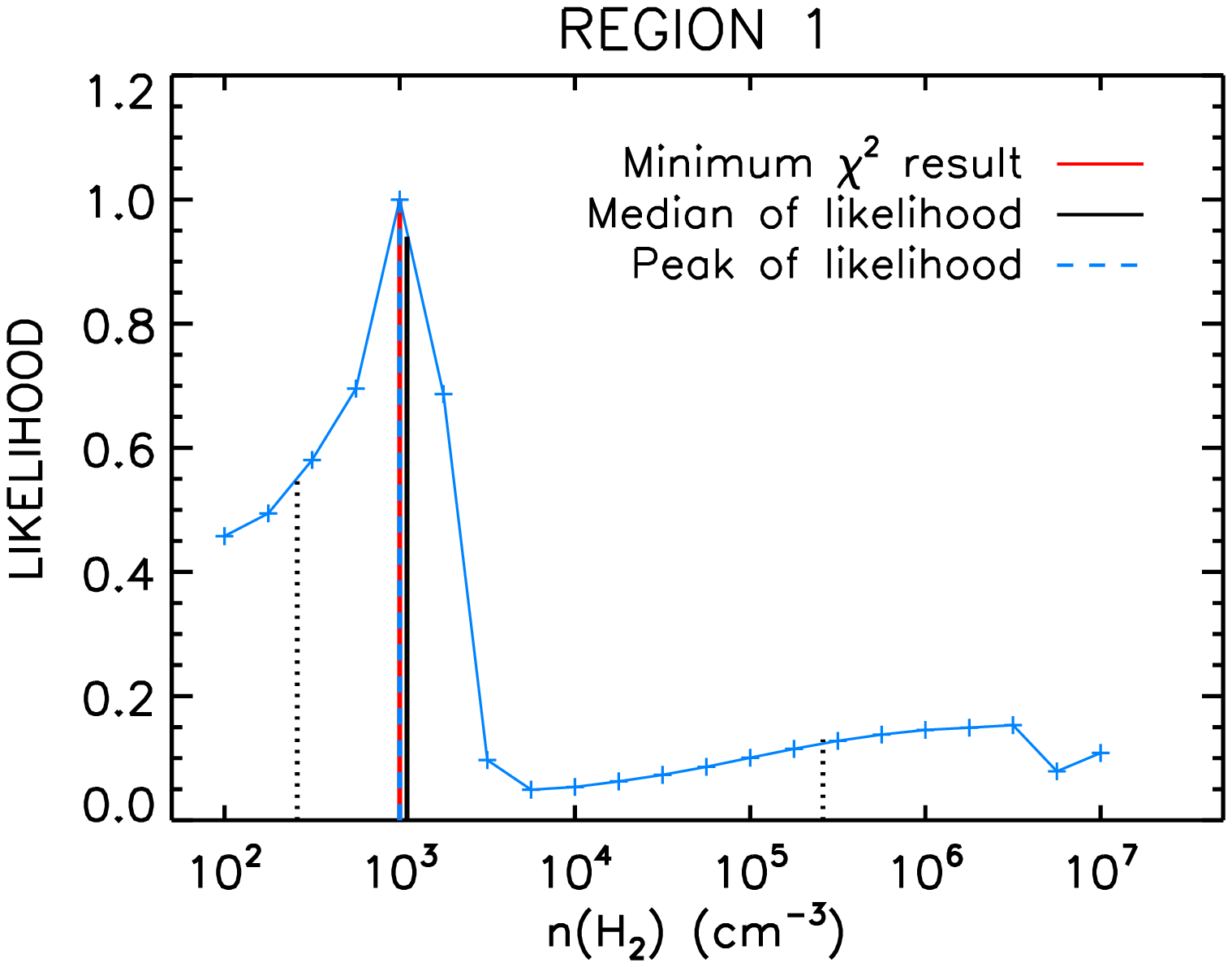}
  \includegraphics[width=5.5cm,clip=]{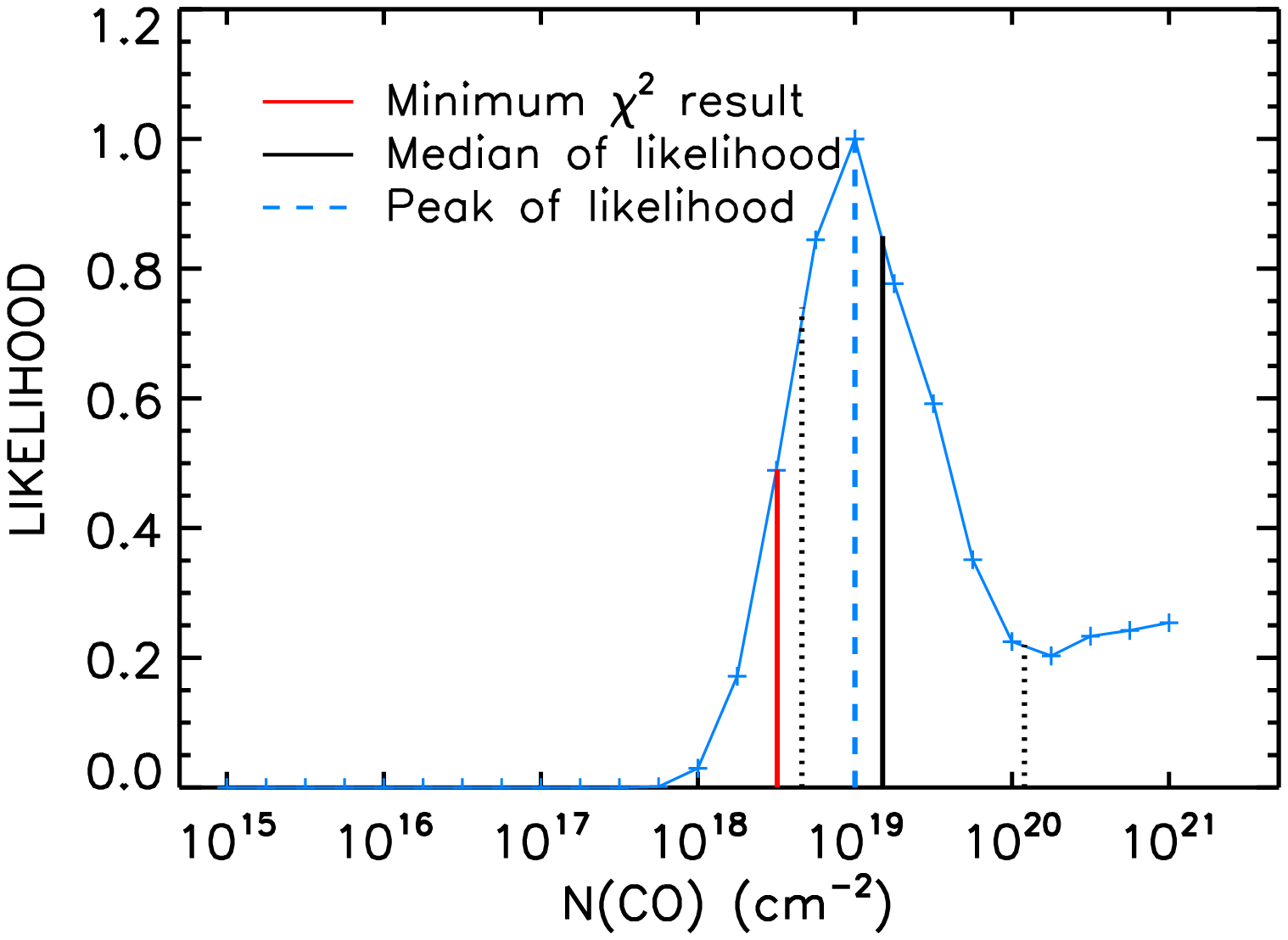}\\
  \includegraphics[width=5.5cm,clip=]{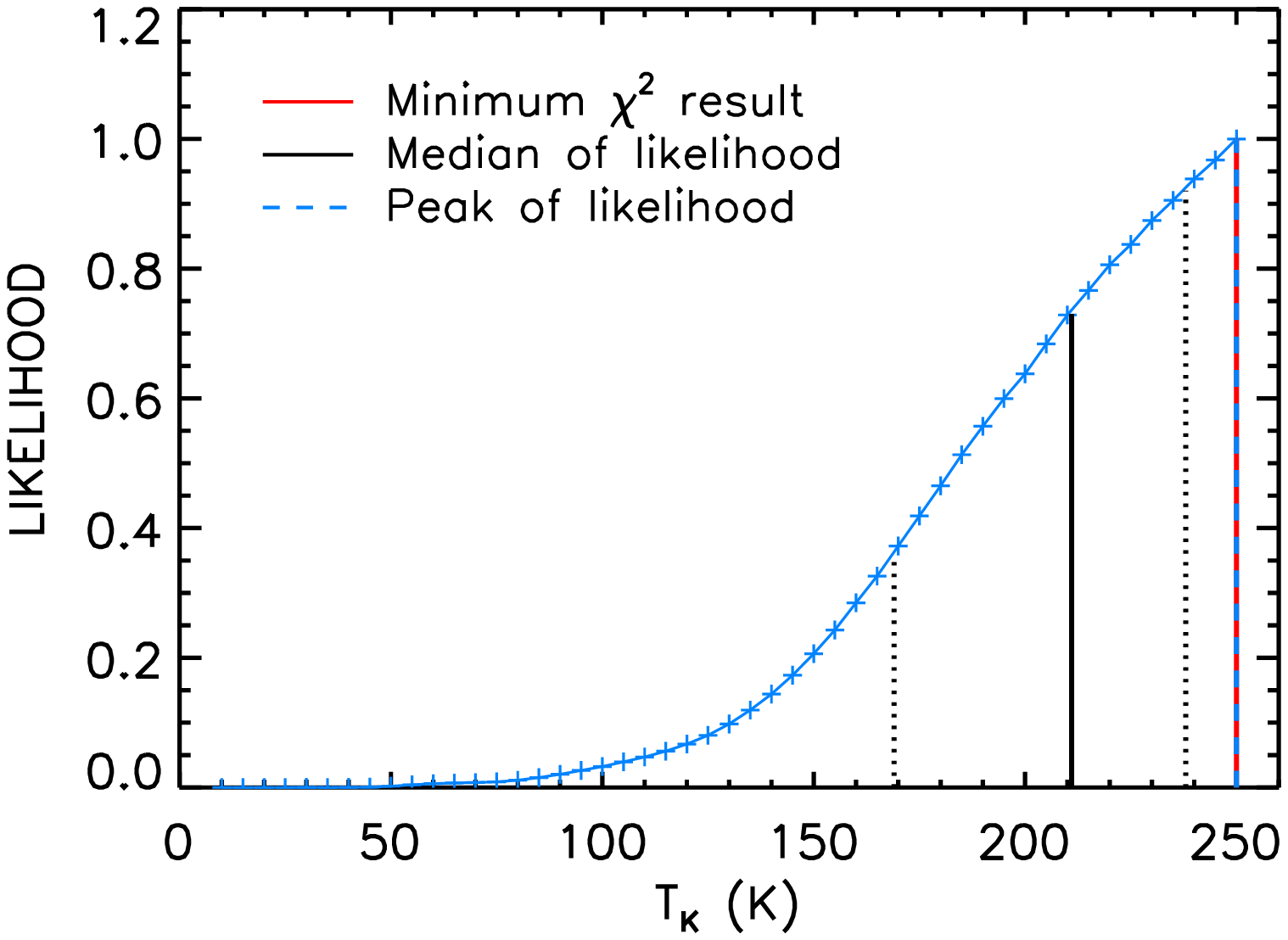}
  \includegraphics[width=5.5cm,clip=]{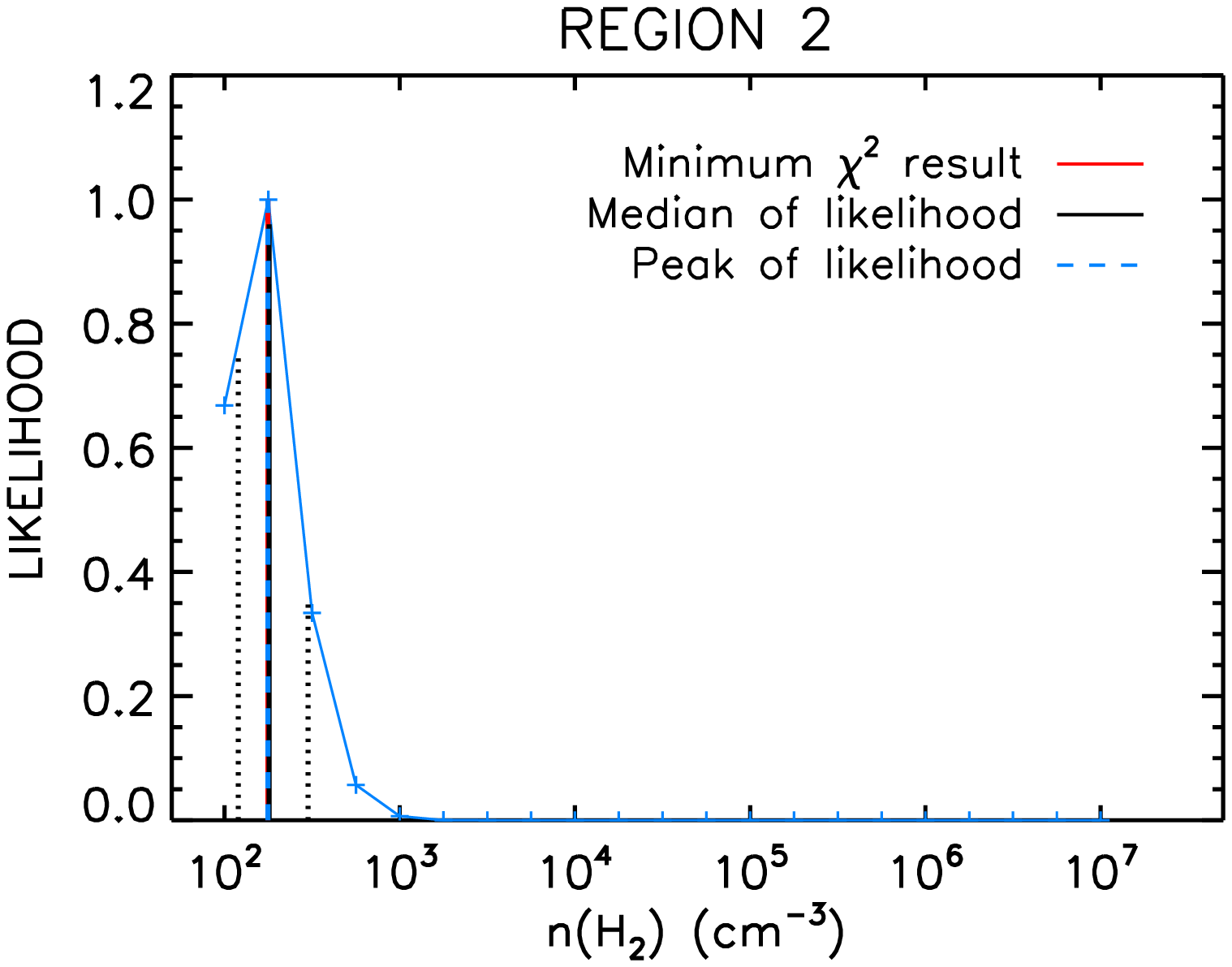}
  \includegraphics[width=5.5cm,clip=]{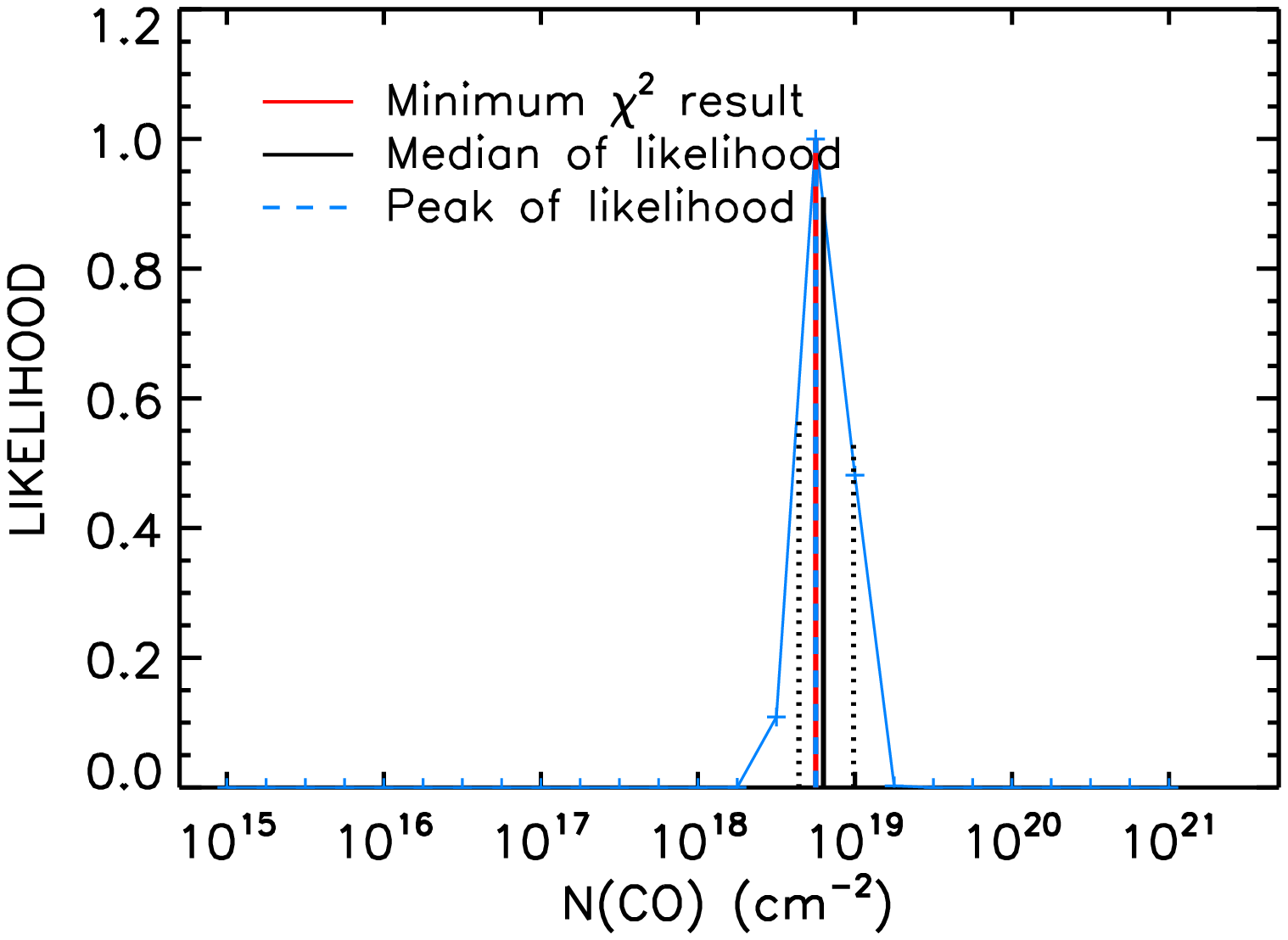}\\
  \includegraphics[width=5.5cm,clip=]{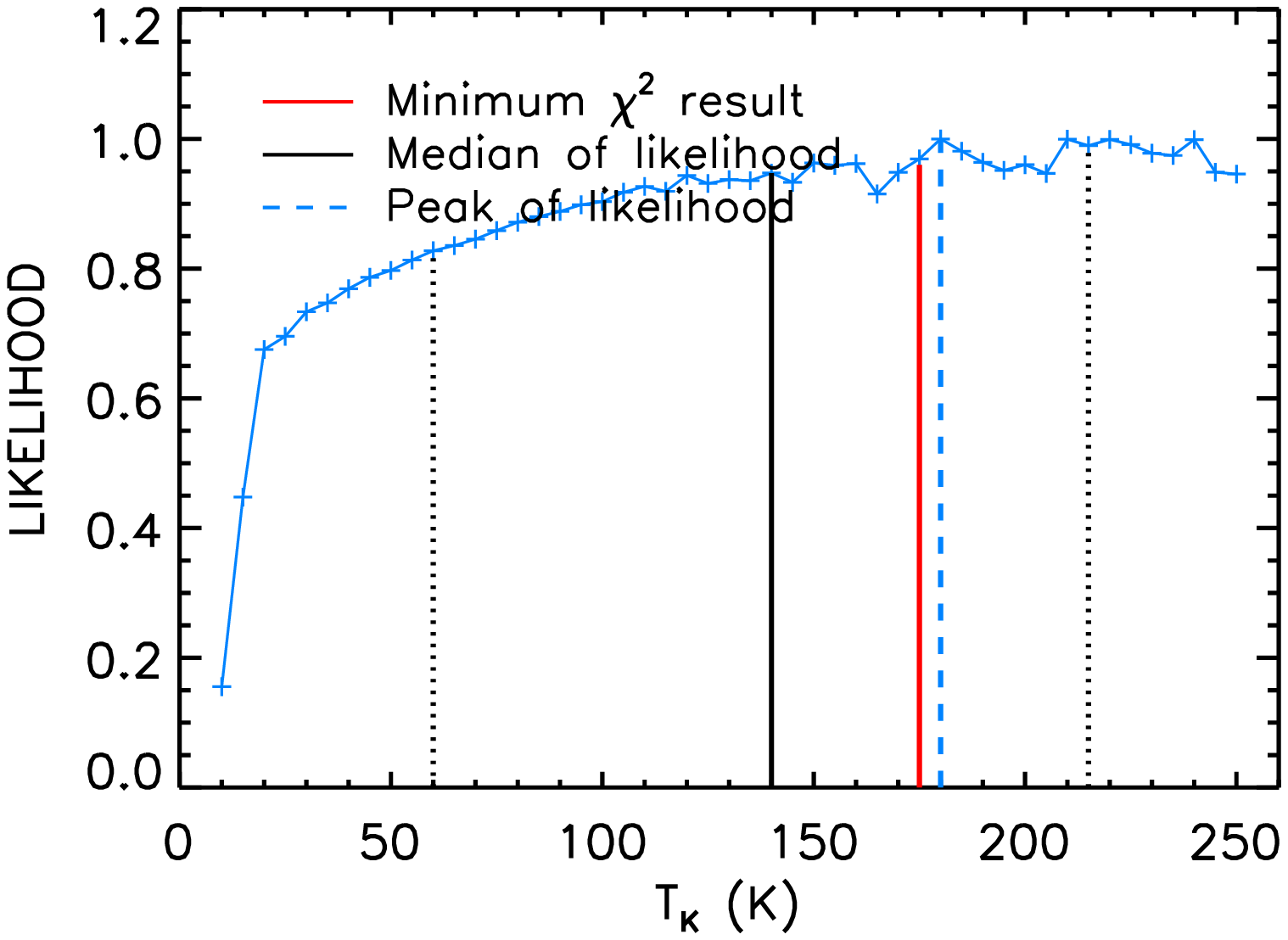}
  \includegraphics[width=5.5cm,clip=]{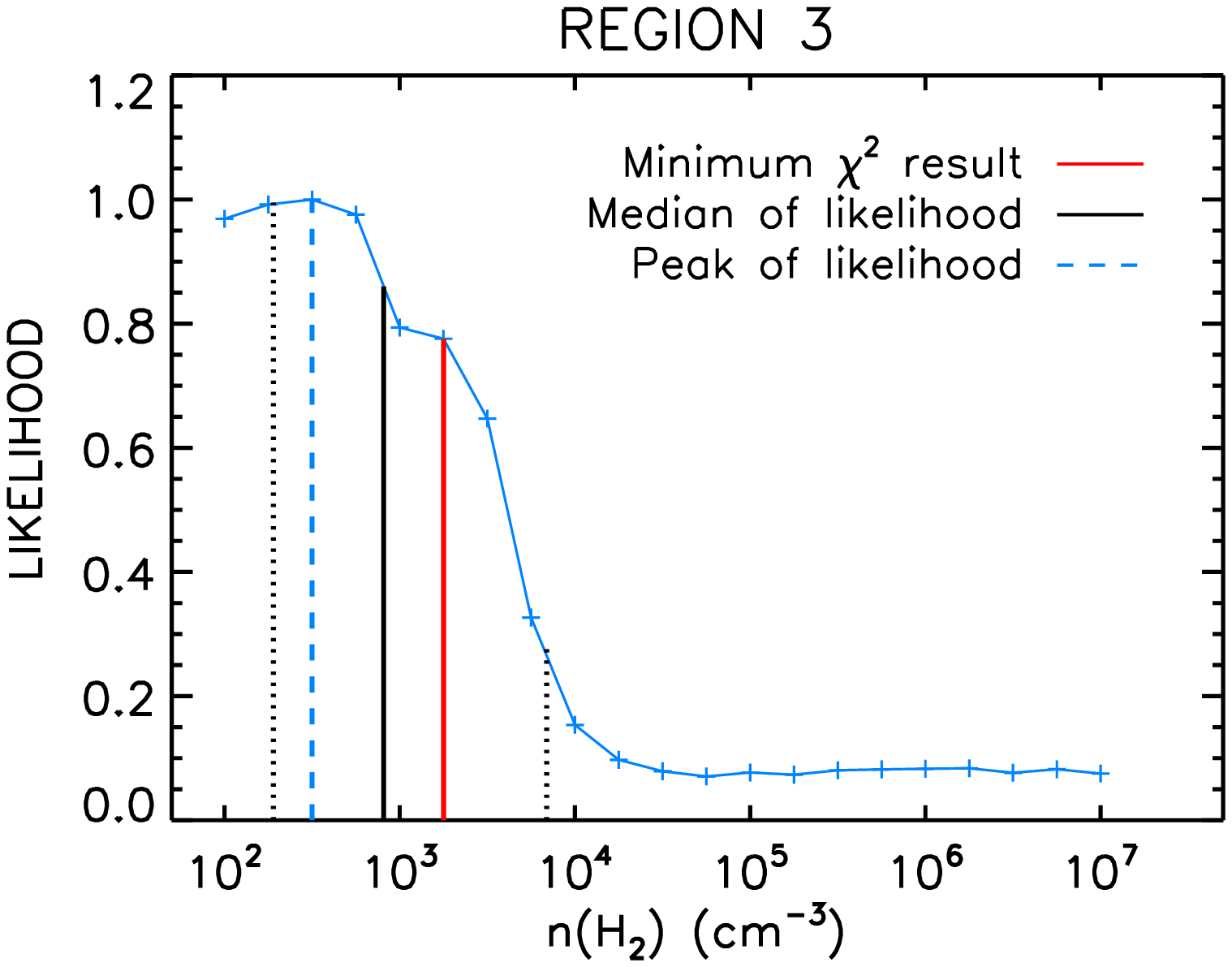}
  \includegraphics[width=5.5cm,clip=]{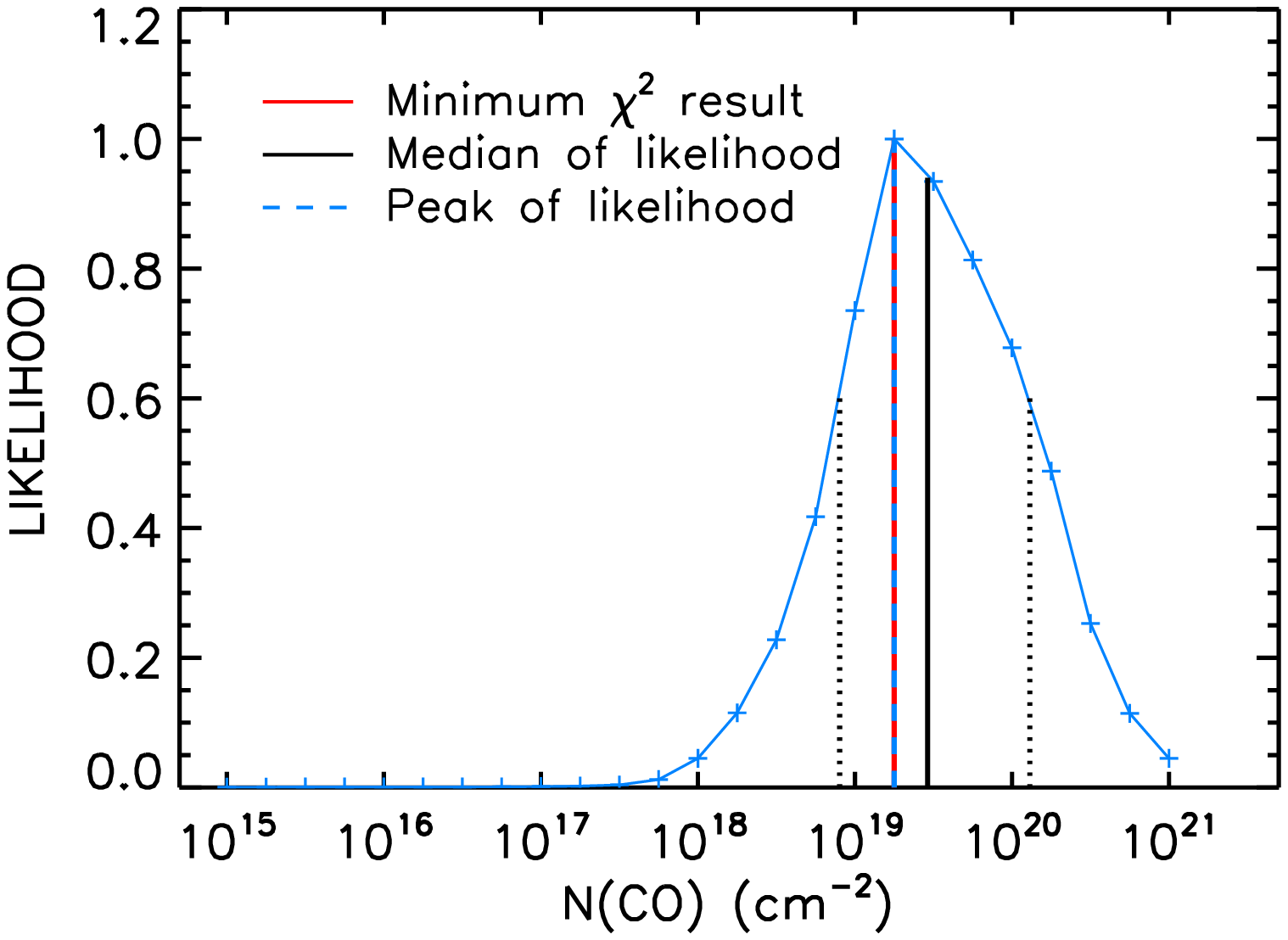}\\
  \includegraphics[width=5.5cm,clip=]{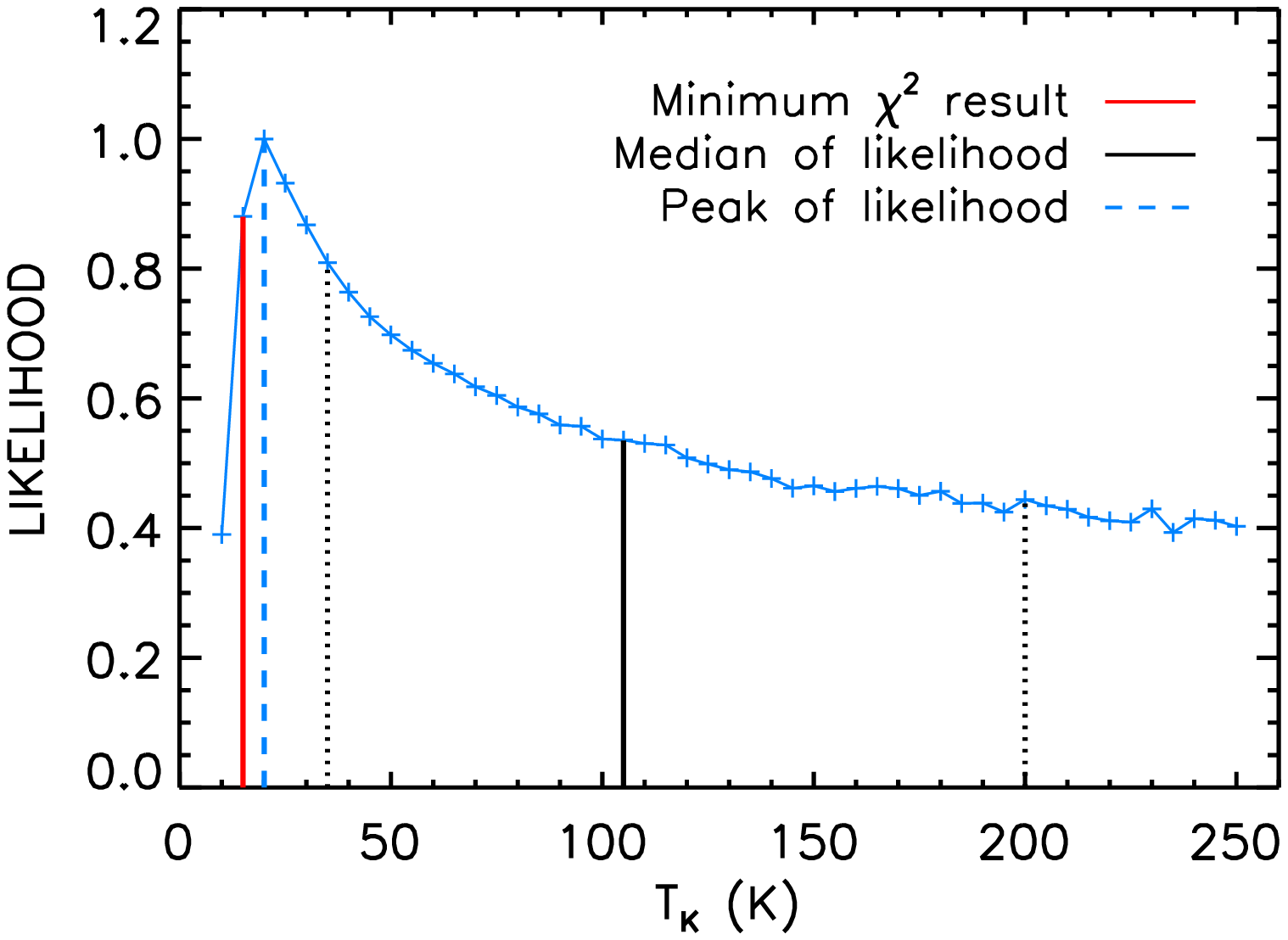}
  \includegraphics[width=5.5cm,clip=]{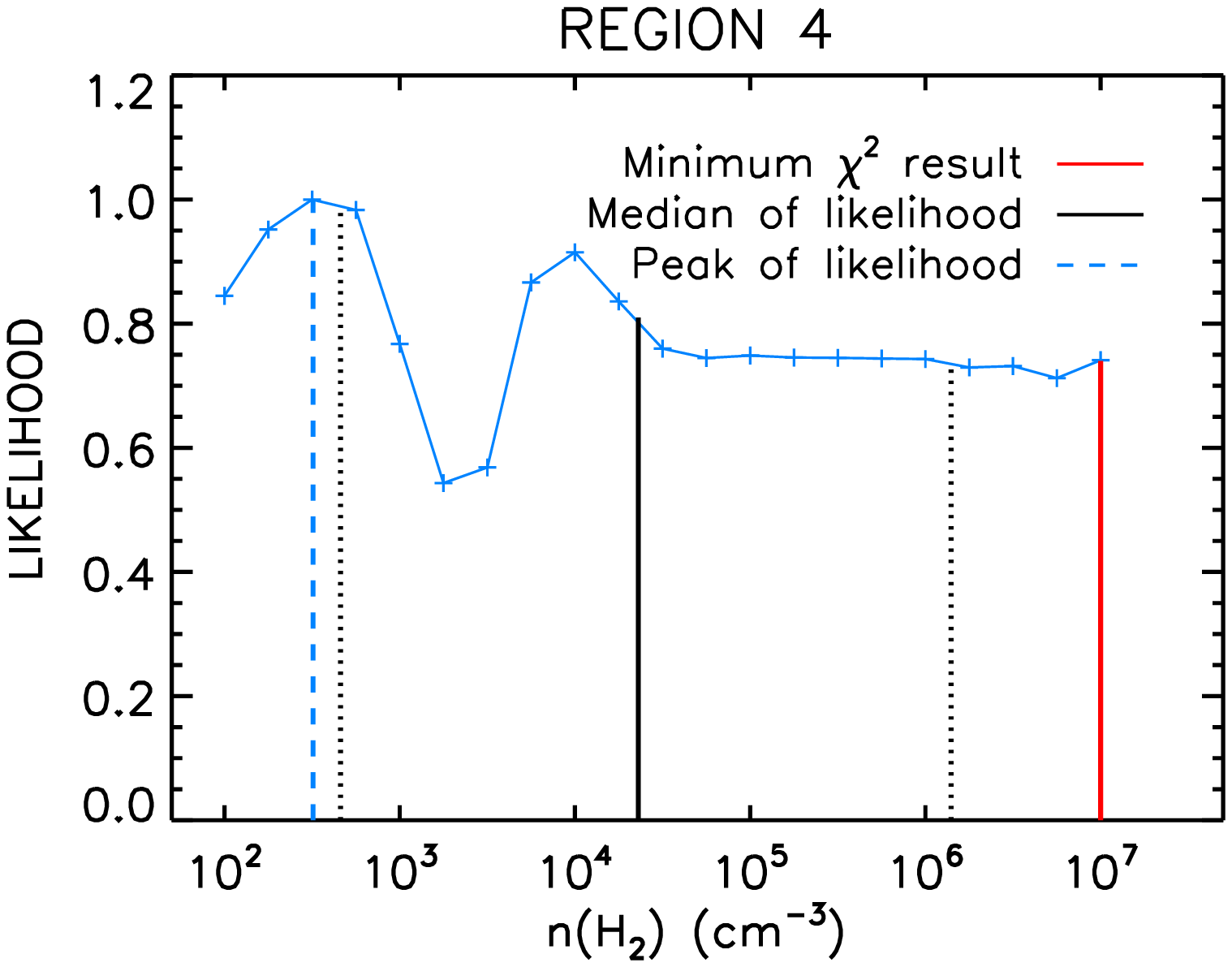}
  \includegraphics[width=5.5cm,clip=]{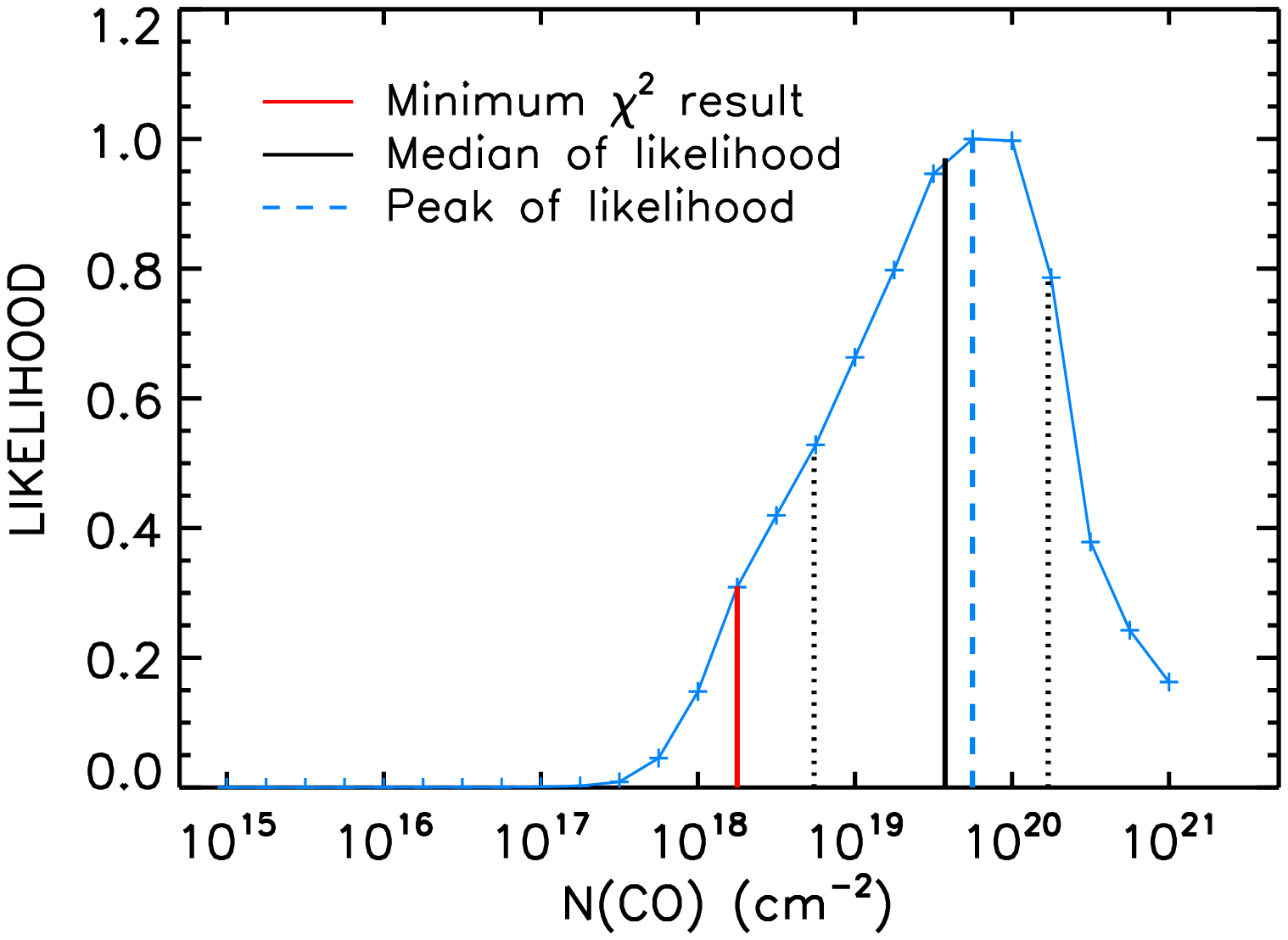}
  \caption{Marginalised probability distribution functions. For each
    region, the PDF of each model parameter marginalised over the
    other two is shown. The peak (most likely) and median values
    within the model grid are identified with dashed blue and solid
    black lines, respectively. The $68$~per cent ($1\sigma$)
    confidence levels around the median are indicated by dotted
    black lines. The best-fit model from $\chi^2$ minimisation is
    indicated by a solid red line.}
  \label{fig:likelihood}
\end{figure*}
\addtocounter{figure}{-1}
\begin{figure*}
  \includegraphics[width=5.5cm,clip=]{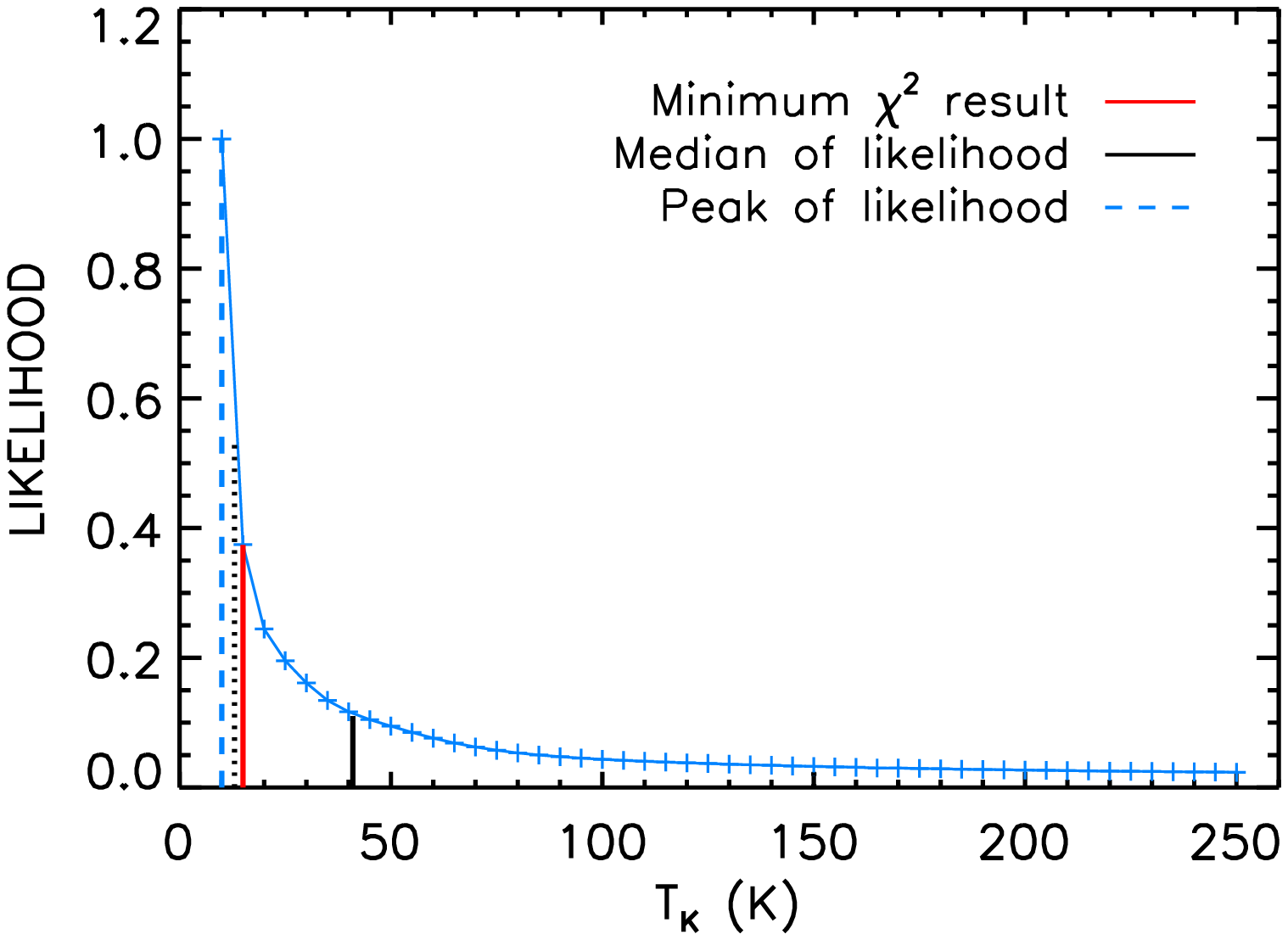}
  \includegraphics[width=5.5cm,clip=]{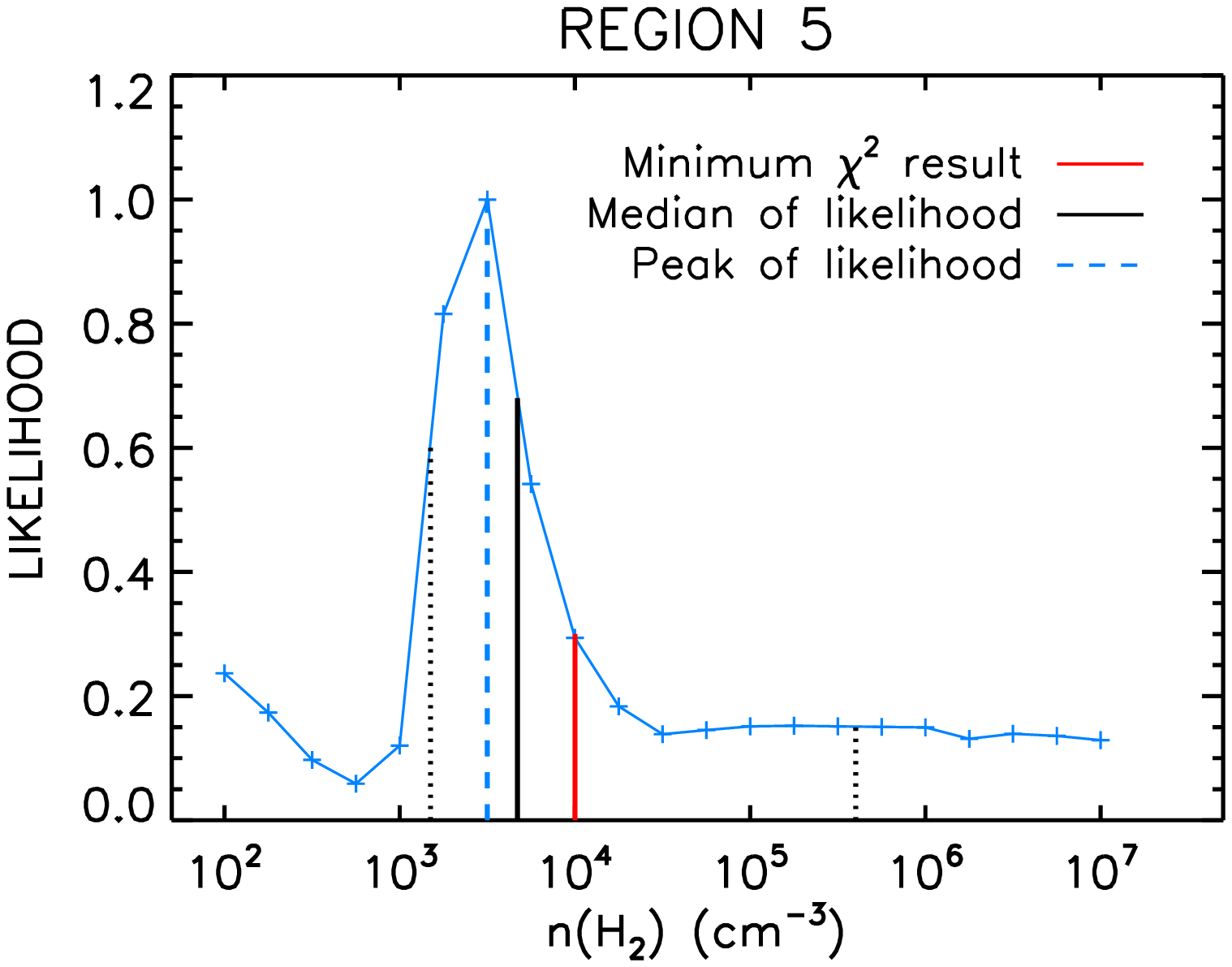}
  \includegraphics[width=5.5cm,clip=]{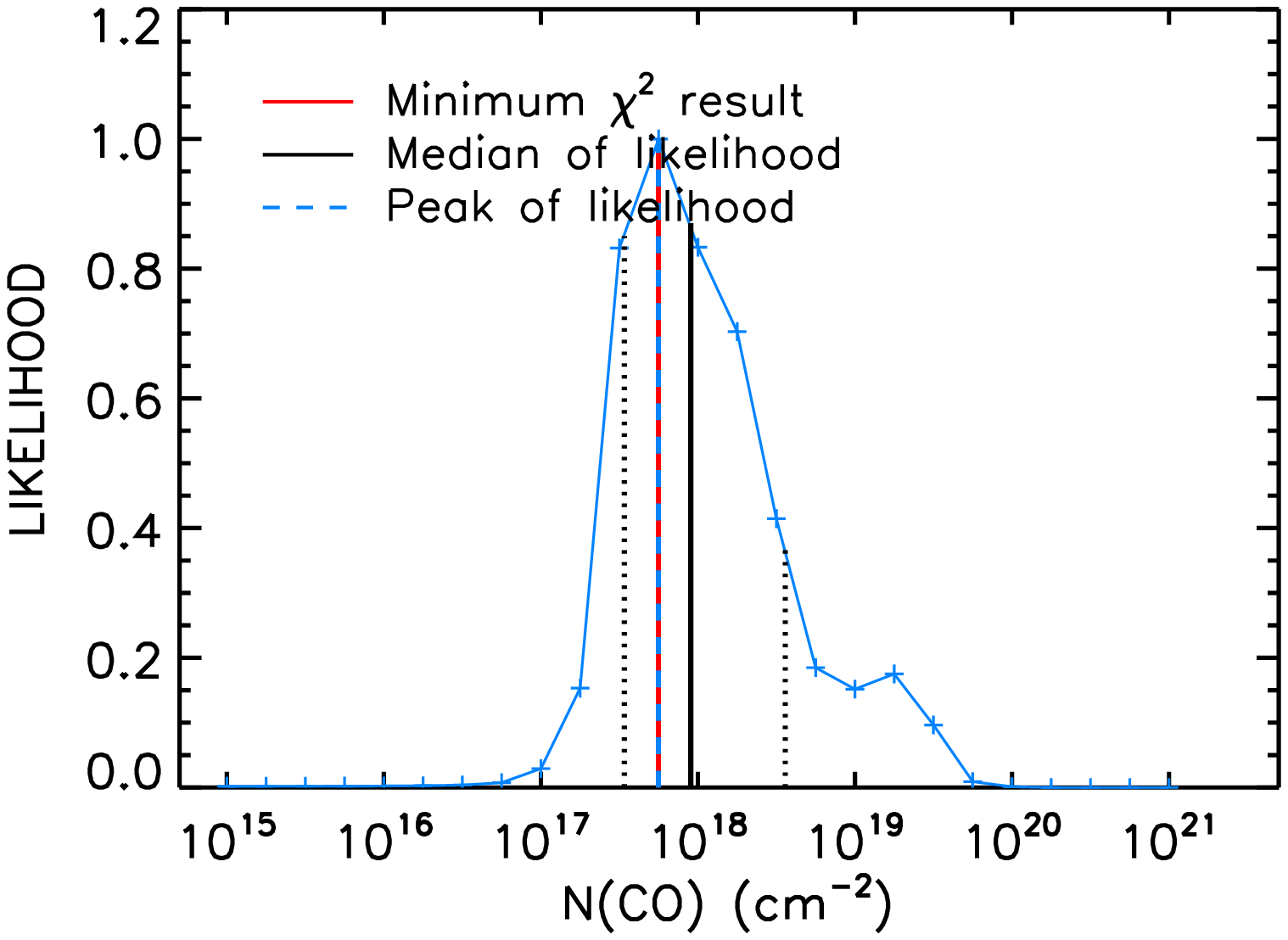}\\
  \includegraphics[width=5.5cm,clip=]{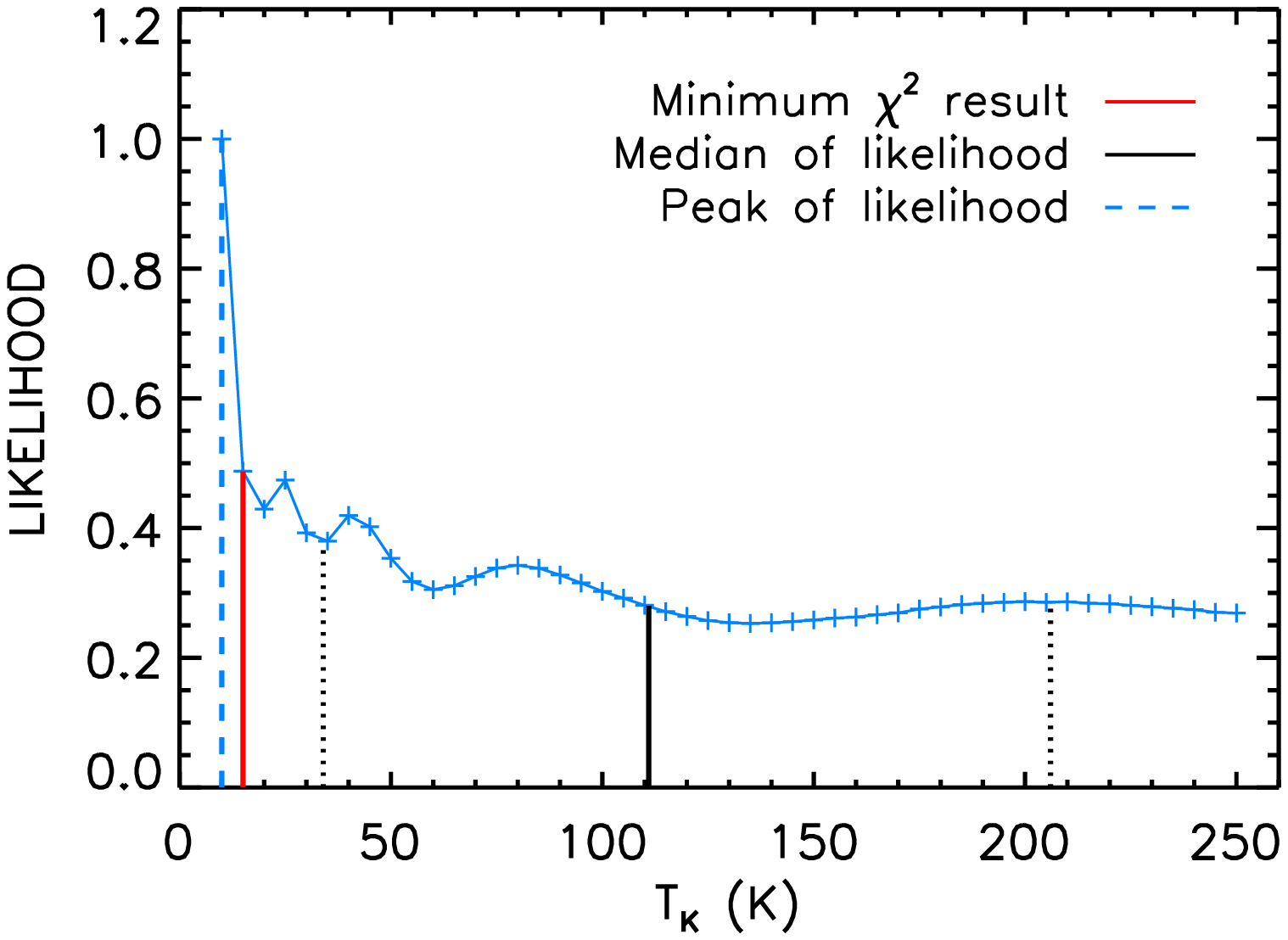}
  \includegraphics[width=5.5cm,clip=]{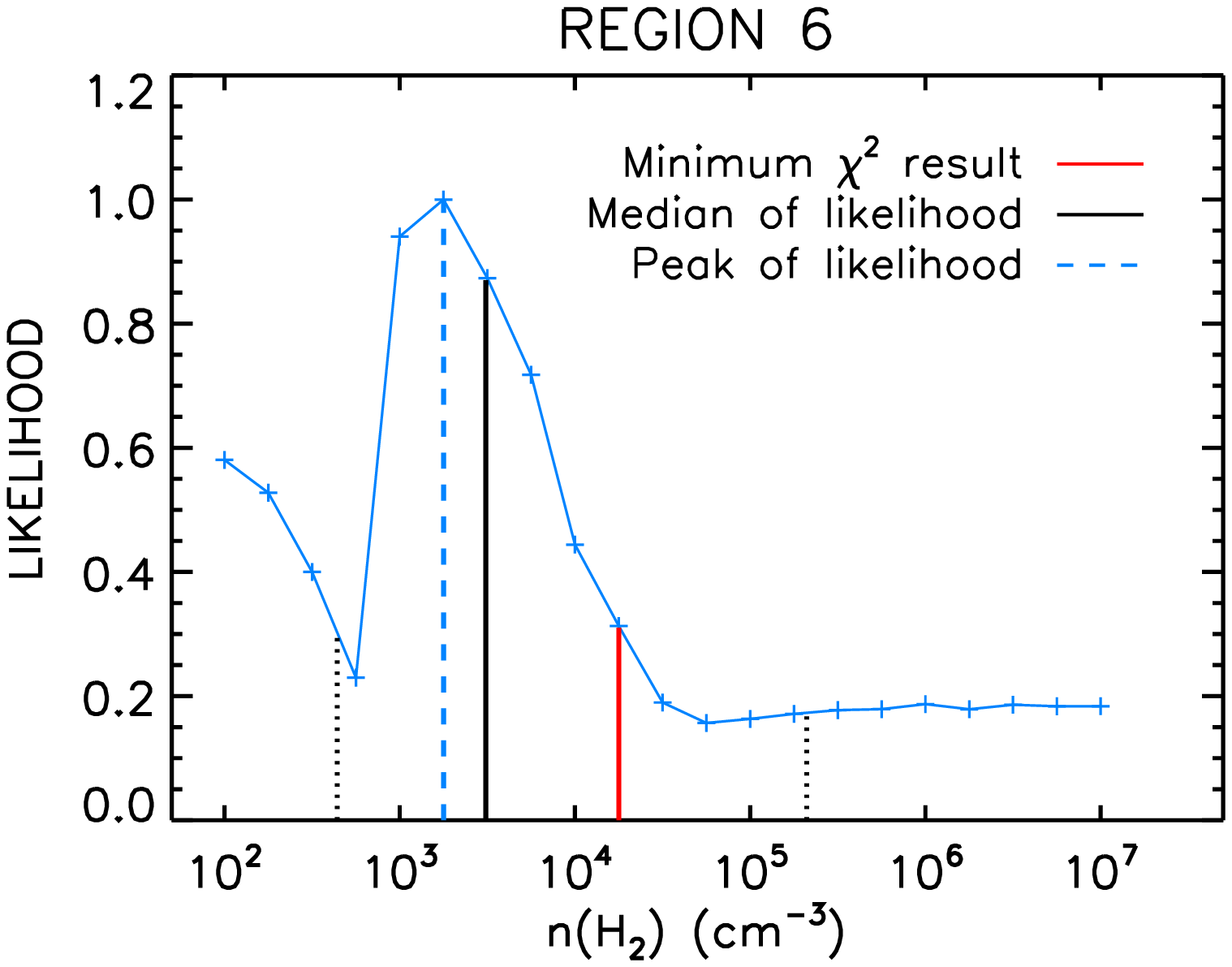}
  \includegraphics[width=5.5cm,clip=]{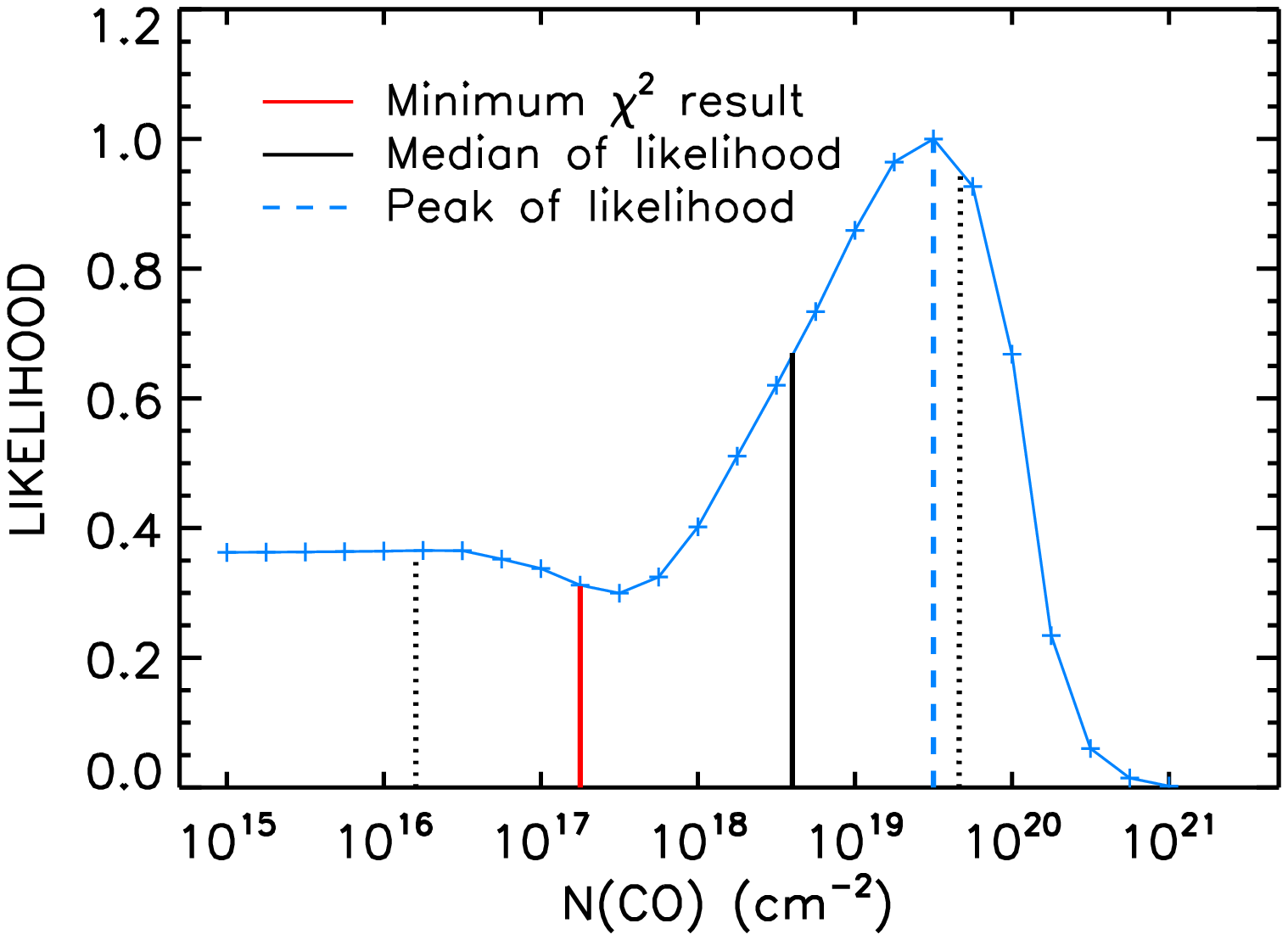}\\
  \includegraphics[width=5.5cm,clip=]{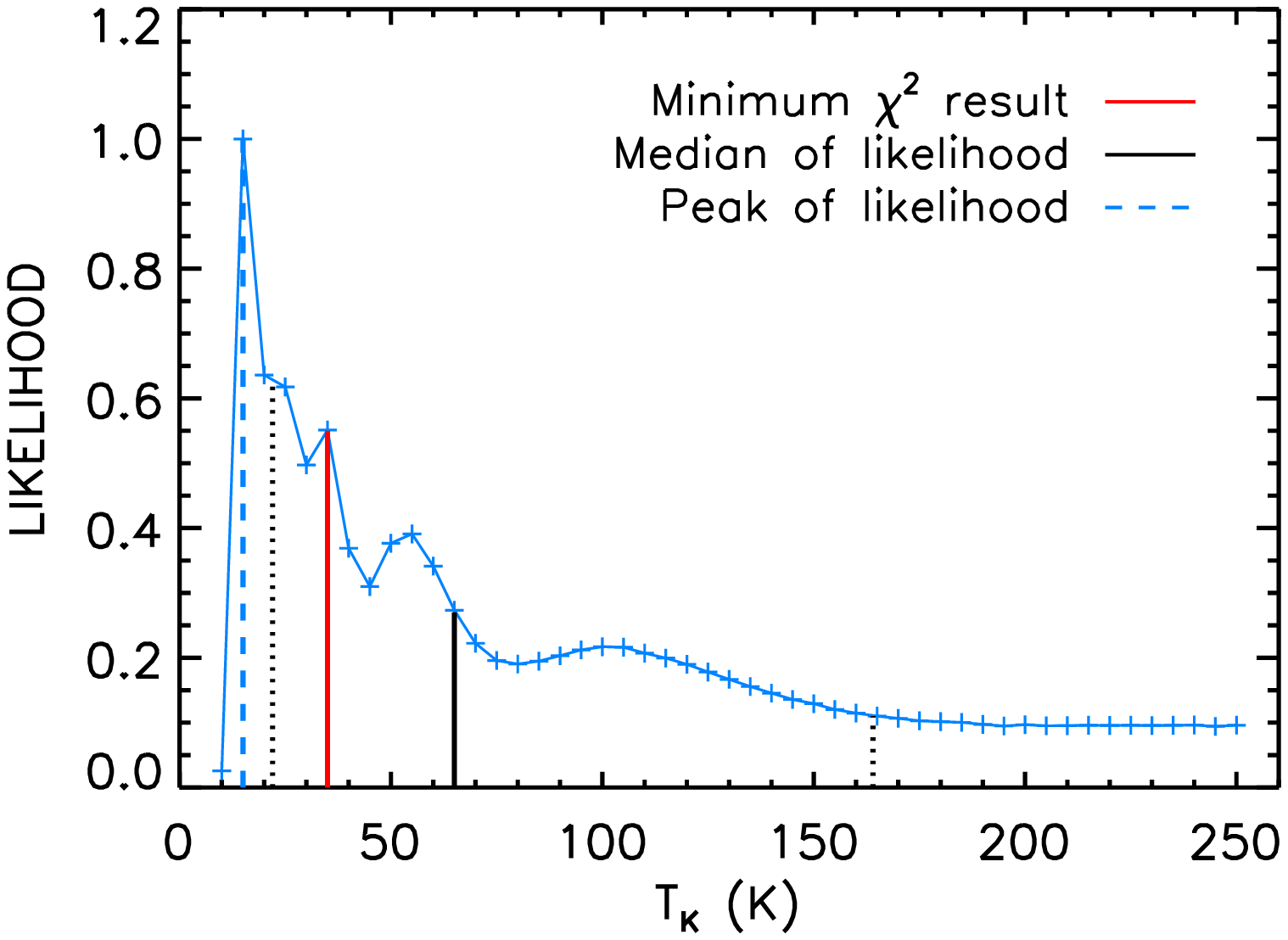}
  \includegraphics[width=5.5cm,clip=]{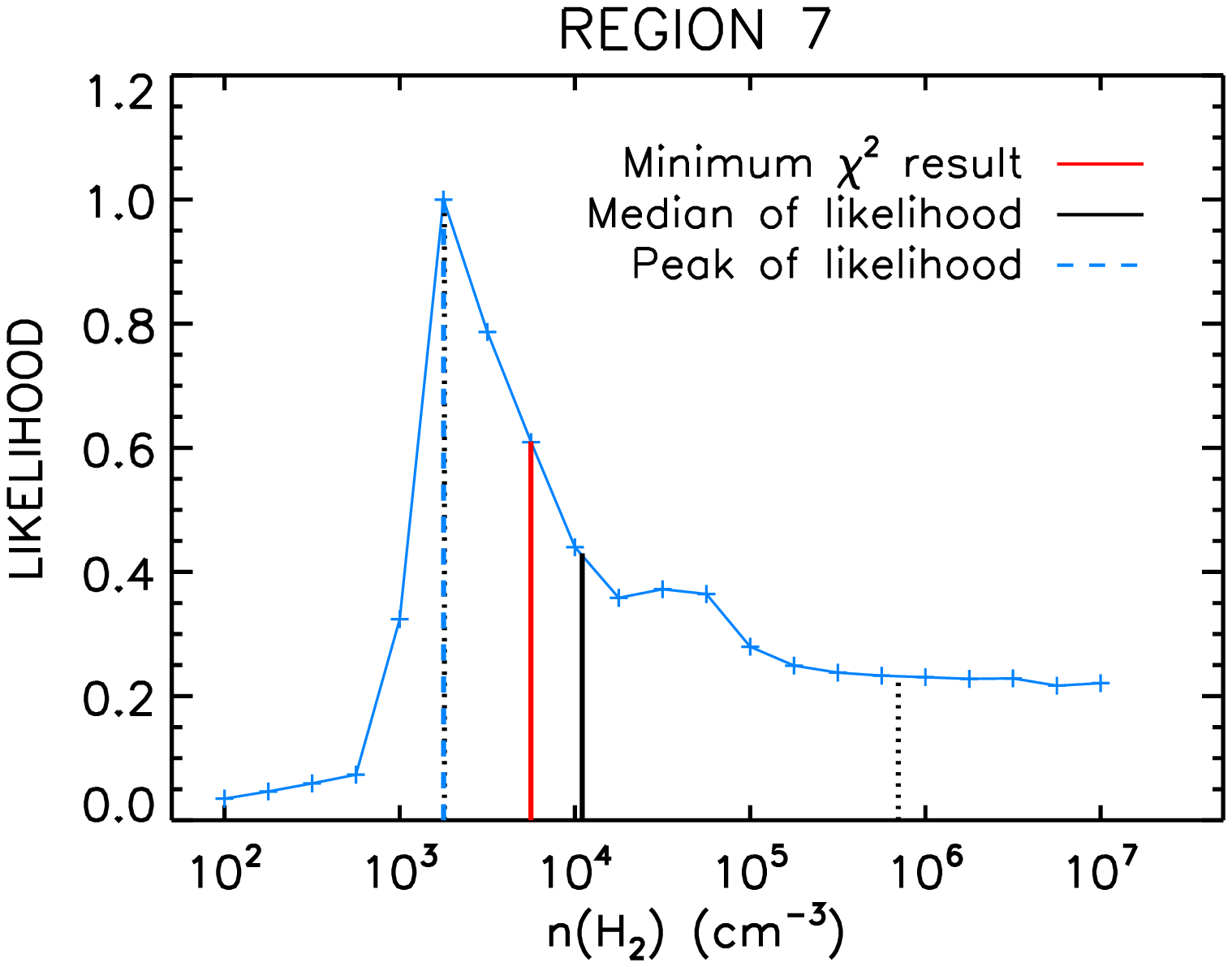}
  \includegraphics[width=5.5cm,clip=]{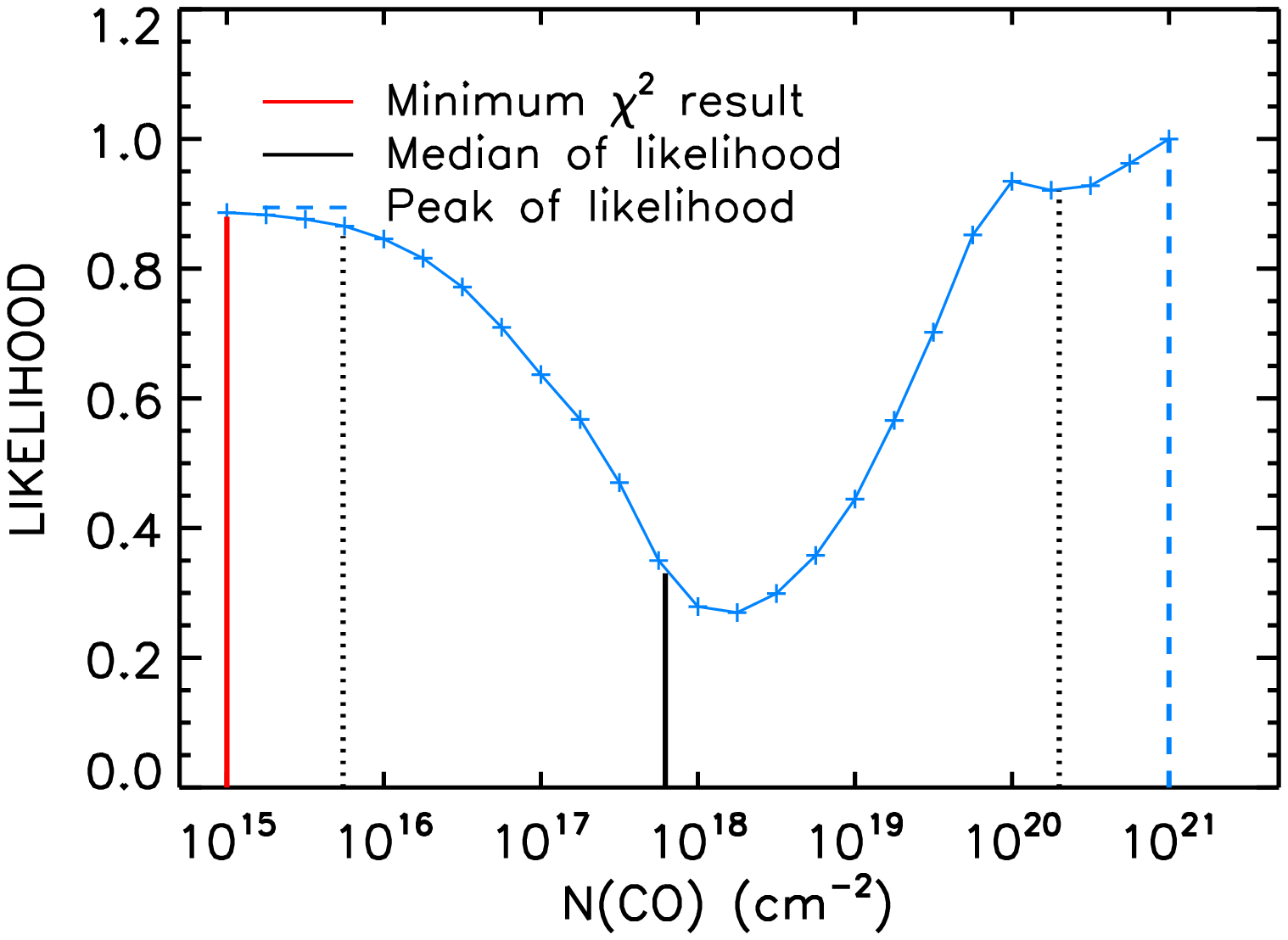}\\
  \includegraphics[width=5.5cm,clip=]{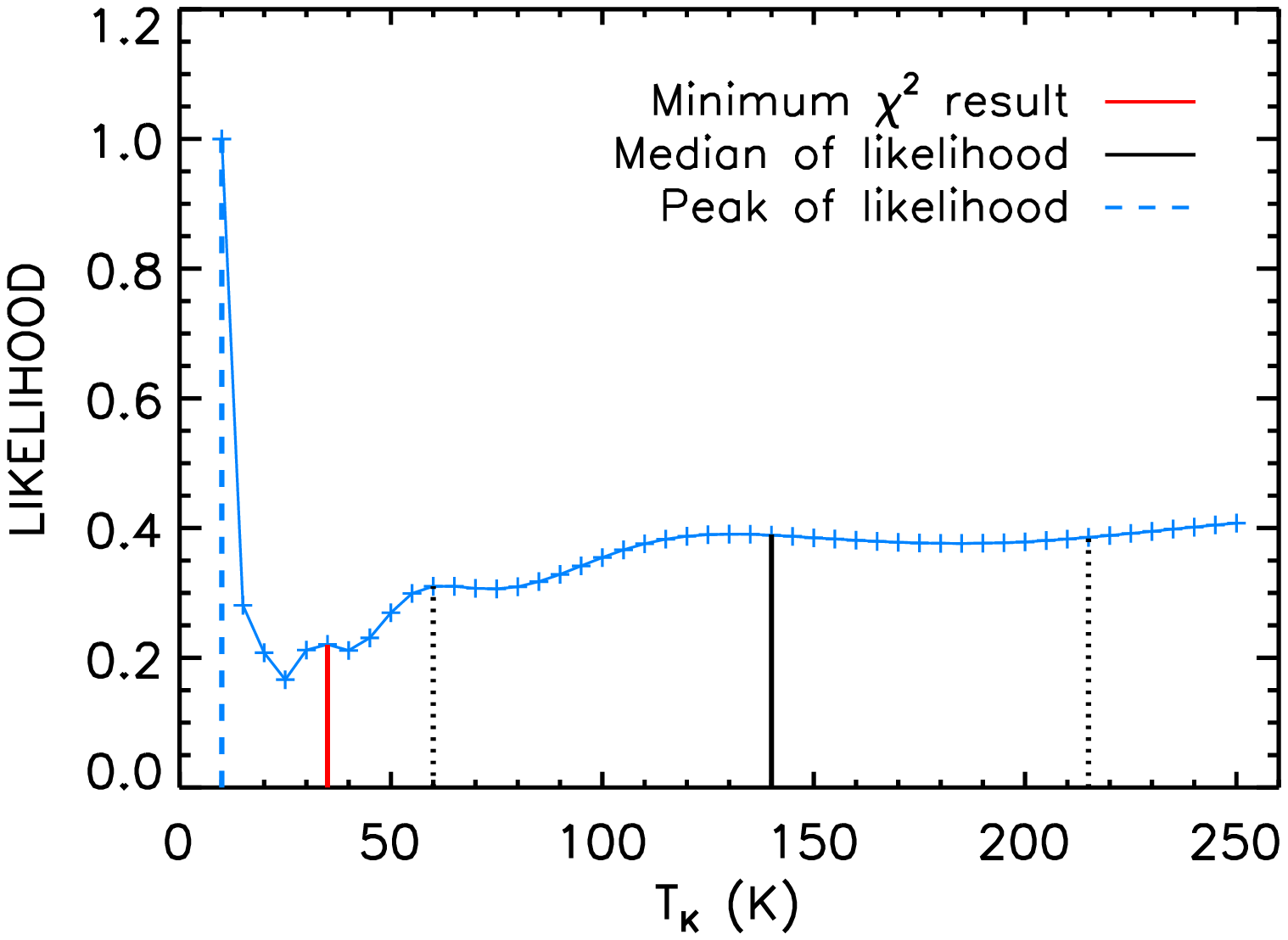}
  \includegraphics[width=5.5cm,clip=]{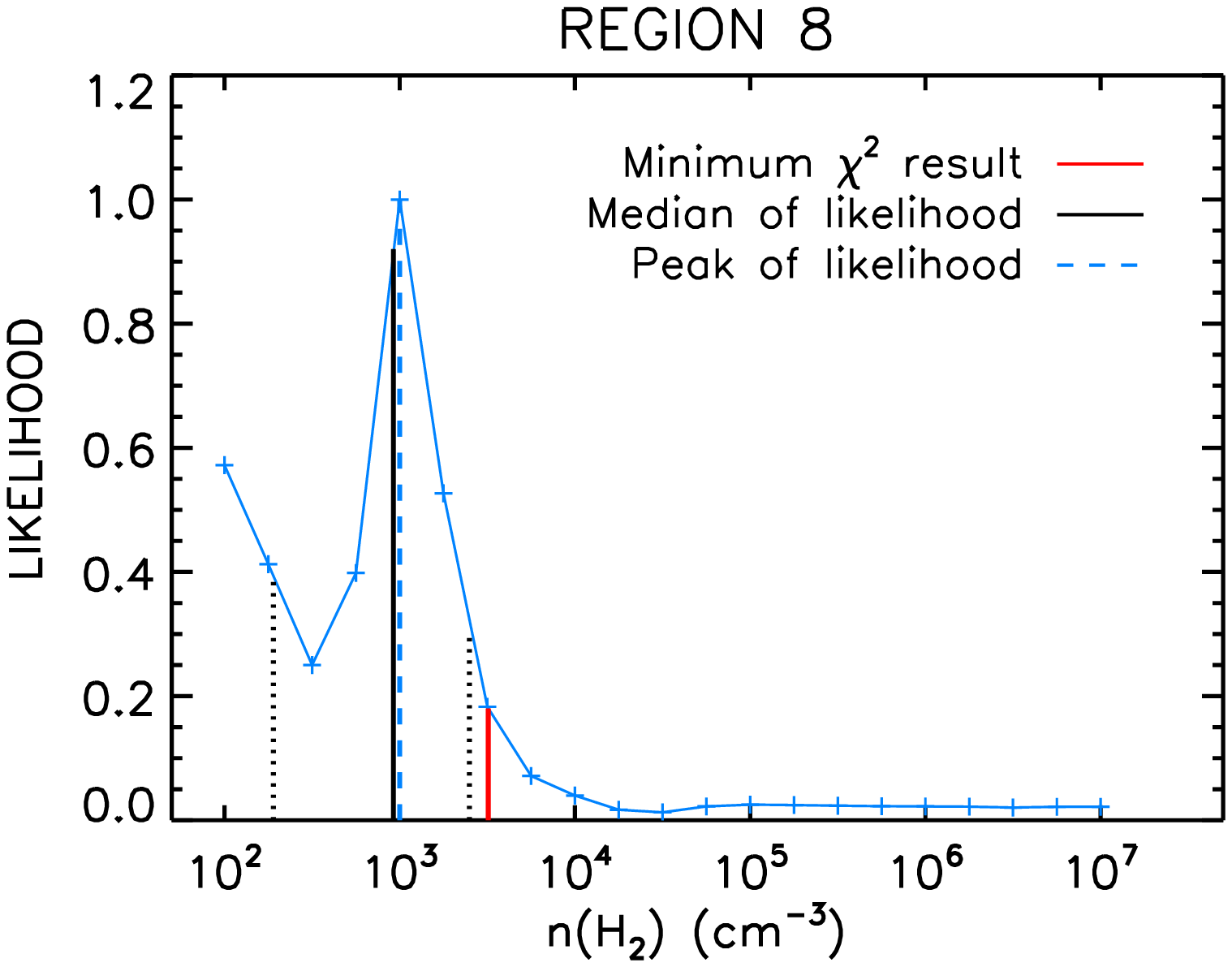}
  \includegraphics[width=5.5cm,clip=]{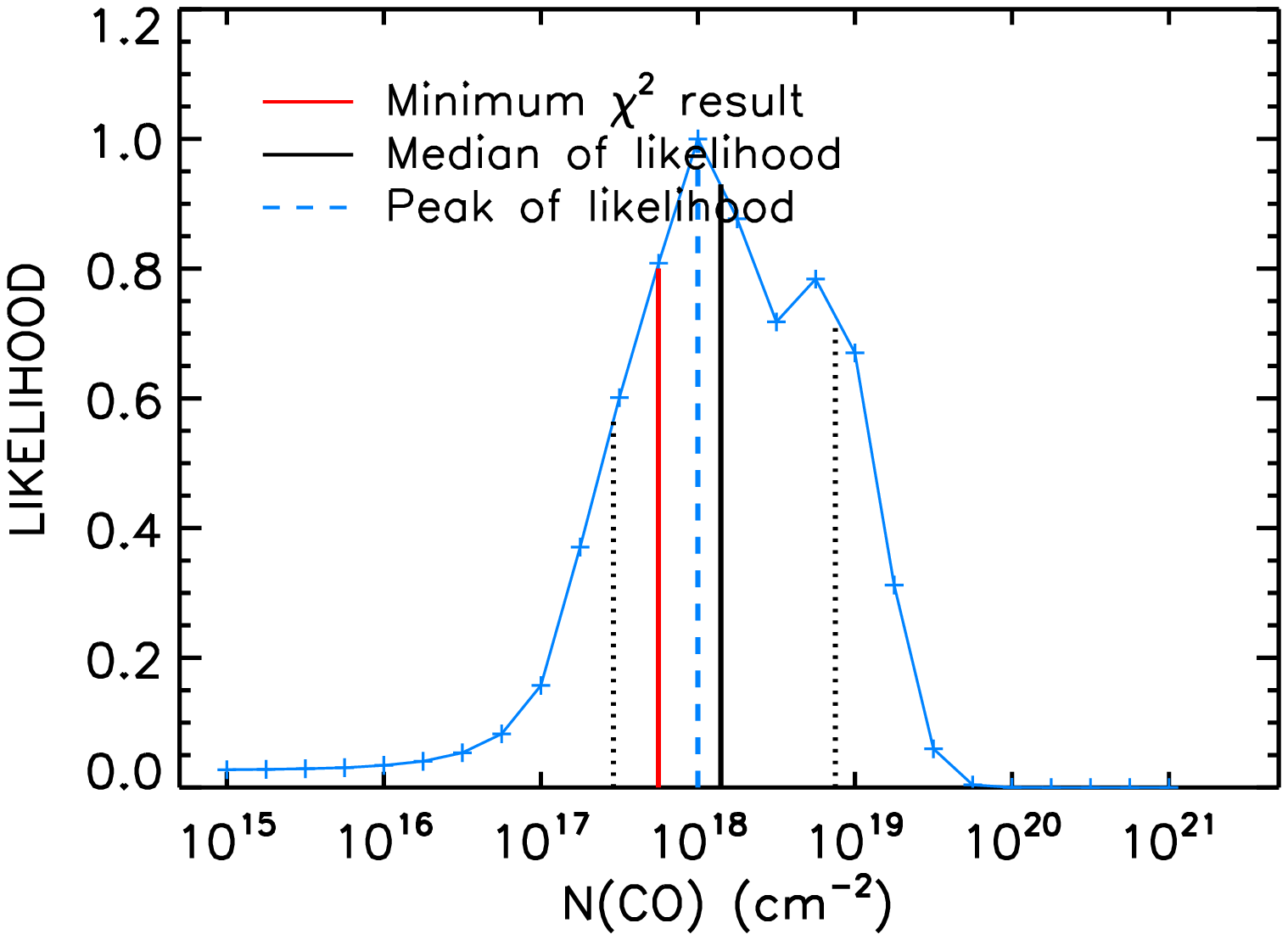}
  \caption{Continued.}
\end{figure*}
%
% Results
%
\section{Results}
\label{sec:results}
%
% Empirical results
%
\subsection{Empirical results}
\label{sec:empirical}
Figure~\ref{fig:CO2-1} shows our new CO(2-1) integrated intensity and
mean velocity maps of NGC~6946, some of the most extensive in the
literature. Molecular gas traced by CO(2-1) emission is pervasive in
NGC~6946, albeit concentrated along the spiral arms, and the velocity
field is regular. Although our coverage is patchy, CO(3-2) emission
(tracing slightly more excited gas) is particularly strong around
regions~1 and 2, located in the prominent East/North-East spiral
arms. These two regions are also very bright at H$\alpha$ \citep{k04},
$24$~$\micron$ \citep{k03}, $850$~$\micron$ \citep{b00} and $6.2$~cm
\citep{bh96}, suggesting intense SF in a dusty ISM. Regions~4, 5, 6
and 7 have weaker H\,{\small I} emission than that of the other
regions (see Fig.~\ref{fig:regions}).

Figure~\ref{fig:radial} shows the radial distribution of the molecular
gas using the multiple tracers available. Two different components
dominate: a highly concentrated central molecular core with a radius
$\la1$~kpc, and a nearly flat lower surface brightness disc extending
to large radii.

Of the $134$ GMC candidates identified by \citet{do12} in the central
$5$~kpc of NGC~6946, only $30$ are fully resolved in their data, and
the others could be blends of GMCs smaller than the beam. \citet{do12}
found that the most massive clouds ($>10^{7}$~$M_\odot$) are located
at the very centre of NGC~6946, and that clouds with masses
$>5\times10^5$~$M_\odot$ are preferentially found where the cloud
density is highest (i.e.\ in the spiral arms). They also showed that
the GMC sizes range from $40$ to $200$~pc in the central $5$~kpc,
similar to the sizes found by \citet{reb12} for $64$ GMCs in the
eastern part of the galaxy ($50$ to $150$~pc). As our adopted common
beam size of $31\farcs5$ corresponds to a linear diameter of
$\approx840$~pc at the distance of NGC~6946, our beam will usually
encompass an ensemble of (giant) molecular clouds. For each region, we
thus estimate the total H$_2$ mass enclosed within our beam ($M_{{\rm
    H}_2}$) using the CO-to-H$_2$ conversion factor ($X_{\rm CO}$ = $N$(H$_2$)~/~$S_{\rm
  1-0}$) calculated by \citet{do12} from their $30$ spatially-resolved GMCs in the centre of
NGC~6946 (see Fig.~\ref{fig:regions}): $X_{\rm CO}$ = $1.2\times10^{20}$~cm$^{-2}$ (K~km~s$^{-1}$)$^{-1}$, consistent with the value assumed for the eastern GMCs 
by \citet{reb12} and the average for the NGC~6946 disc reported by \citet{sa12}. \citet{sa12} 
do report a depression of the $X_{\rm CO}$ value in the very centre of NGC~6946 (i.e. within a radius
of $1.2$~kpc), but since all our regions are well outside of this our calculations should be unaffected.
Considering the adopted beam, distance, and $X_{\rm CO}$ factor, we thus obtain
\begin{equation}
  \frac{M_{{\rm H}_2}}{M_\odot}=3.6\times10^5\,\bigg(\frac{S_{\rm 1-0}}{{\rm K}\,{\rm km}\,{\rm s}^{-1}}\bigg)\,.
  \label{eq:mass}
\end{equation}
The resulting total H$_2$ mass and beam-averaged H$_2$ column density
of each region are listed in Table~\ref{tab:mass}. 
%
% Table: Calculated Gas mass
%	
\begin{table}
  \caption{Total H$_2$ masses and beam-averaged H$_2$ column densities.}
  \centering
  \begin{tabular}{c c c}
    \hline
    Region & $M_{{\rm H}_2}$ & $N$(H$_2$) \\
           & ($M_\odot$) & (cm$^{-2}$) \\
    \hline
    1 & $3.1\pm0.1\times10^{6}$ & $1.1\pm0.1\times10^{21}$ \\
    2 & $2.9\pm0.2\times10^{6}$ & $9.9\pm0.5\times10^{20}$ \\
    3 & $2.7\pm0.2\times10^{6}$ & $9.0\pm0.5\times10^{20}$ \\
    4 & $1.7\pm0.2\times10^{6}$ & $5.7\pm0.5\times10^{20}$ \\
    5 & $2.1\pm0.1\times10^{6}$ & $6.9\pm0.4\times10^{20}$ \\
    6 & $1.7\pm0.2\times10^{6}$ & $5.7\pm0.5\times10^{20}$ \\
    7 & $1.3\pm0.2\times10^{6}$ & $4.4\pm0.5\times10^{20}$ \\
    8 & $2.6\pm0.2\times10^{6}$ & $8.8\pm0.5\times10^{20}$ \\
    \hline
  \end{tabular}
  \label{tab:mass}
\end{table}		

The gas complex in region~1 has the highest CO(1-0) flux and thus
total H$_2$ mass ($3.1\pm0.1\times10^{6}$~M$_\odot$), while the gas in
the inter-arm region~7 (farthest from the galaxy centre) has the
lowest mass ($1.3\pm0.2\times10^{6}$~M$_\odot$). Having said that, the
total gas mass does not vary much from region to region (factor
$\approx3$), and for the $8$ regions studied it does not vary linearly
with distance from the galaxy centre (as hinted from
Fig.~\ref{fig:radial}). All regions have total H$_2$ masses
$>5\times10^{5}$~$M_\odot$.

As shown in Table~\ref{tab:line}, the beam-corrected
velocity-integrated line intensity of the CO(2-1) transition is higher
than that of all other transitions observed in all regions (including
region~8 within the uncertainties). While $R_{\rm 21}$ (the
CO(2-1)/CO(1-0) line ratio) is higher than unity in all regions, all
other line ratios considered (with respect to the ground state) are
less than unity. Interestingly, $R_{\rm 31}$ (the CO(3-2)/CO(1-0) line
ratio) is the only line ratio that is very similar in all
regions. Where we have measurements, the isotopologue transitions
$^{13}$CO(1-0) and $^{13}$CO(2-1) are $5$--$20$ times fainter than
their parent lines. Anticipating some of our modeling results (see
Table~\ref{tab:results} and \S~\ref{sec:disc}), one also notices that
the line ratios of regions~1, 2 and 3, those of regions~4, 5, and 6,
and lastly those of regions~7 and 8 are similar (although not
identical) to each other, suggesting similar molecular gas physical
conditions in each group of regions.

\begin{figure}
  \includegraphics[width=8.5cm,clip=]{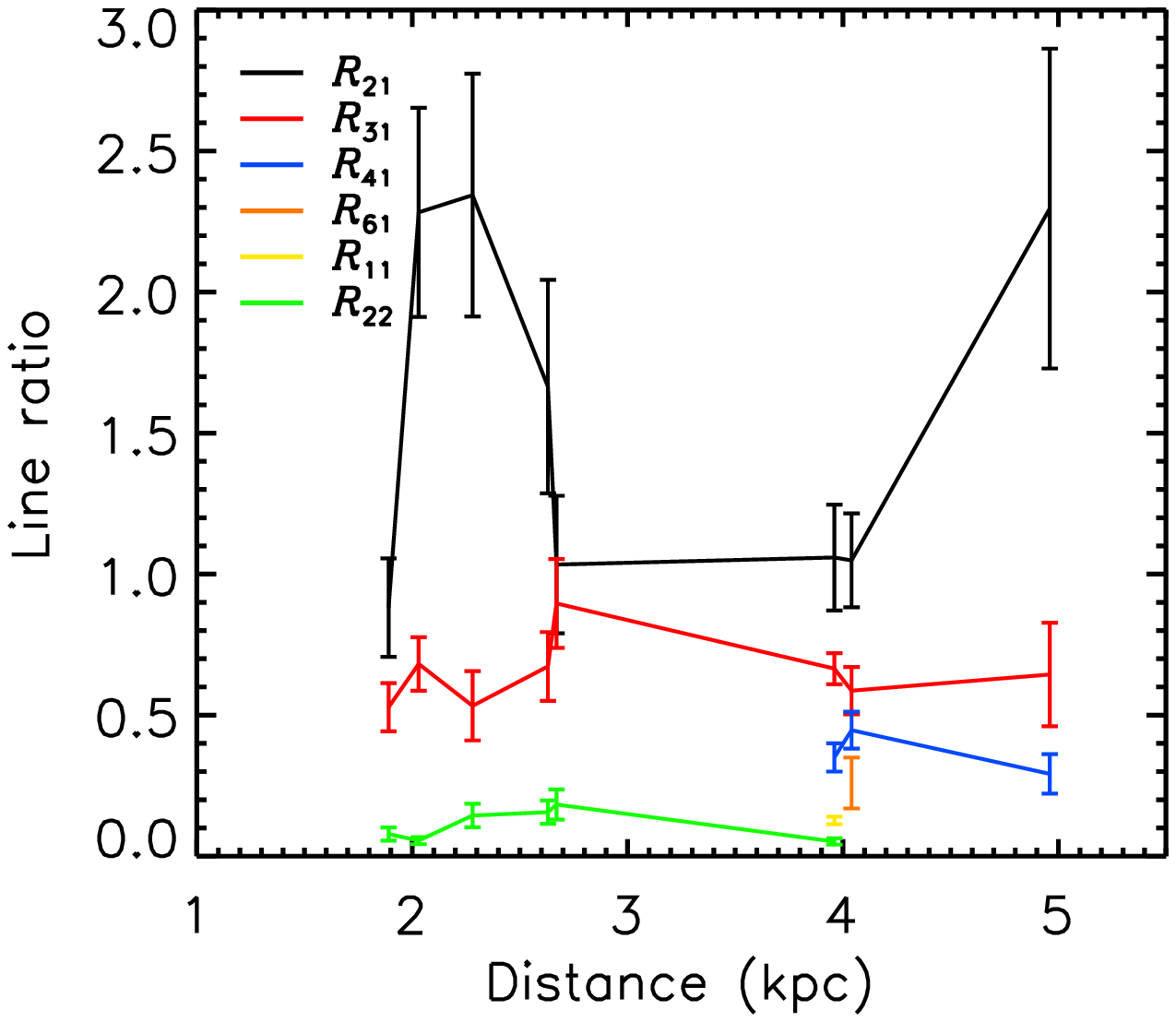}
  \caption{Beam-corrected line ratios as a function of radius in
    NGC~6946. The line ratios are color-coded and defined in the text
    (see Section~\ref{sec:empirical}).}
  \label{fig:ratio}
\end{figure}

Figure~\ref{fig:ratio} shows the beam-corrected velocity-integrated
line ratios (as listed in Table~\ref{tab:line}) as a function of
radius, from $\approx2$ (region~8) to $5$~kpc (region~7). The
well-sampled radial profiles of $R_{\rm 31}$, $R_{\rm 41}$ (the
CO(4-3)/CO(1-0)) line ratio) and $R_{\rm 22}$ (the
$^{13}$CO(2-1)/CO(2-1) line ratio) show little variation with radius
($R_{\rm 31}\approx0.7$, $R_{\rm 41}\approx0.4$ and $R_{\rm
  22}\approx0.1$). $R_{\rm 61}$ (the CO(6-5)/CO(1-0) line ratio) and
$R_{\rm 11}$ (the $^{13}$CO(1-0)/CO(1-0) line ratio) were measured in
two different regions only (see Fig.\ref{fig:ratio}). $R_{\rm 31}$ is
less than that at the centre of our own Galaxy, but higher than in the
Milky Way arms \citep{o07}.

The $R_{\rm 21}$ ratio does vary across the radial range probed
($R_{\rm 21}\approx1.0$--$2.5$), but it is always greater than
$1.0$. \citet{k12} studied the spatial variations of $R_{\rm 21}$ in
both the arms and inter-arm regions of M~51 (with a spatial resolution
comparable to ours), and found that $R_{\rm 21}>0.7$ in the spiral
arms (often $0.8$--$1.0$). They also showed that $R_{\rm 21}$
generally increases with SF activity (as traced by $f_{\rm 24\mu m}$
and $f_{\rm 24\mu m}/S_{\rm CO(1-0)}$). This is opposite to what we
observe in NGC~6946, since regions~4, 5 and 6 (and 7) have the highest
$R_{\rm 21}$ ratios but relatively low SF activity (see
Table~\ref{tab:line} and \S~\ref{sec:disc}). However, \citet{k12} do
note that high values of $R_{\rm 21}$ are also present with low
$f_{\rm 24\mu m}$, $S_{\rm CO(1-0)}$ and $f_{\rm 24\mu m}/S_{\rm
  CO(1-0)}$ values, at the upstream edges of the $24$~\micron\ spiral
arms (indicating dense gas that could potentially form massive OB
stars). As regions~4, 5 and 6 (and 7) in NGC~6946 are located in the
more flocculent spiral arms and inter-arm regions, the CO(1-0)
velocity-integrated line intensity is lower there compared to other
regions, and this could be the cause of the high $R_{\rm 21}$ ratios
observed. These regions may therefore have dense but not very warm
pre-SF gas (as suggested later by our models; see
Table~\ref{tab:results}).
%
% Figure: Beam-corrected line ratios as a function of radius.
%
\subsection{Modeling results}
\label{sec:modeling}
\subsubsection{Molecular ISM physical conditions}
The physical parameters estimated for each region are listed in
Table~\ref{tab:results}, from both the $\chi^{2}$ minimisation and
likelihood methods. Overall, across the $8$ regions considered, the
modeled beam-averaged kinetic temperature $T_{\rm K}$ ranges from $15$
to $250$~K (the whole range of models considered), while the molecular
hydrogen number volume density $n$(H$_2$) and the CO number colum
density $N$(CO) vary by a factor of $\approx100$ (although regions~4
and 7 possibly have more extreme values of $n$(H$_2$) and $N$(CO),
respectively).

In some regions, one of the best-fit model parameters lies at the edge of the model grid
(temperature in regions~1 and 2, H$_2$ number density in region~4 and CO column density in region~7; see Fig.~\ref{fig:chi2} and Table~\ref{tab:results}). However, as also shown in Figure~\ref{fig:chi2}, those $\chi^2$ minima exist within a much broader range of good models, and the likelihood method (see Fig.~\ref{fig:likelihood} and Table~\ref{tab:results}) yields reasonable ranges of probable parameters values. Those parameters are thus simply ill-constrained. We come back to this issue below.

The hottest and most tenuous molecular ISM is found in regions~1 and
2, with the highest numbers of H\,{\small II} regions and GMCs among
the regions studied (see Fig.~\ref{fig:regions}). Interestingly,
\citet{reb12} concluded that while the clouds located in the eastern
part of NGC~6946 have properties independent of the location in the
disc, some arm clouds are more massive, have a broader line width and
have a higher SFR. In particular, they found three such clouds located
within our regions~1 and 2 (regions~6, 7 and 11 in their work; see
also their Fig.~2). Regions~1 and 2 therefore appear to have
particularly extreme properties. Our model results also suggest a
fairly hot and tenuous ISM in region~3 (similar to regions~1 and 2),
with the highest number of SNRs (Fig.~\ref{fig:regions}).

The coldest and densest molecular ISM is located in regions~4, 5 and
6. Interestingly, the physical conditions in the inter-arm regions~7
and 8 are intermediate, although more similar to those of regions~4, 5
and 6 (if somewhat warmer) than those of regions~1, 2 and 3.

\begin{figure*}
  \includegraphics[width=5.7cm,clip=]{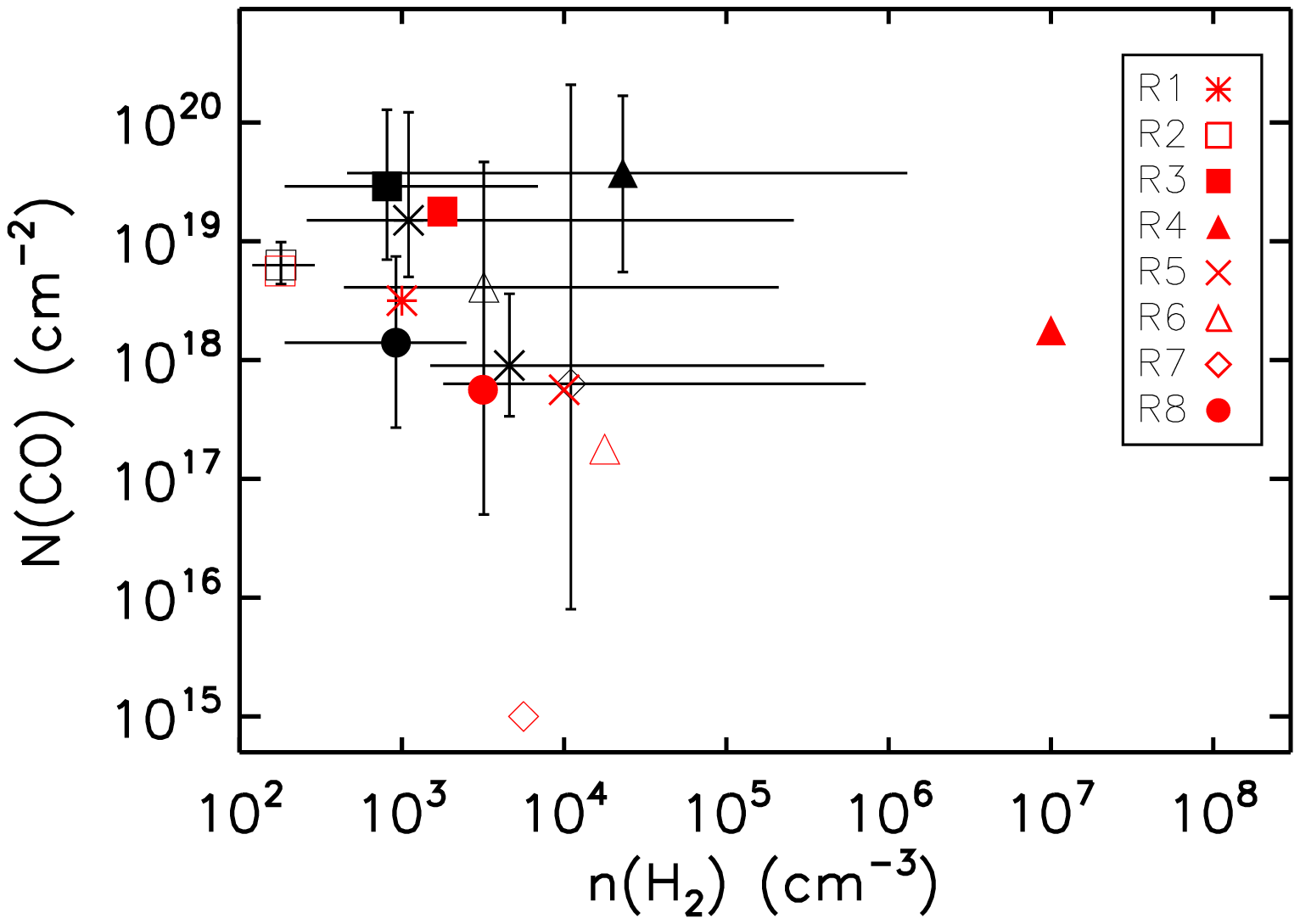}
  \includegraphics[width=5.7cm,clip=]{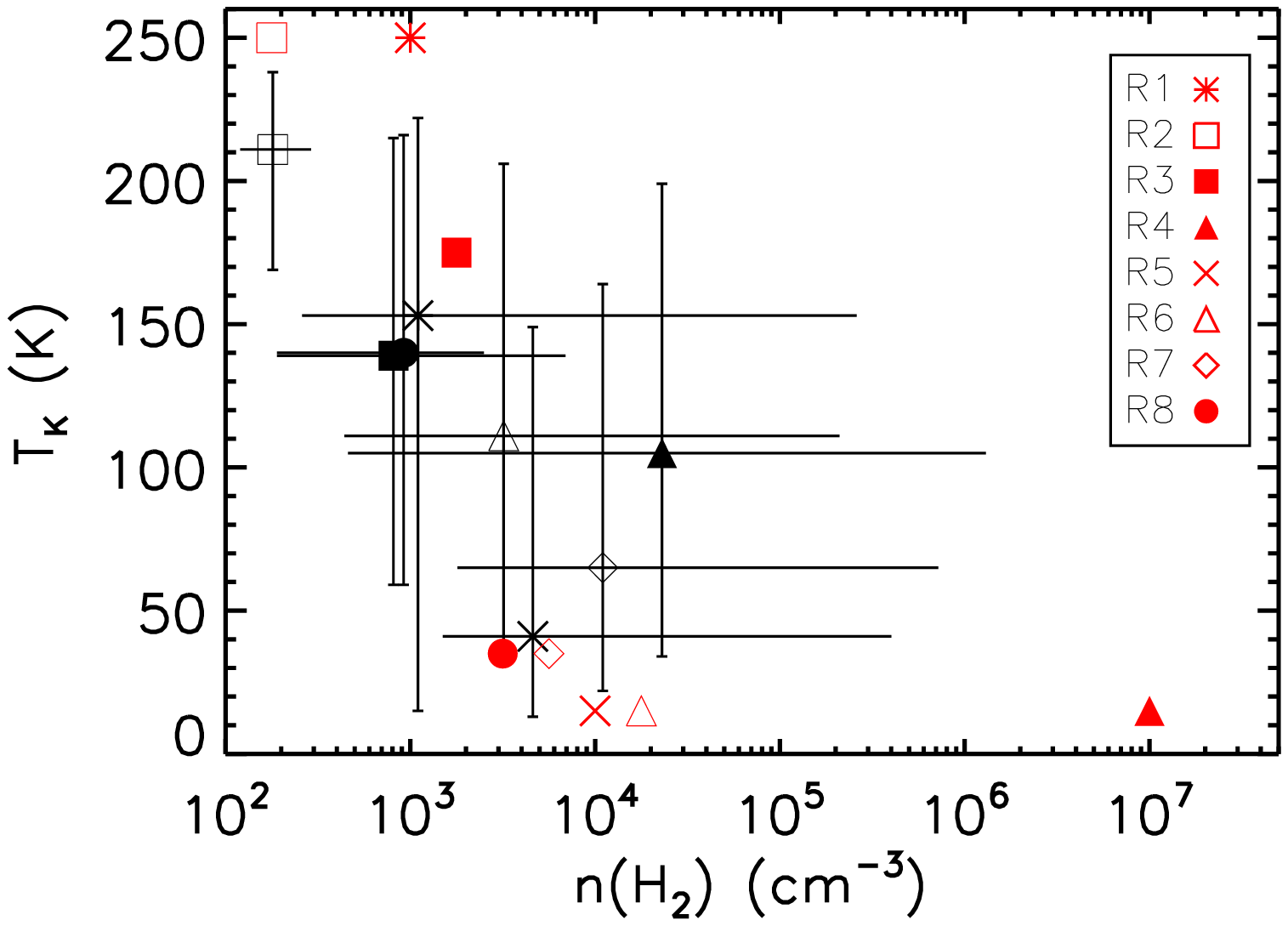}
  \includegraphics[width=5.7cm,clip=]{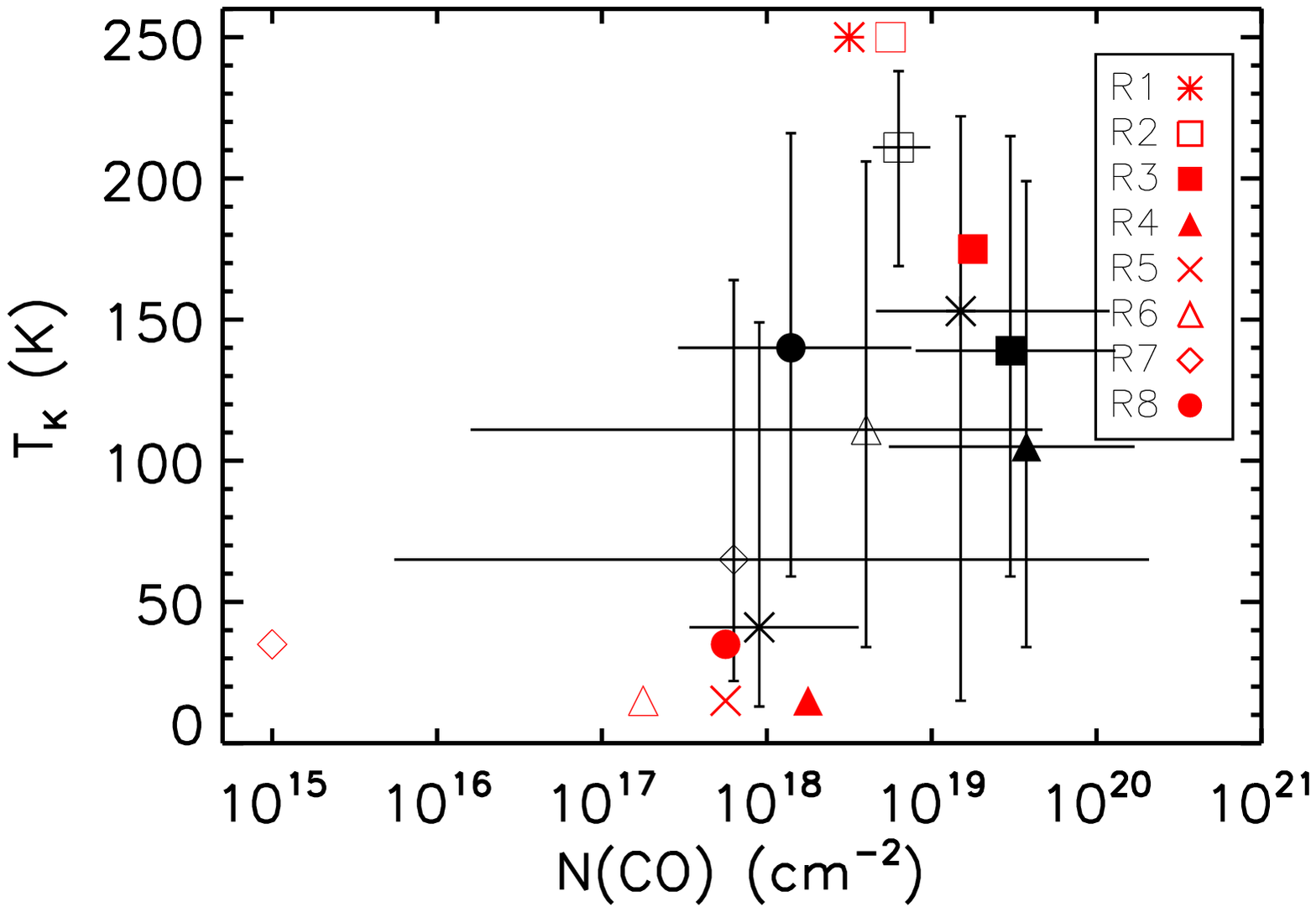}\\
  \includegraphics[width=5.7cm,clip=]{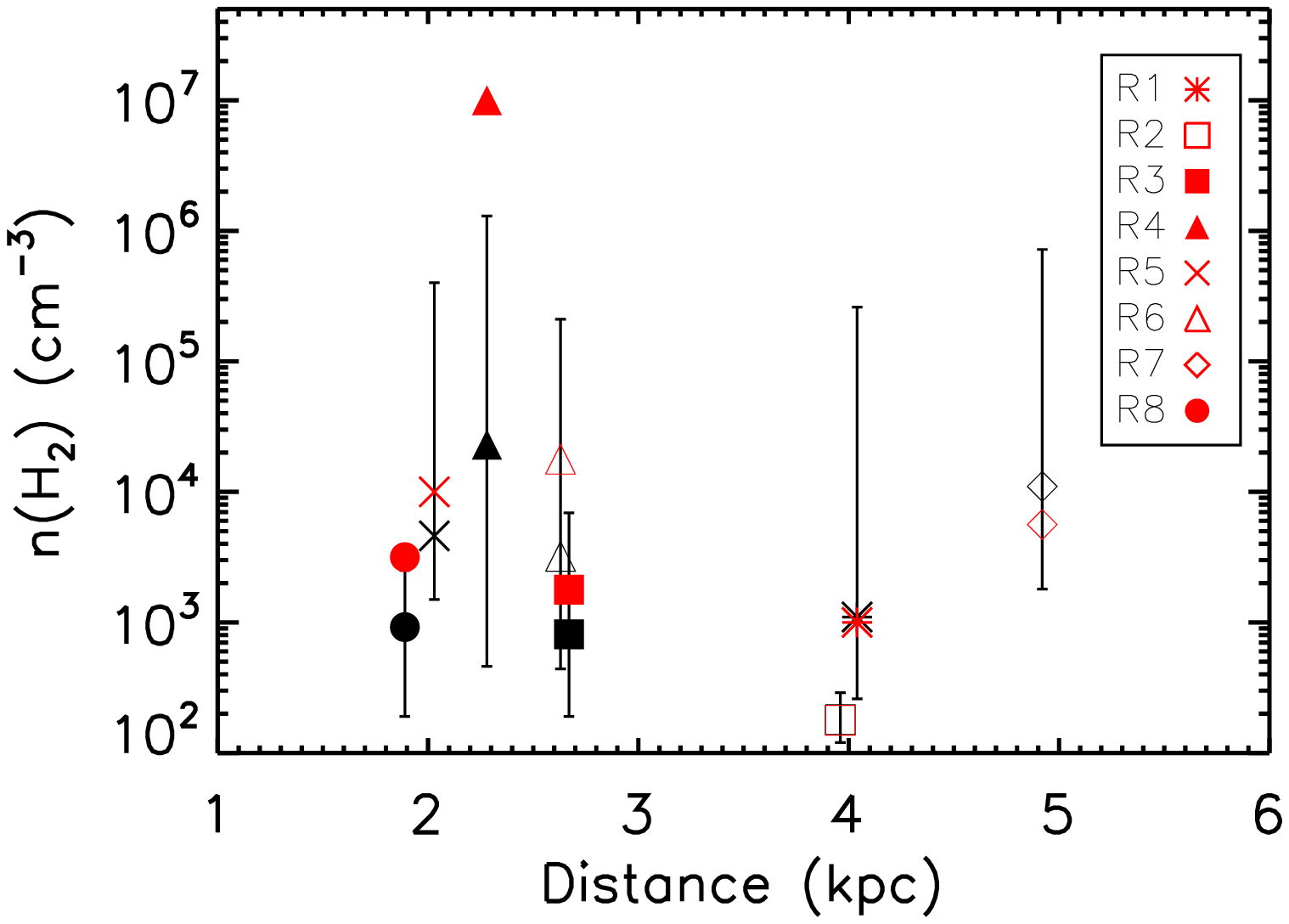}
  \includegraphics[width=5.7cm,clip=]{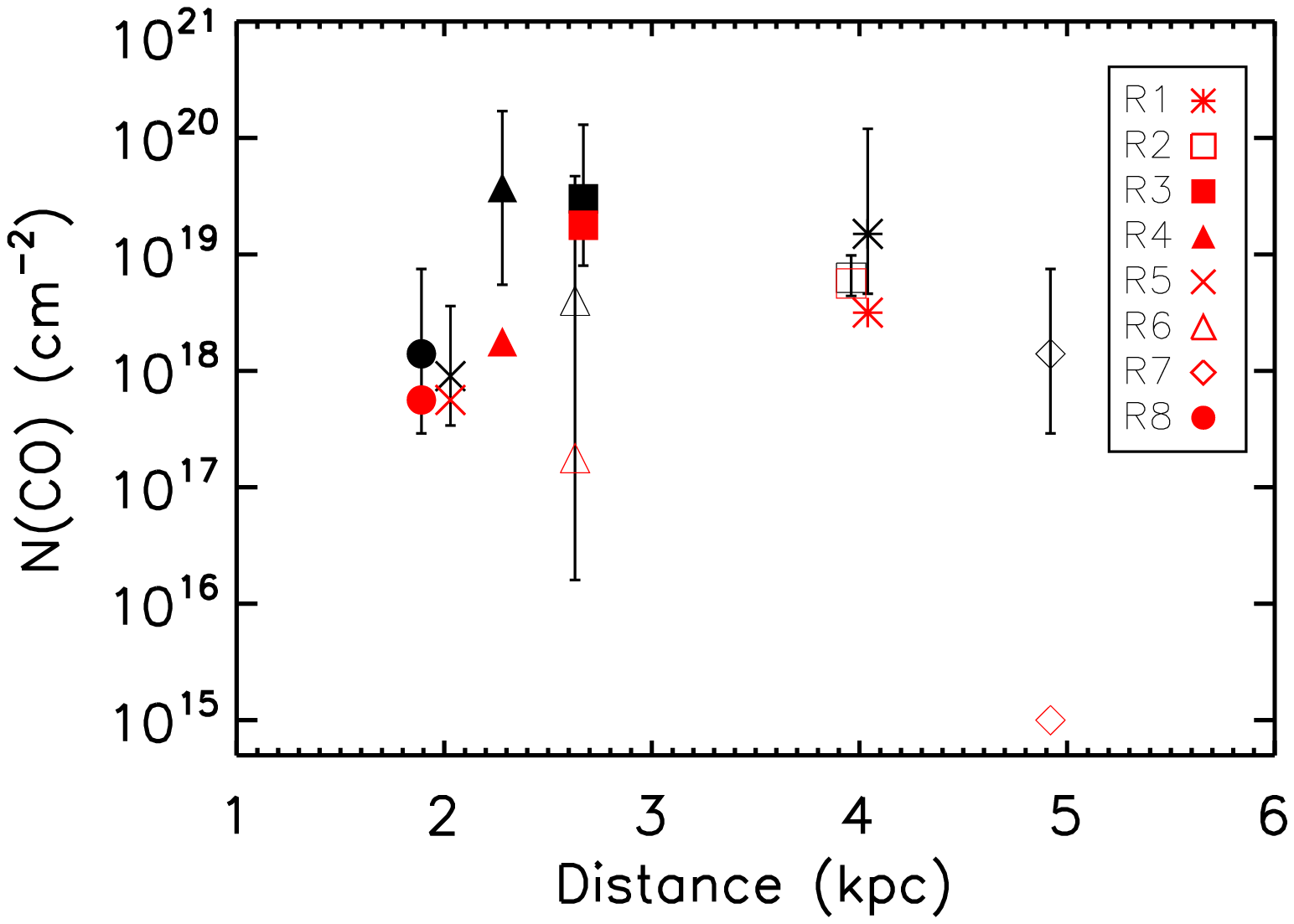}
  \includegraphics[width=5.7cm,clip=]{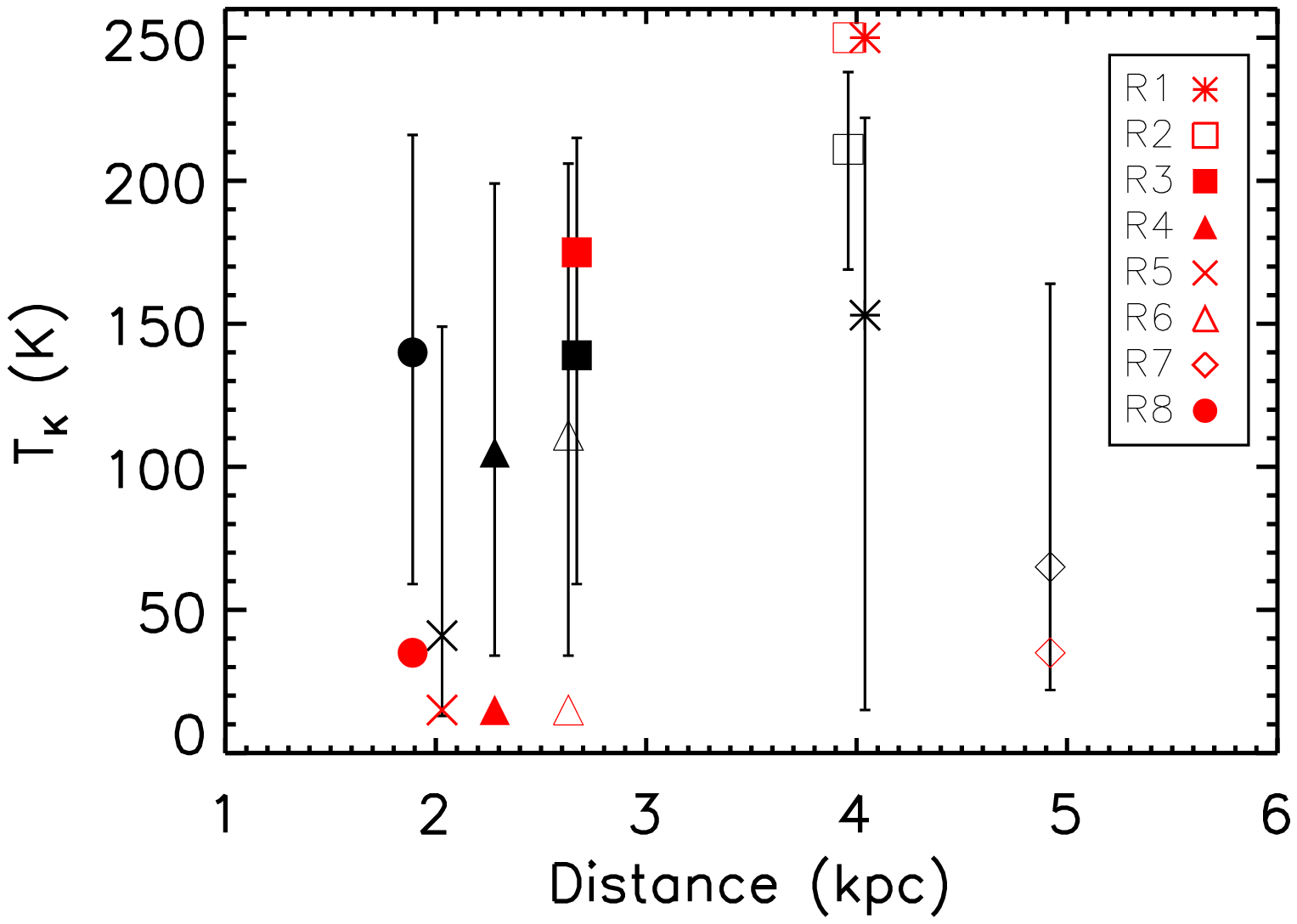}\\
  \includegraphics[width=5.7cm,clip=]{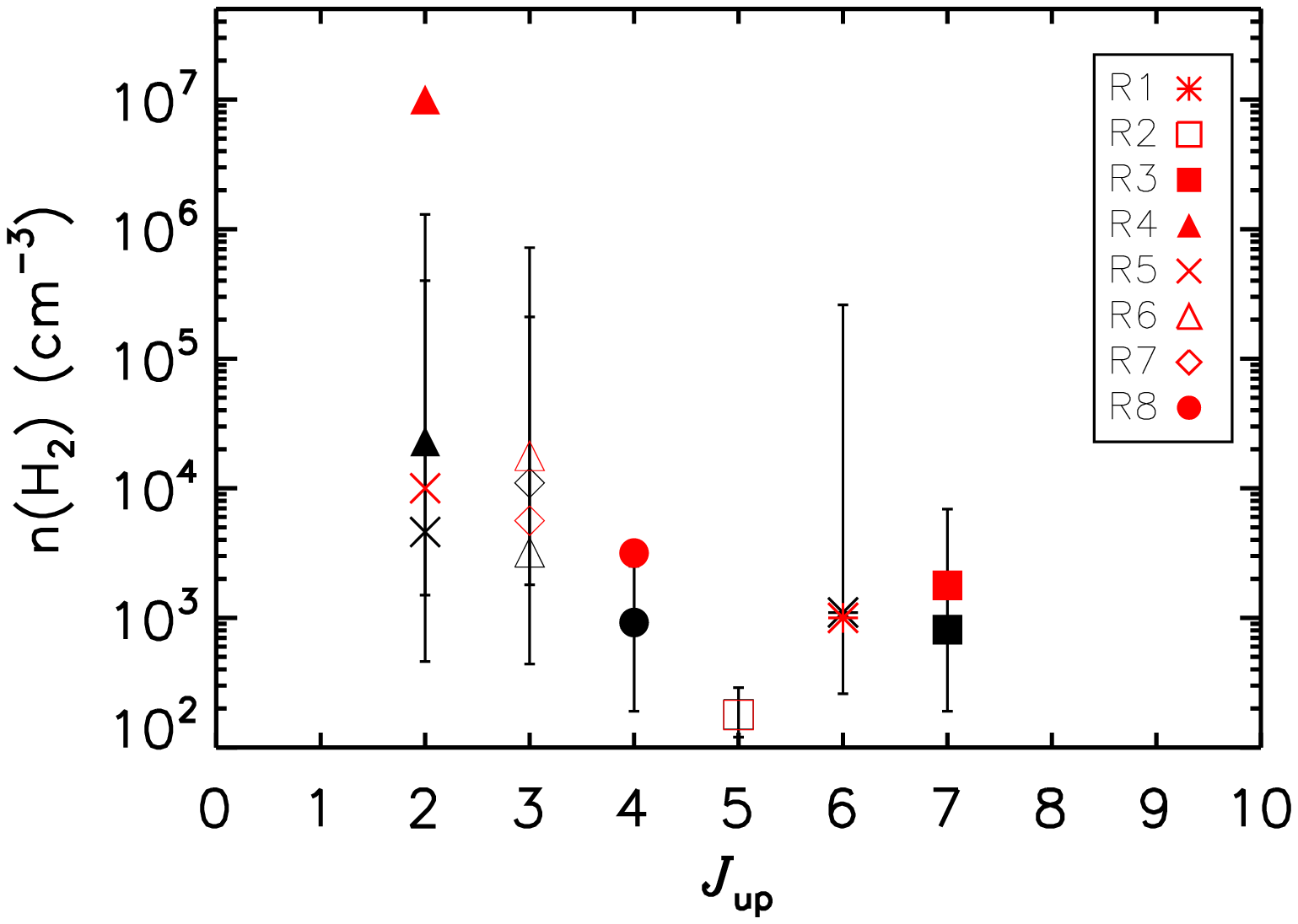}
  \includegraphics[width=5.7cm,clip=]{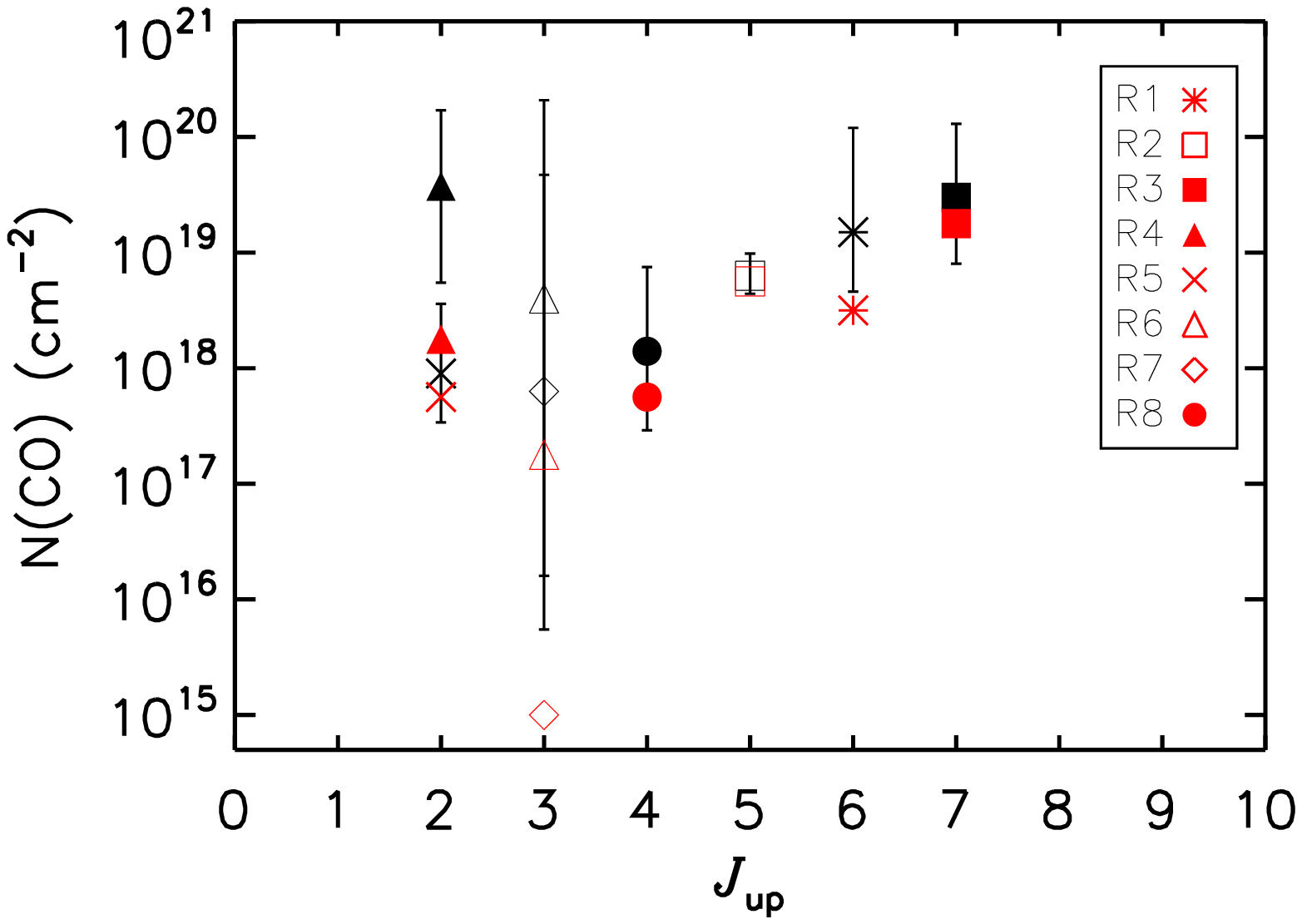}
  \includegraphics[width=5.7cm,clip=]{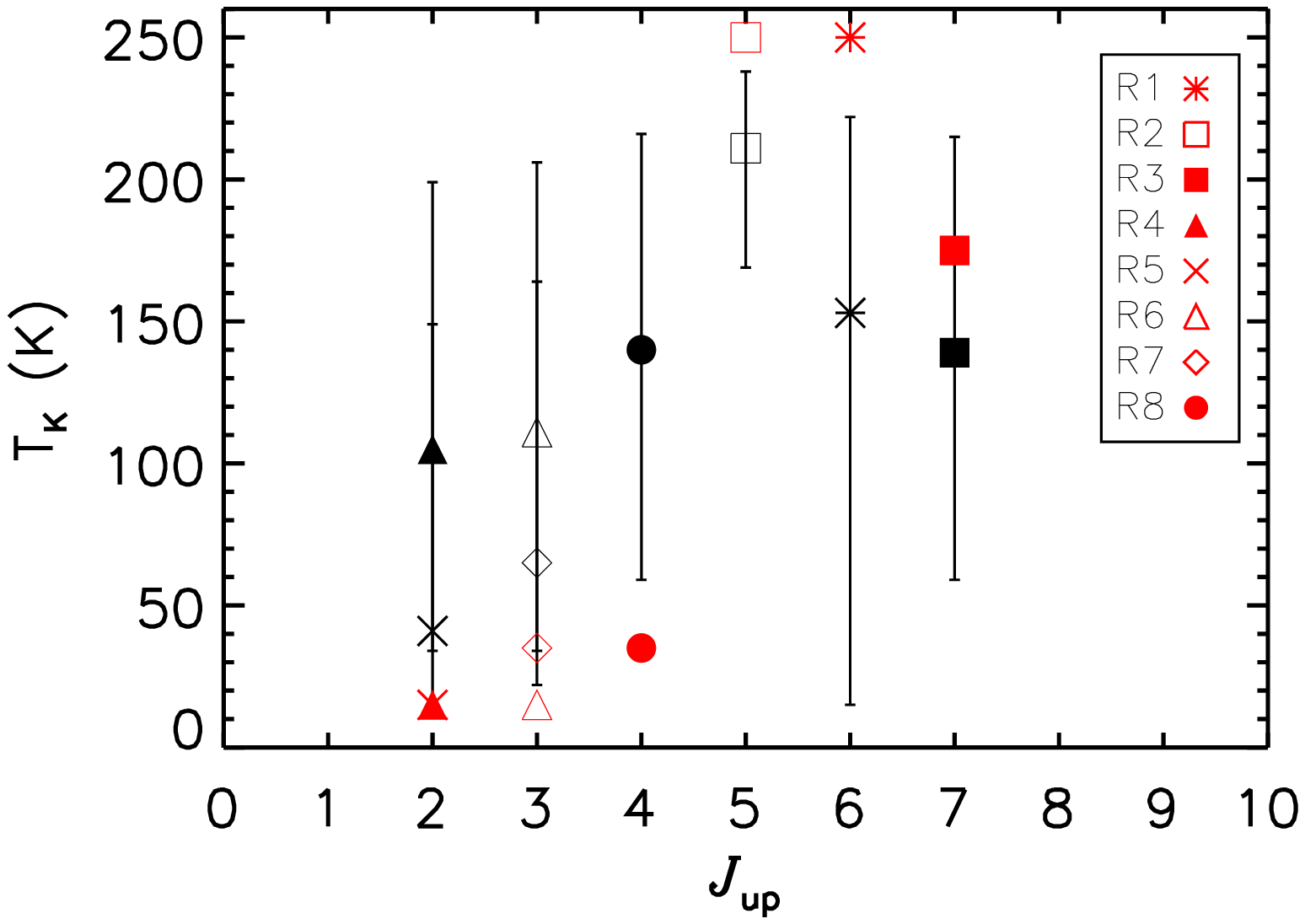}
  \caption{Top: Correlations between the best-fit parameters ($T_{\rm
      K}$, $n$(H$_2$) and $N$(CO)) of the $8$ regions studied. Middle:
    Same physical parameters as a function of galactocentric
    radius. Bottom: Same physical parameters as a function of the
    predicted $J_{\rm max}$. Red symbols show the results from the
    $\chi^2$ analysis. Black symbols with error bars represent the
    median of the marginalised probability distribution function of
    each model parameter, along with the $68$~per cent ($1\sigma$)
    confidence levels (see Table~\ref{tab:results}).}
  \label{fig:corr}
\end{figure*}

As shown in Figure~\ref{fig:corr}, the best-fit CO column density and
total volume density are not clearly correlated. If anything,
excluding region~4, there is a slight anti-correlation. A stronger
anti-correlation exists between $T_{\rm K}$ and $n$(H$_2$) (again
excluding region~4), implying a weak correlation between $N$(CO) and
$T_{\rm K}$. These trends suggest a slight decrease of the CO
abundance when the temperature decreases and the volume density
increases. Although speculative, this could be interpreted as the
effects of depletion processes and freeze-out onto grains during cloud
collapse. However, without confirmation from additional 
  molecules, such as N$_{2}$H$^{+}$ and/or methanol, known for tracing 
  well gas phases experiencing or having experienced freeze-out processes, 
  we only point out the reader a potential
  trend that regions~5, 6 and 8 may have more collapsing clouds/cores
   included in its observed beam than the other regions.

As each region is located at a different distance from the galactic
centre, we can probe radial variations of the three main derived
physical parameters in NGC~6946 (Fig.~\ref{fig:corr}). As might
naively be expected, the total volume density $n$(H$_2$) tends to
decrease with radius (excluding region~4). However, somewhat
counter-intuitively (but expected from the aforementioned
correlations), $N$(CO) and $T_{\rm K}$ tend to increase with radius
(excluding region~7).

Having said that, since we only probed particular star-forming regions
in the spiral arms and inter-arms of NGC~6946 where observations of
multiple lines of CO are available, the aforementioned relationships
among the molecular gas physical parameters can not be generalised to
the entire disc of the galaxy, even less to spirals in the nearby
universe generally. In fact, excluding regions~1, 2 and 3 (with the
highest SF activity and highest $T_{\rm K}$ and $N$(CO)) eradicates
all the correlations (see Fig.~\ref{fig:corr}). The relationships
among the physical parameters presented in Figure~\ref{fig:corr} could
therefore be biased by selection effects. Only a homogenous survey of
multiple molecular lines across the entire disc of NGC~6946 could
provide demonstratively unbiased results.
%
% Figure: Best-fit parameter correlations.
%

As hinted above, it is interesting to compare the probability distribution functions of the
three main physical parameters (marginalised over the other two) with
the simple $\chi^2$ minimisation results, as done in
Figure~\ref{fig:likelihood}. $N$(CO) is generally the best constrained
parameter, with single-peaked Gaussian PDFs and thus good agreement
between the peak and median of the likelihood. Similarly, the best-fit
model (in a $\chi^2$ sense) is generally contained within the $68$~per
cent ($1\sigma$) confidence levels (it is in all but one region at the
$2\sigma$ level). The situation is less satisfactory for $n$(H$_2$)
and particularly for $T_{\rm K}$, due to their degeneracy (i.e.\ the
characteristic ``banana''-shaped $\Delta\chi^2$ countours or inverse
$n$(H$_2$)--$T_{\rm K}$ relationship; see Fig.~\ref{fig:chi2} and
Appendix~C of \citealt{vbsjv07}). The PDFs for $n$(H$_2$) and $T_{\rm
  K}$ typically have a small peak within a rather broad and flat
likelihood distribution. Unsurprisingly, the best-fit model is thus
again generally contained within the $68$~per cent confidence levels
(except in a few cases where the best-fit model is at the egde of the
grid and thus unreliable). However, this also highlights the fact
that, although generally well-defined, the best-fit model is often
only slightly more likely (i.e.\ better fitting) than a host of other
models spanning a wide range of physical parameters. This questions
the wisdom of considering the best-fit model alone as a reasonable
approximation of the true physical conditions, a practice adopted in
many studies. The confidence intervals listed in
Table~\ref{tab:results} are a first step at concisely quantifying the
uncertainties.

\begin{figure}
  \includegraphics[width=8.5cm,clip=]{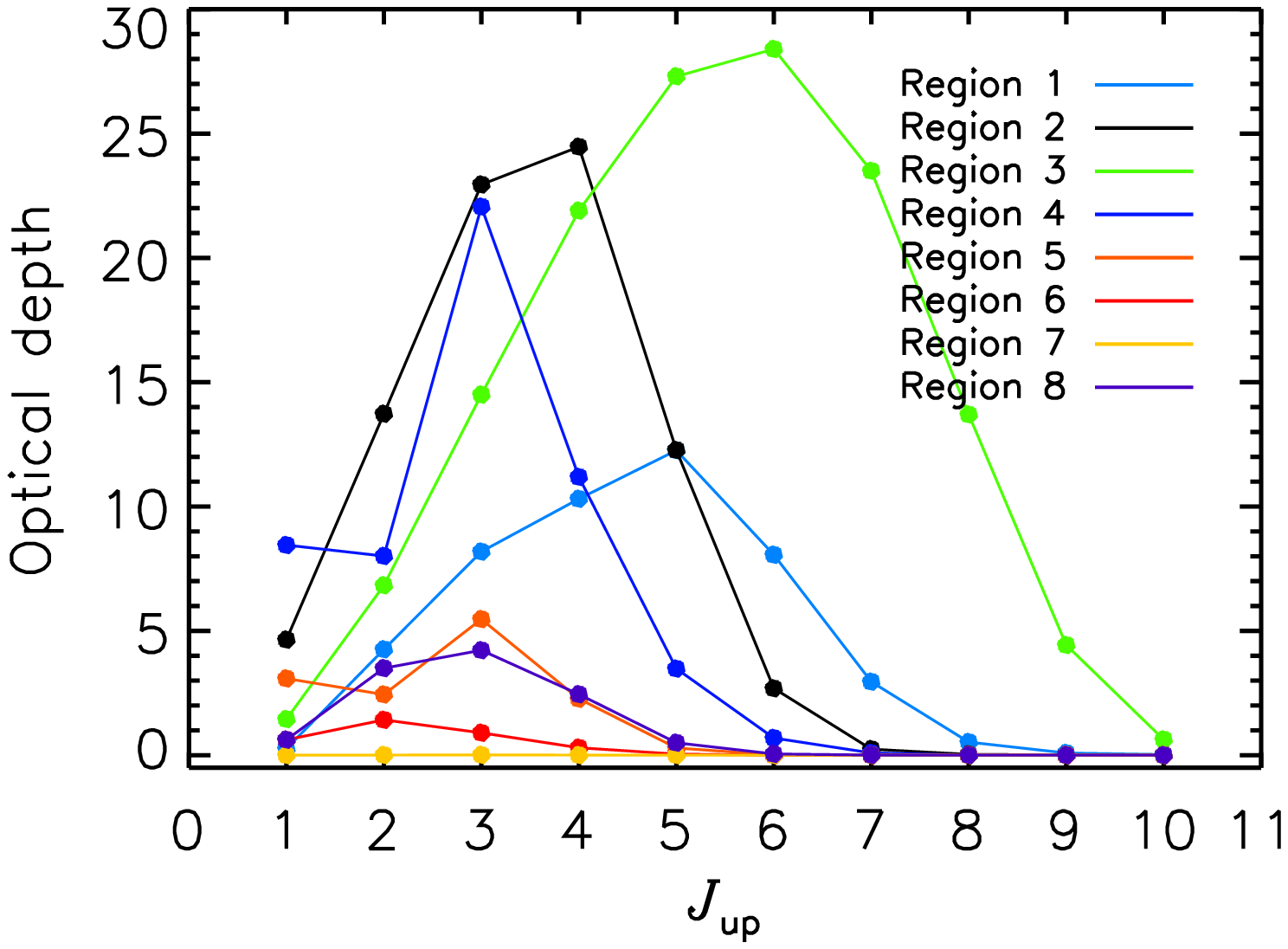}
  \caption{Optical depth distributions. For each region
    (colour-coded), the optical depth of the CO line for the best-fit
    model is shown as a function of the upper $J$ level of the
    transition (up to $J$=10-9).}
  \label{fig:tau}
\end{figure}

In Fig.~\ref{fig:tau}, we show the modeled optical depth $\tau$ of the
CO lines as a function of the upper $J$ level of the transition
$J_{\rm up}$, this for the best-fit model of each region. All regions
except region~7 (which also has the lowest best-fit CO number column
density) have $\tau>1$ for at least one transition, indicating that we
only see the outermost layers of the clouds in those transitions. The
optical depth distributions are qualitatively similar to the
beam-averaged total intensity distributions, in the sense that both
peak at roughly the same transitions. Indeed the optical depth is
maximum at $J_{\rm up}=5$--$7$ in regions~1, 2 and 3, while it peaks
at $J_{\rm up}=2$--$4$ in the other regions. A similar trend has also
been observed for high density tracers by \citet{gl99}.
%
% Figure: Optical depths.
%

The beam-averaged total CO cooling rate of each region was also
estimated from the best-fit model, summing all CO transition
intensities up to $J_{\rm up}=10$. The results are listed in
Table~\ref{tab:results}. As expected, regions~1, 2 and 3 have CO
cooling rates $\approx10$ times higher than those of the other
regions. They are the most powerful emitting regions, encompassing the
highest number of H\,{\small II} regions and GMCs (regions~1 and 2)
and the highest number of SNRs (region~3), all indicating young
stellar populations (see Fig.~\ref{fig:regions}). The other regions,
where the ISM is colder and denser, have lower CO cooling rates.
\subsubsection{Spectral line energy distributions}
\label{sec:sled}
We plot the CO spectral line energy distribution (SLED) of the
best-fit model of each region in Figure~\ref{fig:sleds}, overlaid with
the observed data points. Also shown are the SLED ranges allowed by
the $1\sigma$ confidence level on the best-fit model, as well as the
SLED of the most likely model. By construction, the models are in good
agreement with the data. The modeled integrated line intensities
generally decrease with increasing upper energy level $J_{\rm up}$,
but the beam-averaged total intensities peak at intermediate
transitions.

The location of the SLED turnover ($J_{\rm max}$) is interesting, as
it provides information on the SF activity level, and it is listed in
Table~\ref{tab:results} for each region. In the heavily star-forming
galaxy APM~08279+5255 at redshift $z\approx4$, $J_{\rm max}=9$
\citep{w07}. In the centre of local starbursting galaxies such as
Henize~$2$-$10$, NGC~253 and M~82, $J_{\rm max}\approx6$--$7$
\citep{b04, w05}, while it is $\approx3$ in molecular
streamers/outflows around M~82's central molecular disc \citep{w05}. $J_{\rm
  max}\approx4$ in the centre of our own Galaxy, whereas it is
$\approx3$ in the Galactic plane at longitudes $2\fdg5<l<32\fdg5$,
where CO emission is much weaker than in the central regions
\citep{fbm99}. On the other hand, the Orion Bar photodissociation
region in the Milky Way disc shows extreme star formation activity,
with $J_{\rm max}\approx13$ \citep{h10}.

In NGC~6946, the predicted turnover takes place at $J_{\rm
  max}=5$--$7$ in regions~1, 2 and 3 (hot and tenuous ISM), while
$J_{\rm max}=2$--$4$ in the other regions (colder and denser
ISM). From the above discussion, this suggests more intense SF
activity in regions~1, 2 and 3 compared to the other regions. In fact,
SF in regions~1, 2 and 3 appears more extreme than that in the Galaxy
centre.

As visible in Figure~\ref{fig:sleds}, an inflection point is present
at $J_{\rm up}=3$ in the SLED of region~4, where the CO(3-2)
integrated line intensity is slightly smaller than that of both
CO(2-1) and CO(4-3). Although we refer the reader to \citet{vbsjv07}
for more details on the relation between opacity and line intensity,
we note that the opacity in region~4 peaks at $J_{\rm up}=3$ and is
more than twice that for $J_{\rm up}=2$ and $4$ (see
Fig.~\ref{fig:tau}), causing the apparent reduction of the CO(3-2)
integrated line intensity. Since the low-$J$ trend is more consistent
with the observed lines and the best-fit physical parameters (see
Fig.\ref{fig:sleds}), we adopt $J_{\rm max}=2$ for region~4.

Perhaps unsurprisingly, the strongest correlation of $J_{\rm max}$
with the molecular ISM physical conditions is that with the kinetic
temperature $T_{\rm K}$ (see Fig.~\ref{fig:corr}). Indeed, $J_{\rm
  max}$ is maximum for regions~1, 2 and 3, with the highest $T_{\rm
  K}$, while it is lowest for regions~4, 5 and 6, with the lowest
$T_{\rm K}$ (regions~7 and 8 are, again, intermediate). As seen in
Figure~\ref{fig:corr} and expected from the aforementioned
correlations among the physical parameters, $N$(CO) also shows a
correlation with $J_{\rm max}$ but it is relatively weak (and
region~7 is an outlier), whereas $n$(H$_2$) shows an anti-correlation
(and region~4 is an outlier). It therefore seems that in the regions
where the turnover takes place at high $J_{\rm max}$ ($>4$), the gas
tends to be hot (high $T_{\rm K}$) and diffuse (low $n$(H$_2$)) with a
high CO column number density, while in regions where the turnover
takes place at low $J_{\rm max}$ ($\le4$), the gas is cold and dense
with a relatively low CO column number density.
\begin{figure*}
  \includegraphics[height=5.0cm,clip=]{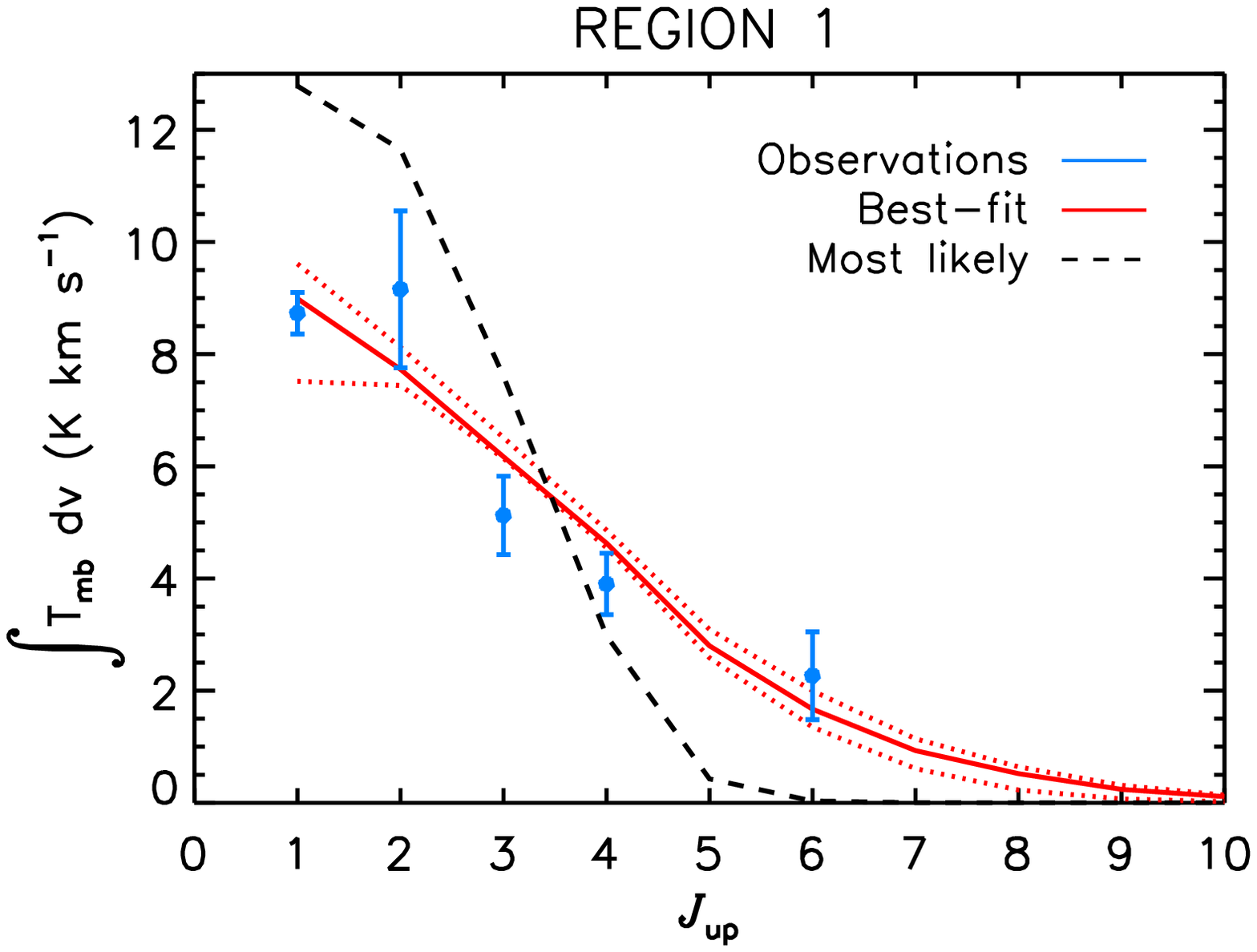}
  \includegraphics[height=5.0cm,clip=]{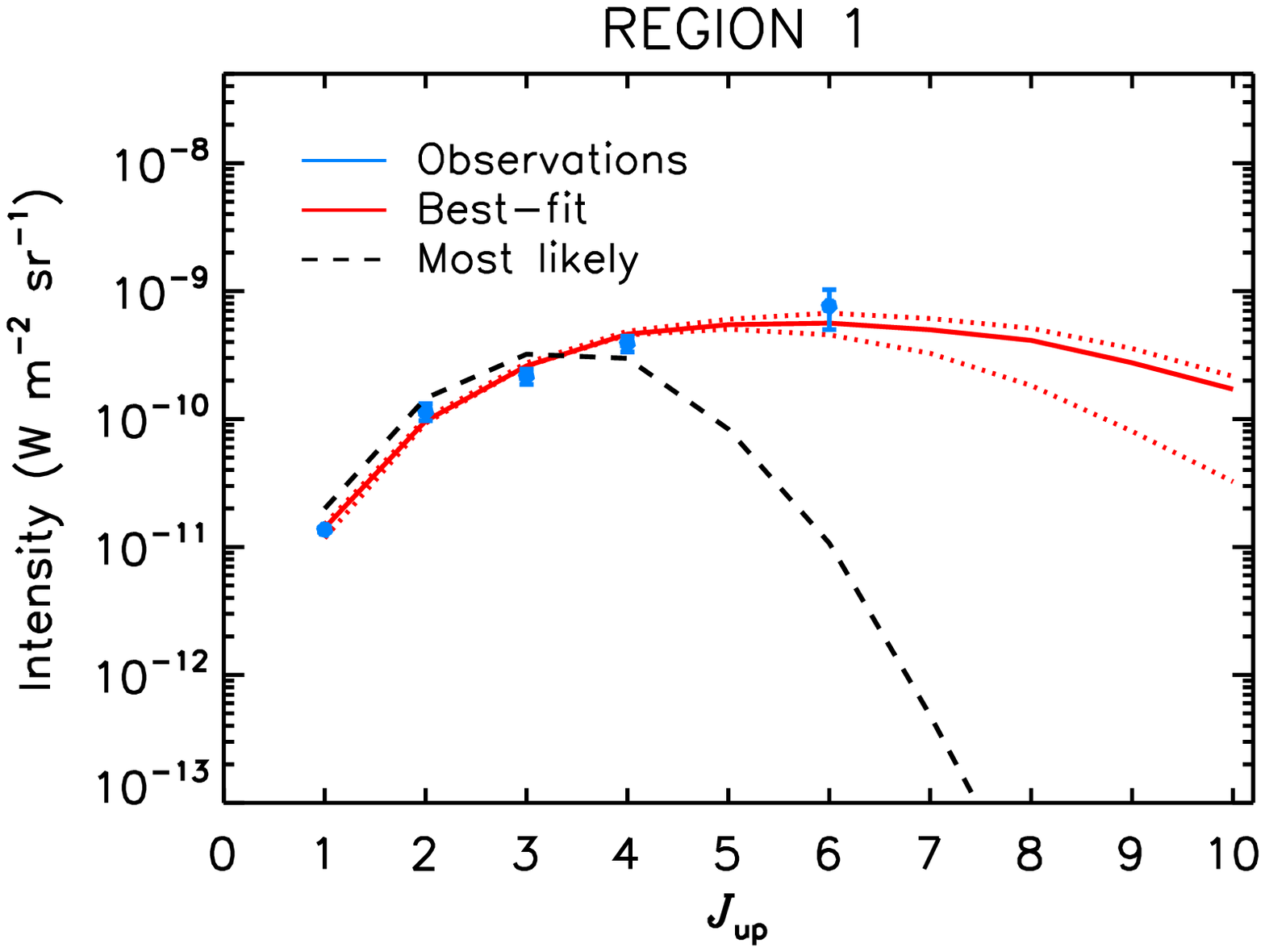}\\
  \includegraphics[height=5.0cm,clip=]{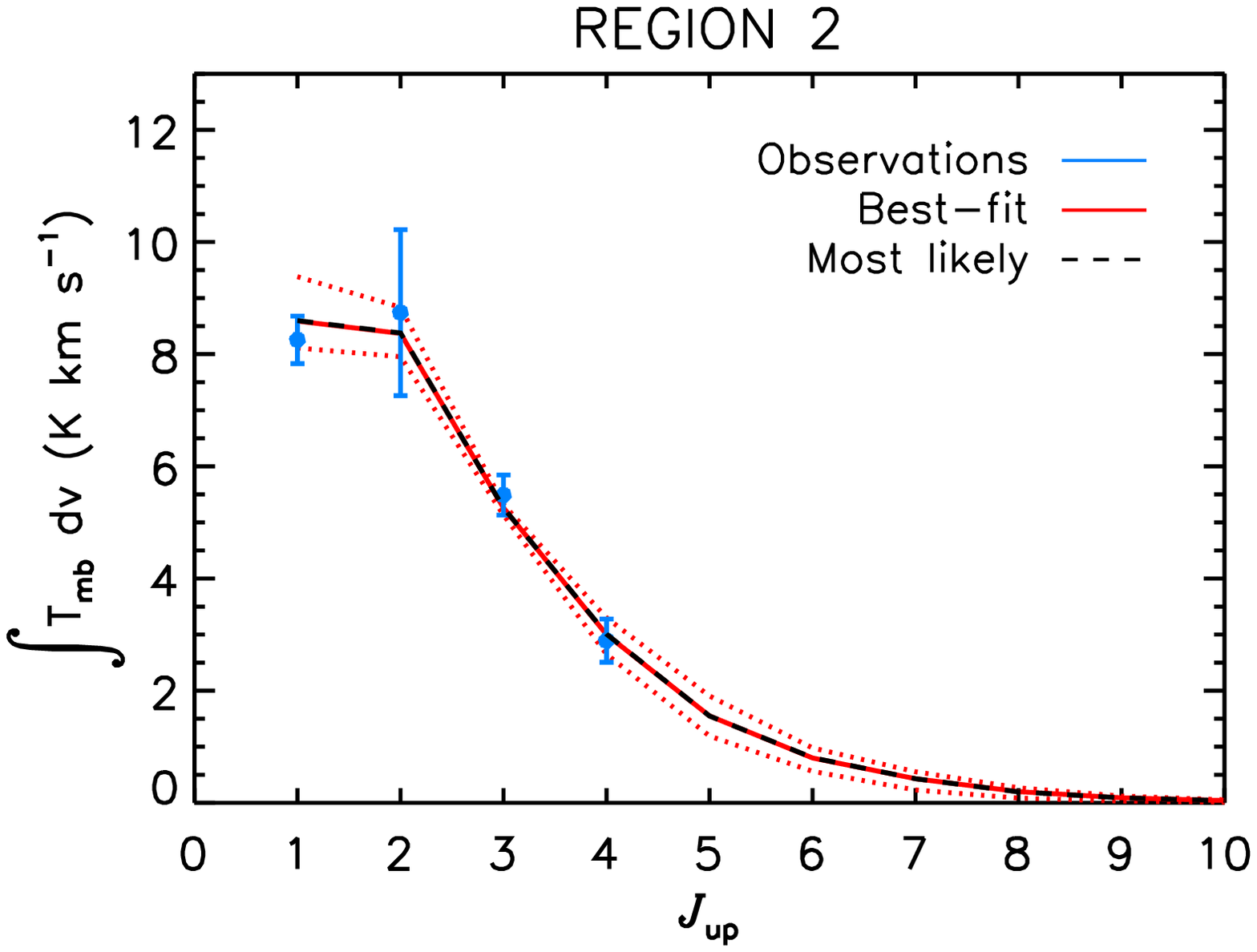}
  \includegraphics[height=5.0cm,clip=]{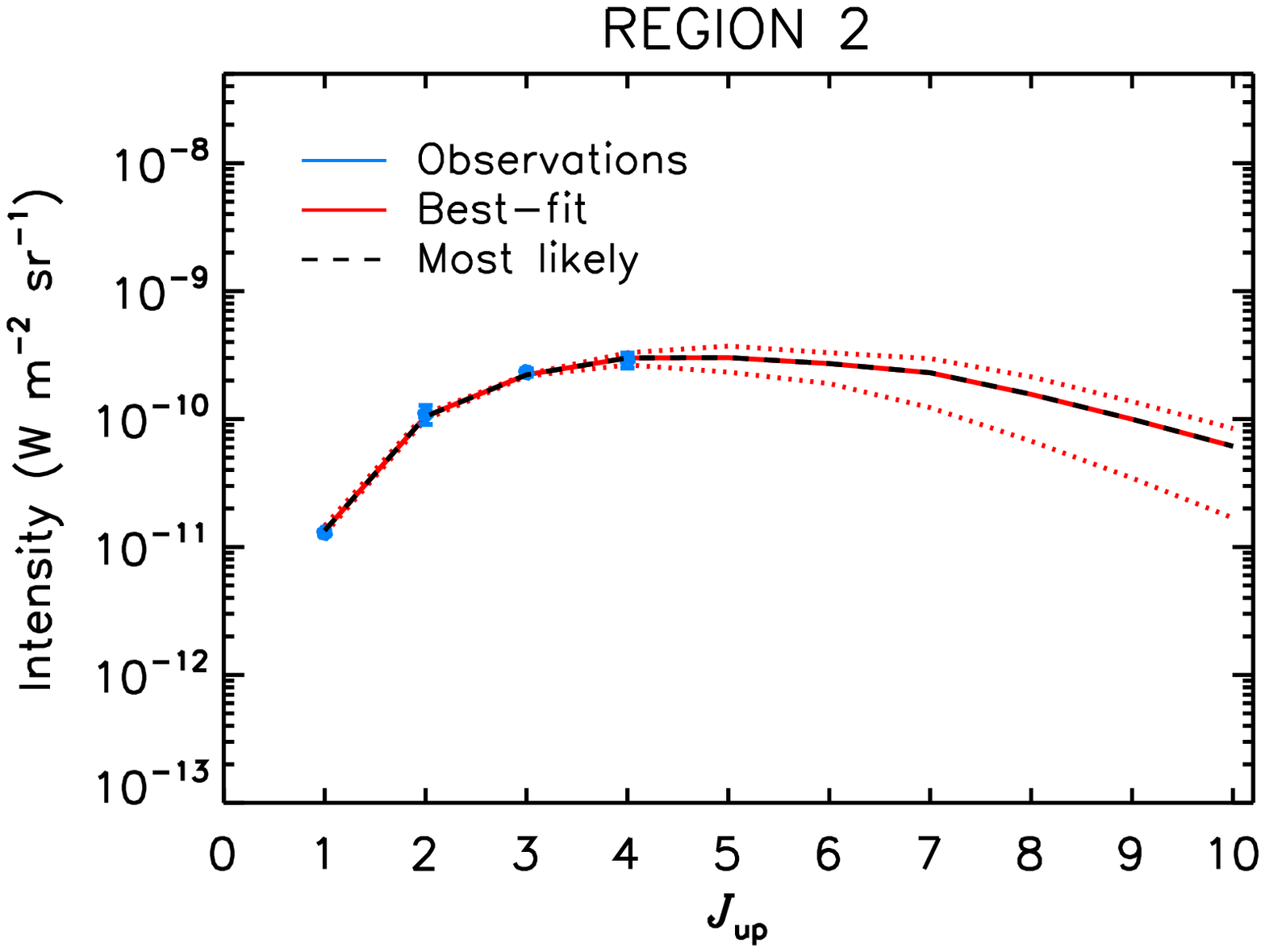}\\
  \includegraphics[height=5.0cm,clip=]{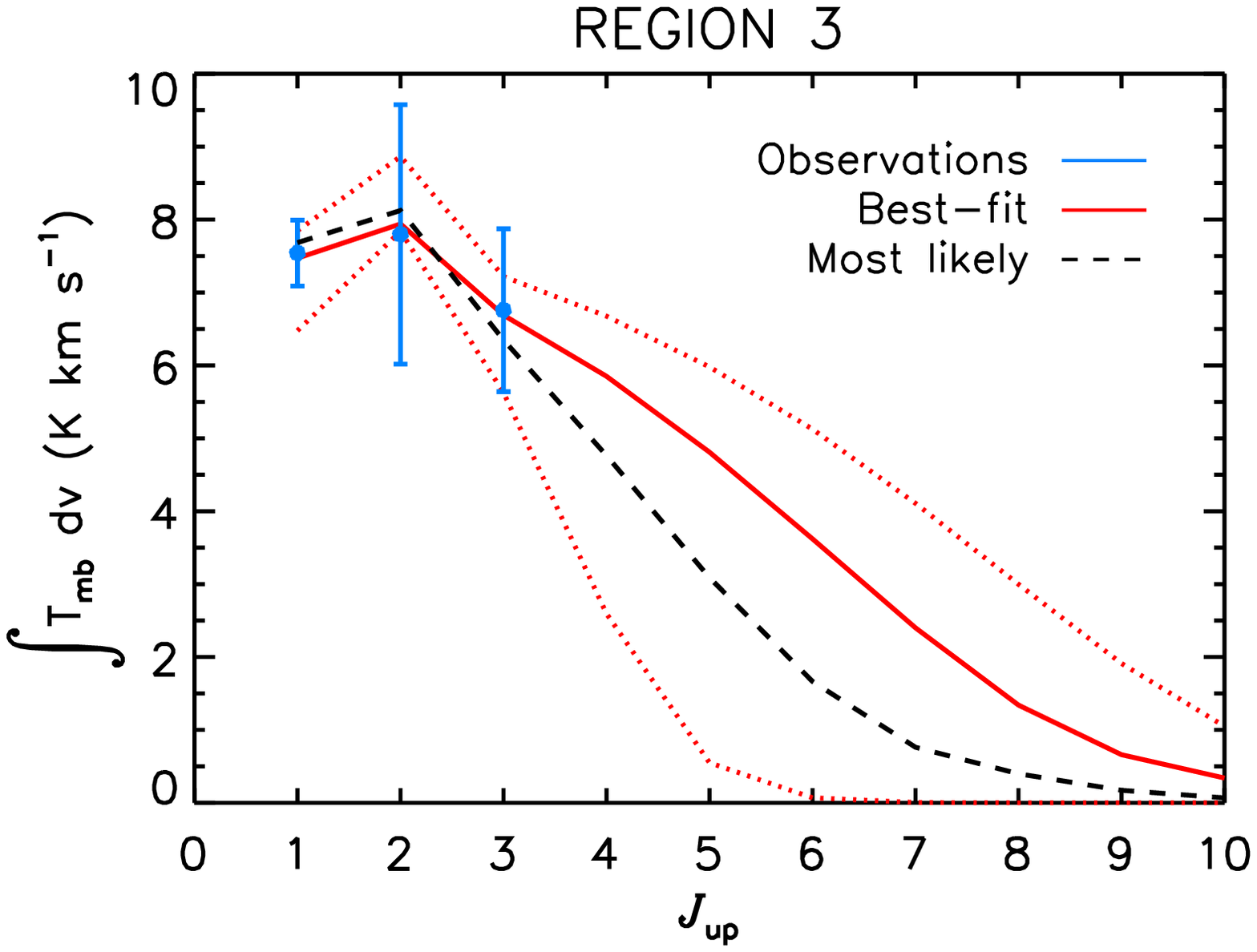}
  \includegraphics[height=5.0cm,clip=]{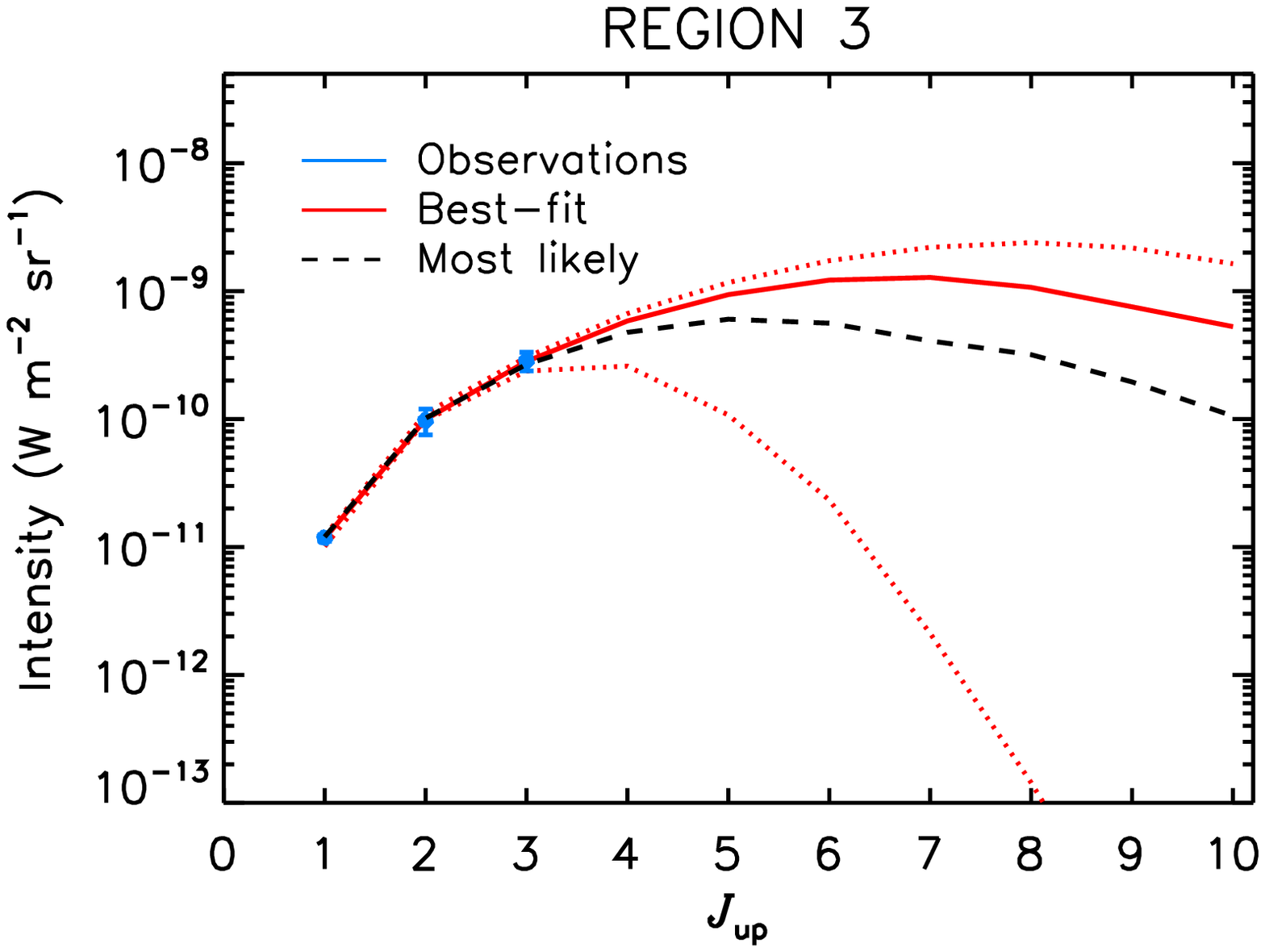}\\
  \includegraphics[height=5.0cm,clip=]{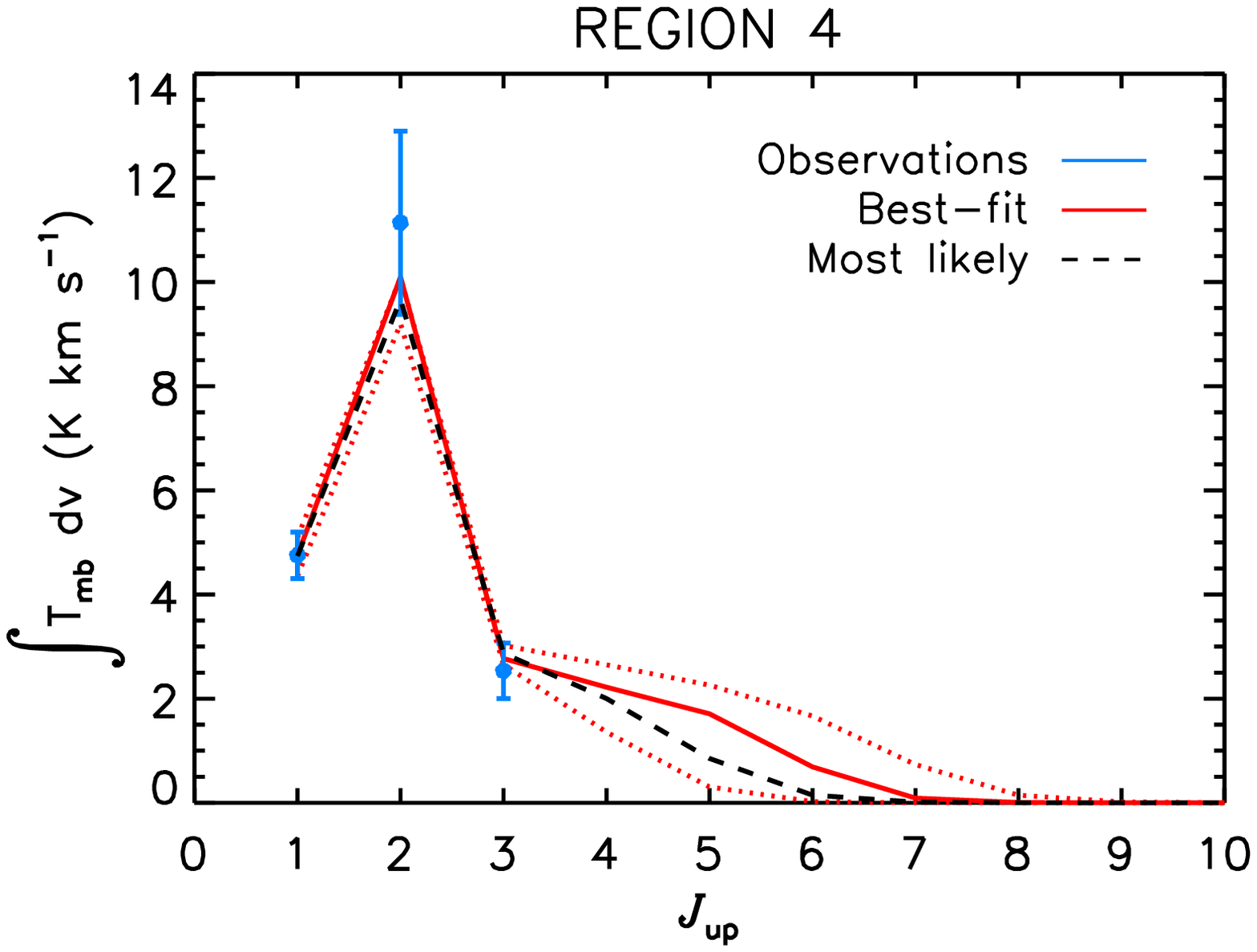}
  \includegraphics[height=5.0cm,clip=]{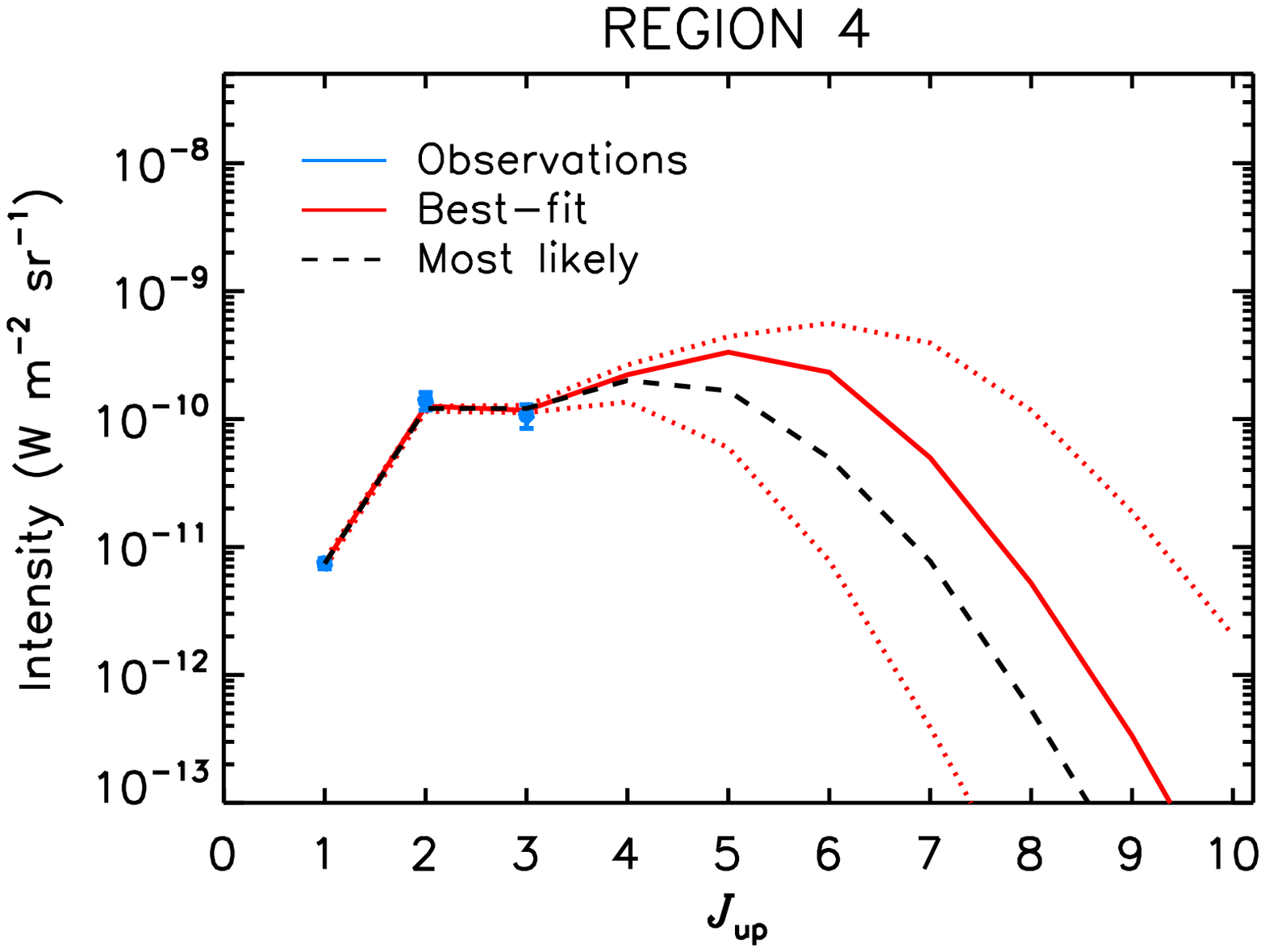}
  \caption{Spectral line energy distributions. For each region, the
    integrated line intensity $S$ (left) and beam-averaged total
    intensity per unit area $I$ (right) are shown as a function of the
    upper $J$ level of the transition (up to $J$=10-9). The solid red
    line shows the best-fit model, while the dotted red lines
    delineate the range of possible SLEDs encompassed by the $1\sigma$
    confidence level on the best-fit model (the darkest zone in the
    $\Delta\chi^2$ contour maps of Fig.~\ref{fig:chi2}). The dashed
    black line shows the SLED corresponding to the most likely
    model. Blue circles with error bars are our observations.}
  \label{fig:sleds}
\end{figure*}
\addtocounter{figure}{-1}
\begin{figure*}
  \includegraphics[height=5.0cm,clip=]{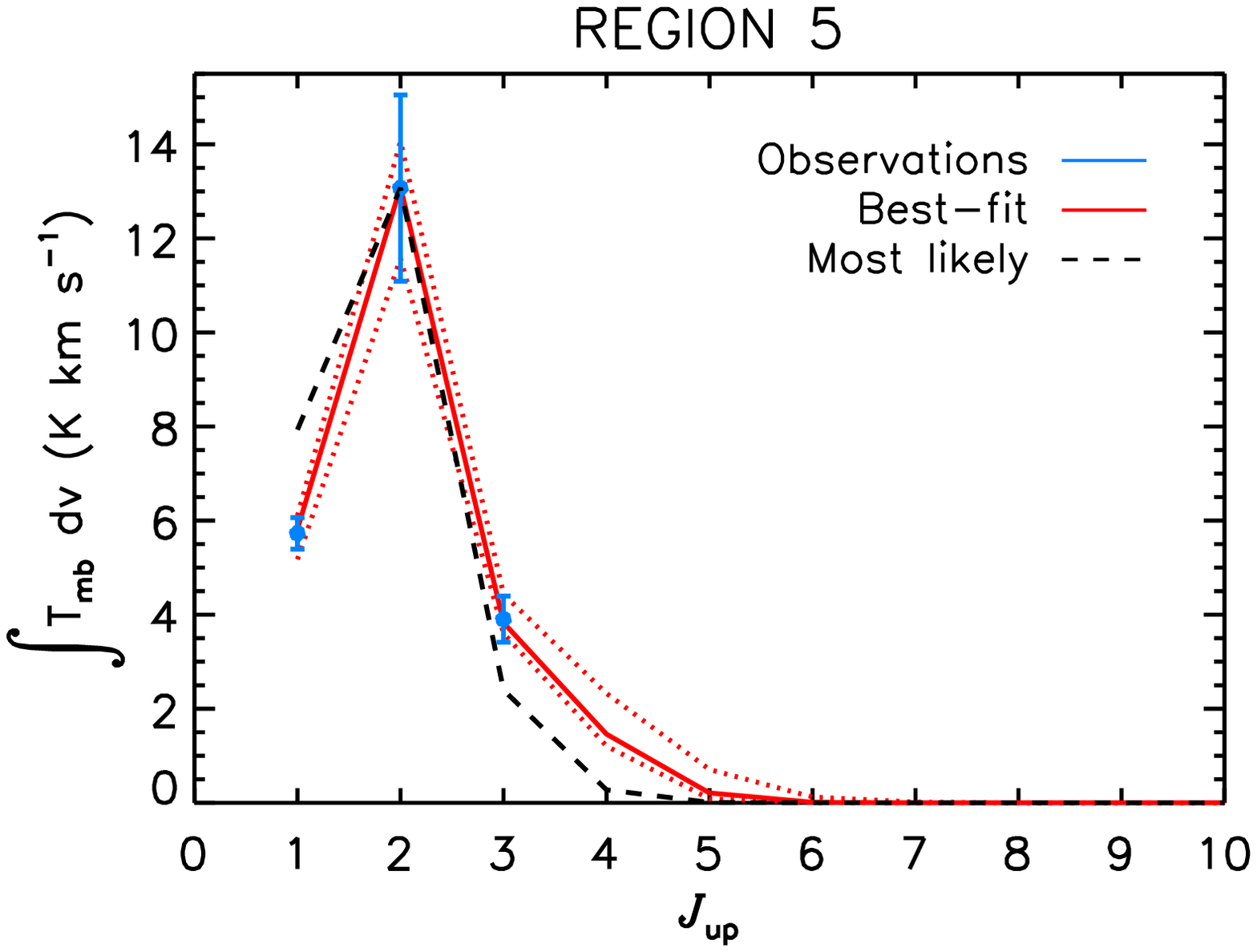}
  \includegraphics[height=5.0cm,clip=]{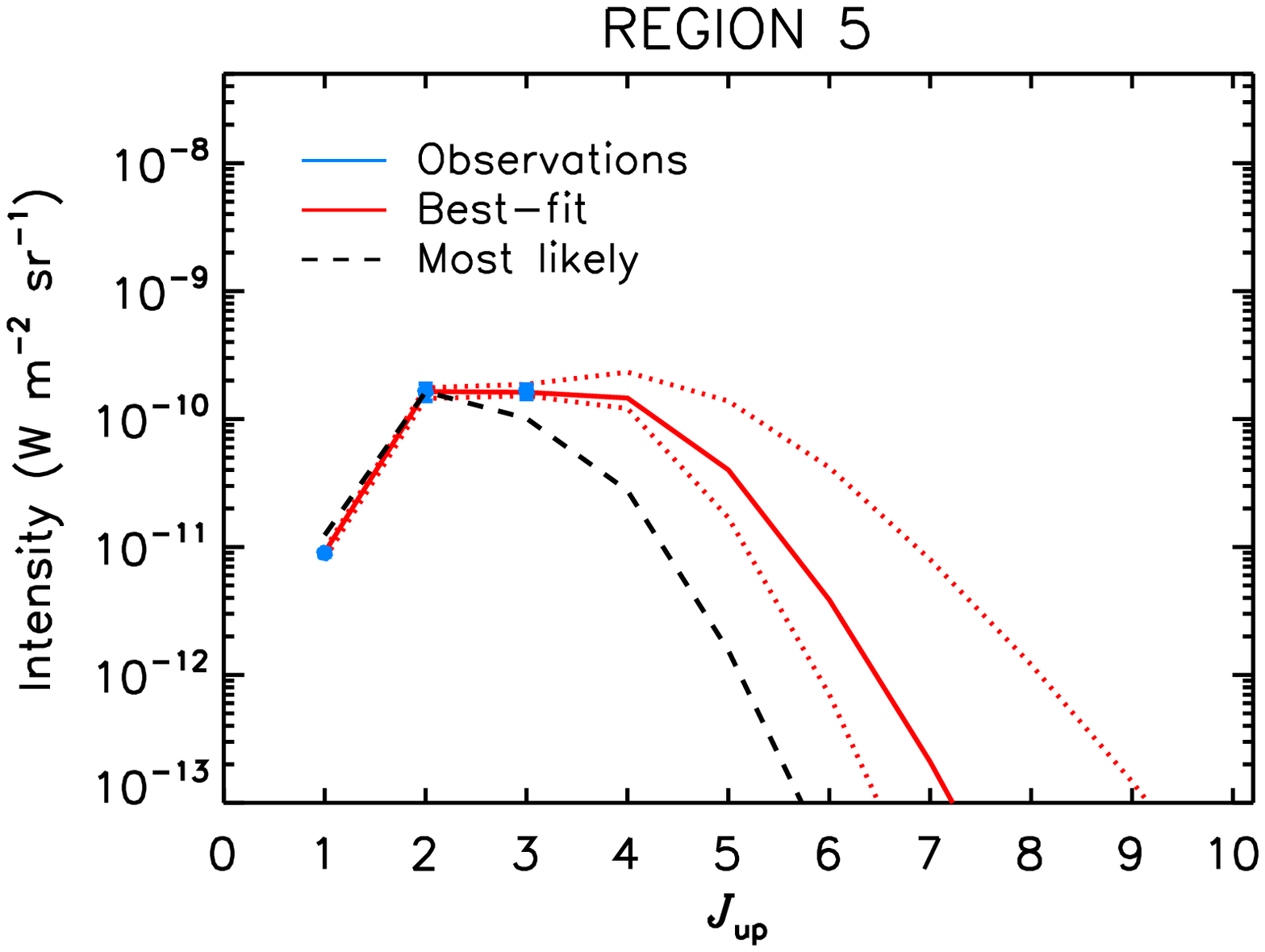}\\
  \includegraphics[height=5.0cm,clip=]{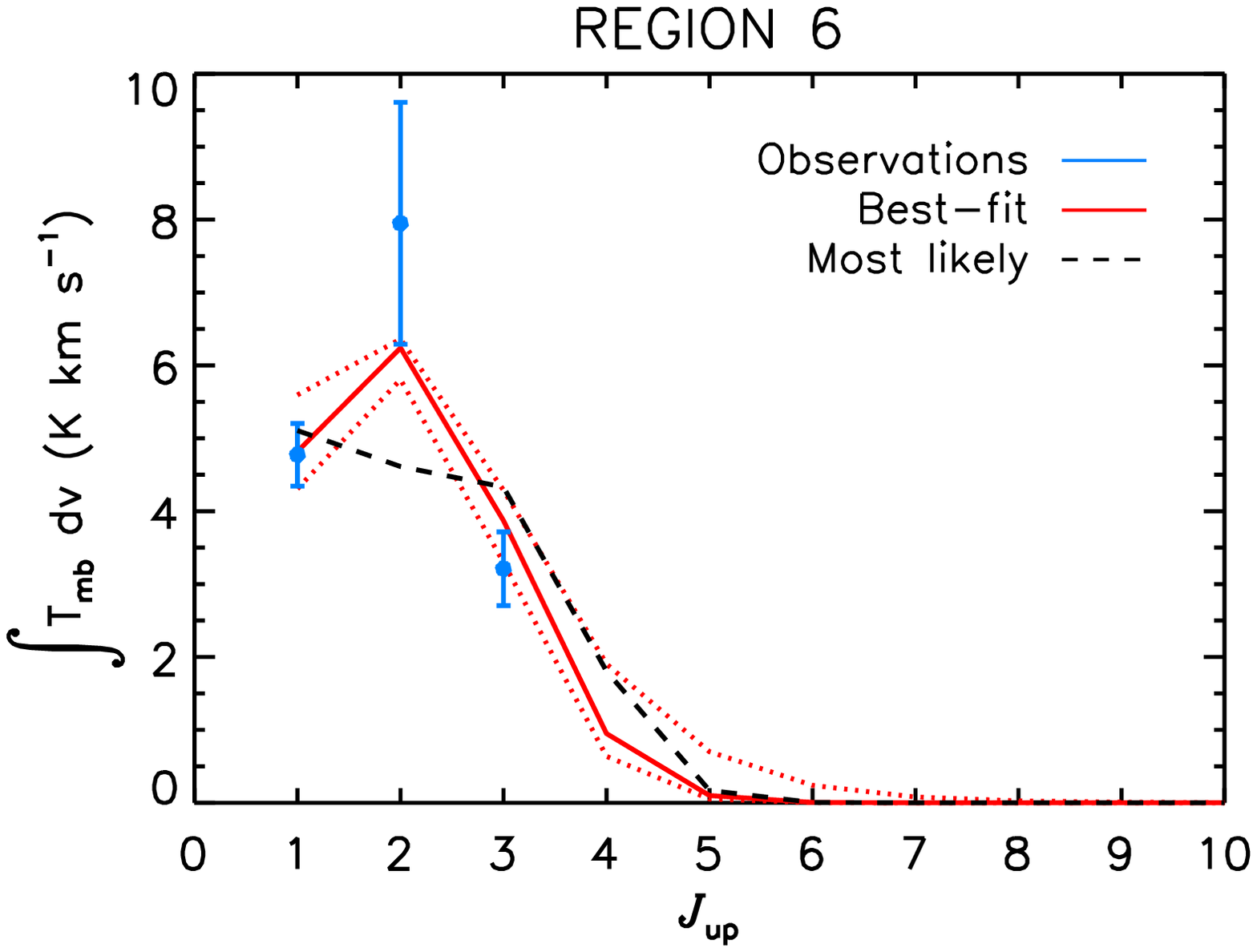}
  \includegraphics[height=5.0cm,clip=]{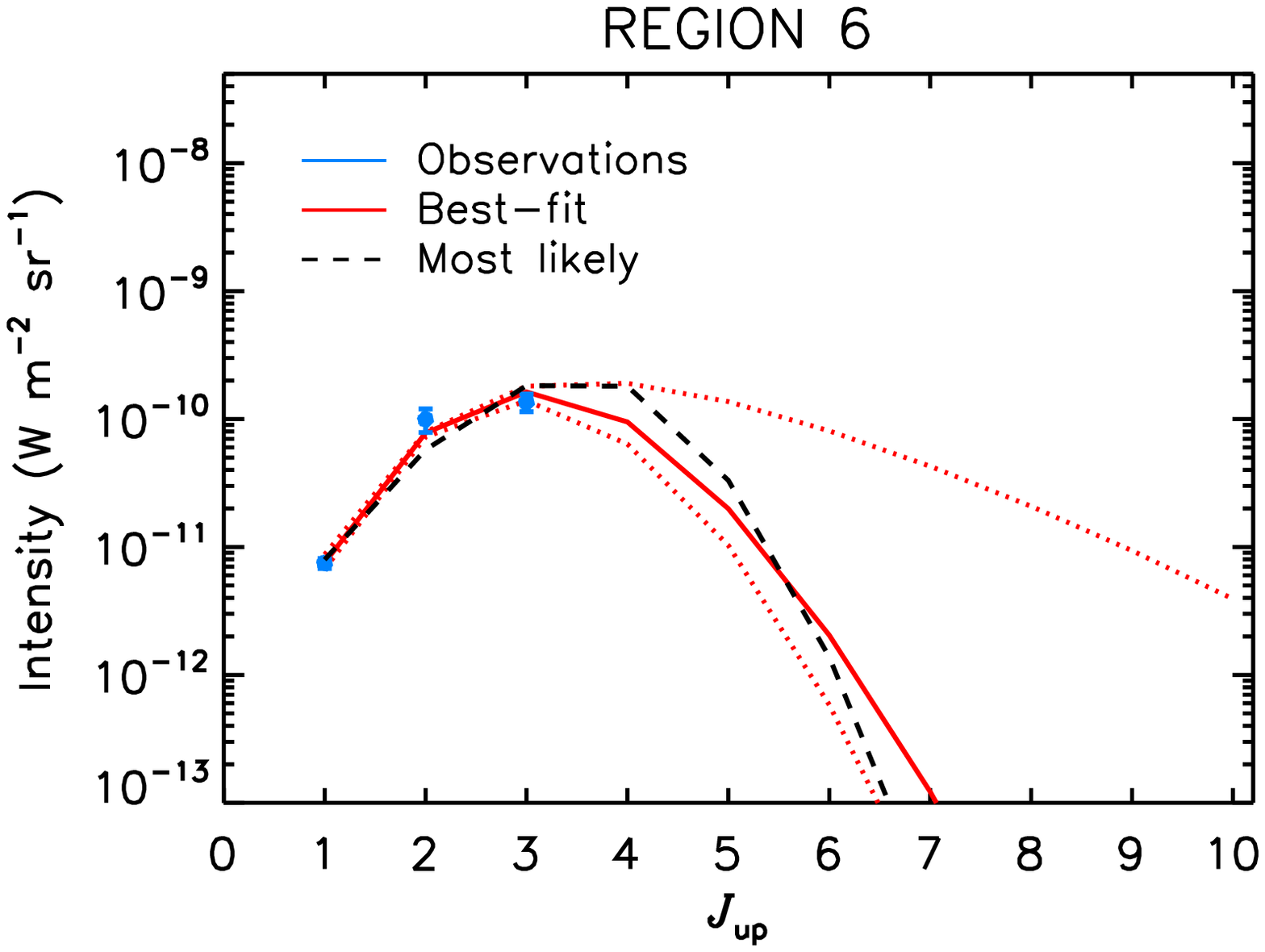}\\
  \includegraphics[height=5.0cm,clip=]{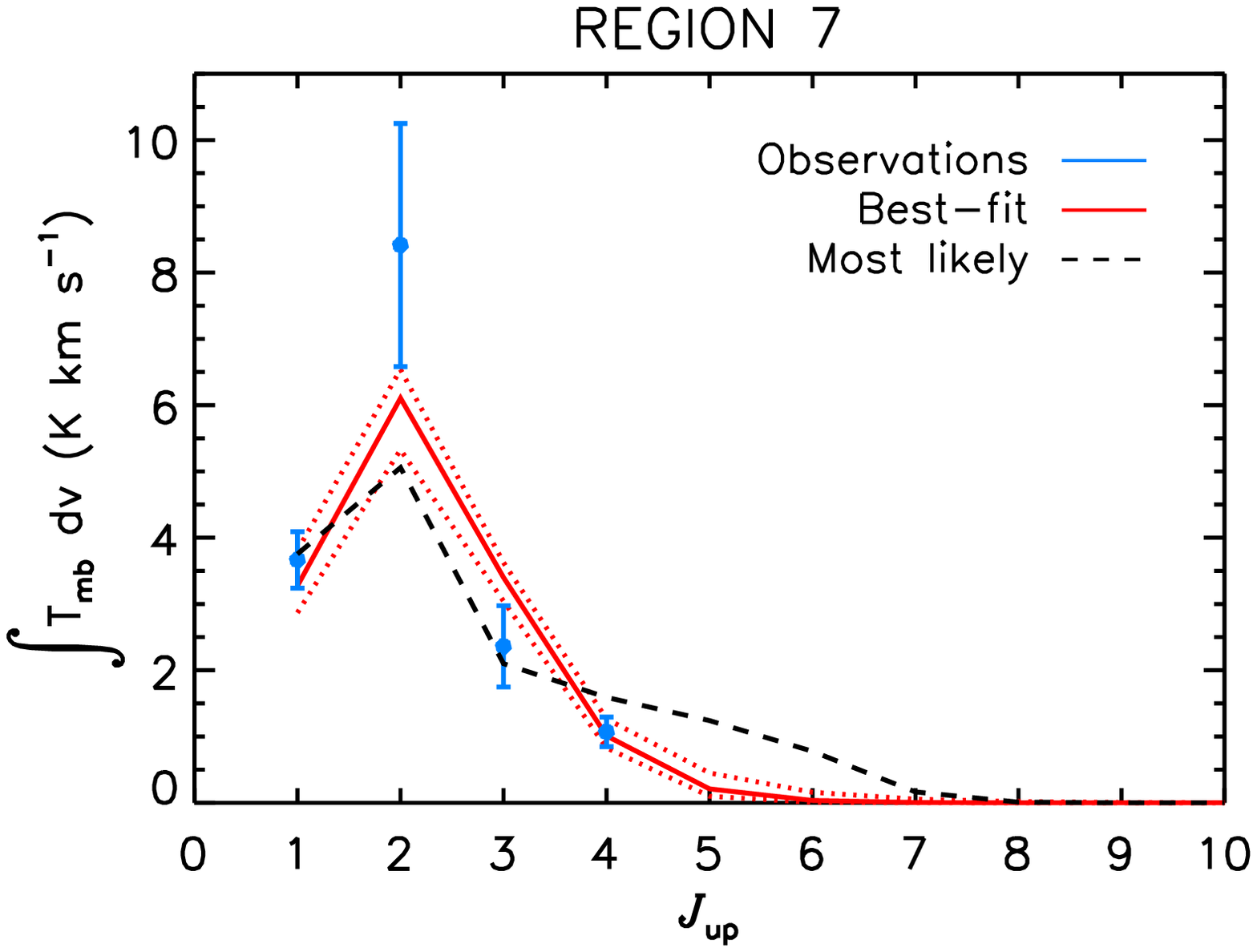}
  \includegraphics[height=5.0cm,clip=]{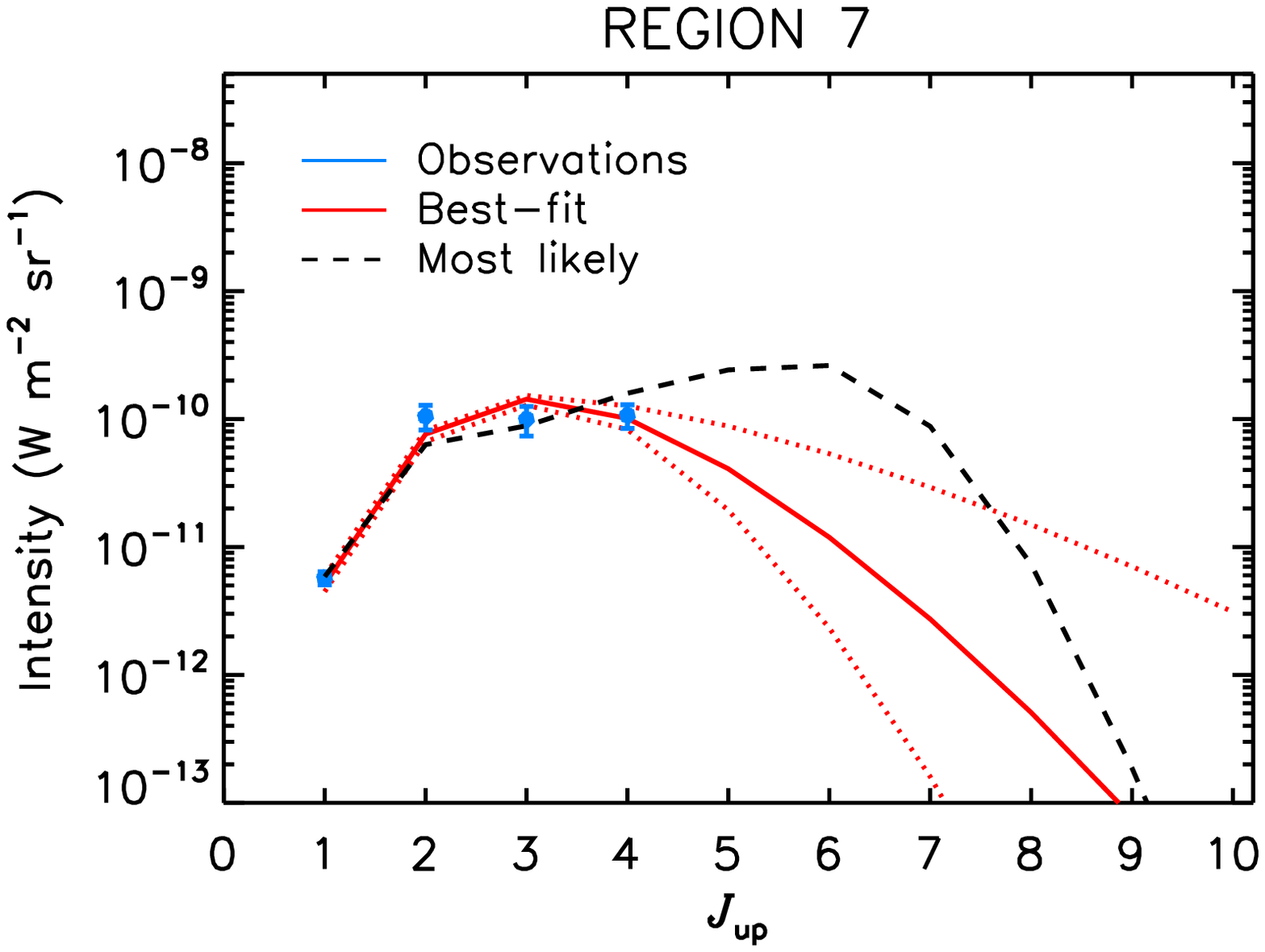}\\
  \includegraphics[height=5.0cm,clip=]{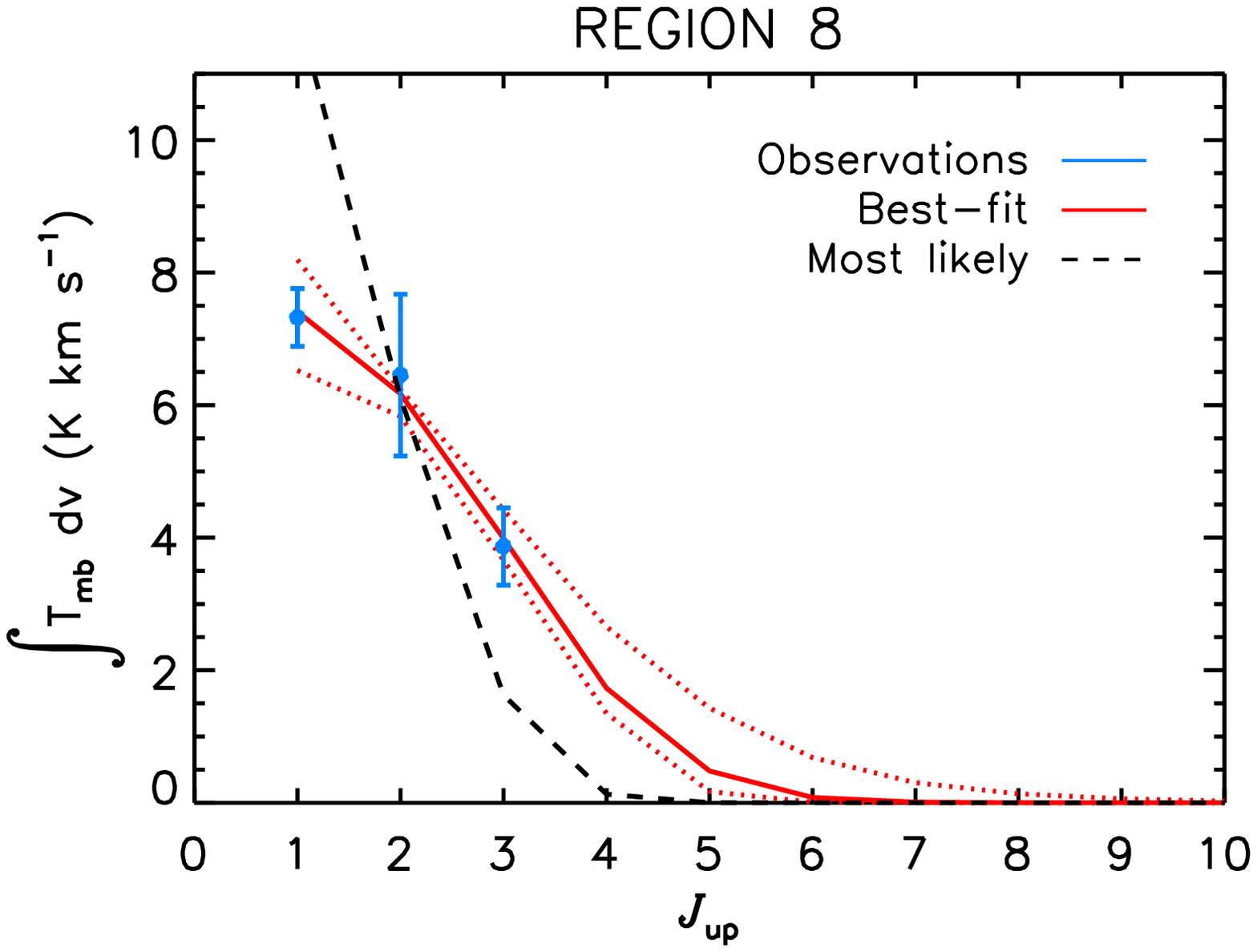}
  \includegraphics[height=5.0cm,clip=]{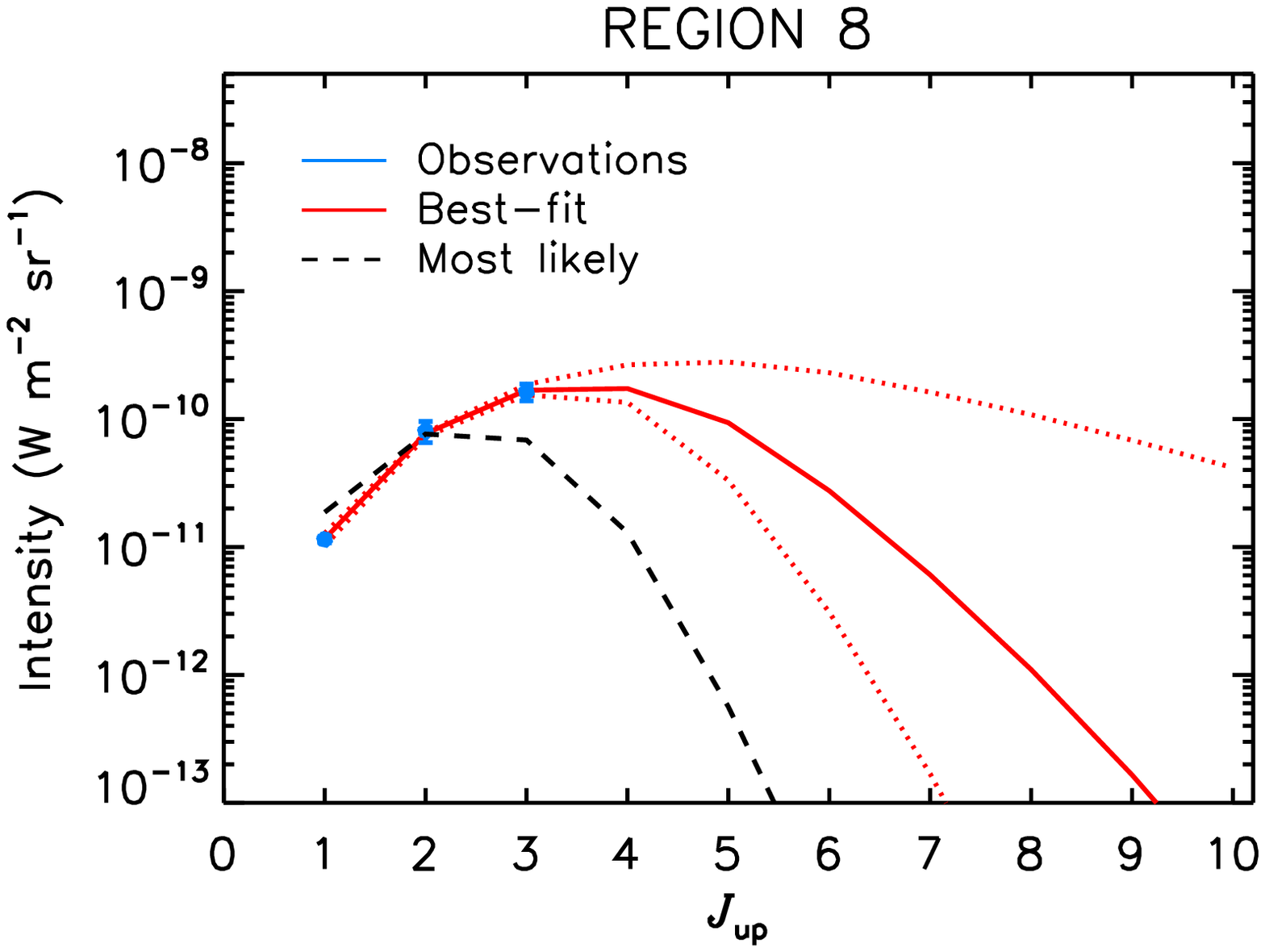}
  \caption{Continued.} 
\end{figure*}
%
% Discussion
%
\section{Discussion}
\label{sec:disc}
%
% Discussion: Star formation activity
%
\subsection{Star formation activity}
\label{sec:sf}
A variety of heating mechanisms of the molecular ISM are present in
galaxies, including far-ultraviolet (FUV) radiation (from massive
stars and active galactic nuclei), dust-gas collisions, cosmic rays,
mechanical heating (e.g.\ turbulence) and chemical heating \citep[see,
e.g.,][]{r07}. But which heating mechanisms dominate in the regions
studied? More specifically, how does SF feedback regulate the physical
conditions of the ISM? From both observational and theoretical
arguments, there must be a connection between the gas physical
conditions and the immediate environment of the regions, such as the
presence SNRs, H\,{\small II} regions, and H\,{\small I}
holes. In this section, we therefore discuss the physical properties
derived from our models in relation to the environmental
characteristics of the regions studied, and compare our results in the
spiral arms and inter-arm regions with previous results for the centre
of NGC~6946.

Overlaid on H$\alpha$ and H\,{\small I} images, some of the features
present in NGC~6946 were illustrated in Figure~\ref{fig:regions}:
H\,{\small I} holes \citep{bo08}, GMCs \citep{do12, reb12}, H\,{\small
  II} regions \citep{bbm86}, giant H\,{\small II} complexes
\citep{v77}, CO complexes \citep{reb12} and SNRs
\citep{mf97,ld01}. All the regions studied in this paper are
associated with one or more of these features.

Of the $62$ SNRs detected in NGC~6946 (both optically and in the
radio), $40$ are associated with spiral arms while $22$ are located in
inter-arm regions. \citet{ld01} argued that the arm SNRs (mostly
radio-bright) have massive progenitors (Population~I stars) and
explode as Type~II/Ib/Ic supernovae. Indeed, with short lifetimes,
many of the progenitors will explode inside or close to their parent
molecular cloud. \citet{ld01} in fact showed that the radio-bright
SNRs are likely associated with H\,{\small II} and star-forming
regions, an association borne out by Figure~\ref{fig:regions}. On the
other hand, the inter-arm SNRs (mostly optically-selected) are thought
to have Population~II progenitors \citep{ld01}.

In the galaxy M33, most radio-emitting SNRs have a diameter ranging
from $20$ to $80$~pc \citep{d95}. In the Galaxy, the bulk of the $47$
SNRs with a distance determination in the catalogue of \citet{g12}
have a diameter between $10$ and
$100$~pc\footnote{http://www.mrao.cam.ac.uk/projects/surveys/snrs/}. However,
stellar winds and/or diffuse ionised gas can help create shells with
diameters $\ga100$~pc \citep{mf97}. As cosmic rays produced by SN
explosions and shocks can travel large distances in low density
environments (weak absorption), where they heat up their surroundings
\citep{mj11}, we can expect that the average temperature will be
higher in low density environments than in high density ones.

Of course, UV radiation from short-lived massive stars is generally
believed to be the main stellar feedback mechanism affecting GMCs
(e.g.\ \citealt{c03} and references therein). Indeed,
simulations show that massive stars ($M_{\star}\ga~20$~$M_{\sun}$) can
locally disrupt their surrounding clouds via photoionisation, creating
H\,{\small II} regions and potentially triggering SF at new locations
in the clouds.

Spiral arms clearly lead to an increase in the number of cloud collisions \citep{dob08}. 
Classical grand-design spiral structures are thought to form from perturbations
due to bars or companions, while flocculent structures develop from 
local gravitational instabilities (see; e.g.,\ \citealt{dobb08}). NGC~6946 is a flocculent galaxy, 
with characteristic broken spiral arms. For the formation of GMCs, and thus of stars, 
the local gravitational instabilities seen in flocculent galaxies are therefore likely to be more important than global spiral dynamics.

In the following sub-sections, our results are therefore discussed in
view of the environment and SF activity in each region, grouping
regions with similar characteristics. Specifically, we consider
regions~1, 2 and 3 (intense SF activity), regions~4, 5 and 6 (low SF
activity), and regions~7 and 8 (inter-arm regions and intermediate SF
activity). We also compare the gas physical properties previously
found in the centre of NGC~6946 with our own results in the arm and
inter-arm regions.
%
% Regions 1+2+3
%
\subsection{Regions~1, 2 and 3}
As shown in Figure~\ref{fig:regions}, regions~1 and 2 are associated
with giant H\,{\small II} complexes, encompassing many H\,{\small II}
regions that each indicate the presence of OB stars and thus FUV
radiation. In particular, as mentioned before, region~1 has the
highest number of H\,{\small II} regions of all the regions
studied. The H\,{\small I} hole associated with region~2 may well have
been created by supernova explosions, as three SNRs are located near
the centre of the hole. Regions~1 and 2 are also associated with some
of the main star-forming regions previously studied by
\citet{mhc10,m11} using $33$~GHz free-free emission, a standard tracer
of SF. Although not part of a catalogued giant H\,{\small II} complex,
region~3 nevertheless lies in an important spiral arm and harbours
many H\,{\small II} regions and SNRs (see Fig.~\ref{fig:regions}).

The environment of and features present in regions~1, 2 and 3 are thus
similar, both indicating a high level of SF activity (slightly higher
in region~1 than 2, and in region~2 than 3). Our modeling results
reflect this, as the gas physical conditions are indeed similar and
indicate significant heating of the molecular ISM in all three regions
(see Table~\ref{tab:results}), presumably from FUV radiation and
stellar winds associated with OB stars (and potentially cosmic rays
from supernovae): high kinetic temperature $T_{\rm K}$ and low H$_2$
number volume density $n$(H$_2$) despite a high opacity.

We note that \citet{wbtwwd02} also carried out LVG modeling of
region~2 but found $T_{\rm K}=15$~K, at the opposite end of the range
of possible values and firmly ruled out here ($n$(H$_2$) and $N$(CO)
are also different, but less so). While worrying at face value, this
large temperature discrepancy probably arises from the different
transitions used to determine the best-fit model. In particular,
\citet{wbtwwd02} did not use relatively high-$J$ transitions of CO,
like the CO(4-3) line used here, that require relatively high
temperatures to be excited.
%
% Regions 4+5+6
%
\subsection{Regions~4, 5 and 6}
\label{sec:reg456}
Regions~4 and 5 are very close to each other and overalp significantly
(see Fig.~\ref{fig:regions}), so as expected the physical conditions
suggested by our modeling are very similar (see
Table~\ref{tab:results}). The best-fit $n$(H$_2$) is substantially
higher for region~4 (at the edge of our model grid), but the
likelihood shows that this parameter is rather ill-constrained. In
addition, the molecular ISM physical conditions derived in region~6
are very similar to those of regions~4 and 5. This is easily
understood, as all three regions are located in rather similar
environments (flocculent spiral arms; see Fig.~\ref{fig:regions}).

Compared to regions~1--3, regions~4--6 have much lower temperatures
and higher densities. This can be again explained by considering their
environment and observed features. Regions~4 and 5 do encompass some
H\,{\small II} regions (and region~4 has one catalogued SNR), but not
as many as regions~1--3. Similarly, region~6 has a small number of
H\,{\small II} regions, SNRs and GMCs at its edge, but none within our
adopted beam (see Fig.~\ref{fig:regions}). It is also largely
co-spatial with a large H\,{\small I} hole ($\approx900$~pc
diameter). It is thus likely that, while SF likely took place in the
past and the potential for future SF is high, current SF activity is
relatively mild in regions~4--6 (compared to regions~1--3), and this
is reflected in the current molecular gas physical conditions.
%
% Regions 7+8
%
\subsection{Regions 7 and 8}
Regions~7 and 8 are located in the inter-arm regions of NGC~6946,
respectively to the NE and SW of the galaxy centre (see
Fig.~\ref{fig:regions}). Interestingly, while one would naively expect
the lowest SF activity of all our regions, and correspondingly the
lowest ISM temperatures and highest densities, this is not the
case. Our models reveal physical conditions ($T_{\rm K}$, $n$(H$_2$)
and $N$(CO)) that are intermediate between those of regions~1--3 and
those of regions~4--6 (see Table~\ref{tab:results}).

While we do not have a truly satisfying explanation for this, we
suspect that SF activity is simply not as low as expected in those two
regions. Indeed, region~7 lies at the edge of a major spiral arm and
has some H\,{\small II} regions and GMCs, while region~8 harbours one
particularly large GMC (spatially) and a few H\,{\small II} regions
(at the centre of the GMC; see Fig.~\ref{fig:regions}). As there is no
SNR in either region, it is possible that the SF activity in those
regions is simply lower or is just beginning.

In any case, SF activity in regions~7 and 8 may well be higher than
naive expectations for the general inter-arm medium by construction,
as the regions studied here were necessarily chosen to contain some
molecular clouds.
%
% Spiral arms
%
\subsection{Spiral arms vs. galaxy centre}

Many studies have probed the molecular gas physical conditions at the
centre of NGC~6946, where there is a small bar
\citep{bssls85,wcc88,bbg88,iki90,rw95} and starburst activity
\citep{errl96}. \citet{ib01} concluded that there are at least two
molecular components in the central region, a warm ($T_{\rm
  K}=30$--$60$~K) and dense ($n$(H$_2$)$=3$--$10\times10^3$~cm$^{-3}$)
component and a more tenuous ($n$(H$_{2}$)$\le10^3$~cm$^{-3}$) and
hotter ($T_{\rm K}=100$--$150$~K) component. \citet{wbtwwd02} and
\citet{bgpc06} also probed the gas physical properties at the centre
using LVG modeling. \citet{wbtwwd02} found $T_{\rm K}=40$~K for the
gas kinetic temperature at the centre, while \citet{bgpc06} found
$T_{\rm K}=130$~K. The higher temperature found by \citet{bgpc06} may
be caused by the high-$J$ transition of CO used (CO(6-5)), requiring a
higher temperature to be excited. The densities derived are however
very similar ($n$(H$_2$)$=10^{3.2}$--$10^{3.3}$~cm$^{-3}$).

It is interesting to compare the properties of the molecular gas
present at the centre of the galaxy with those of the gas in the
spiral arms and inter-arm regions. As discussed above (see also
Table~\ref{tab:results}), our results indicate three different media
across the regions studied; hot and tenuous (regions~1--3), cold and
dense (regions~4--6), and intermediate (warm; inter-arm
regions~7--8). Regions~1--3 have hot and relatively tenuous molecular
gas similar to the hot and tenuous central component identified by
\citet{ib01}. This is likely related to intense SF activity and
associated shocks from SNRs (both heating up and sweeping up the gas).
The inter-arm regions~7--8 have warm and dense molecular gas similar
to that of the central warm and dense component identified by
\citet{ib01}. Unsurprisingly, the cold and dense gas in regions~4--6
does not match any of the central molecular gas properties.
%
% Conclusions
%
\section{Conclusions}
\label{sec:conc}
In this paper, we have provided the most complete CO survey of the
nearby galaxy NGC~6946 ever performed, including for each of the $8$
regions studied at least $4$ transitions of CO. The data presented in
Table~\ref{tab:ourdata1} and Appendix~\ref{sec:spec} further cover a
much broader area of NGC~6946, and should be useful for diverse
projects investigating its SF activity. From the observations and gas
physical parameters derived for each region, our main results can be
summarised as follows:
\begin{enumerate}
\item The CO(2-1) map indicates that molecular gas is ubiquitous
  across NGC~6946 but is concentrated along the spiral arms, with a
  regular velocity field. The CO(3-2) data reveal two compact
  'hotspots' (very active star-forming regions) in the eastern spiral
  arms (regions~1 and 2), where the gas is also bright at H$\alpha$,
  $24$~$\micron$, $850$~$\micron$ and $6.2$~cm.
\item The radial profiles of the different CO transitions probed
  (eastward from the centre up to $\approx4$~kpc) reveal the likely
  presence of two components: a highly-peaked molecular core and a
  nearly flat lower surface brightness extended disc.
\item While the beam-corrected line ratio $R_{\rm 21}$ varies
  significantly across NGC~6946 ($\approx1$ to $2.5$), the other
  ratios are more constant: $R_{\rm 31}\approx0.7$, $R_{\rm
    41}\approx0.4$ and $R_{\rm 22}\approx0.1$.
\item According to LVG modeling, selected regions in NGC~6946 have a
  large variety of physical conditions (see
  Table~\ref{tab:results}). Overall, the gas kinetic temperature
  $T_{\rm K}$ ranges from $15$ to $250$~K, the H$_2$ number volume
  density $n$(H$_2$) ranges from $10^{2.3}$ to $10^{7.0}$~cm$^{-3}$,
  and the CO number column density $N$(CO) ranges from $10^{15.0}$ to
  $10^{19.3}$~cm$^{-2}$.
\item Comparing these physical properties with star formation-related
  environmental characteristics such as the presence of SNRs,
  H\,{\small I} holes, H\,{\small II} regions/complexes and GMCs, the
  regions probed can naturally be grouped into $3$ broad
  categories. First, regions~1, 2 and 3 are hotter and more tenuous
  than the other regions. They also show the strongest signs of SF
  activity, most likely driven by young massive OB stars. FUV
  radiation and stellar winds therefore likely dominate the heating
  and excitation of the molecular ISM in these regions, with a
  possible contribution from SNR-related cosmic rays in
  region~3. Second, regions~4, 5 and 6 are colder and denser than the
  other regions. They also likely harbour less intense SF activity,
  with fewer SF-related characteristic features. Third, the inter-arm
  regions~7 and 8 reveal intermediate gas physical properties,
  suggesting (somewhat surprisingly) equally intermediate SF activity.
\item The turnovers of the predicted SLEDs indicate that in regions
  where the turnover takes place at high $J$ ($J_{\rm max}>4$),
  $T_{\rm K}$ and $N$(CO) are high whereas $n$(H$_2$) is relatively
  low. On the other hand, in regions where the turnover takes place at
  low $J$ ($J_{\rm max}\le4$), $T_{\rm K}$ and $N$(CO) are low whereas
  $n$(H$_2$) is relatively high.
\item Based on the predicted SLEDs, some extra-nuclear regions (i.e.\
  regions~1, 2 and 3) have a SF activity comparable to that seen at
  the centre of NGC~6946, much higher than that at the centre of our
  own Galaxy. The SF activity in the other regions studied is however
  comparable to that at the centre of the Milky Way.
\end{enumerate}
%
% Acknowledgements
%
\section*{Acknowledgements}
The authors thank S.\ Kaviraj for useful discussions on the likelihood
method and F.\ van der Tak for very useful discussions on the
radiative transfer code RADEX. ST gives special thanks to David
Rebolledo and Jennifer Donovan Meyer for providing their GMC data. ST
was supported by the Republic of Turkey, Ministry of National
Education, and The Philip Wetton Graduate Scholarship at Christ
Church. EB and MB were partially supported by an Oxford Fell Fund
award during the course of this work, and by STFC rolling grant
Astrophysics at Oxford ST/H002456/1. ST and MB thank the Department of
Physics at Nagoya University for its hospitality during a sabbatical
visit. The research leading to these results has received funding from
the European Community's Seventh Framework Programme (/FP7/2007-2013/)
under grant agreement No 229517. This research also made use of the
NASA/IPAC Extragalactic Database (NED), which is operated by the Jet
Propulsion Laboratory, California Institute of Technology, under
contract with the National Aeronautics and Space Administration.
%
% Bibliography
%

%
% Appendices
%
\appendix
%
% CSO Observations
%
\section{CSO Observations}
\label{sec:cso}
\begin{table*}\footnotesize
  \setlength{\tabcolsep}{4pt}
  \caption{CSO detections.}
  \begin{tabular}{rcrr@{\hskip 1cm}rcrr}
    \hline
    Transition & Offset & $\int T_{\rm mb}$ d$v$ & FWHM & Transition &
    Offset & $\int T_{\rm mb}$ d$v$ & FWHM \\
    & (arcsec) & (K~km~s$^{-1}$) & (km~s$^{-1}$) & & (arcsec) &  (K~km~s$^{-1}$) & (km~s$^{-1}$) \\ \hline 
    C\,{\small I} ($^3$P$_1$-$^3$P$_0$) & $\phantom{0}\phantom{0}(0,\phantom{0}\phantom{0}0)$ & $29.9\,\pm\,2.8$ & $148.5\,\pm\,15.2$ & CO(2-1) & $\phantom{0}($-$90,\phantom{0}$-$30)$ & $7.9\,\pm\,\phantom{0}1.5$ & $29.8\,\pm\,\phantom{0}5.4$ \\
    & $\phantom{0}(20,\phantom{0}\phantom{0}0)$ & $25.1\,\pm\,2.9$ & $87.3\,\pm\,11.9$ && $\phantom{0}($-$90,\phantom{0}$-$60)$ & $7.9\,\pm\,\phantom{0}1.8$ & $25.4\,\pm\,\phantom{0}8.6$ \\
    & $\phantom{0}(40,\phantom{0}\phantom{0}0)$ & $5.6\,\pm\,2.3$ & $37.9\,\pm\,27.3$ && $($-$120,\phantom{0}\phantom{0}0)$ & $2.5\,\pm\,\phantom{0}0.9$ & $14.4\,\pm\,\phantom{0}5.3$ \\
    & $\phantom{0}(80,\phantom{0}\phantom{0}0)$ & $18.9\,\pm\,2.5$ & $155.8\,\pm\,22.6$ && $($-$120,120)$ & $6.5\,\pm\,\phantom{0}1.3$ & $25.1\,\pm\,\phantom{0}6.4$ \\
    & $(100,\phantom{0}\phantom{0}0)$ & $4.1\,\pm\,1.3$ & $93.3\,\pm\,38.2$ && $($-$120,\phantom{0}$-$30)$ & $8.6\,\pm\,\phantom{0}1.4$ & $30.3\,\pm\,\phantom{0}5.3$\\
    & $(110,100)$ & $4.4\,\pm\,0.9$ & $29.9\,\pm\,\phantom{0}5.7$ && $($-$120,\phantom{0}$-$60)$ & $10.8\,\pm\,\phantom{0}2.6$ & $46.6\,\pm\,17.0$\\
    & $(120,\phantom{0}\phantom{0}0)$ & $4.0\,\pm\,1.3$ & $41.7\,\pm\,18.7$ && $($-$150,\phantom{0}30)$ & $3.2\,\pm\,\phantom{0}0.8$ & $12.1\,\pm\,\phantom{0}5.5$\\
    & $(150,\phantom{0}$-$20)$ & $3.8\,\pm\,0.9$ & $20.2\,\pm\,\phantom{0}4.7$ && $($-$150,\phantom{0}60)$ & $3.9\,\pm\,\phantom{0}1.5$ & $18.6\,\pm\,14.1$\\
    CO(2-1) & $\phantom{0}\phantom{0}(0,\phantom{0}\phantom{0}0)$ & $102.7\,\pm\,3.0$ & $155.5\,\pm\,\phantom{0}4.8$ && $($-$150,\phantom{0}90)$ & $6.0\,\pm\,\phantom{0}1.6$ & $41.6\,\pm\,12.7$\\ 
    & $\phantom{0}\phantom{0}(0,\phantom{0}30)$ & $43.7\,\pm\,2.4$ & $106.6\,\pm\,\phantom{0}6.3$ && $($-$150,120)$ & $11.1\,\pm\,\phantom{0}2.6$ & $73.8\,\pm\,26.2$\\
    & $\phantom{0}\phantom{0}(0,\phantom{0}60)$ & $20.3\,\pm\,2.7$ & $88.9\,\pm\,14.3$ && $($-$150,\phantom{0}$-$30)$ & $5.7\,\pm\,\phantom{0}1.5$ & $30.6\,\pm\,\phantom{0}8.5$\\
    & $\phantom{0}\phantom{0}(0,\phantom{0}90)$ & $6.6\,\pm\,1.9$ & $49.5\,\pm\,17.2$ && $($-$150,\phantom{0}$-$60)$ & $8.9\,\pm\,\phantom{0}2.0$ & $57.8\,\pm\,17.8$ \\
    & $\phantom{0}\phantom{0}(0,120)$ & $5.5\,\pm\,0.9$ & $35.2\,\pm\,\phantom{0}7.4$ && $($-$180,\phantom{0}$-$30)$ & $4.4\,\pm\,\phantom{0}1.2$ & $25.9\,\pm\,\phantom{0}8.1$\\
    & $\phantom{0}\phantom{0}(0,\phantom{0}$-$30)$ & $18.4\,\pm\,2.5$ & $87.2\,\pm\,12.8$ && $($-$210,\phantom{0}\phantom{0}0)$ & $3.0\,\pm\,\phantom{0}1.1$ & $18.4\,\pm\,11.1$ \\
    & $\phantom{0}\phantom{0}(0,\phantom{0}$-$90)$ & $7.0\,\pm\,1.5$ & $31.8\,\pm\,\phantom{0}9.1$ && $($-$210,\phantom{0}90)$ & $6.5\,\pm\,\phantom{0}2.6$ & $43.6\,\pm\,27.7$\\
    & $\phantom{0}\phantom{0}(0, $-$120)$ & $9.4\,\pm\,2.4$ & $44.9\,\pm\,15.7$ && $($-$210,120)$ & $16.2\,\pm\,\phantom{0}2.7$ & $114.1\,\pm\,18.8$  \\
    & $\phantom{0}(30,\phantom{0}\phantom{0}0)$ & $31.7\,\pm\,3.6$ & $98.6\,\pm\,17.7$ && $($-$210,180)$ & $13.1\,\pm\,\phantom{0}2.3$ & $94.8\,\pm\,17.5$\\
    & $\phantom{0}(30,\phantom{0}30)$ & $31.4\,\pm\,2.4$ & $96.6\,\pm\,\phantom{0}9.2$ && $($-$210,\phantom{0}$-$60)$ & $8.4\,\pm\,\phantom{0}2.2$ & $64.6\,\pm\,19.3$\\
    & $\phantom{0}(30,\phantom{0}90)$ & $10.9\,\pm\,1.7$ & $48.3\,\pm\,11.3$ && $($-$240,\phantom{0}60)$ & $3.6\,\pm\,\phantom{0}1.1$ & $23.6\,\pm\,\phantom{0}8.2$\\
    & $\phantom{0}(30,120)$ & $4.6\,\pm\,1.3$ & $30.5\,\pm\,10.3$ && $($-$240,180)$ & $11.8\,\pm\,\phantom{0}2.5$ & $93.9\,\pm\,20.4$\\
    & $\phantom{0}(30,\phantom{0}$-$30)$ & $13.2\,\pm\,2.7$ & $81.3\,\pm\,18.2$ && $($-$240,240)$ & $34.6\,\pm\,\phantom{0}4.2$ & $217.5\,\pm\,28.2$\\
    & $\phantom{0}(30,\phantom{0}$-$60)$ & $7.9\,\pm\,1.4$ & $39.1\,\pm\,\phantom{0}7.3$ & CO(3-2) & $\phantom{0}\phantom{0}(0,\phantom{0}\phantom{0}0)$ & $111.6\,\pm\,\phantom{0}2.5$ & $151.0\,\pm\,\phantom{0}3.7$ \\
    & $\phantom{0}(30,\phantom{0}$-$90)$ & $4.3\,\pm\,1.2$ & $29.5\,\pm\,\phantom{0}9.1$ && $\phantom{0}\phantom{0}(0,\phantom{0}10)$ & $100.8\,\pm\,\phantom{0}2.2$ & $155.9\,\pm\,\phantom{0}3.6$\\
    & $\phantom{0}(60,\phantom{0}30)$ & $15.0\,\pm\,2.2$ & $67.6\,\pm\,11.9$ && $\phantom{0}\phantom{0}(0,\phantom{0}20)$ & $66.7\,\pm\,\phantom{0}5.2$ & $138.0\,\pm\,12.3$ \\
    & $\phantom{0}(60,\phantom{0}60)$ & $9.5\,\pm\,1.7$ & $47.7\,\pm\,11.1$ && $\phantom{0}\phantom{0}(0,\phantom{0}$-$10)$ & $89.1\,\pm\,\phantom{0}3.0$ & $133.1\,\pm\,\phantom{0}5.3$\\
    & $\phantom{0}(60,\phantom{0}90)$ & $7.9\,\pm\,1.2$ & $25.5\,\pm\,\phantom{0}4.4$ && $\phantom{0}(10,\phantom{0}\phantom{0}0)$ & $131.2\,\pm\,\phantom{0}4.2$ & $155.0\,\pm\,\phantom{0}5.3$\\
    & $\phantom{0}(60,120)$ & $4.5\,\pm\,1.2$ & $24.9\,\pm\,\phantom{0}7.3$ && $\phantom{0}(10,\phantom{0}10)$ & $54.9\,\pm\,\phantom{0}7.9$ & $107.3\,\pm\,26.6$ \\
    & $\phantom{0}(60,\phantom{0}$-$60)$ & $3.7\,\pm\,1.0$ & $20.7\,\pm\,\phantom{0}6.5$ && $\phantom{0}(10,\phantom{0}$-$10)$ & $79.5\,\pm\,\phantom{0}5.8$ & $169.9\,\pm\,13.6$\\
    & $\phantom{0}(90,\phantom{0}\phantom{0}0)$ & $11.4\,\pm\,2.5$ & $61.4\,\pm\,22.5$ && $\phantom{0}(20,\phantom{0}\phantom{0}0)$ & $29.8\,\pm\,\phantom{0}2.6$ & $75.1\,\pm\,\phantom{0}9.7$\\
    & $\phantom{0}(90,\phantom{0}30)$ & $9.2\,\pm\,1.5$ & $25.3\,\pm\,\phantom{0}6.5$ && $\phantom{0}(20,\phantom{0}$-$20)$ & $7.3\,\pm\,\phantom{0}1.9$ & $20.1\,\pm\,\phantom{0}5.1$\\
    & $\phantom{0}(90,\phantom{0}60)$ & $5.6\,\pm\,1.2$ & $22.2\,\pm\,\phantom{0}6.3$ && $\phantom{0}(40,\phantom{0}\phantom{0}0)$ & $13.0\,\pm\,\phantom{0}1.2$ & $57.2\,\pm\,\phantom{0}9.2$\\
    & $\phantom{0}(90,\phantom{0}90)$ & $12.4\,\pm\,1.6$ & $46.4\,\pm\,\phantom{0}7.8$ && $\phantom{0}(60,\phantom{0}\phantom{0}0)$ & $3.9\,\pm\,\phantom{0}1.1$ & $20.8\,\pm\,\phantom{0}6.6$\\
    & $\phantom{0}(90,120)$ & $8.4\,\pm\,1.3$ & $36.8\,\pm\,\phantom{0}5.6$ && $\phantom{0}(80,\phantom{0}\phantom{0}0)$ & $3.9\,\pm\,\phantom{0}0.9$ & $11.4\,\pm\,\phantom{0}3.7$\\
    & $\phantom{0}(90,150)$ & $5.1\,\pm\,1.5$ & $31.9\,\pm\,11.1$ && $\phantom{0}(90,100)$ & $9.2\,\pm\,\phantom{0}1.7$ & $35.9\,\pm\,\phantom{0}8.5$\\
    & $(120,\phantom{0}\phantom{0}0)$ & $4.7\,\pm\,1.3$ & $24.0\,\pm\,\phantom{0}6.9$ && $(100,\phantom{0}\phantom{0}0)$ & $2.0\,\pm\,\phantom{0}0.6$ & $14.3\,\pm\,\phantom{0}3.7$\\
    & $(120,\phantom{0}30)$ & $7.8\,\pm\,1.4$ & $25.4\,\pm\,\phantom{0}4.9$ && $(100,\phantom{0}90)$ & $7.0\,\pm\,\phantom{0}1.4$ & $25.0\,\pm\,\phantom{0}5.2$\\
    & $(120,\phantom{0}60)$ & $6.5\,\pm\,1.1$ & $22.4\,\pm\,\phantom{0}3.9$ && $(100,110)$ & $6.7\,\pm\,\phantom{0}1.3$ & $14.7\,\pm\,\phantom{0}3.6$\\
    & $(120,\phantom{0}90)$ & $9.1\,\pm\,1.6$ & $52.4\,\pm\,10.7$ && $(110,100)$ & $11.2\,\pm\,\phantom{0}0.7$ & $34.1\,\pm\,\phantom{0}2.4$\\
    & $(120,$-$120)$ & $6.6\,\pm\,2.2$ & $42.5\,\pm\,21.8$ && $(110,110)$ & $3.4\,\pm\,\phantom{0}0.9$ & $16.5\,\pm\,\phantom{0}4.5$\\
    & $(150,\phantom{0}\phantom{0}0)$ & $8.7\,\pm\,1.2$ & $29.8\,\pm\,\phantom{0}5.1$ && $(120,\phantom{0}90)$ & $2.8\,\pm\,\phantom{0}0.9$ & $8.4\,\pm\,\phantom{0}3.4$\\
    & $(150,\phantom{0}60)$ & $9.3\,\pm\,2.1$ & $43.4\,\pm\,15.0$ && $(140,\phantom{0}$-$20)$ & $8.8\,\pm\,\phantom{0}0.8$ & $25.2\,\pm\,\phantom{0}2.3$ \\
    & $(150,\phantom{0}90)$ & $8.4\,\pm\,1.8$ & $43.0\,\pm\,12.9$ && $(150,\phantom{0}\phantom{0}0)$ & $6.1\,\pm\,\phantom{0}0.6$ & $27.9\,\pm\,\phantom{0}3.1$\\
    & $(150,\phantom{0}$-$30)$ & $9.7\,\pm\,1.6$ & $25.2\,\pm\,\phantom{0}4.1$ && $(150,\phantom{0}$-$10)$ & $6.6\,\pm\,\phantom{0}0.6$ & $22.1\,\pm\,\phantom{0}2.4$\\
    & $(150,\phantom{0}$-$60)$ & $3.4\,\pm\,1.1$ & $23.4\,\pm\,\phantom{0}7.9$ && $(150,\phantom{0}$-$20)$ & $10.8\,\pm\,\phantom{0}1.5$ & $26.8\,\pm\,\phantom{0}4.3$\\
    & $(180,\phantom{0}\phantom{0}0)$ & $7.7\,\pm\,1.5$ & $38.8\,\pm\,\phantom{0}9.2$ && $(150,\phantom{0}$-$30)$ & $6.2\,\pm\,\phantom{0}0.7$ & $21.1\,\pm\,\phantom{0}2.5$ \\
    & $(180,\phantom{0}30)$ & $5.6\,\pm\,1.5$ & $28.8\,\pm\,\phantom{0}9.3$ && $(150,\phantom{0}$-$40)$ & $3.7\,\pm\,\phantom{0}0.6$ & $14.4\,\pm\,\phantom{0}2.4$\\
    & $(180,\phantom{0}60)$ & $7.7\,\pm\,2.4$ & $44.7\,\pm\,17.6$ && $(160,\phantom{0}$-$20)$ & $9.2\,\pm\,\phantom{0}1.6$ & $23.5\,\pm\,\phantom{0}4.7$\\
    & $(180,\phantom{0}$-$30)$ & $7.3\,\pm\,1.9$ & $32.5\,\pm\,\phantom{0}8.9$ && $\phantom{0}($-$10,\phantom{0}\phantom{0}0)$ & $70.3\,\pm\,\phantom{0}2.9$ & $129.8\,\pm\,\phantom{0}6.2$\\
    & $(210,\phantom{0}30)$ & $5.9\,\pm\,1.4$ & $34.9\,\pm\,\phantom{0}9.8$ && $\phantom{0}($-$10,\phantom{0}10)$ & $37.9\,\pm\,\phantom{0}2.9$ & $125.9\,\pm\,11.0$\\
    & $(240,\phantom{0}$-$30)$ & $9.4\,\pm\,2.3$ & $76.8\,\pm\,22.9$ && $\phantom{0}($-$10,\phantom{0}$-$10)$ & $36.9\,\pm\,\phantom{0}3.7$ & $116.0\,\pm\,15.9$\\
    & $\phantom{0}($-$30,\phantom{0}\phantom{0}0)$ & $13.1\,\pm\,2.7$ & $97.7\,\pm\,26.2$ && $\phantom{0}($-$20,\phantom{0}\phantom{0}0)$ & $8.9\,\pm\,\phantom{0}2.7$ & $14.2\,\pm\,\phantom{0}6.0$\\
    & $\phantom{0}($-$30,\phantom{0}30)$ & $25.1\,\pm\,2.9$ & $170.1\,\pm\,21.6$ && $\phantom{0}($-$20,\phantom{0}20)$ & $3.0\,\pm\,\phantom{0}1.7$ & $14.5\,\pm\,\phantom{0}6.2$\\
    & $\phantom{0}($-$30,\phantom{0}90)$ & $12.1\,\pm\,1.9$ & $86.2\,\pm\,14.1$ && $\phantom{0}($-$40,\phantom{0}\phantom{0}0)$ & $5.0\,\pm\,\phantom{0}0.4$ & $36.1\,\pm\,\phantom{0}3.5$\\
    & $\phantom{0}($-$30,\phantom{0}$-$60)$ & $11.7\,\pm\,2.2$ & $54.1\,\pm\,13.3$ & CO(4-3) & $\phantom{0}\phantom{0}(0,\phantom{0}\phantom{0}0)$ & $187.1\,\pm\,\phantom{0}5.5$ & $144.9\,\pm\,\phantom{0}4.6$ \\
    & $\phantom{0}($-$30,\phantom{0}$-$90)$ & $8.6\,\pm\,1.4$ & $31.6\,\pm\,\phantom{0}5.9$ && $\phantom{0}\phantom{0}(0,\phantom{0}10)$ & $40.0\,\pm\,\phantom{0}7.9$ & $105.2\,\pm\,21.7$\\
    & $\phantom{0}($-$30,$-$120)$ & $9.1\,\pm\,2.1$ & $68.3\,\pm\,21.2$ && $\phantom{0}(10,\phantom{0}\phantom{0}0)$ & $167.4\,\pm\,\phantom{0}9.9$ & $137.4\,\pm\,\phantom{0}8.8$\\
    & $\phantom{0}($-$60,\phantom{0}\phantom{0}0)$ & $11.4\,\pm\,1.8$ & $39.5\,\pm\,\phantom{0}8.6$ && $\phantom{0}(20,\phantom{0}\phantom{0}0)$ & $41.9\,\pm\,\phantom{0}3.6$ & $141.3\,\pm\,13.0$\\
    & $\phantom{0}($-$60,\phantom{0}60)$ & $10.2\,\pm\,1.7$ & $46.9\,\pm\,\phantom{0}9.5$ && $(110,100)$ & $9.4\,\pm\,\phantom{0}1.3$ & $27.1\,\pm\,\phantom{0}3.9$ \\
    & $\phantom{0}($-$60,\phantom{0}$-$30)$ & $9.5\,\pm\,1.4$ & $33.2\,\pm\,\phantom{0}4.7$ && $(150,\phantom{0}$-$20)$ & $13.4\,\pm\,\phantom{0}1.9$ & $27.5\,\pm\,\phantom{0}5.1$\\     
    & $\phantom{0}($-$60,\phantom{0}$-$60)$ & $7.3\,\pm\,1.6$ & $30.2\,\pm\,\phantom{0}9.9$ && $($-$110,\phantom{0}$-$30)$ & $6.5\,\pm\,\phantom{0}1.1$ & $19.0\,\pm\,\phantom{0}4.6$\\
    & $\phantom{0}($-$60,$-$150)$ & $7.1\,\pm\,1.9$ & $56.4\,\pm\,19.0$ & CO(6-5) & $\phantom{0}\phantom{0}(0,\phantom{0}\phantom{0}0)$ & $70.1\,\pm\,10.4$ & $101.1\,\pm\,18.5$\\
    & $\phantom{0}($-$90,\phantom{0}60)$ & $5.6\,\pm\,1.5$ & $31.3\,\pm\,\phantom{0}9.2$ && $\phantom{0}\phantom{0}(5,\phantom{0}\phantom{0}0)$ & $90.7\,\pm\,\phantom{0}9.8$ & $129.5\,\pm\, 13.9$\\
    & $\phantom{0}($-$90,\phantom{0}90)$ & $4.3\,\pm\,1.3$ & $30.0\,\pm\,\phantom{0}8.1$ && $\phantom{0}(10,\phantom{0}\phantom{0}0)$ & $152.4\,\pm\,23.7$ & $191.3\,\pm\,30.8$\\
    & $\phantom{0}($-$90,120)$ & $19.2\,\pm\,3.8$ & $136.0\,\pm\,37.0$ && $(150,\phantom{0}$-$20)$ & $13.9\,\pm\,\phantom{0}4.8$ & $20.2\,\pm\,\phantom{0}8.3$\\
    & $\phantom{0}($-$90,\phantom{0}\phantom{0}0)$ & $3.3\,\pm\,1.0$ & $12.1\,\pm\,\phantom{0}3.3$ && & &\\
    \hline
  \end{tabular}
  \label{tab:ourdata1}
  \parbox[t]{1.5\textwidth}{\textit{Notes:} Offsets are measured with
    respect to the galaxy centre.}    
\end{table*}
%
%
% Observed spectra
%
\section{Integrated spectra of the 8 regions studied}
\label{sec:spec}
%
% Spectra of regions 1+2
%
\begin{figure*}
  \includegraphics[height=6.75cm,clip=,angle=270]{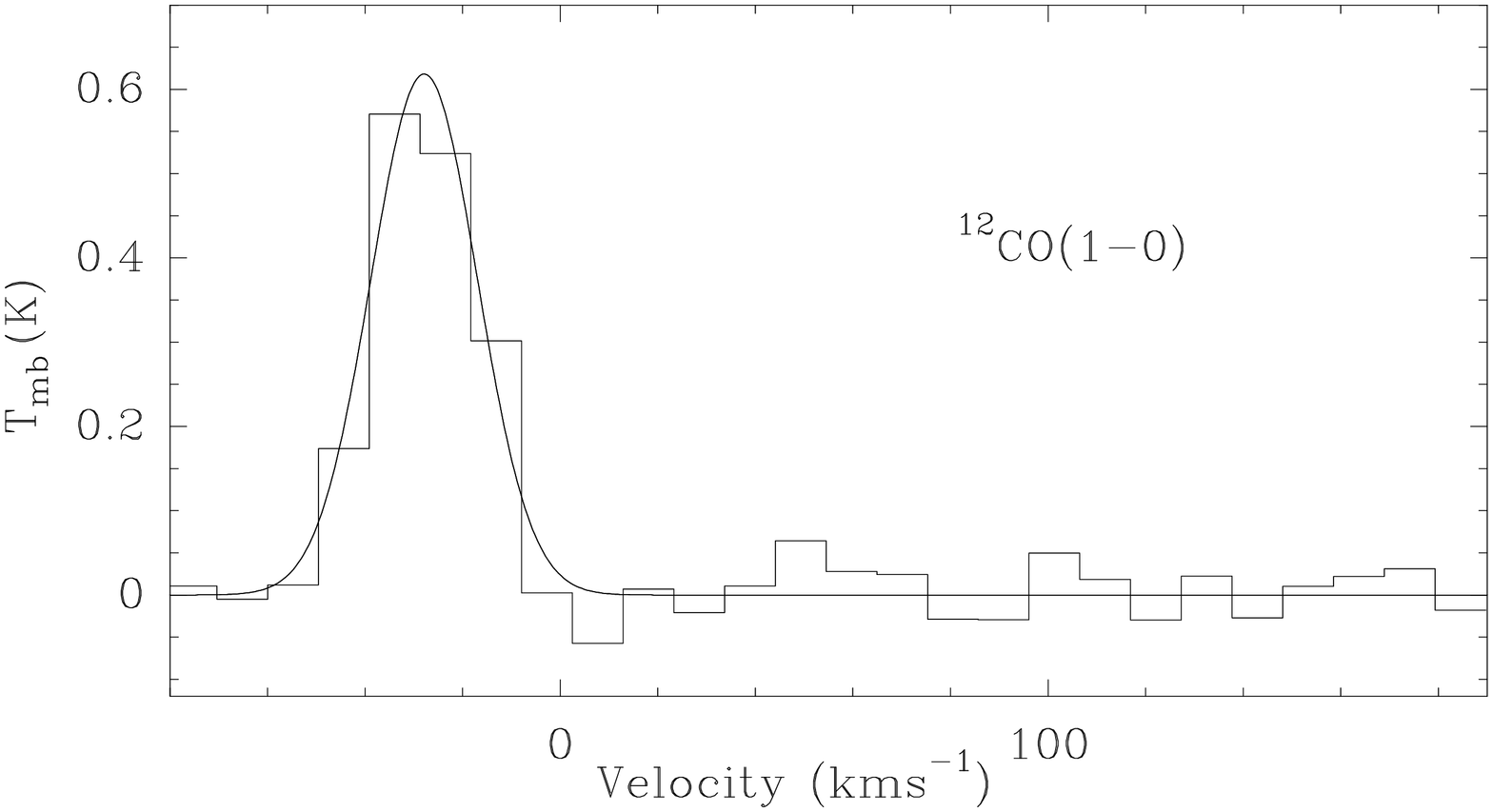}
  \includegraphics[height=6.75cm,clip=,angle=270]{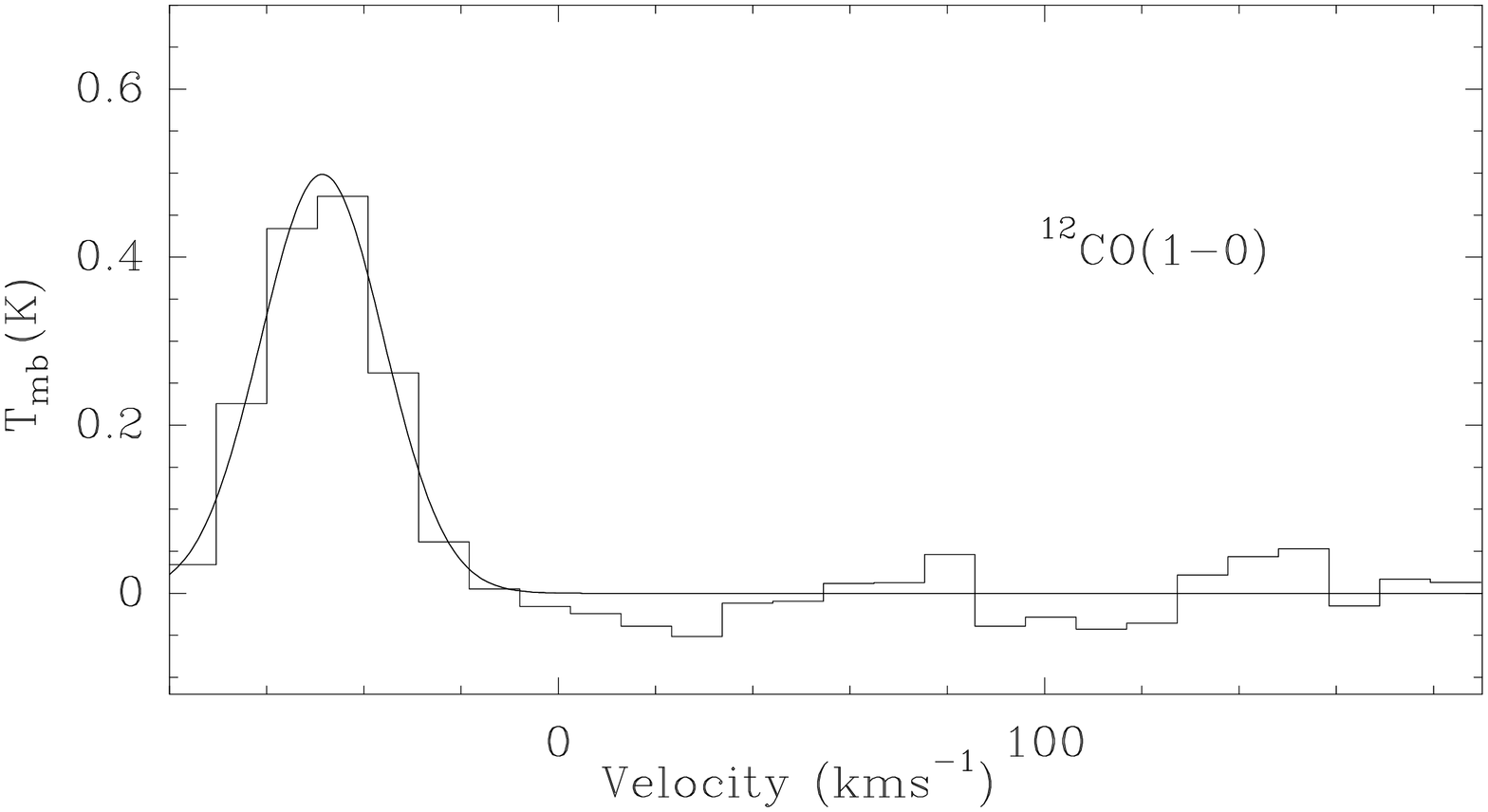}\\
  \includegraphics[height=6.75cm,clip=,angle=270]{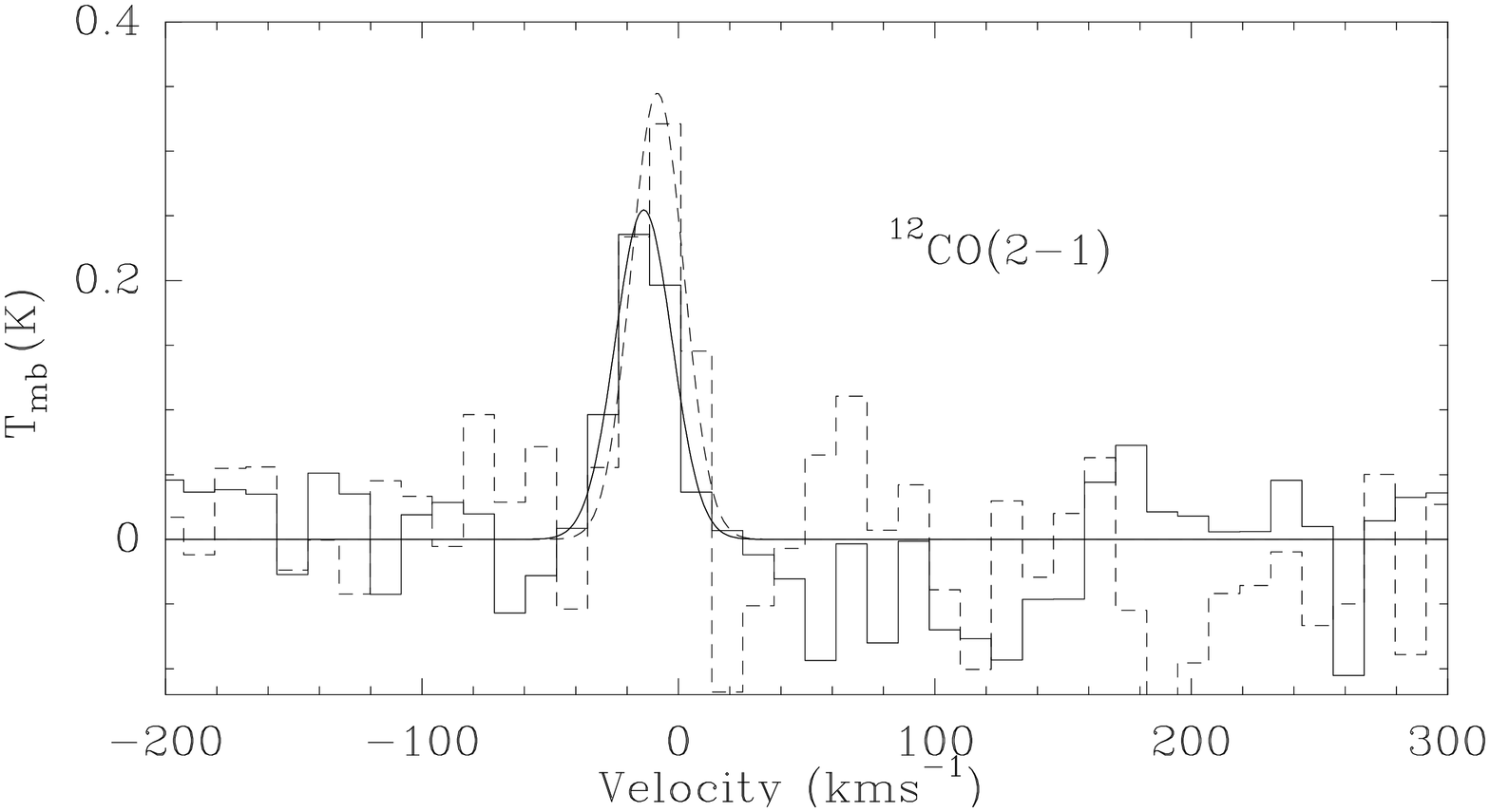}
  \includegraphics[height=6.75cm,clip=,angle=270]{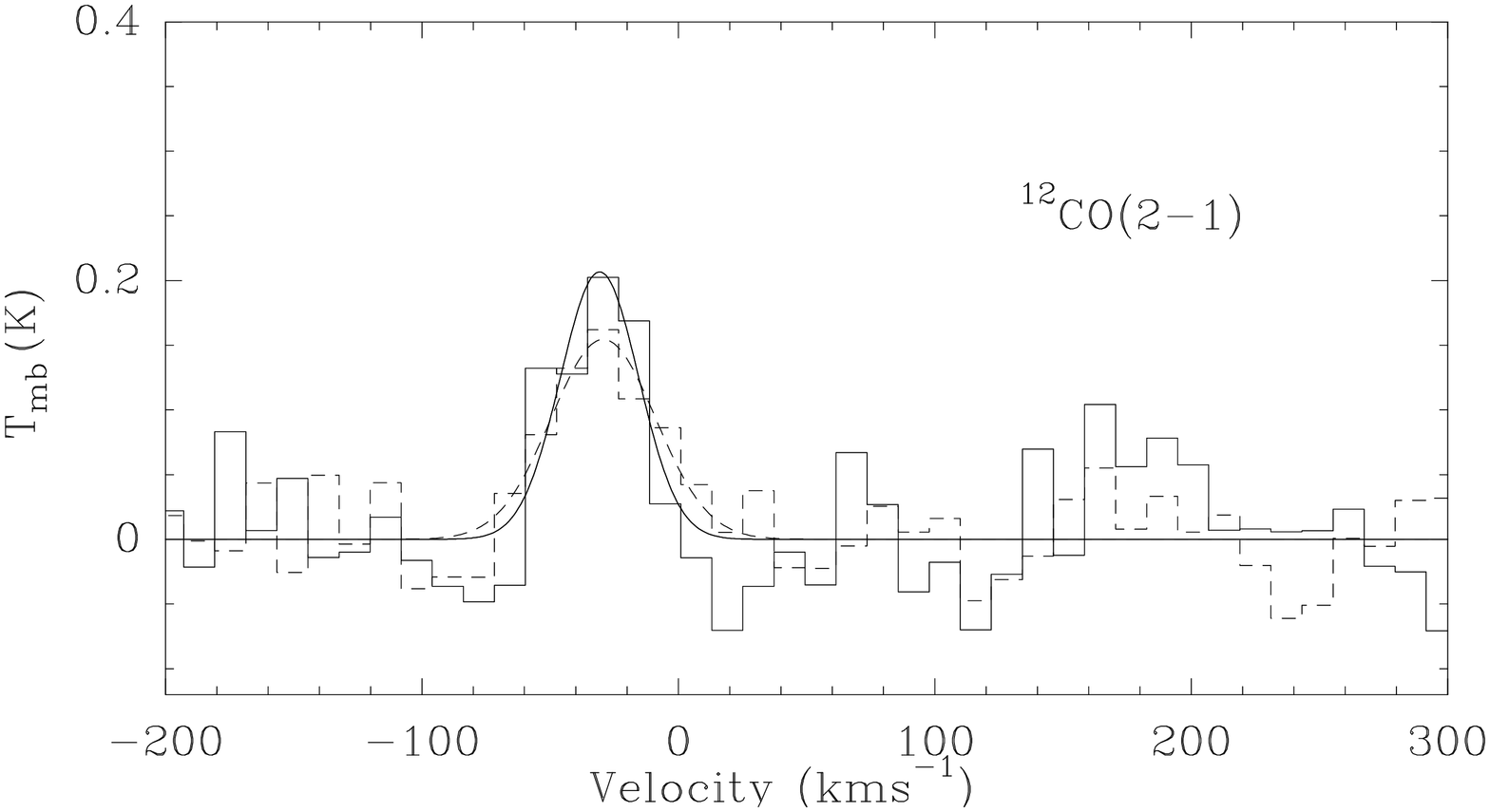}\\
  \includegraphics[height=6.75cm,clip=,angle=270]{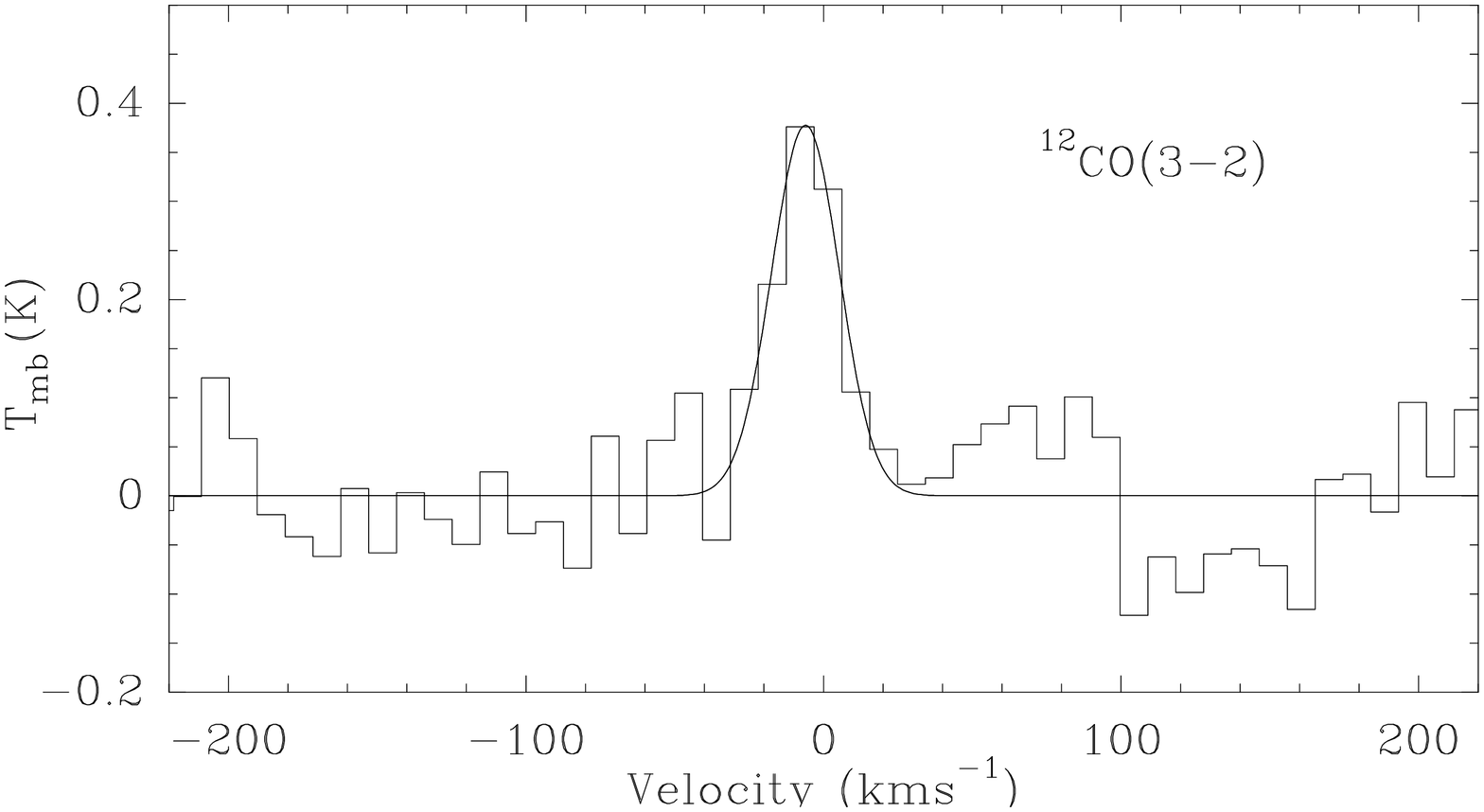}
  \includegraphics[height=6.75cm,clip=,angle=270]{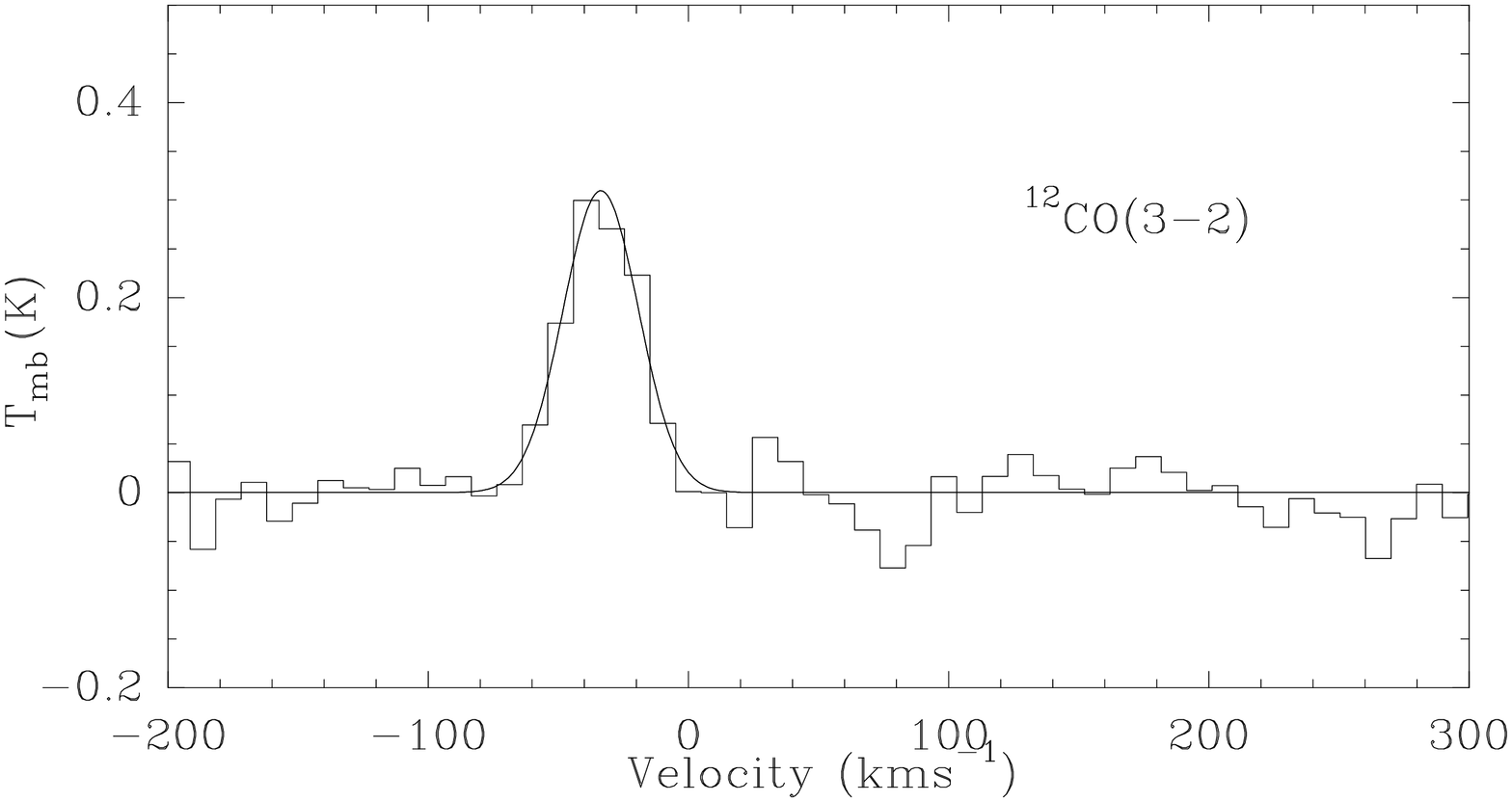}\\
  \includegraphics[height=6.75cm,clip=,angle=270]{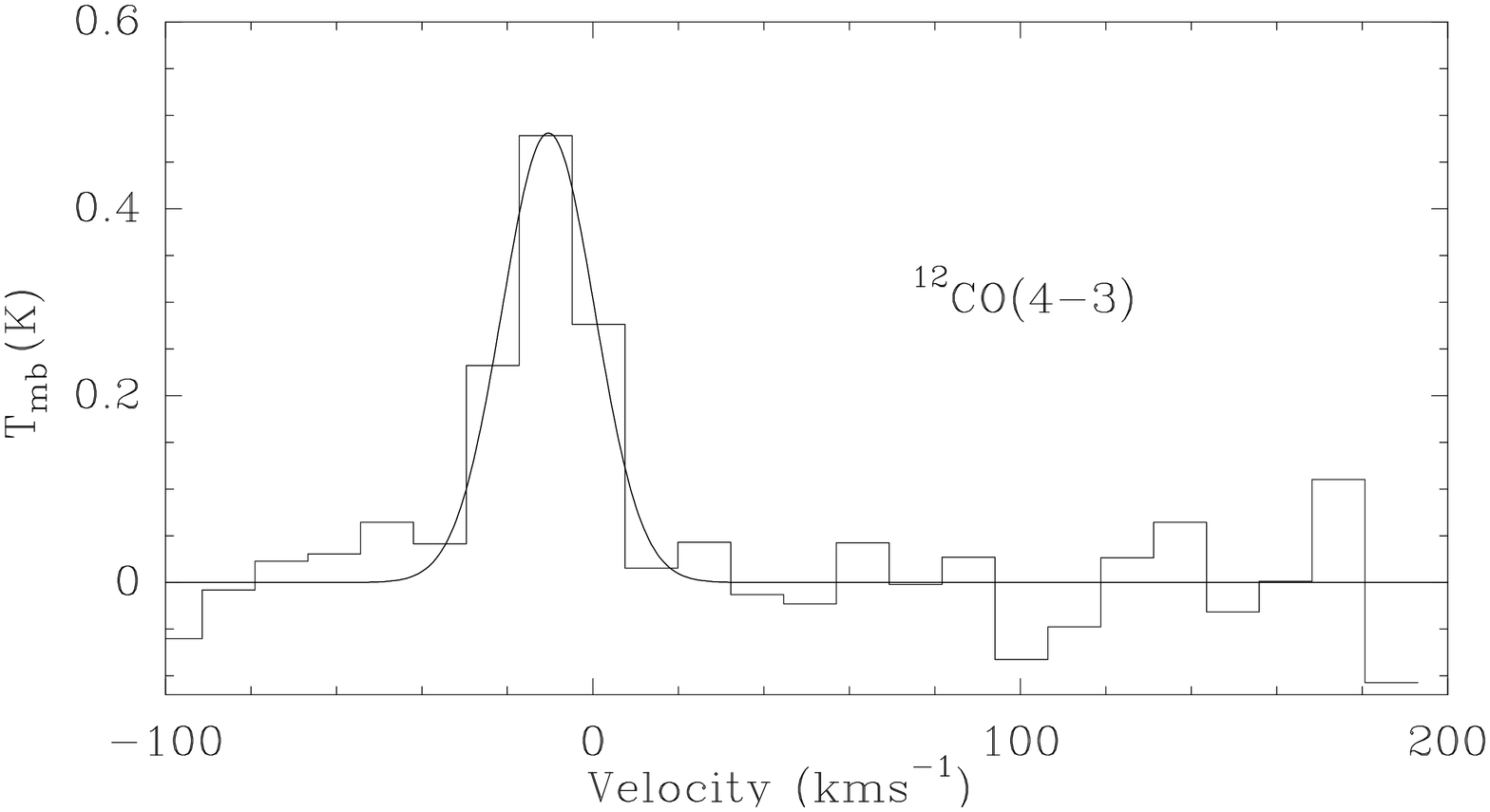}
  \includegraphics[height=6.75cm,clip=,angle=270]{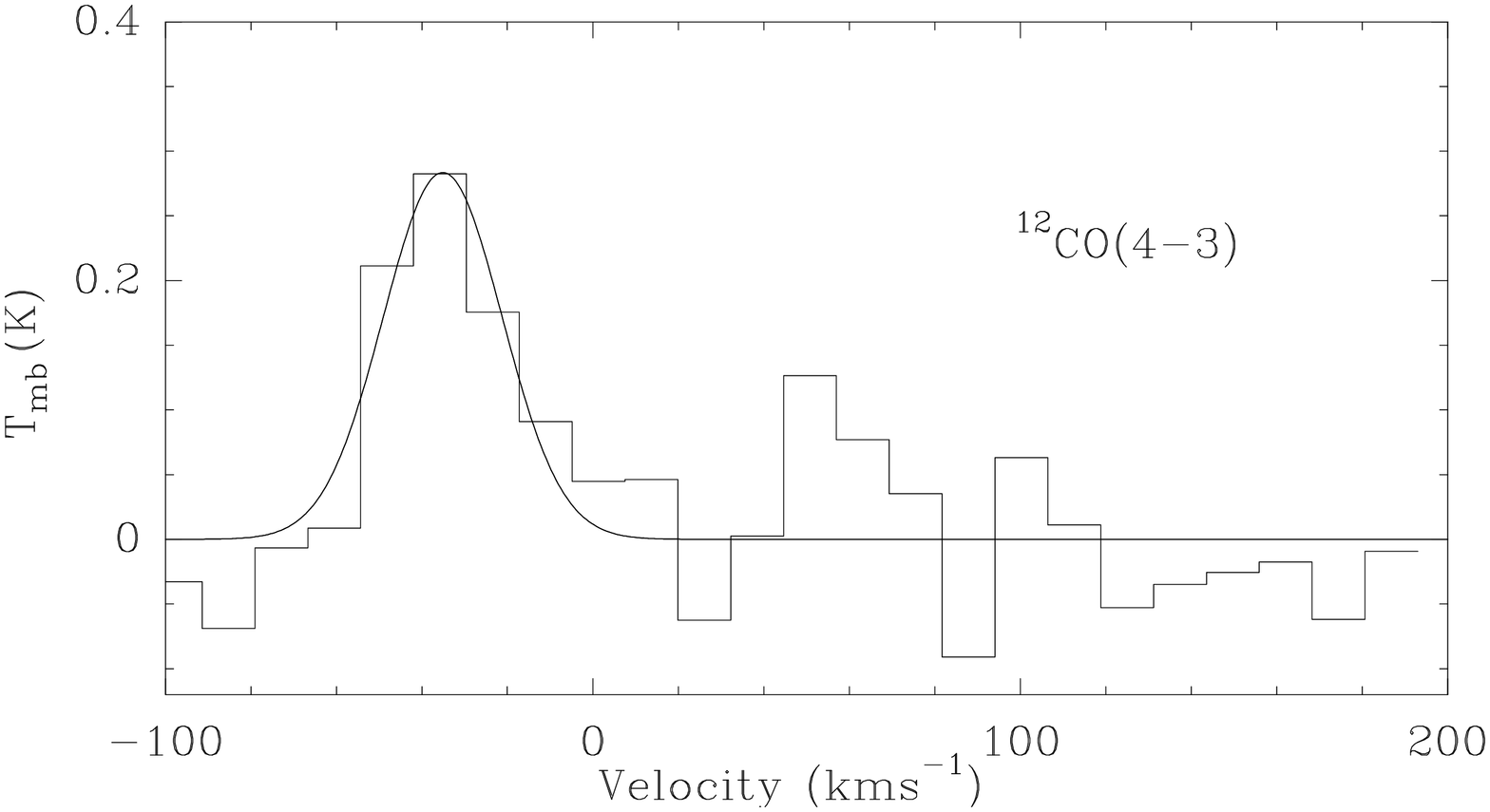}\\
  \includegraphics[height=6.75cm,clip=,angle=270]{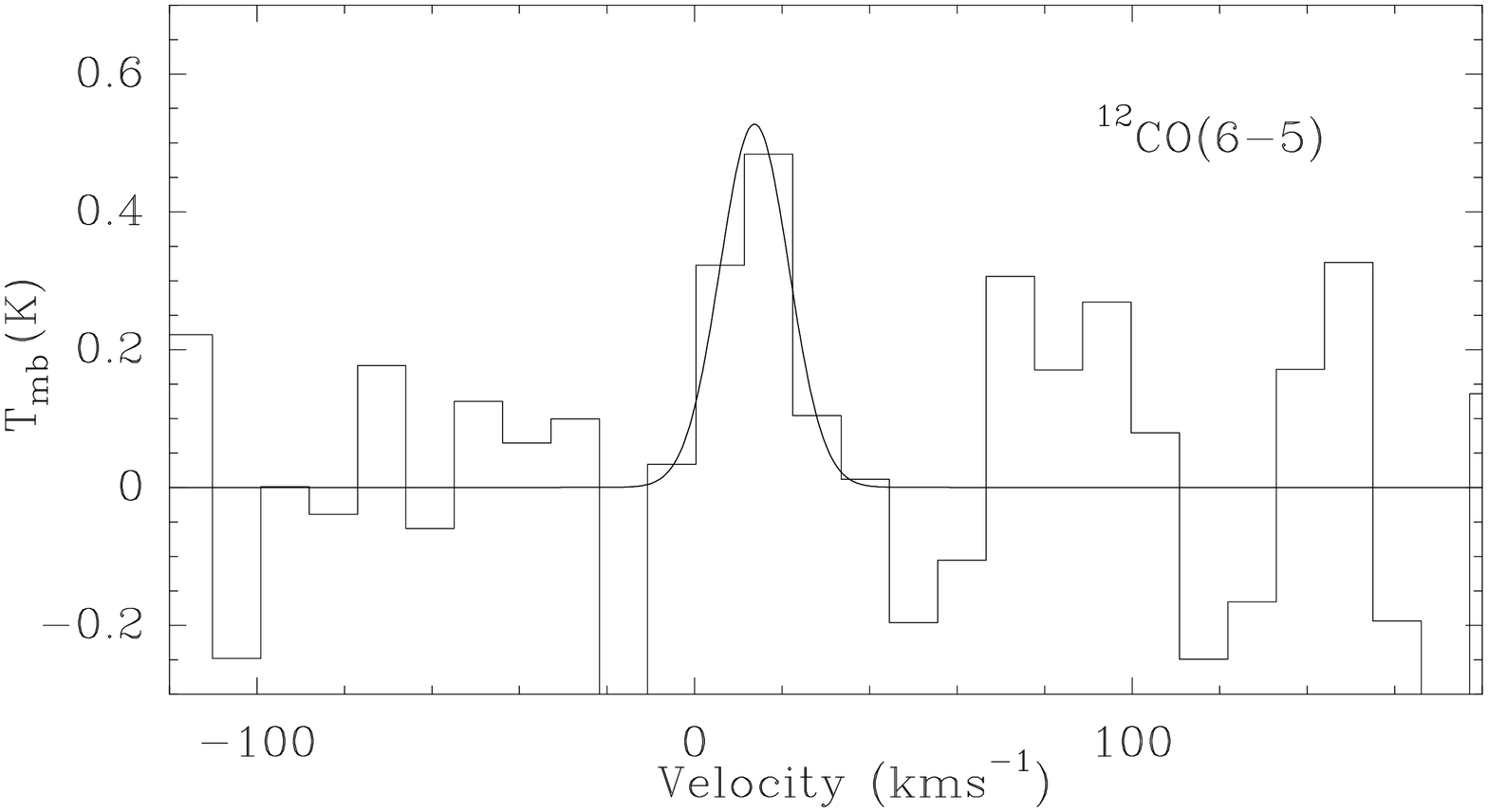}
  \includegraphics[height=6.75cm,clip=,angle=270]{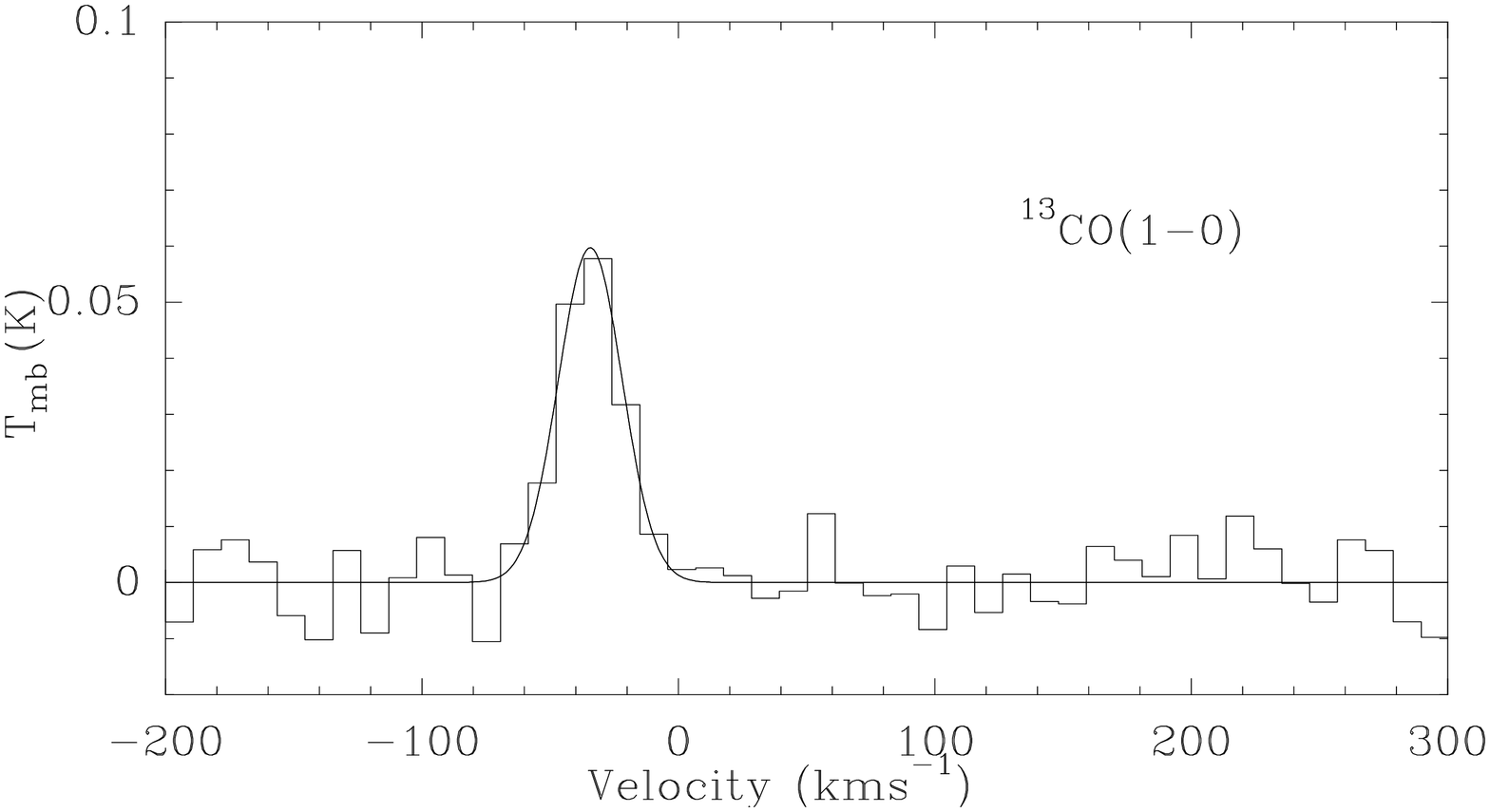}\\
  \includegraphics[height=13.6cm,clip=,angle=270]{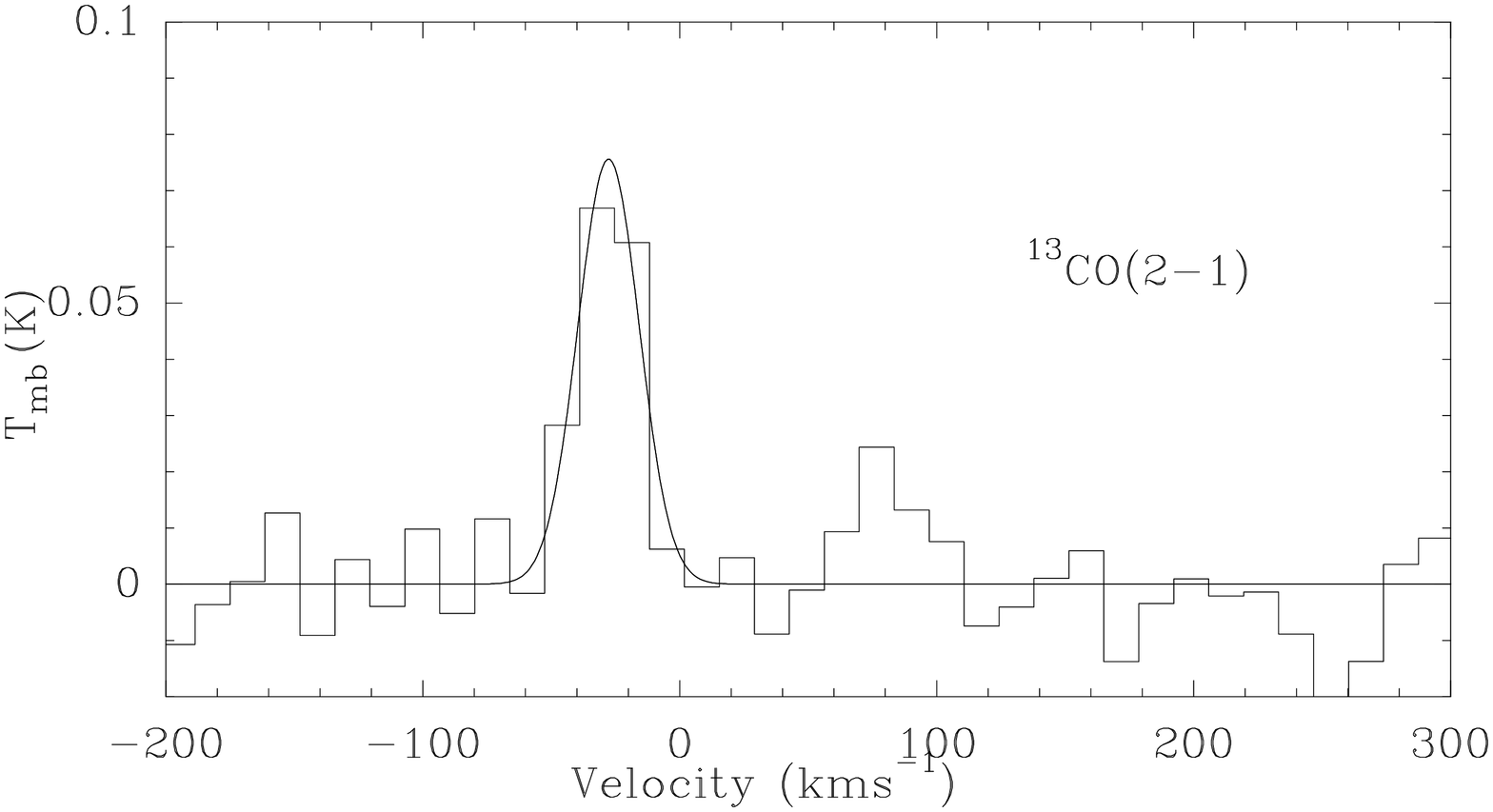}
  \caption{Integrated spectra used in the line ratio analysis for
    region~1 (left) and 2 (right). Gaussian fits are overlaid. The
    $^{12}$CO(2-1) line integrated intensity and line width in each region was
    obtained by interpolating those of the two closest detections of the
    large-scale mapping. These are shown as dashed and solid lines in
    the $^{12}$CO(2-1) panel.}
    \label{fig:spec12}
\end{figure*}

% 
% Spectra of regions 3+4
%
\addtocounter{figure}{-1}
\begin{figure*}
  \includegraphics[height=6.75cm,clip=,angle=270]{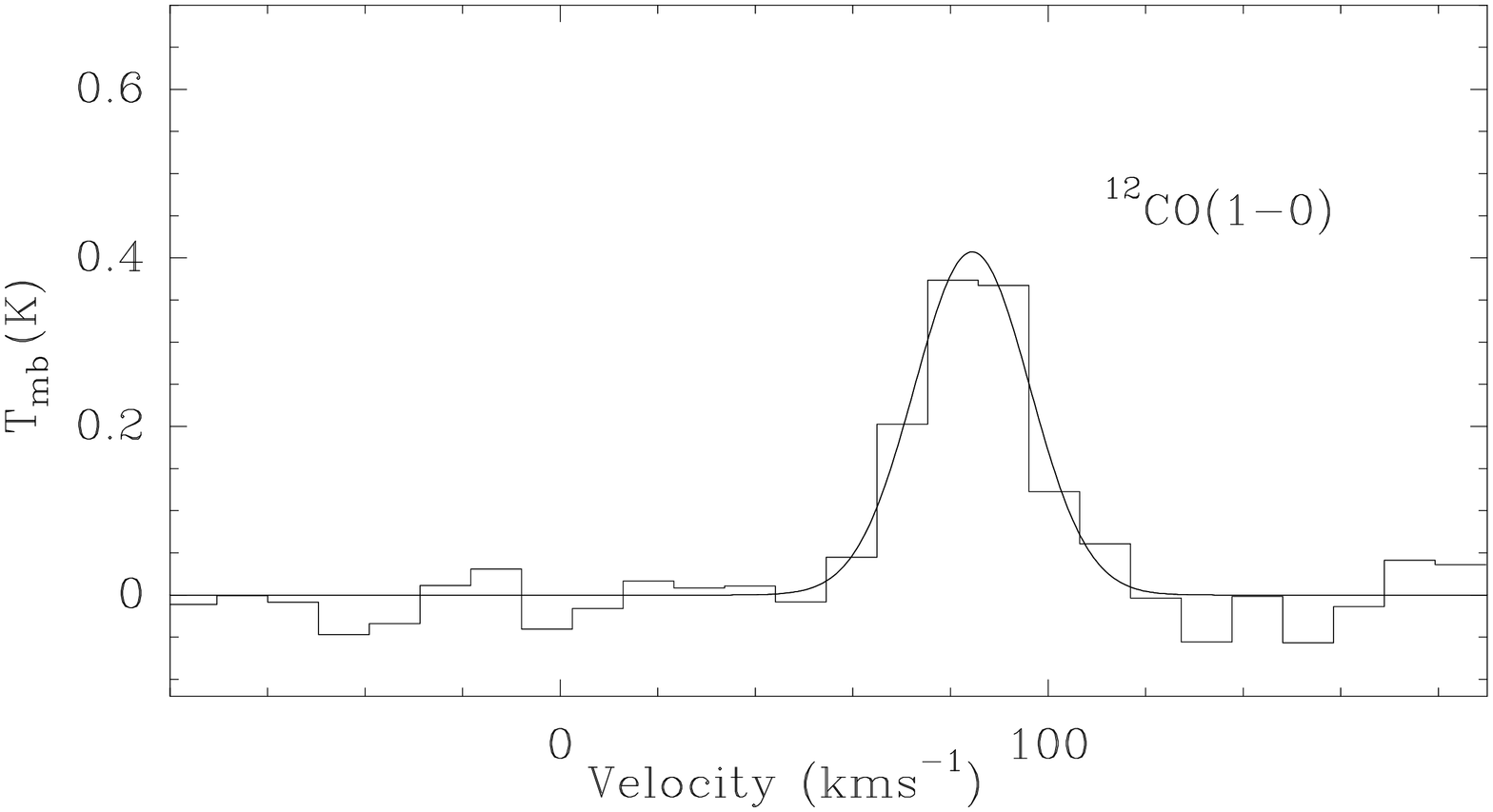}
  \includegraphics[height=6.75cm,clip=,angle=270]{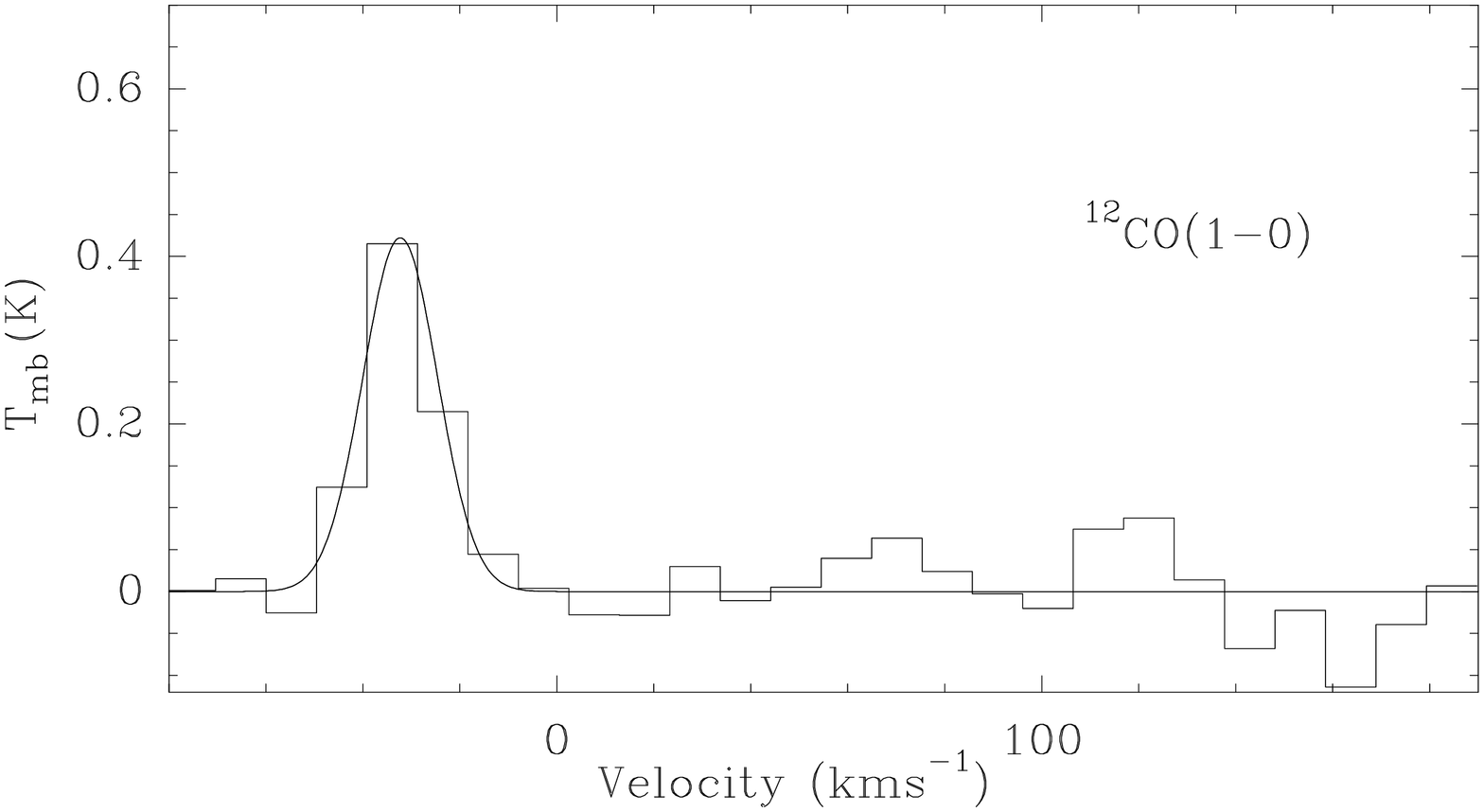}\\
  \includegraphics[height=6.75cm,clip=,angle=270]{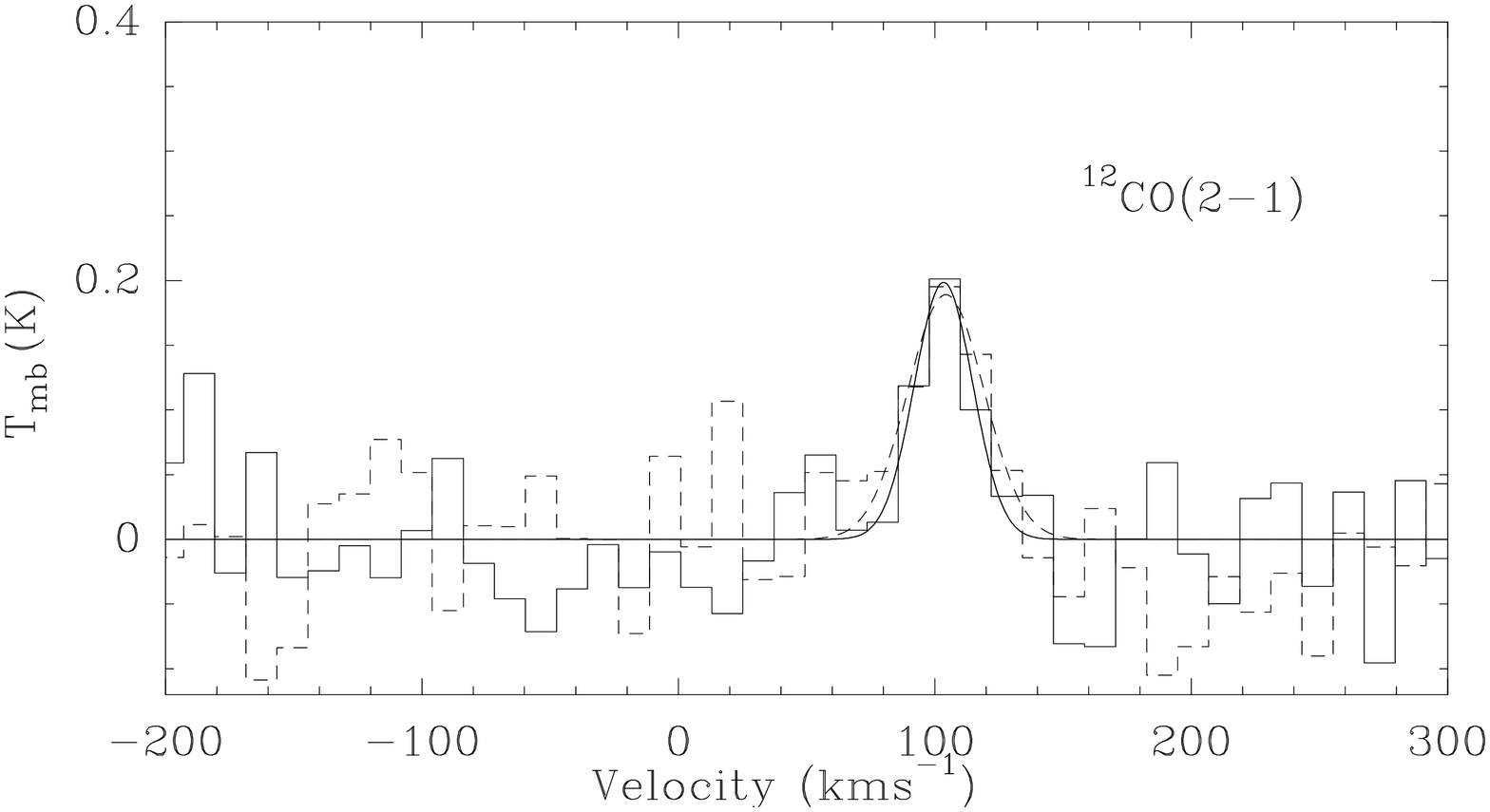}
  \includegraphics[height=6.75cm,clip=,angle=270]{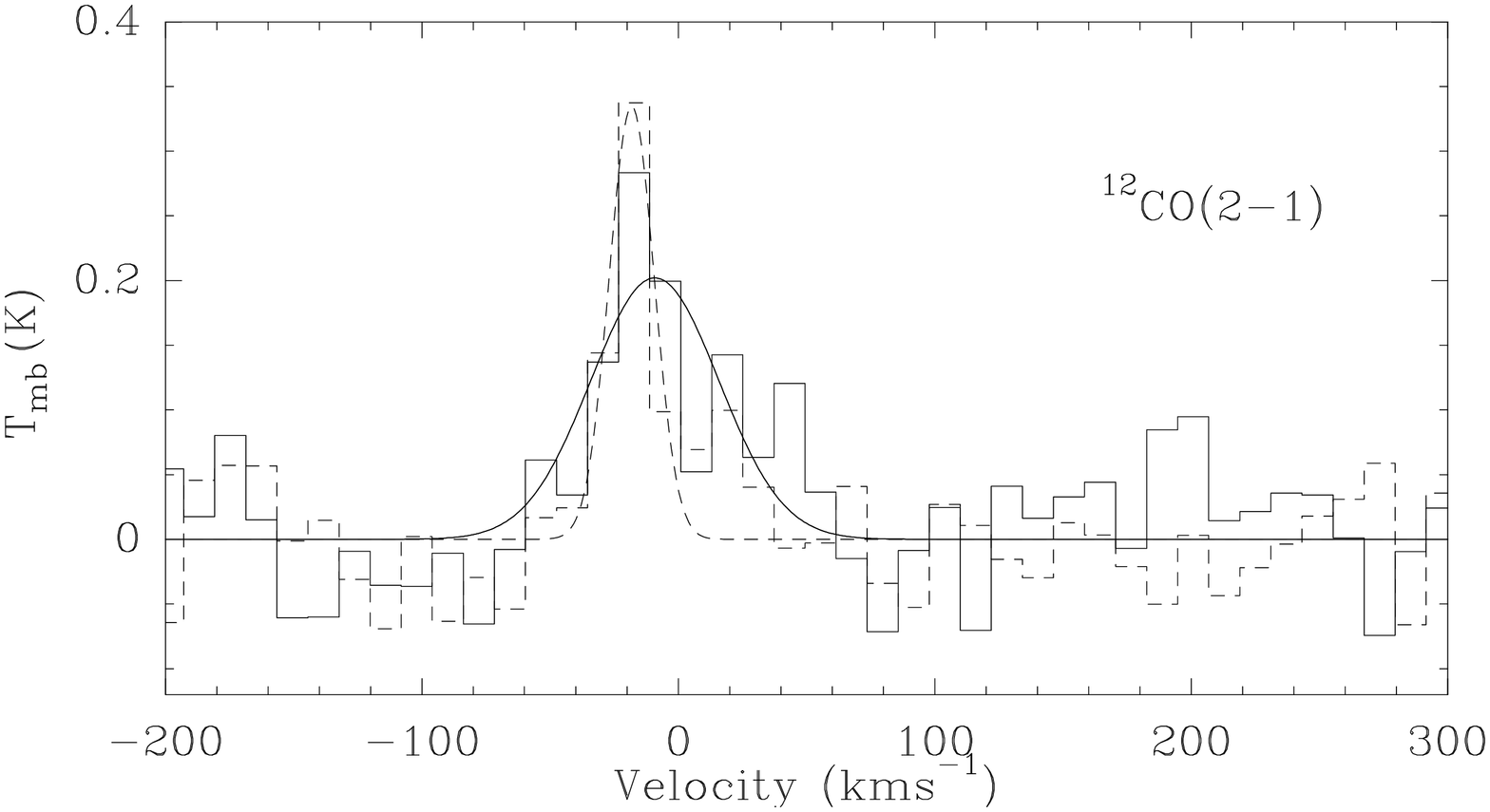}\\
  \includegraphics[height=6.75cm,clip=,angle=270]{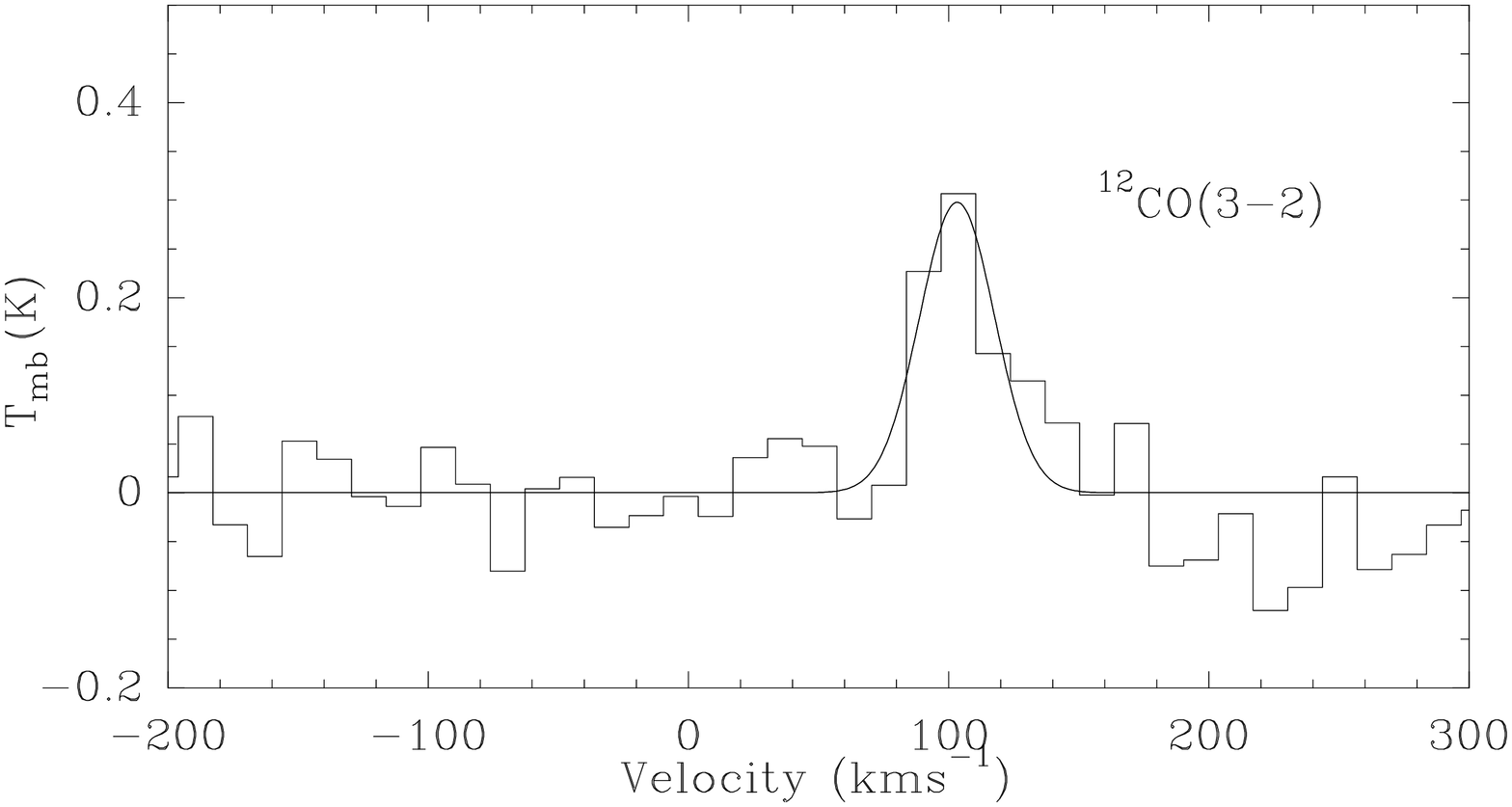}
  \includegraphics[height=6.75cm,clip=,angle=270]{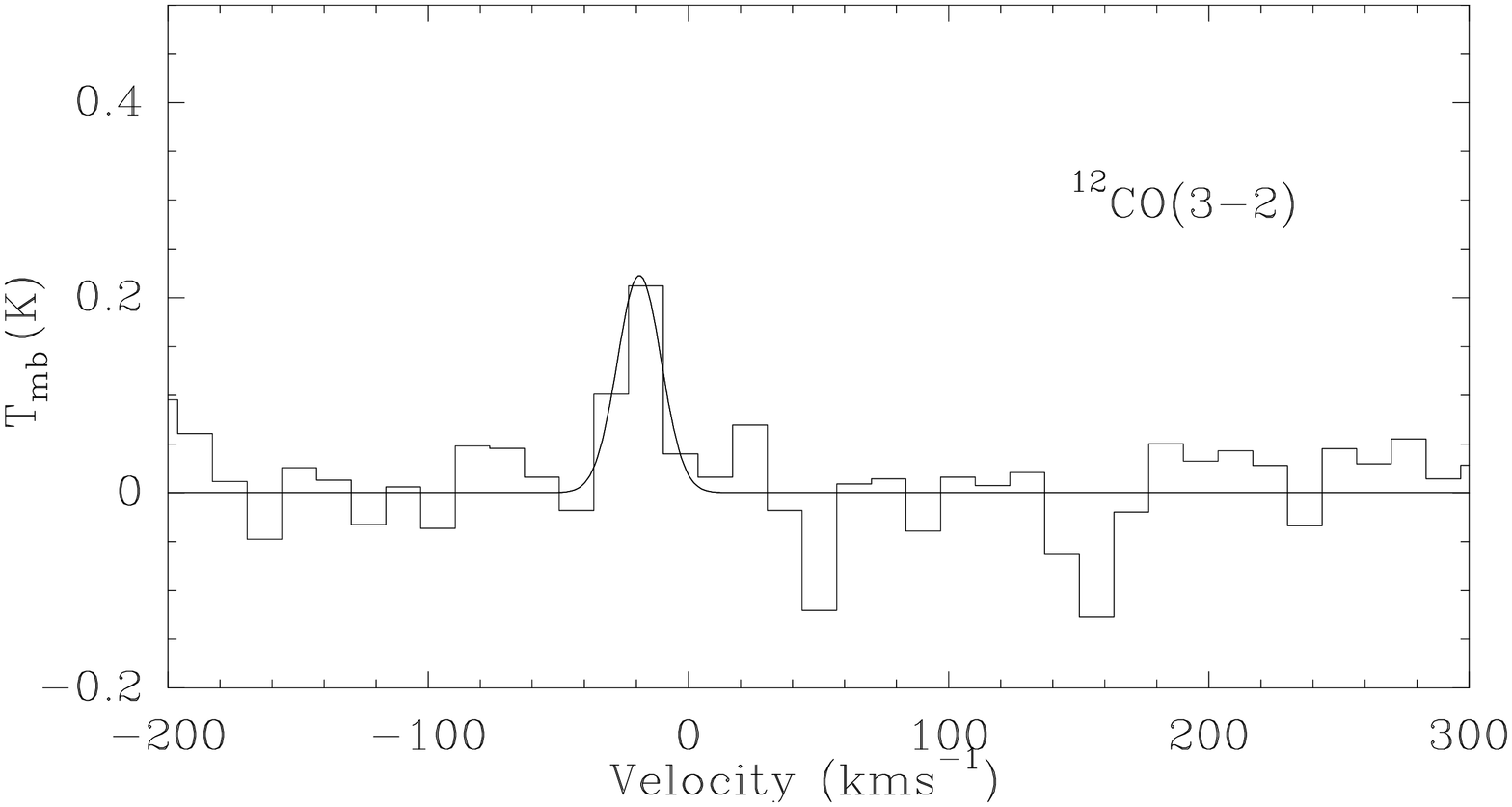}\\
  \includegraphics[height=6.75cm,clip=,angle=270]{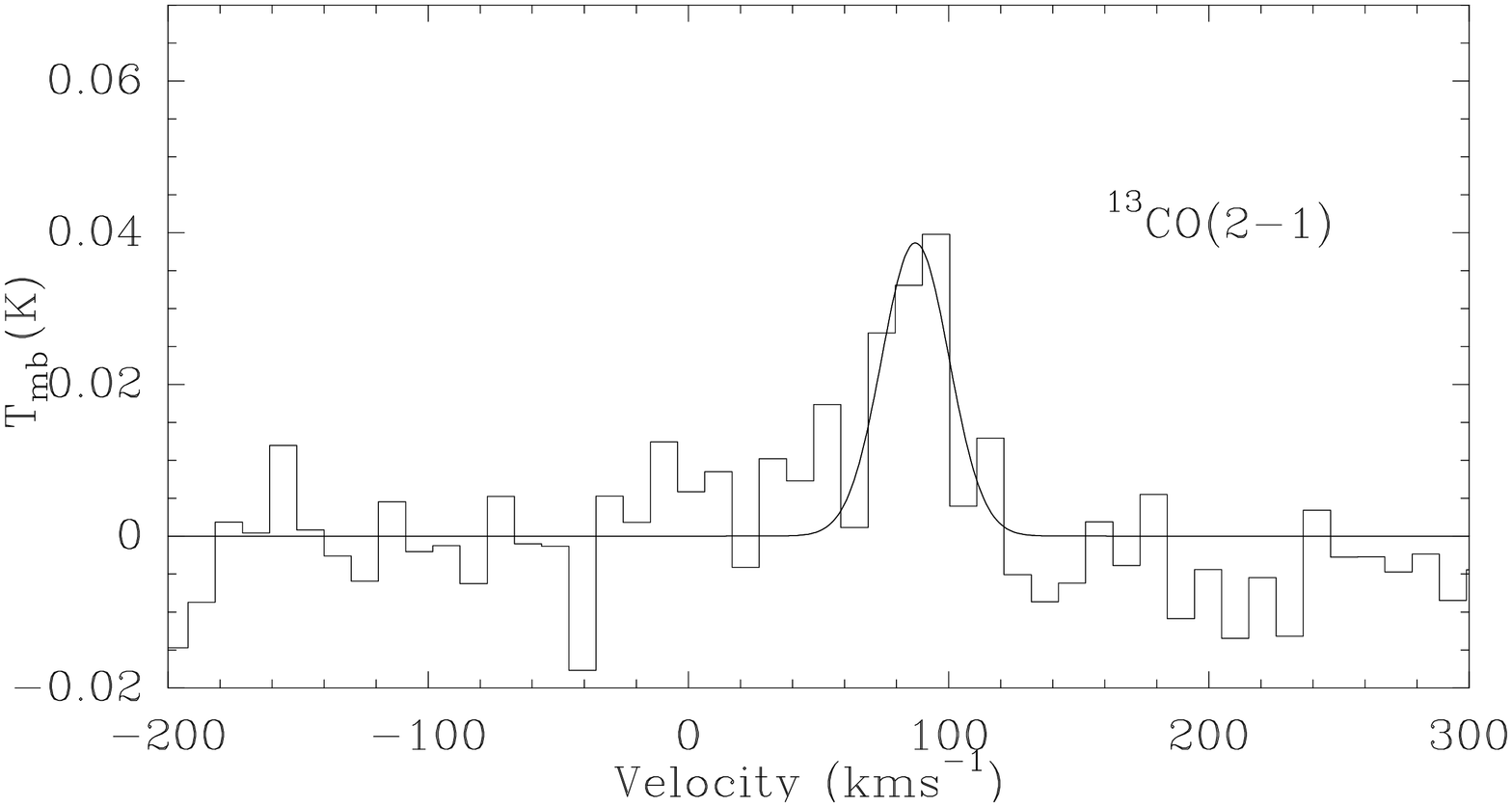}
  \includegraphics[height=6.75cm,clip=,angle=270]{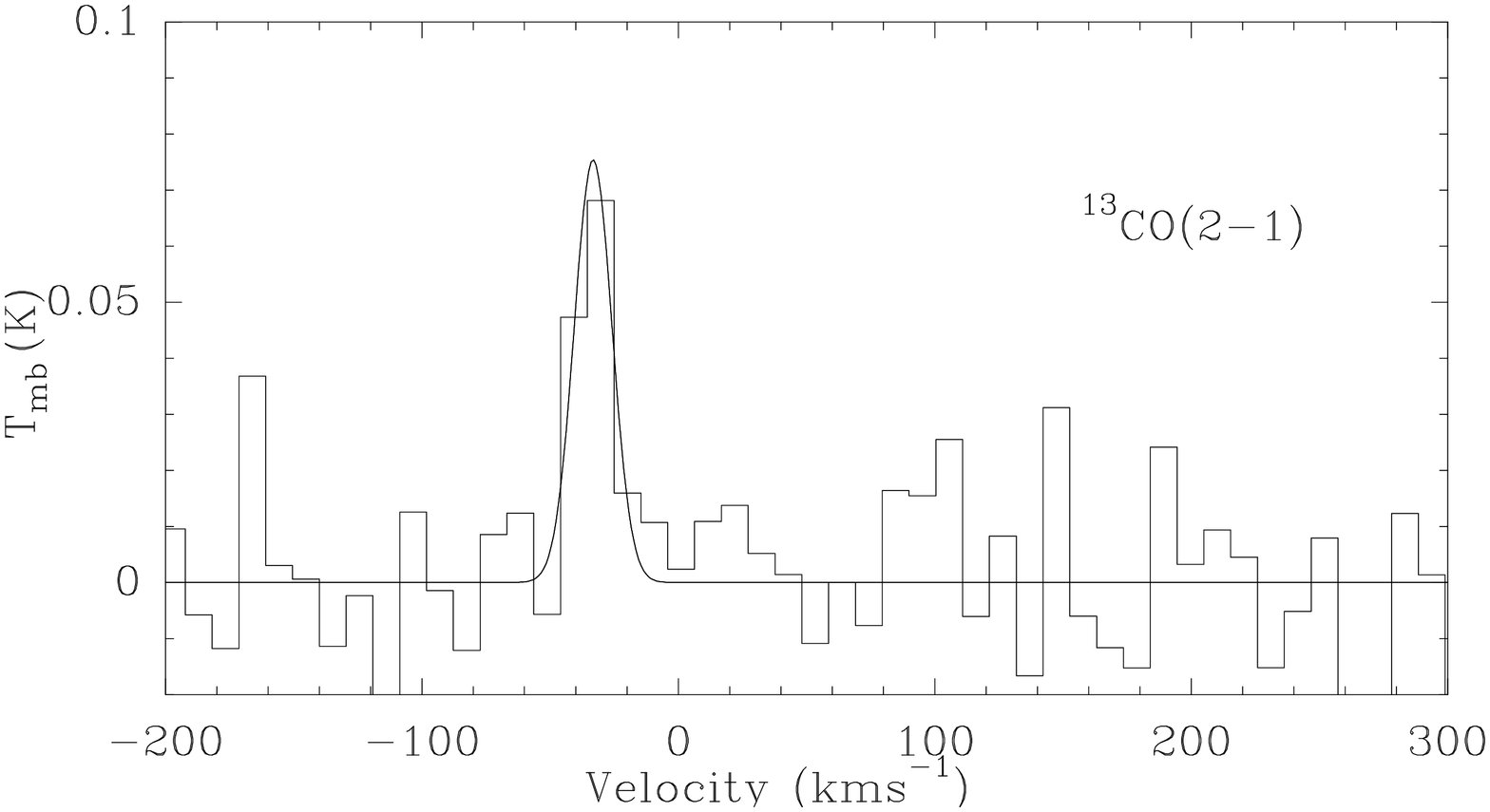}
  \caption{{\bf (Continued.)} Region~3 (left) and 4 (right).}
  \label{fig:spec34}
\end{figure*}

% 
% Spectra of regions 5+6
%
\addtocounter{figure}{-1}
\begin{figure*}
  \includegraphics[height=6.75cm,clip=,angle=270]{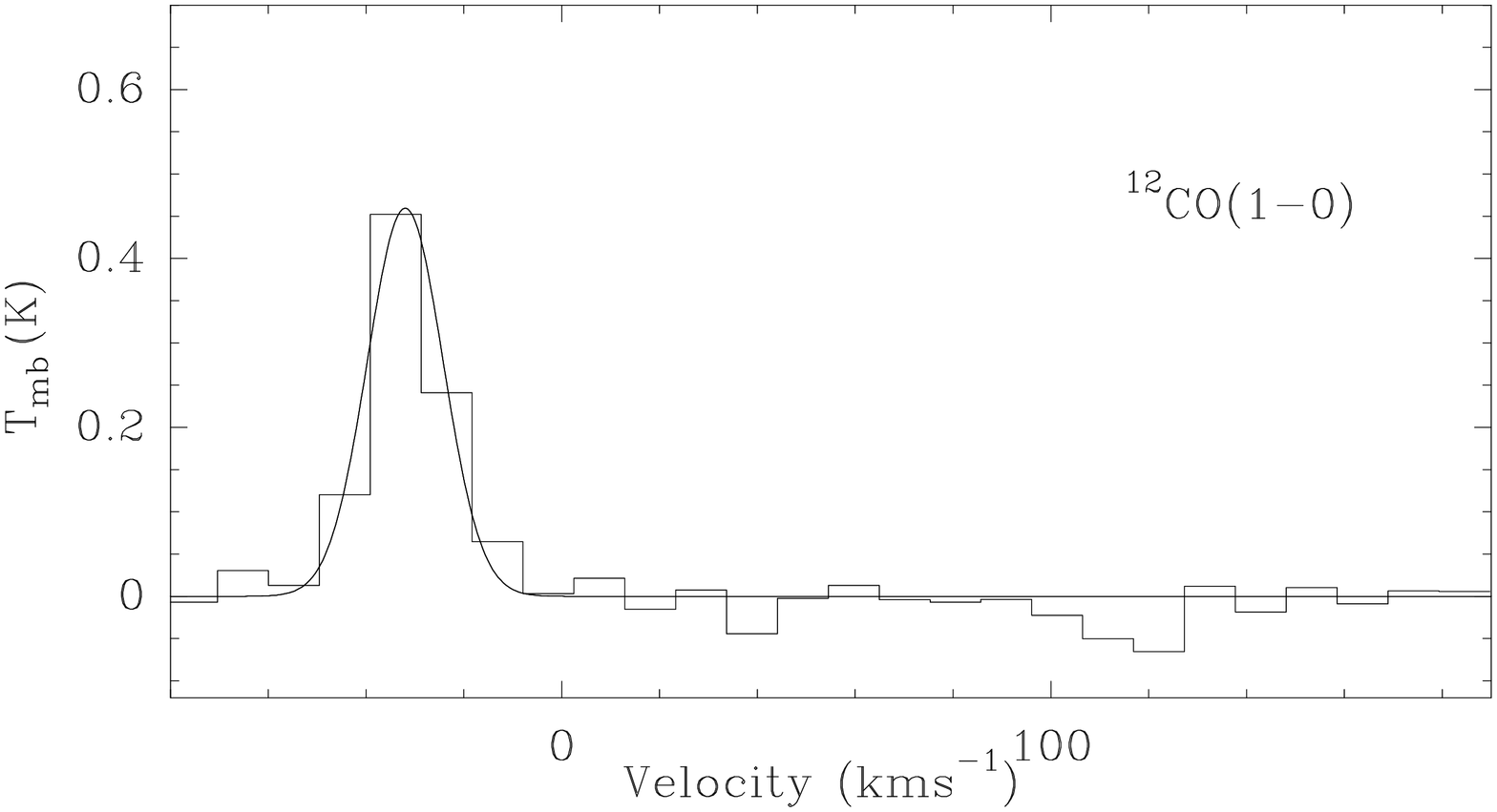}
  \includegraphics[height=6.75cm,clip=,angle=270]{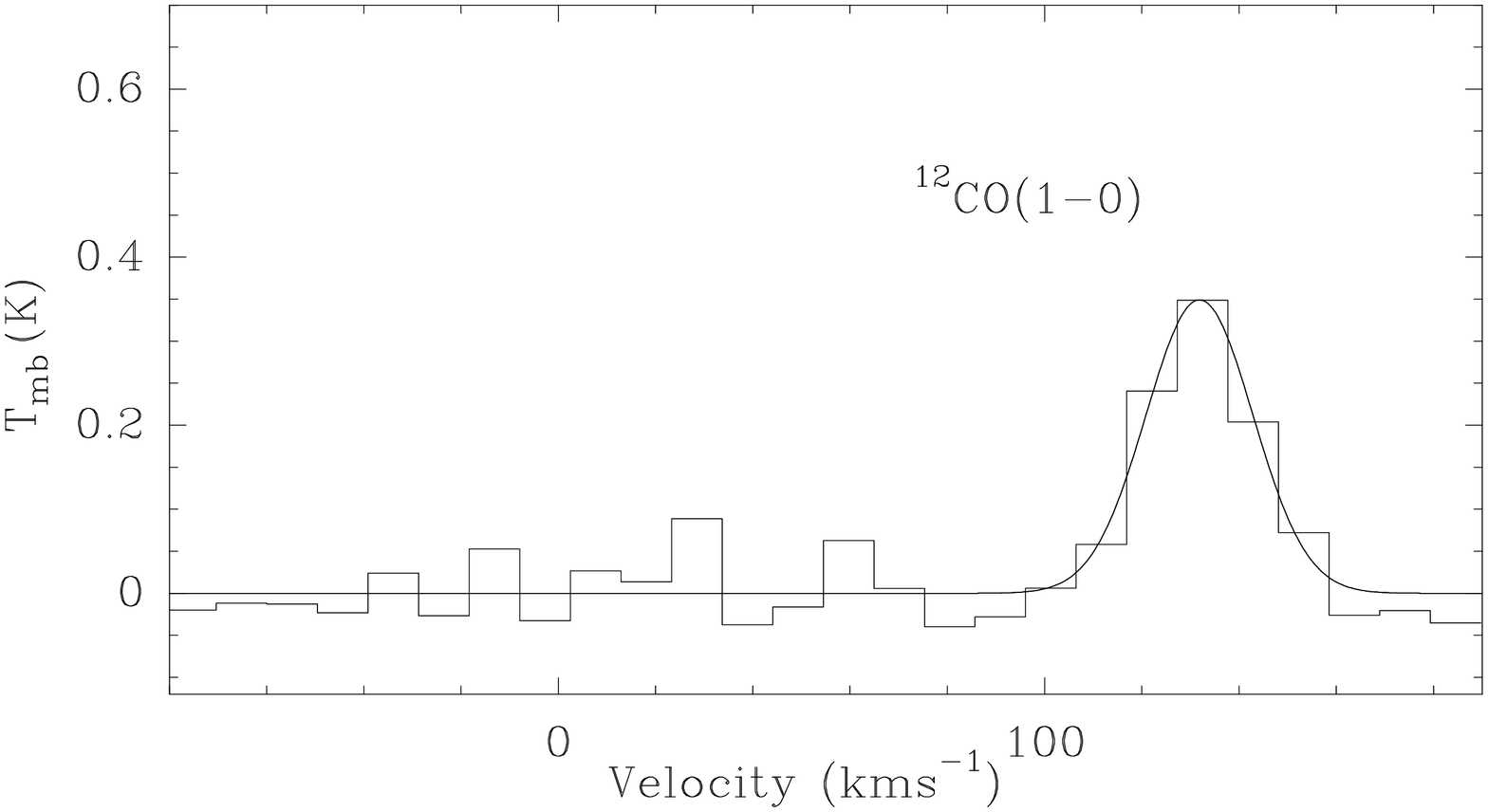}\\
  \includegraphics[height=6.75cm,clip=,angle=270]{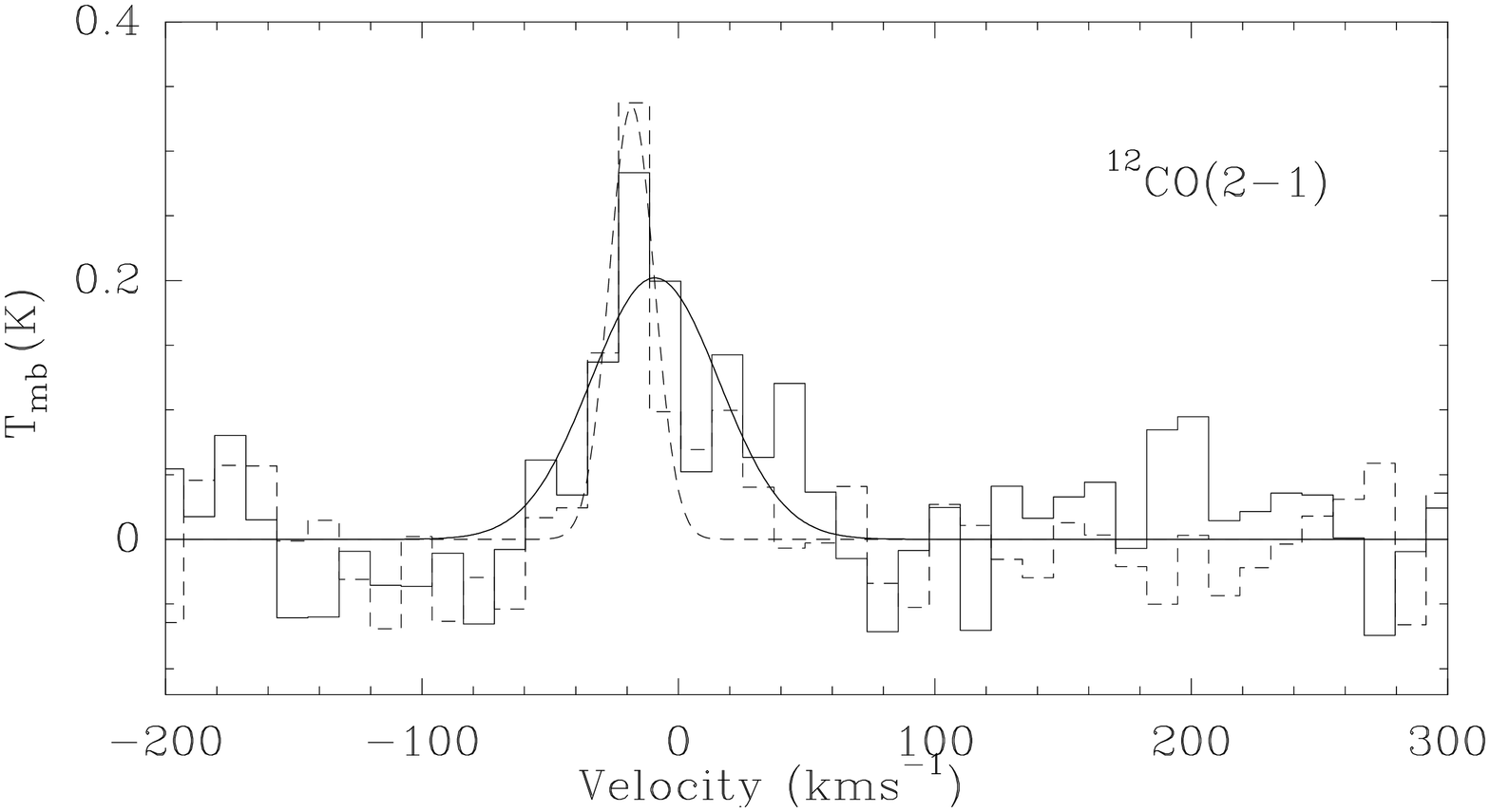}
  \includegraphics[height=6.75cm,clip=,angle=270]{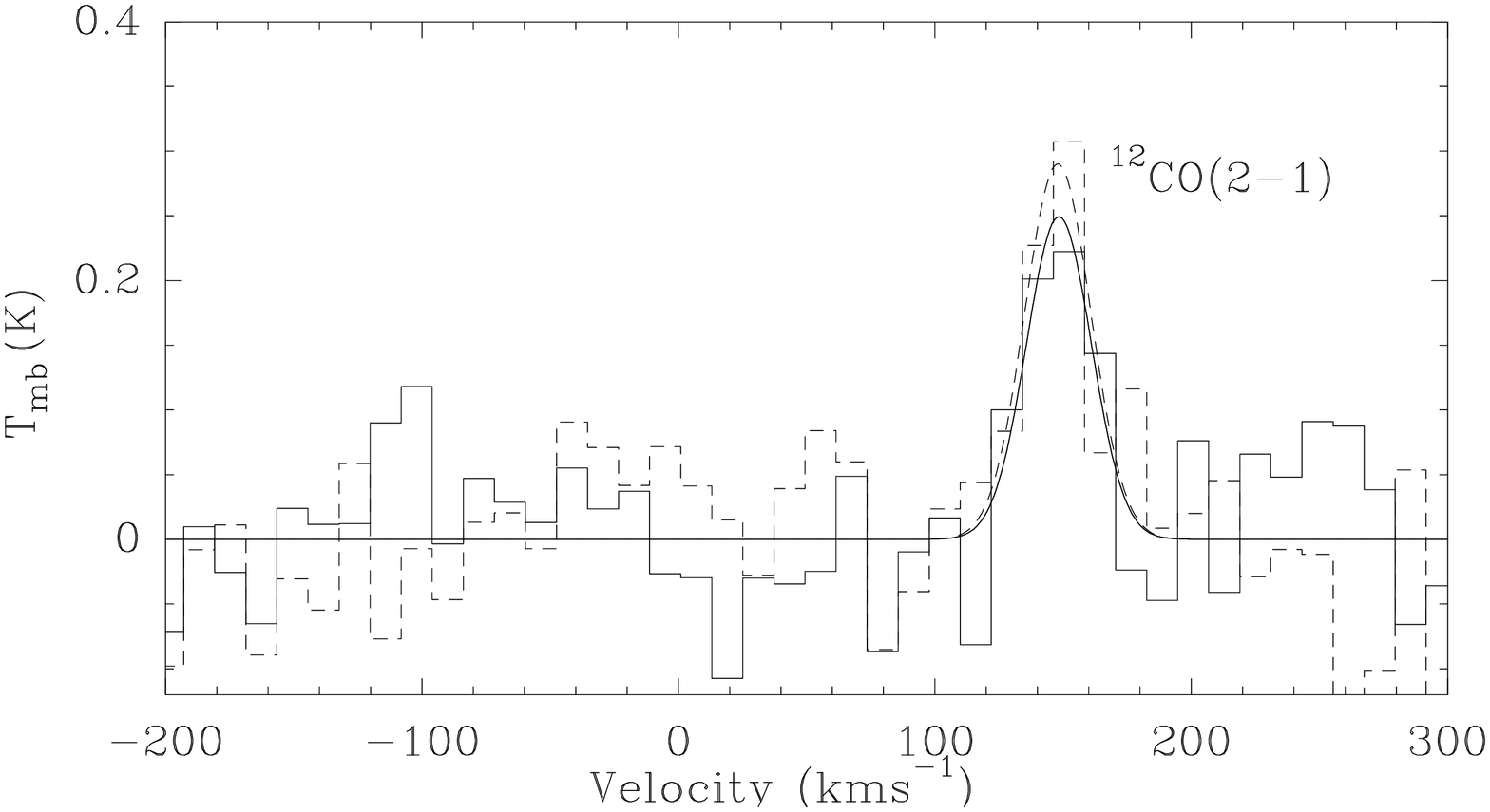}\\
  \includegraphics[height=6.75cm,clip=,angle=270]{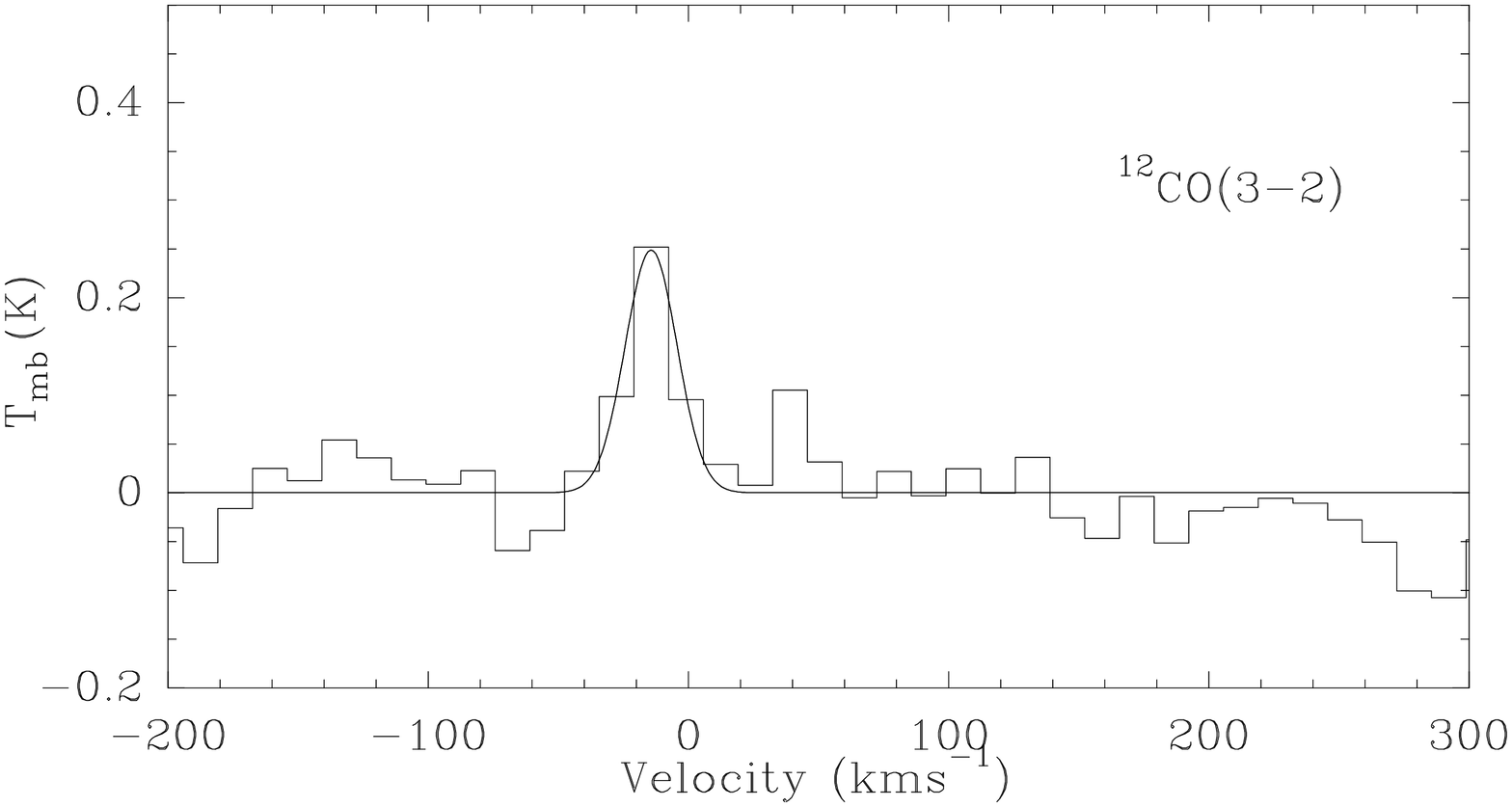}
  \includegraphics[height=6.75cm,clip=,angle=270]{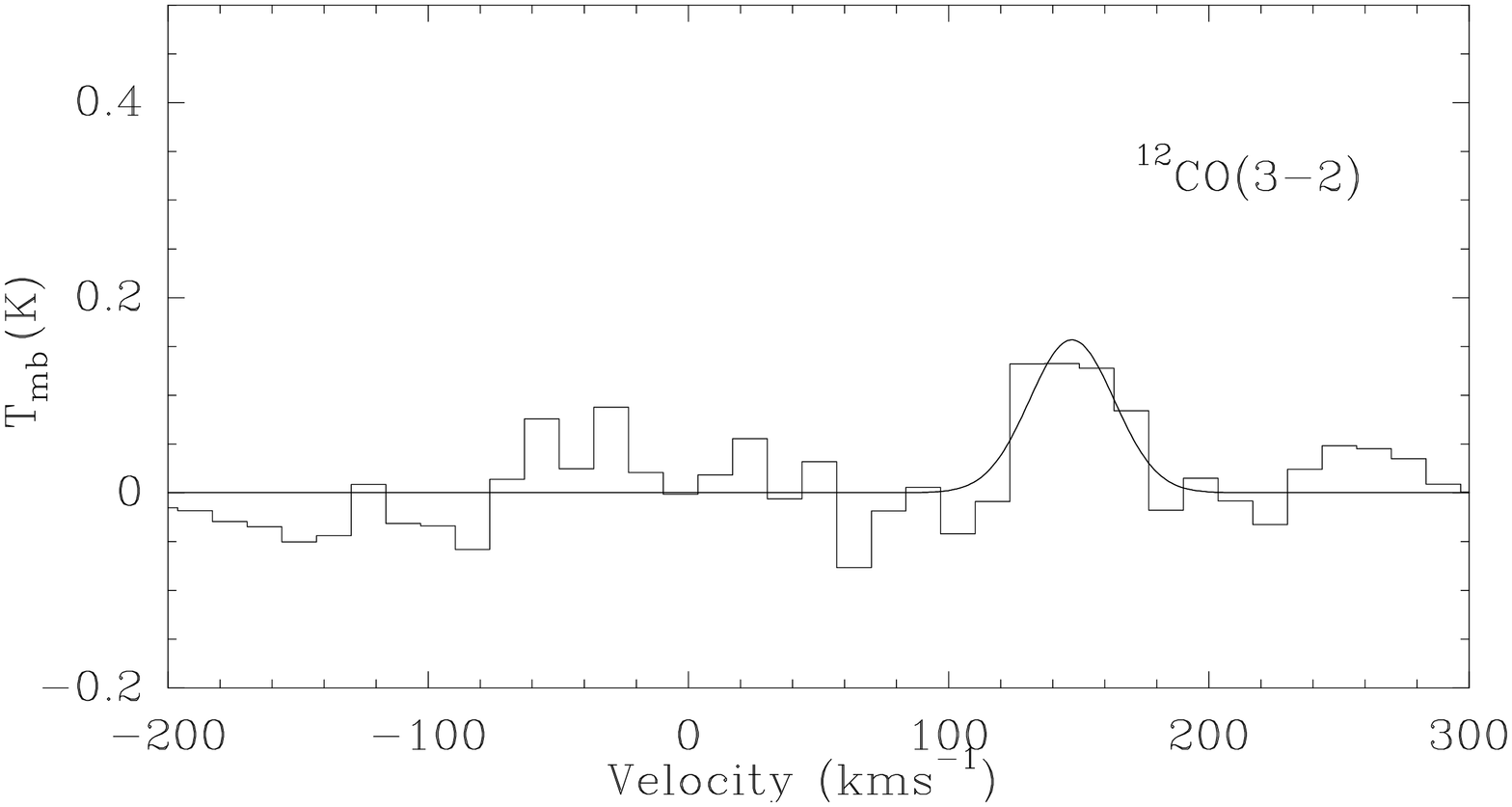}\\
  \includegraphics[height=6.75cm,clip=,angle=270]{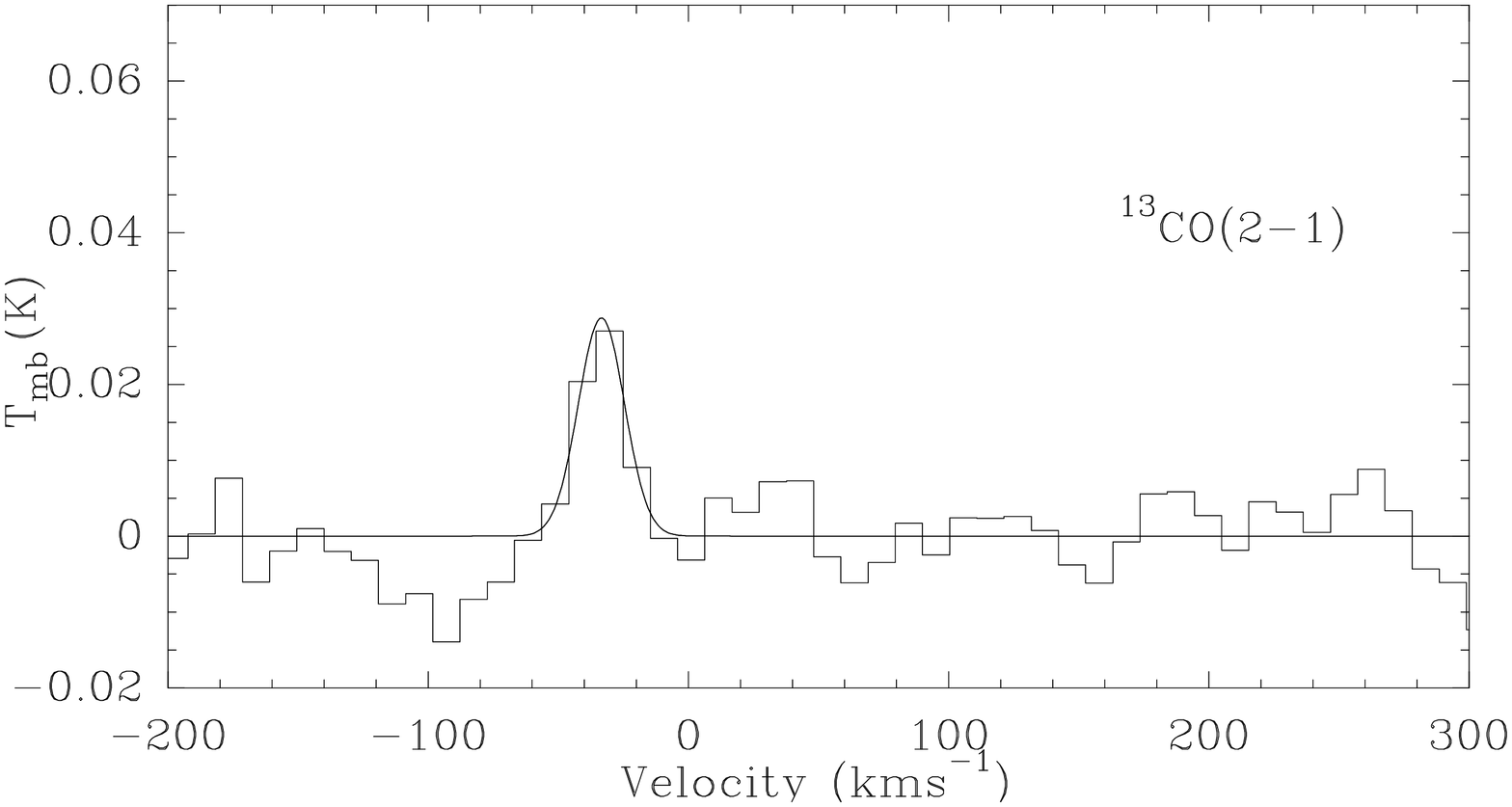}
  \includegraphics[height=6.75cm,clip=,angle=270]{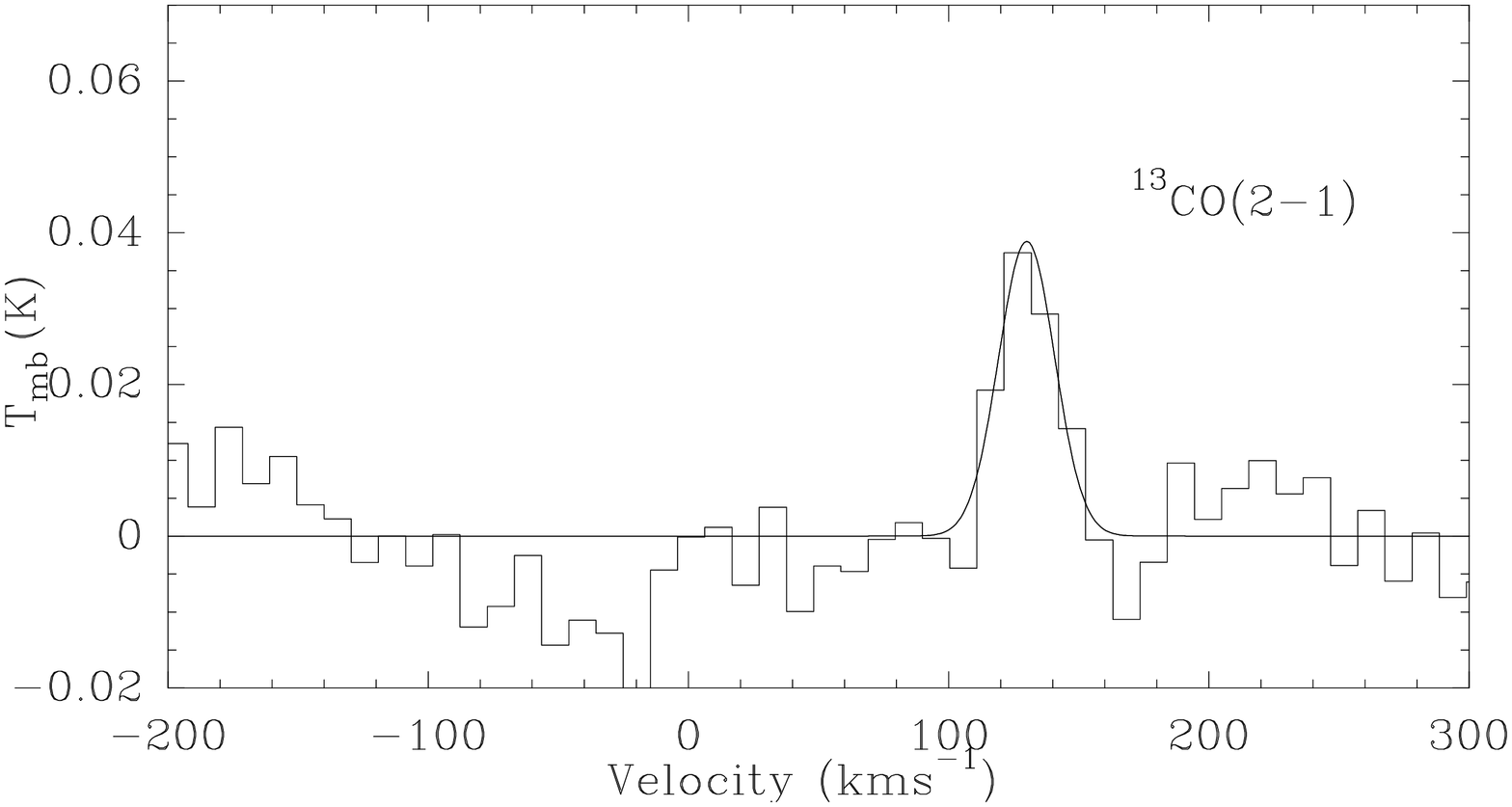}
  \caption{{\bf (Continued.)} Region~5 (left) and 6 (right).}
  \label{fig:spec56}
\end{figure*}

% 
% Spectra of regions 7+8
%
\addtocounter{figure}{-1}
\begin{figure*}
  \includegraphics[height=6.75cm,clip=,angle=270]{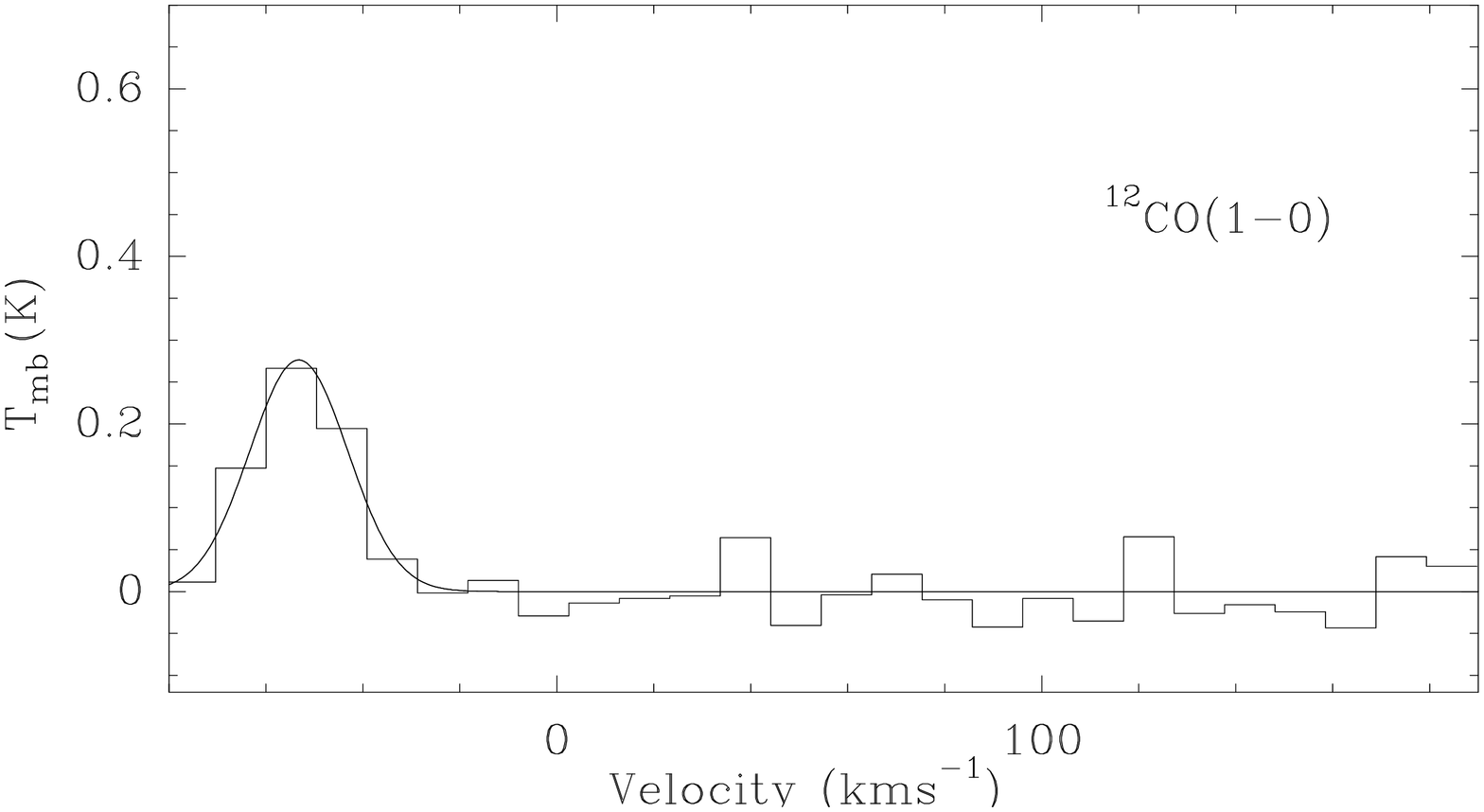}
  \includegraphics[height=6.75cm,clip=,angle=270]{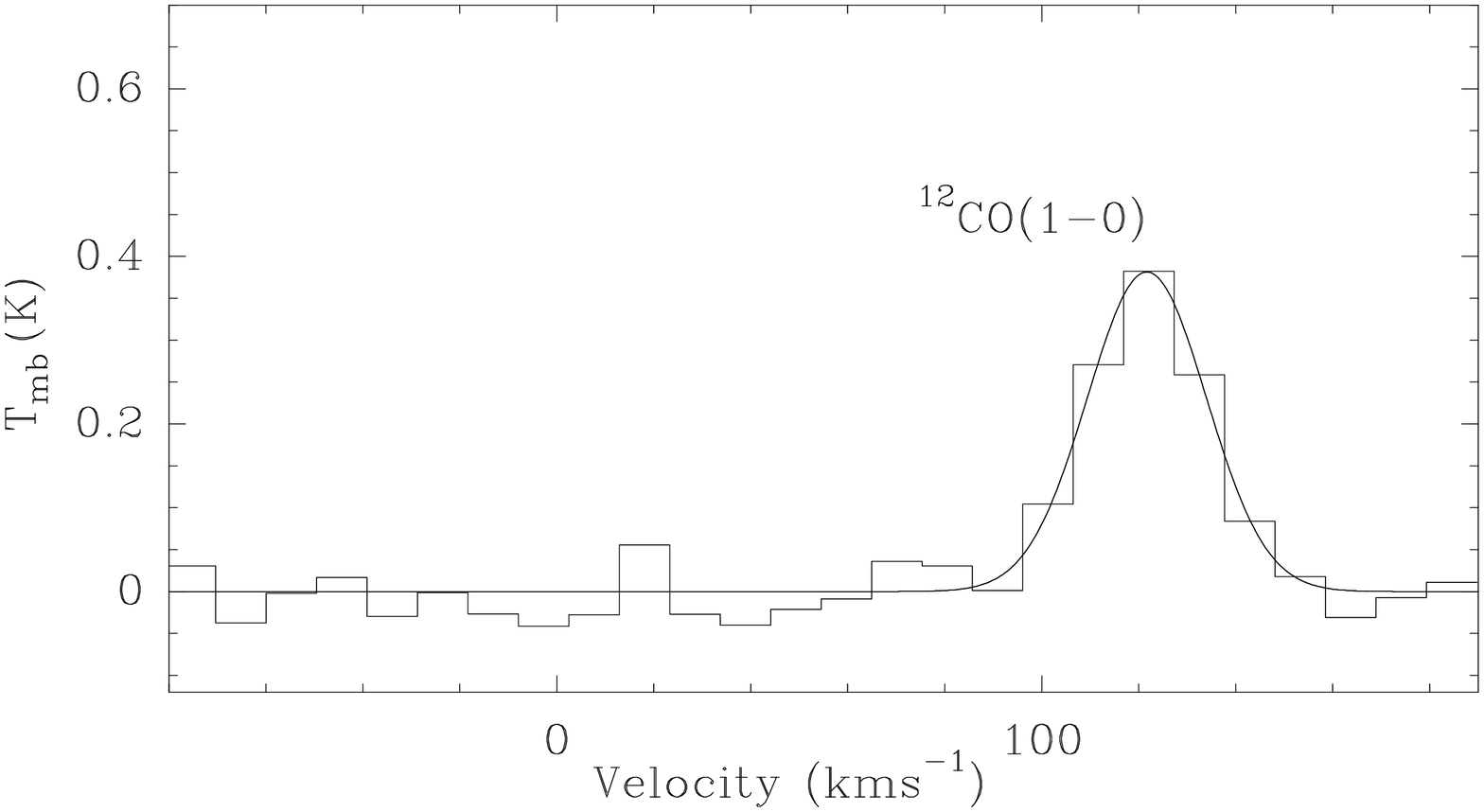}\\
  \includegraphics[height=6.75cm,clip=,angle=270]{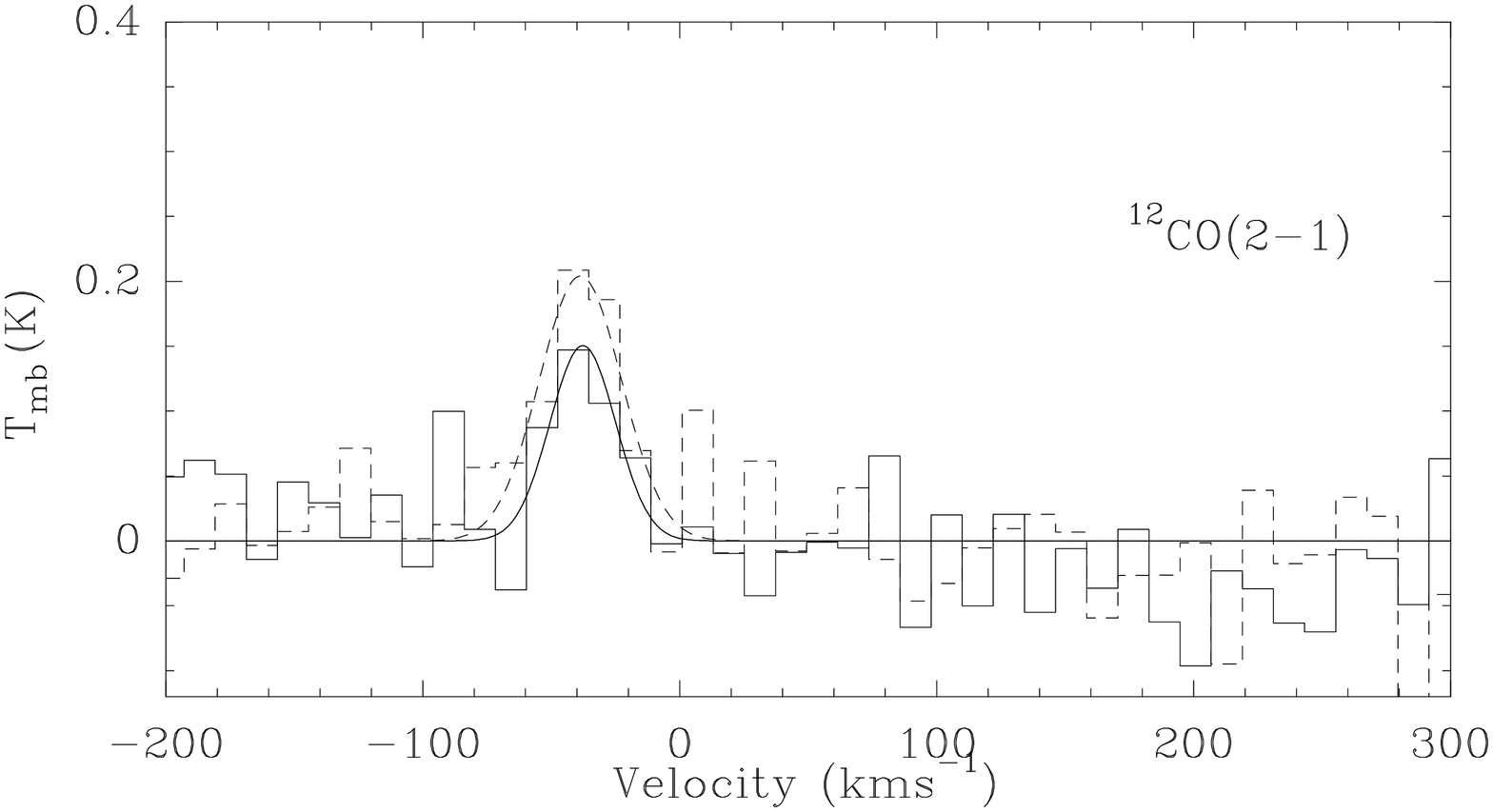}
  \includegraphics[height=6.75cm,clip=,angle=270]{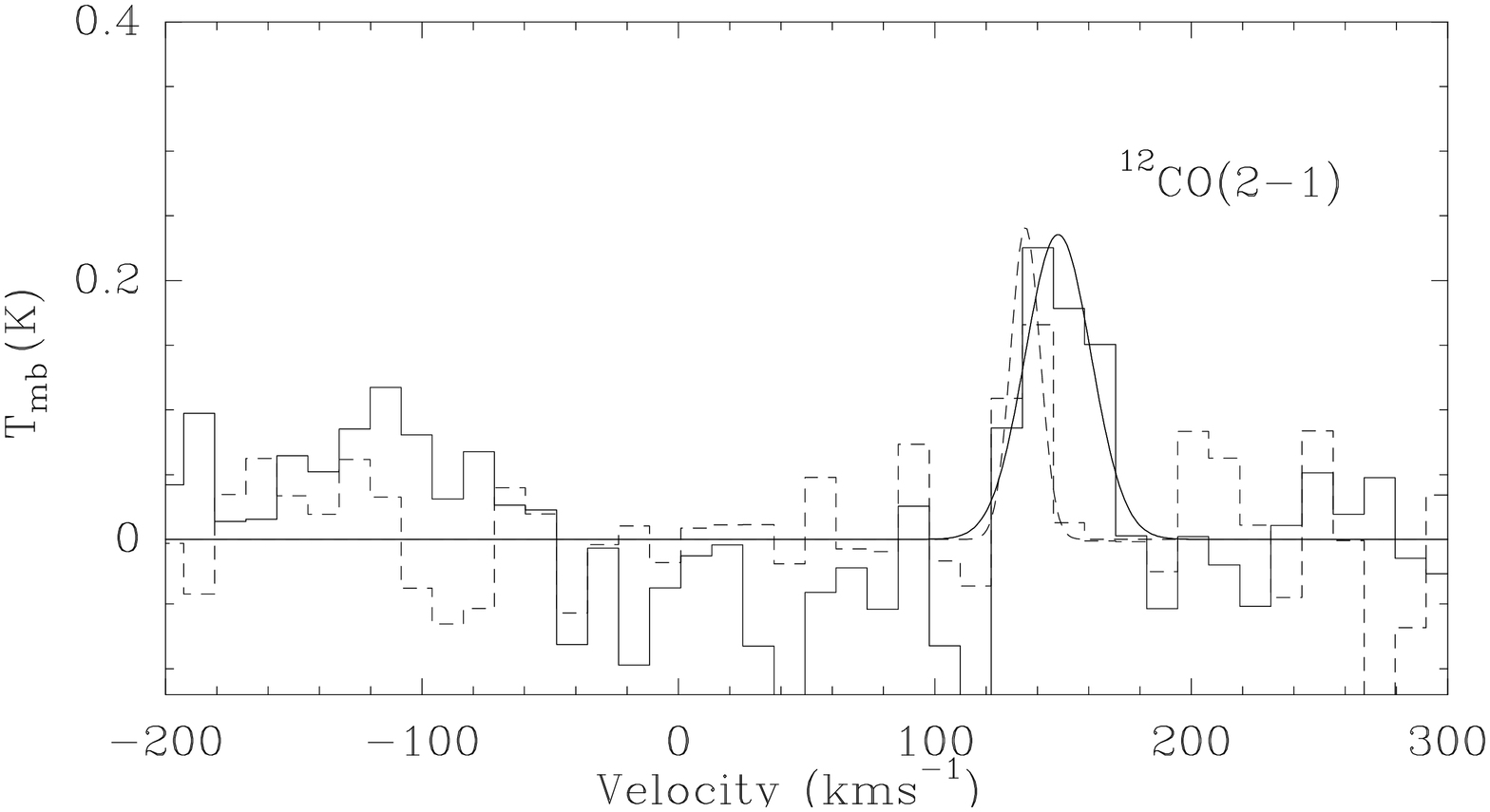}\\
  \includegraphics[height=6.75cm,clip=,angle=270]{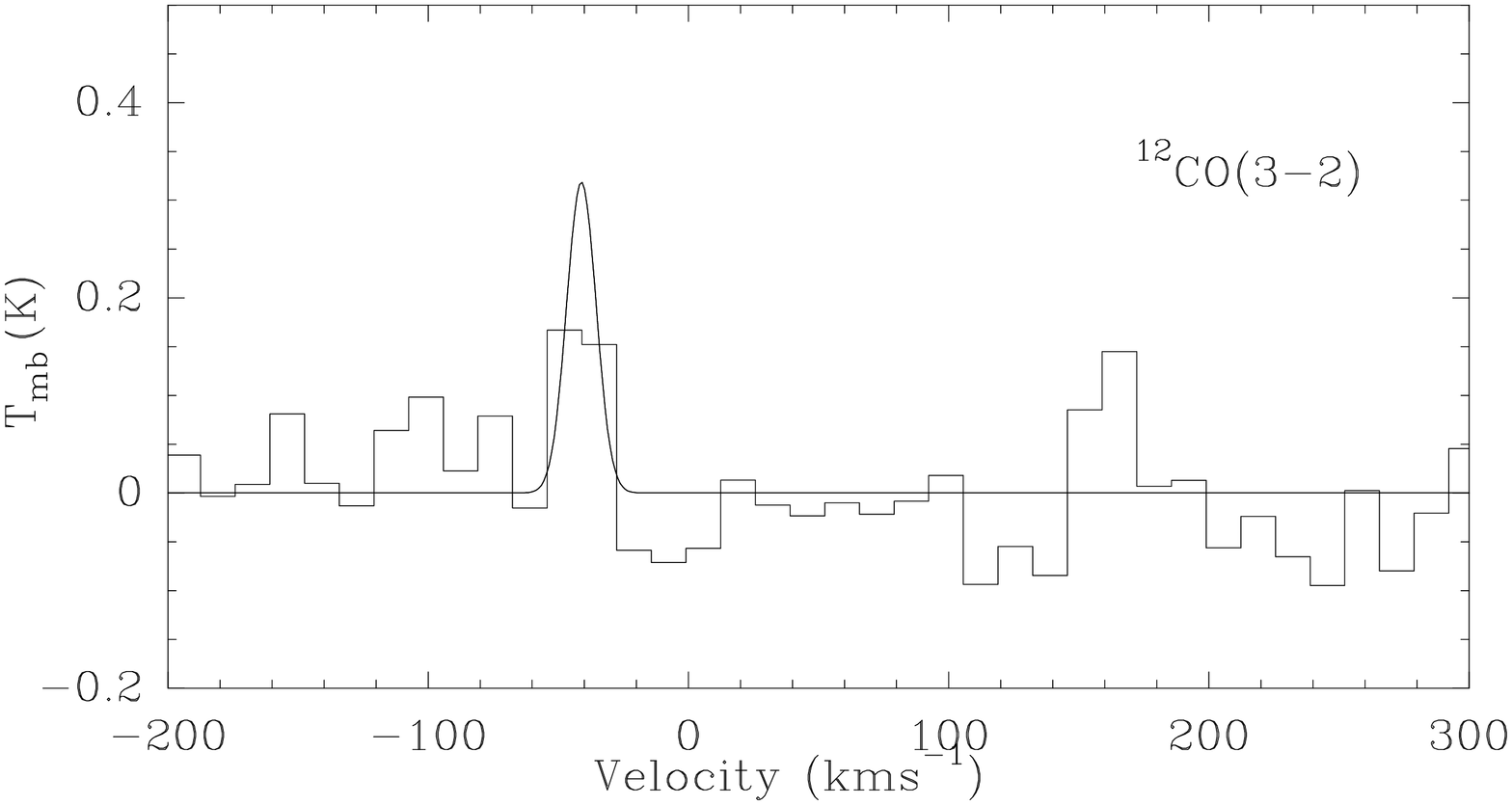}
  \includegraphics[height=6.75cm,clip=,angle=270]{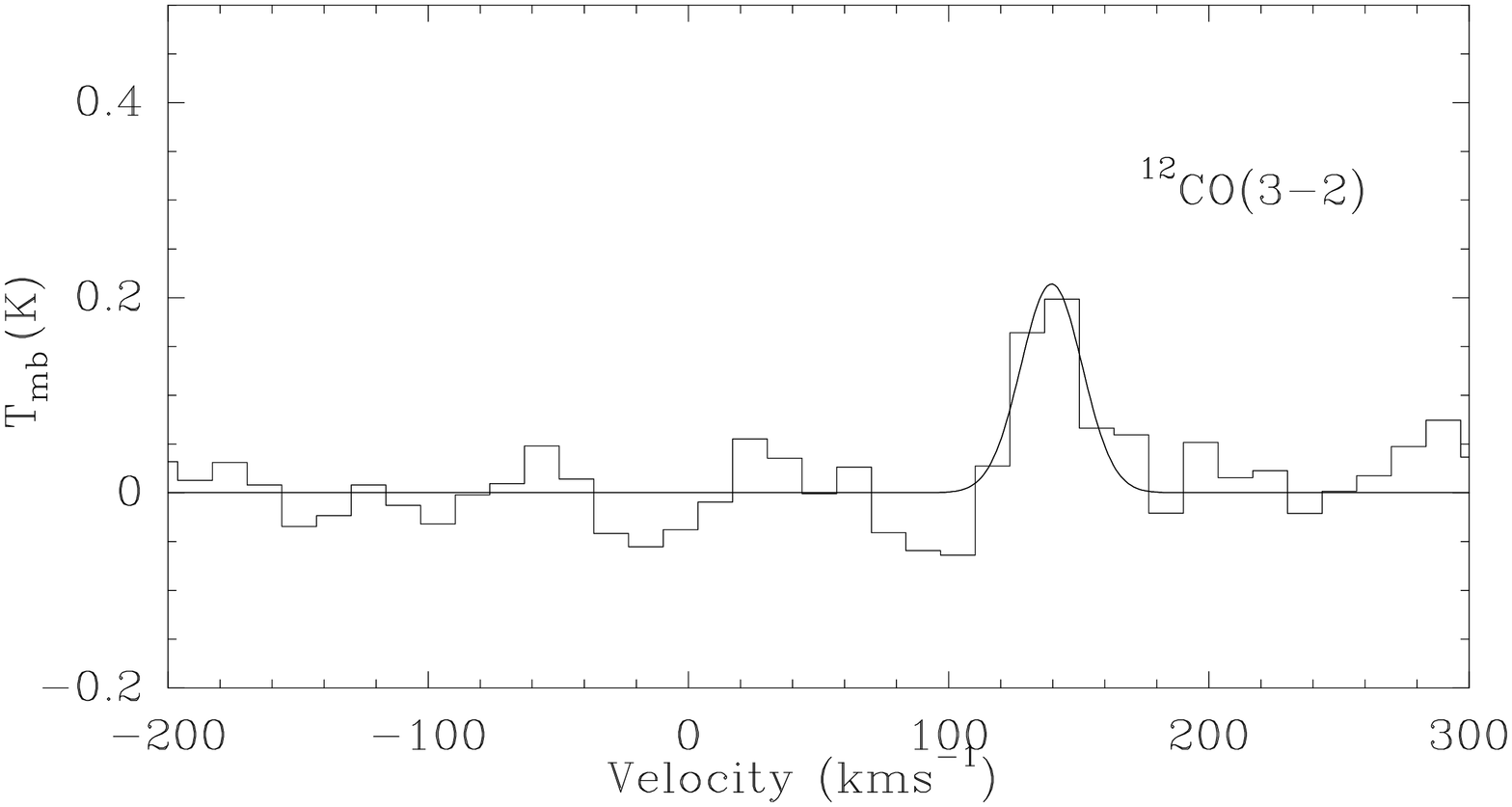}\\
  \includegraphics[height=6.75cm,clip=,angle=270]{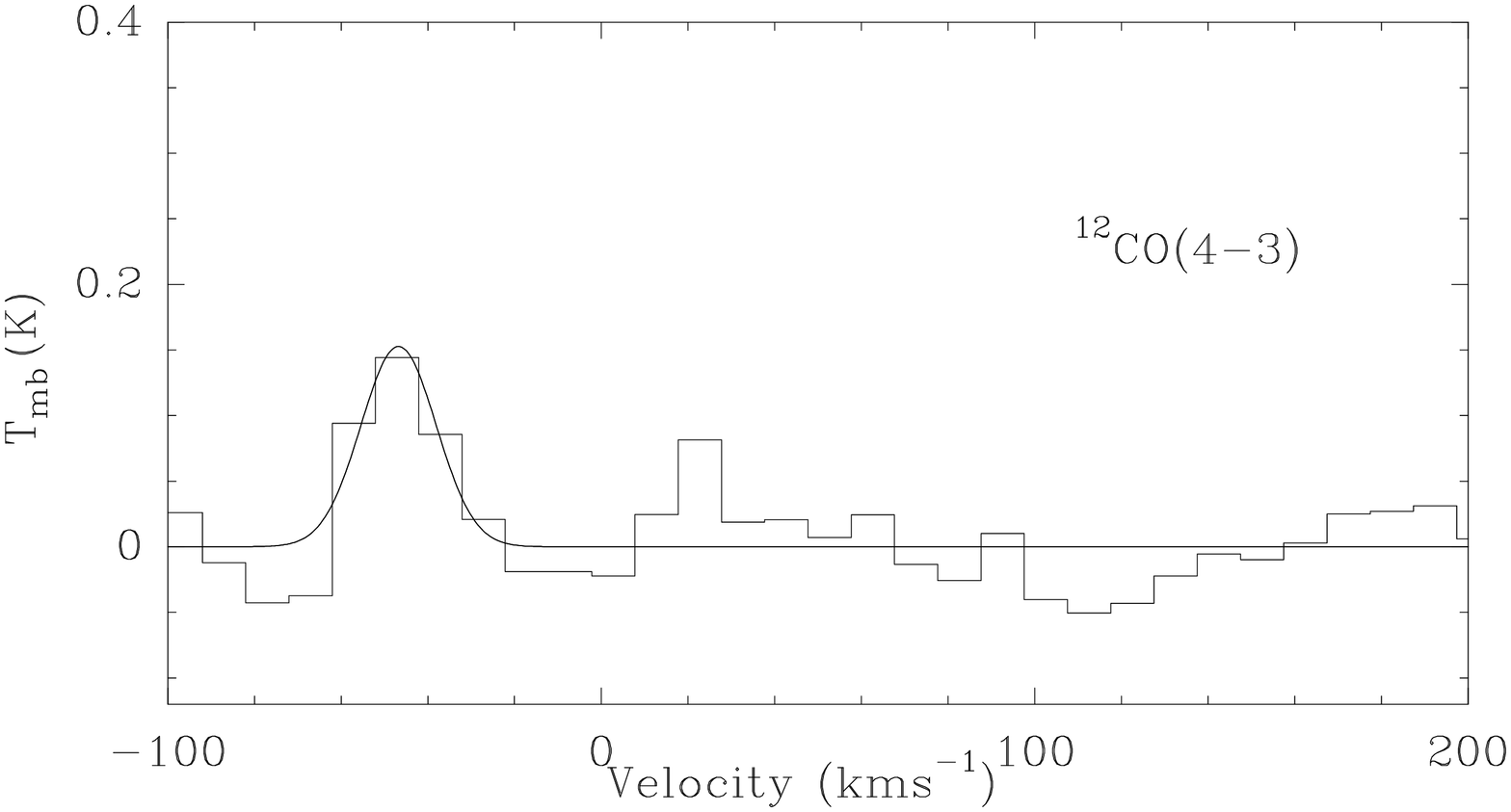}
  \includegraphics[height=6.75cm,clip=,angle=270]{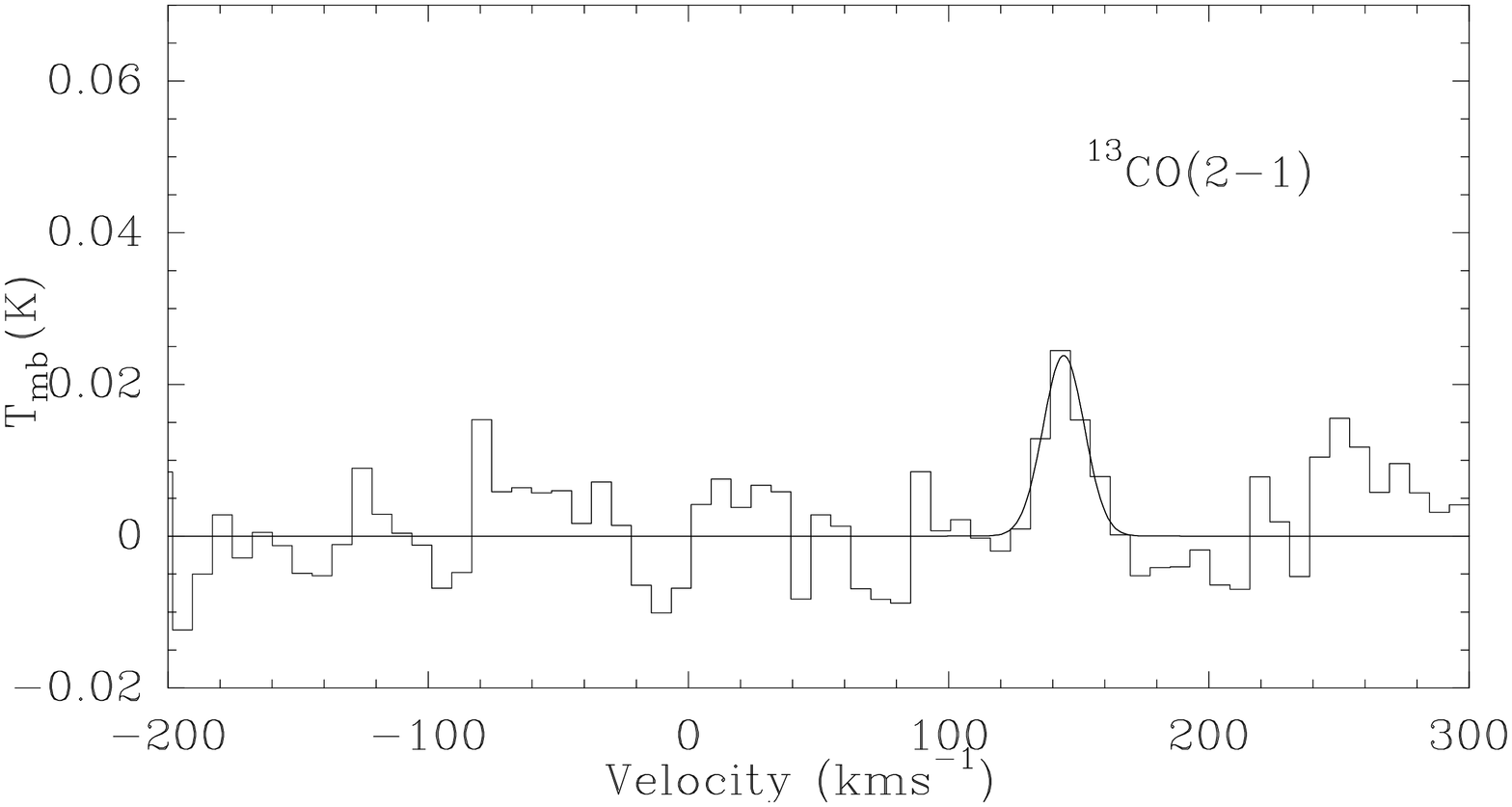}
  \caption{{\bf (Continued.)} Region~7 (left) and 8 (right).}
  \label{fig:spec78}
\end{figure*}

\label{lastpage}
\end{document}